\documentclass[floats,floatfix,amssymb,prd,twocolumn,superscriptaddress,nofootinbib,preprintnumbers]{revtex4-1}

\usepackage{subcaption}
\usepackage{ragged2e}
\DeclareCaptionJustification{justified}{\justifying}
\makeatletter
\newcommand{\subsetsim}{\mathrel{\mathpalette\subset@sim\relax}}
\newcommand{\subset@sim}[2]{%
  \vtop{\offinterlineskip\m@th
    \ialign{\hfil##\cr
      $#1\subset$\cr\noalign{\kern0.5pt}\scalebox{0.9}{$#1\sim$}\cr
    }%
  }%
}
\makeatother

\usepackage{amssymb,amsmath,verbatim,mathtools,needspace,enumitem,etoolbox,graphicx,physics,microtype,afterpage,bm}
\usepackage[dvipsnames, usenames]{xcolor}
\definecolor{linkcolor}{rgb}{0.0,0.3,0.5}
\usepackage{booktabs}
\usepackage[unicode, colorlinks=true, linkcolor=linkcolor, citecolor=linkcolor, filecolor=linkcolor,urlcolor=linkcolor, pdfusetitle]{hyperref}
\usepackage[all]{hypcap}
\usepackage[T1]{fontenc}
\usepackage[utf8]{inputenc}
\usepackage{tabularx}
\usepackage{float}
\renewcommand{\arraystretch}{1.4}

\usepackage{multirow}
\usepackage{pifont}
\newcommand{\cmark}{\textcolor{Green}{\ding{51}}}%
\newcommand{\xmark}{\textcolor{Red}{\ding{55}}}%
\usepackage{lmodern}

\allowdisplaybreaks
\usepackage{tikz}
\usepackage{color}
\usepackage{framed}
\usepackage{hyperref}
\hypersetup{colorlinks, citecolor=bluscuro, linkcolor=black, urlcolor=bluscuro}
\definecolor{rossos}{cmyk}{0,1,1,0.55}
\definecolor{bluscuro}{rgb}{0.15, 0.2, .85}
\definecolor{bluchiaro}{cmyk}{1,.3,0.,0.1}
\definecolor{ForestGreen}{rgb}{0.13, 0.55, 0.13}

\newcommand{\aapr}{"Astronomy and Astrophysics Reviews"}
\newcommand{\aap}{"Astronomy and Astrophysics"}
\newcommand{\mnras}{"Mon. Not. Roy. Astron. Soc."}

\def\ii{{\text{\tiny i}}}
\def\co{{\text{\tiny cut-off}}}
\def\d{{\mathrm{d}}}
\def\PBH{\text{\tiny{PBH}}}

\newcommand{\bs}{\begin{subequations}}
\newcommand{\es}{\end{subequations}}

\newcommand{\be}{\begin{equation}}
\newcommand{\ee}{\end{equation}}
\renewcommand{\d}{{\rm d}}

\newcommand{\llp}{\left [}
\newcommand{\rrp}{\right ]}
\newcommand{\lp}{\left (}
\newcommand{\rp}{\right )}
\newcommand{\tn}{\textnormal}
\def\lsim{\mathrel{\rlap{\lower4pt\hbox{\hskip0.5pt$\sim$}}
    \raise1pt\hbox{$<$}}}         
\def\gsim{\mathrel{\rlap{\lower4pt\hbox{\hskip0.5pt$\sim$}}
    \raise1pt\hbox{$>$}}}         

\newcommand{\jhu}{\affiliation{Department of Physics and Astronomy, Johns Hopkins University, 3400 N. Charles
Street, Baltimore, MD 21218, USA}}

\begin{document}

\title{How to assess the primordial origin of single gravitational-wave events\\
with mass, spin, eccentricity, and deformability measurements
}

\author{Gabriele Franciolini}
\email{gabriele.franciolini@uniroma1.it}
\affiliation{Dipartimento di Fisica, Sapienza Università
di Roma, Piazzale Aldo Moro 5, 00185, Roma, Italy}
\affiliation{INFN, Sezione di Roma, Piazzale Aldo Moro 2, 00185, Roma, Italy}

\author{Roberto Cotesta} 
\jhu

\author{Nicholas Loutrel}
\affiliation{Dipartimento di Fisica, Sapienza Università 
	di Roma, Piazzale Aldo Moro 5, 00185, Roma, Italy}
\affiliation{INFN, Sezione di Roma, Piazzale Aldo Moro 2, 00185, Roma, Italy}

\author{Emanuele Berti} 
\jhu

\author{Paolo Pani}
\affiliation{Dipartimento di Fisica, Sapienza Università 
	di Roma, Piazzale Aldo Moro 5, 00185, Roma, Italy}
\affiliation{INFN, Sezione di Roma, Piazzale Aldo Moro 2, 00185, Roma, Italy}

\author{Antonio Riotto}
\affiliation{D\'epartement de Physique Th\'eorique and Centre for Astroparticle Physics (CAP), Universit\'e de Gen\`eve, 24 quai E. Ansermet, CH-1211 Geneva, Switzerland}

\date{\today}

\begin{abstract}
A population of primordial black holes formed in the early Universe could contribute to at least a fraction of the black-hole merger events detectable by current and future gravitational-wave interferometers. With the ever-increasing number of detections, an important open problem is how to discriminate whether a given event is of primordial or astrophysical origin.
 We systematically present a comprehensive and interconnected list of discriminators that would allow us to rule out, or potentially claim, the primordial origin of a binary by measuring different parameters, including redshift, masses, spins, eccentricity, and tidal deformability.
We estimate how accurately future detectors (such as the Einstein Telescope and LISA) could measure these quantities, and we quantify the constraining power of each discriminator for current interferometers.
We apply this strategy to the GWTC-3 catalog of compact binary mergers. We show that current measurement uncertainties do not allow us to draw solid conclusions on the primordial origin of individual events, but this may become possible with next-generation ground-based detectors. 
\end{abstract}

\preprint{ET-0464A-21}

\maketitle

\begin{figure*}[ht]
\centering
\includegraphics[width=.95\textwidth]{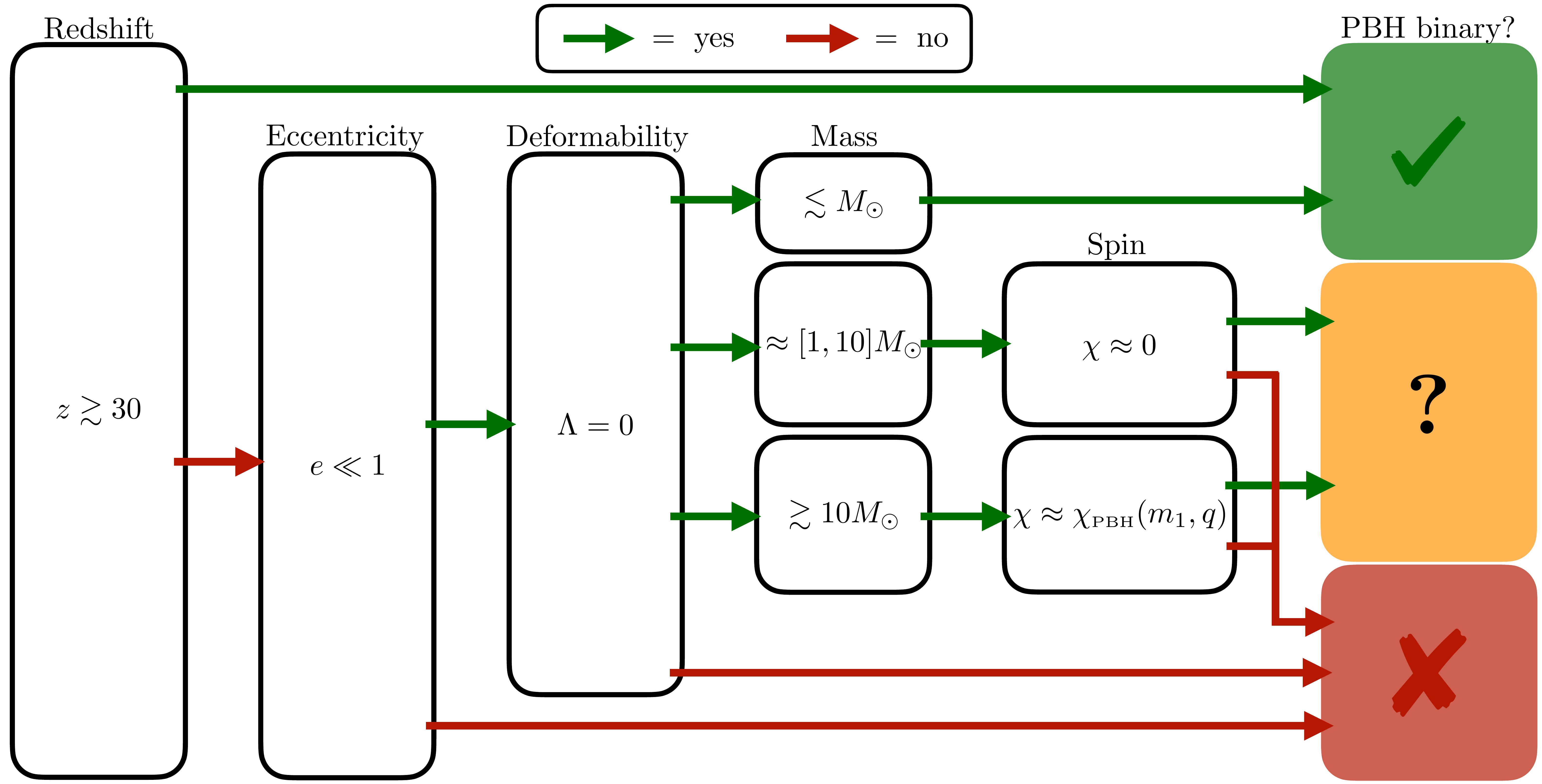}
\caption{ 
Schematic flowchart to systematically rule out or potentially assess the primordial nature of a binary merger based on { measurements of the redshift $z$, eccentricity $e$, tidal deformability $\Lambda$, component masses $m$, and dimensionless spins $\chi$}. 
Each arrow indicates if the condition in the box is met (green) or violated (red).
The various marks indicate: 
\cmark) likely to be a PBH binary; 
\xmark) cannot to be a PBH binary;
\textcolor{Black}{\textbf{?}}) may be a PBH  binary. 
} 
\label{fig:flowchart}
\end{figure*}

\section{Introduction}

The collapse of very large inhomogeneities during the radiation-dominated era could produce primordial black holes~(PBHs)~\cite{Zeldovich:1967lct,Hawking:1974rv,Chapline:1975ojl,Carr:1975qj} across a wide mass range~\cite{Ivanov:1994pa,GarciaBellido:1996qt,Ivanov:1997ia,Blinnikov:2016bxu}. 
Despite several observational constraints on these objects (see~\cite{Carr:2020gox} for a recent review), in certain mass ranges PBHs could comprise the entirety of the dark matter, and could seed supermassive black holes (BHs) at high redshift~\cite{2010A&ARv..18..279V,Clesse:2015wea,Serpico:2020ehh}.
Furthermore, PBHs could contribute to at least a fraction of the BH merger events detected by LIGO-Virgo~\cite{LIGOScientific:2018mvr, LIGOScientific:2020ibl,LIGOScientific:2021djp} so far~\cite{Bird:2016dcv,Sasaki:2016jop,Eroshenko:2016hmn, Wang:2016ana, Ali-Haimoud:2017rtz, Chen:2018czv,Raidal:2018bbj,Liu:2019rnx, Hutsi:2019hlw, Vaskonen:2019jpv, Gow:2019pok,Wu:2020drm,DeLuca:2020qqa, Hall:2020daa,Wong:2020yig,Hutsi:2020sol,Kritos:2020wcl,DeLuca:2021wjr,Deng:2021ezy,Kimura:2021sqz,Franciolini:2021tla,Bavera:2021wmw,Liu:2021jnw}, and to those that will be detected by future gravitational-wave~(GW) instruments~\cite{DeLuca:2021wjr,DeLuca:2021hde,Pujolas:2021yaw}: see Refs.~\cite{LIGOScientific:2021djp,LIGOScientific:2021psn} for the most recent LIGO-Virgo-KAGRA~(LVK) Collaboration catalog and population studies, and Refs.~\cite{Sasaki:2018dmp,Green:2020jor,Franciolini:2021nvv} for reviews on PBHs as GW sources.

{ ``Special'' events such as GW190425 (with a total mass that exceeds that of known galactic neutron star binaries) and the mass-gap events (such as GW190814~\cite{Clesse:2020ghq}, GW190521~\cite{DeLuca:2020sae} and GW190426\_190642) may have a PBH origin.}  Also, a subpopulation of PBHs may be competitive with certain astrophysical population models at explaining a fraction of events detected thus far by the LVK Collaboration~\cite{Franciolini:2021tla}.  However, astrophysical uncertainties make it hard to draw definite conclusions at a population level, and confidently claiming the primordial origin of an individual BH merger is much more challenging.  Indeed, an important problem in the ``PBHs as GW sources'' program is to disentangle a PBH candidate from the astrophysical foreground, thus discriminating between the primordial or astrophysical origin of a given binary.
Attempts have been made for single-event detections using Bayesian model selection based on astrophysically or primordial-motivated different priors~\cite{Bhagwat:2020bzh}, whereas catalog analyses could use the peculiar mass-spin-redshift distributions predicted for PBH binaries~\cite{Raidal:2018bbj,DeLuca:2020qqa} or perform population studies~\cite{Hall:2020daa,Wong:2020yig,Hutsi:2020sol,DeLuca:2021wjr,Franciolini:2021tla}. Given current measurement accuracy, the relatively modest number of GW events, and the 
uncertainties in both PBH and astrophysical models, none of the aforementioned strategies is currently able to give irrefutable evidence for or against the PBH scenario~\cite{Franciolini:2021tla}.

This state of affairs is expected to improve greatly in the era of next-generation detectors, such as the third-generation~(3G) ground-based interferometers Cosmic Explorer~(CE)~\cite{Reitze:2019iox} and Einstein Telescope~(ET)~\cite{Hild:2010id}, and the future space mission LISA~\cite{2017arXiv170200786A}. In particular, 3G detection rates will be orders of magnitude larger than current ones~\cite{Baibhav:2019gxm,Maggiore:2019uih,Kalogera:2021bya}, and much more accurate measurements will be possible for ``golden'' events with high signal-to-noise ratio (SNR).

The main goal of this paper is to present a systematic discussion of the various discriminators that would allow us to either rule out or confidently claim the primordial origin of a GW event by measuring different key binary parameters: the redshift, masses, spins, eccentricity, and tidal deformability (see Sec.~\ref{sec:key}).
A systematic strategy to use these discriminators is summarized in the flowchart of Fig.~\ref{fig:flowchart}, based on the predictions of the standard PBH scenario summarized in Sec.~\ref{sec:key} (where we also discuss some caveats).
In Sec.~\ref{sec:results}  we estimate the measurement errors on the PBH discriminators needed to apply this flowchart, and we quantify their constraining power for current and future detectors.
In Sec.~\ref{sec:nGWe} we apply the strategy to the GWTC-3 event catalog. We conclude in Sec.~\ref{sec:conclusions} with a summary of our findings and a discussion of future research directions. 

We will focus on binaries with individual component masses up to around ${\cal O}(10^2) M_\odot$, which comprise all of the events currently observed by LVK. We do not attempt to assess the primordial nature of more massive BHs, up to the supermassive range potentially detectable by LISA. Accretion throughout the cosmological evolution prior to the reionization epoch is still poorly modelled for those PBHs, and the predictions used in this paper (following Refs.~\cite{DeLuca:2020bjf,DeLuca:2020qqa}) have not been properly extended including feedback effects: see the discussion in Ref.~\cite{Ricotti:2007au}. Therefore we leave this effort for future work. 
Throughout this paper we adopt geometrical units ($G=c=1$).

\section{Key predictions for PBHs} \label{sec:key}

In this section, we review the main properties of PBH binaries, whose
characteristic features will be used in the rest of the paper to address the question: how can we rule out or confirm the primordial origin of a merger signal?
We highlight that throughout this work we will consider the ``standard'' PBH formation scenario, in which PBHs are formed out of large density fluctuations in the radiation dominated Universe~\cite{Sasaki:2018dmp}.
We will comment later on about other possible PBH scenarios, and whether they may lead to different predictions.

To clarify our notation, we consider binaries with masses $m_1$ and $m_2$, mass ratio $q=m_2/m_1\leq1$, total mass $M=m_1+m_2$, and dimensionless spins $\chi_i=J_i/m_i^2$ (with $0\leq \chi_j\leq 1$), located at redshift $z$.
Additionally, an important parameter measurable through GW observations is the effective spin 
\begin{equation}
    \chi_\text{\tiny eff} \equiv 
    \frac{\chi_1 \cos{\alpha_1} + q \chi_2 \cos{\alpha_2}}
    {1+q},
\end{equation} 
which is a function of the mass ratio $q$, of both BH spin magnitudes $\chi_j$ ($j=1,2$), and of their orientation with respect to the orbital angular momentum, parametrized by the tilt angles $\alpha_j$.

\subsection{PBH binary formation vs redshift}

In the standard formation scenario, PBHs are generated from the collapse of large overdensities in the primordial Universe~\cite{Ivanov:1994pa,GarciaBellido:1996qt,Ivanov:1997ia,Blinnikov:2016bxu} (see~\cite{Green:2020jor} for a recent review). 
As the PBH mass is related to the size of the cosmological horizon at the time of collapse, the formation of a PBH of mass $m_\PBH$ takes place  deep in the radiation-dominated era, at a typical redshift~\cite{Sasaki:2018dmp}
\begin{equation}
    z_\ii \approx 2 \times 10^{11} \lp {m_\PBH}\over{M_\odot}\rp ^{-1/2}.
\end{equation}
At that epoch, the standard scenario predicts that PBH locations in space are described by a Poisson distribution~\citep{Ali-Haimoud:2018dau,Desjacques:2018wuu,Ballesteros:2018swv,MoradinezhadDizgah:2019wjf,DeLuca:2020ioi}. 
{ In simple terms, this means that the number of PBHs in a given volume $V$ is described by a Poisson distribution with mean $\lambda =  n_\text{\tiny PBH} \times V$, where $n_\text{\tiny PBH}$ is the average number density of PBHs in the Universe. }
This initial condition is used to compute the properties of the population of PBH binaries formed at high redshift and contributing to the PBH merger rate.

It is important to stress that the merger rate of PBHs is dominated by binaries formed in the early Universe via gravitational decoupling from the Hubble flow before the matter-radiation equality~\cite{Nakamura:1997sm,Ioka:1998nz}. Another binary formation mechanism is possible, i.e., the formation of binaries taking place in present-day halos through gravitational capture. This second possibility was previously considered in the literature, see for example Refs.~\cite{Bird:2016dcv,Cholis:2016kqi}, 
but it was later on shown to produce a largely subdominant contribution to the overall merger rate~\cite{Ali-Haimoud:2017rtz,Raidal:2017mfl,Vaskonen:2019jpv,DeLuca:2020jug}.
We will, therefore, only consider the former mechanism throughout this paper. 
As a consequence, in contrast to the astrophysical channels, primordial binary BHs~(BBHs) have a merger rate density that monotonically increases with redshift as~\citep{Ali-Haimoud:2017rtz,Raidal:2018bbj,DeLuca:2020qqa}
\begin{equation}\label{redevo}
R_\text{\tiny PBH} (z) \propto \left ( \frac{ t(z)}{t (z=0)} \right)^{-34/37}, 
\end{equation}
extending up to redshifts $z\gtrsim{\cal O}(10^3)$. 
Notice that the evolution of the merger rate with time shown in Eq.~\eqref{redevo} is entirely determined by the binary formation mechanism (i.e. how pairs of PBHs decouple from the Hubble flow) before the matter-radiation equality era. 
Eq.~\eqref{redevo} is, therefore, a robust prediction of the PBH model in the standard formation scenario.

On the contrary, astrophysical-origin mergers should not occur at $z\gtrsim 30$.
The redshift corresponding to the epoch of first star formation is poorly known: theoretical calculations and cosmological simulations suggest 
this to fall below $z\sim 40$~\cite{Schneider:1999us,Schneider:2001bu,Schneider:2003em,Bromm:2005ep,deSouza:2011ea,Koushiappas:2017kqm,Mocz:2019uyd,Liu:2020ufc} (but see Refs.~\cite{Trenti:2009cj} and~\cite{Tornatore:2007ds}, where Pop~III star formation was suggested to start at higher or lower redshift).
The time delay between Pop~III star formation and BBH mergers was studied using population synthesis models, and found to be around $\mathcal{O}(10)~\rm Myr$~\cite{Kinugawa:2014zha,Kinugawa:2015nla,Hartwig:2016nde,Belczynski:2016ieo,Inayoshi:2017mrs,Liu:2020lmi,Liu:2020ufc,Kinugawa:2020ego,Tanikawa:2020cca,Singh:2021zah}.  This means that we can conservatively assume BBHs from Pop~III remnants to merge below $z\approx 30$, and consider merger redshifts $z\gtrsim30$ to be smoking guns for primordial binaries~\cite{Koushiappas:2017kqm,DeLuca:2021wjr,Ng:2021sqn}.

\subsection{PBH masses and spins}\label{sec:th_mspins}

The distribution of PBH masses $m_\PBH$ is determined by the characteristic size and statistical properties of the density perturbations, corresponding to curvature perturbations generated during the inflationary epoch. 
As $m_\PBH$ is related to the mass contained in the cosmological horizon at the time of collapse,  a much wider range of masses is accessible compared to astrophysical BHs~\cite{Sasaki:2018dmp}.
In particular, PBHs can have subsolar masses, which are unexpected from standard stellar evolution, and they can also populate the astrophysical mass gaps~\cite{Clesse:2020ghq,DeLuca:2020sae}. 

Given an accurate mass measurement, we can discriminate among three cases:
\begin{itemize}[leftmargin=*]
\item $m_i \lesssim M_\odot$: subsolar compact objects could be PBHs\footnote{See, however, Ref.~\cite{Shandera:2018xkn} for models in which subsolar BHs are born out of dark sector interactions.}, white dwarfs, brown dwarfs, or exotic compact objects~\cite{Cardoso:2019rvt}, e.g. boson stars~\cite{Guo:2019sns}. Distinguishing PBHs from other compact objects requires taking into account tidal disruption and tidal deformability measurements (see Sec.~\ref{sec:tidal} below).
As we will see, less compact objects like brown and white dwarfs are tidally disrupted well before the contact frequency, so detecting the merger of a subsolar compact object would imply new physics, regardless of the nature of the object~\cite{Barsanti:2021ydd}.
\item $1 \lesssim m_i/M_\odot \lesssim 3$: PBHs in this mass range can be confused with neutron stars~(NSs). Once again, tidal deformability measurement can in principle be used to break the degeneracy. Additionally, solar-mass BHs can form out of NS transmutation in certain particle-dark-matter scenarios~\cite{Bramante:2017ulk,Takhistov:2020vxs,Dasgupta:2020mqg,Giffin:2021kgb}.
In the upper half of this mass range ($2 \lesssim m_i/M_\odot \lesssim 3$), the component BHs in the binary may form out of previous NS-NS mergers and then pair again to produce a light binary~\cite{Fasano:2020eum}. In this case, however, the second-generation BH formed as a result of the NS-NS merger is expected to be spinning~\cite{Hofmann:2016yih}. This is in contrast with the prediction for the PBH scenario in this mass range, as we shall see in the following.
\item $m_i>3M_\odot$: PBHs in this mass range must be distinguished from stellar-origin BHs by other means.
\end{itemize}
Obviously, the boundaries between the mass ranges discussed above should be understood as approximate, and taken with a grain of salt.

\begin{figure*}[th]
\centering
\includegraphics[width=0.49\textwidth]{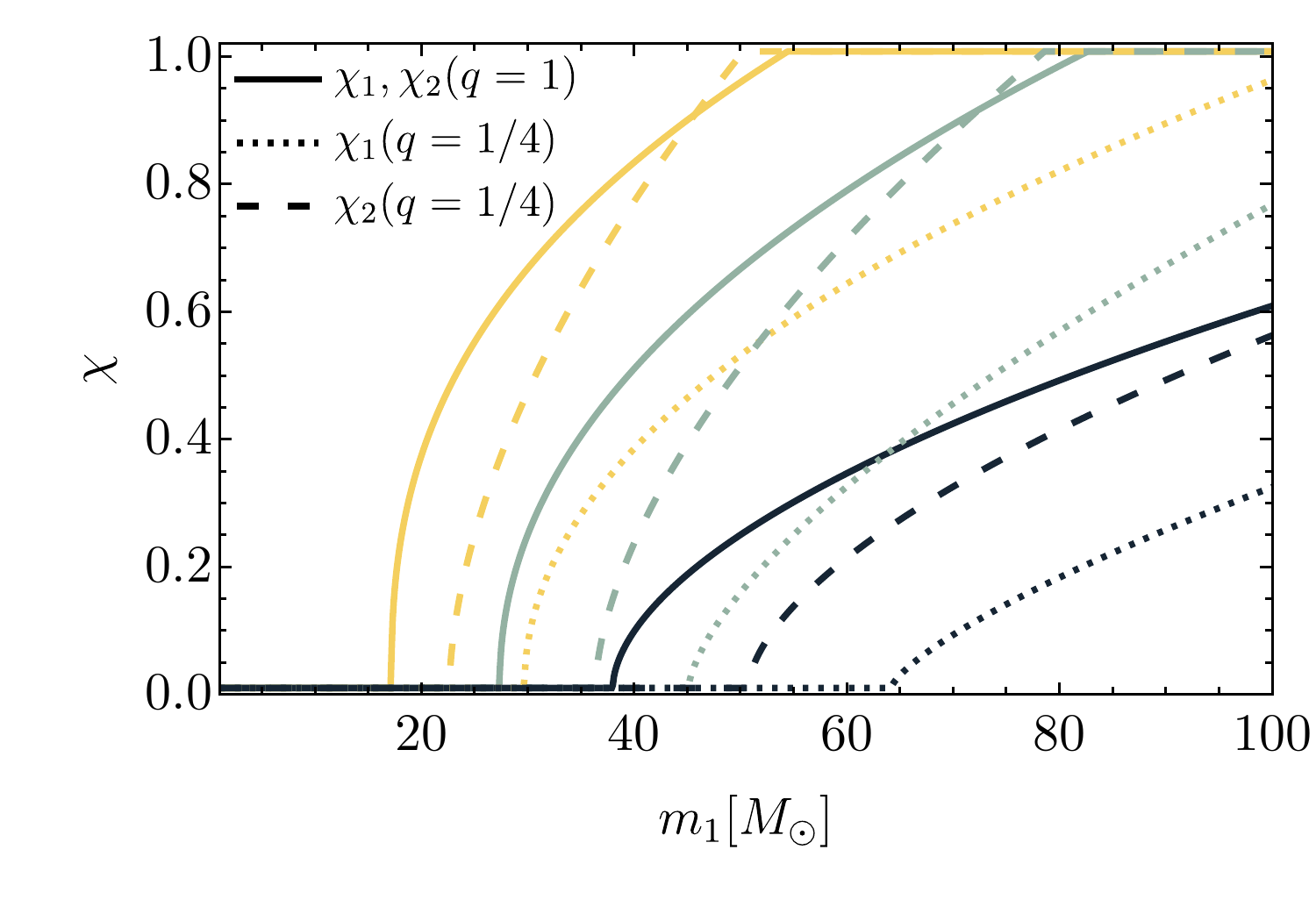}
\includegraphics[width=0.49\textwidth]{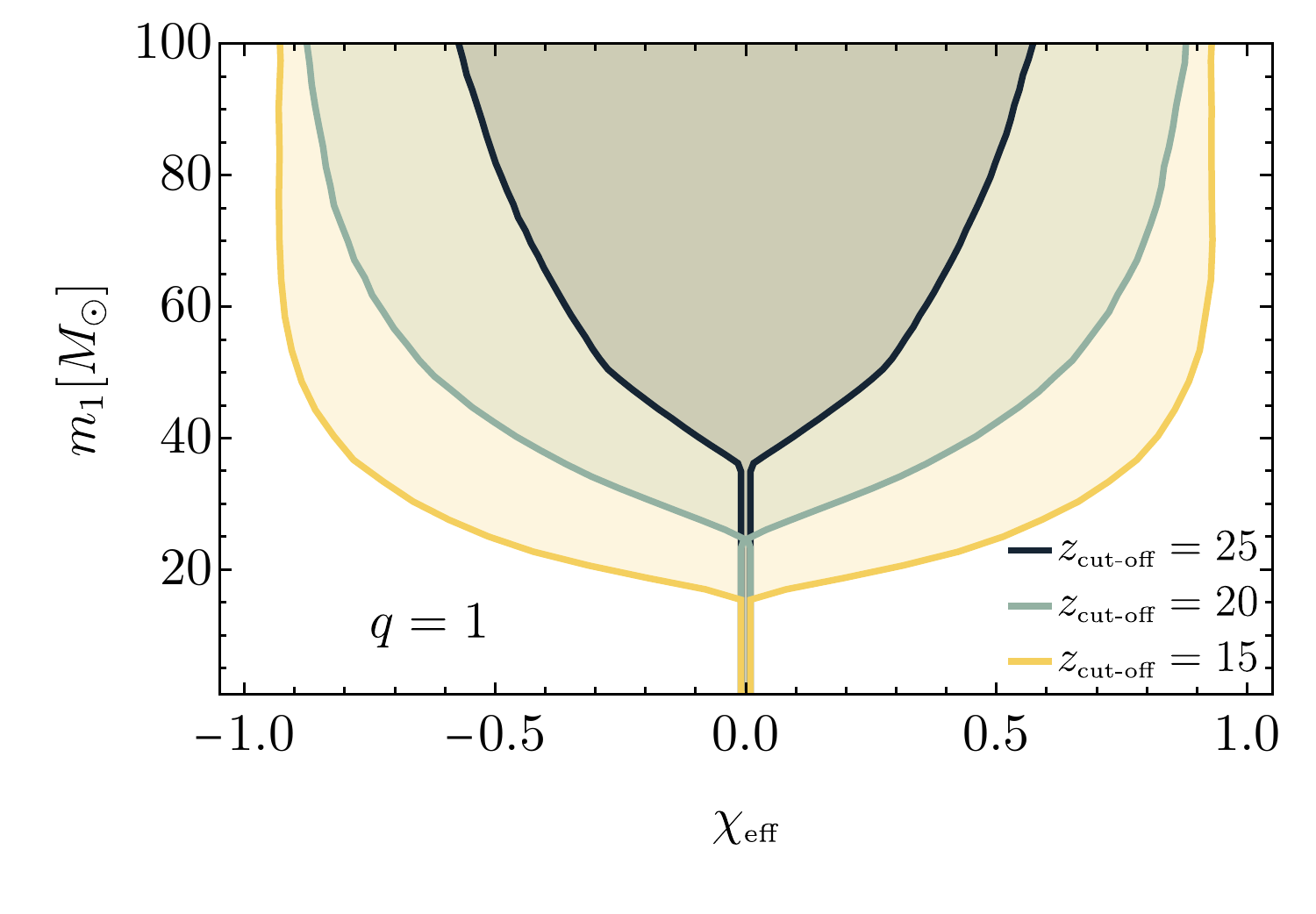}
\caption{ 
\textbf{Left:}
Predicted primary ($\chi_1$) and secondary ($\chi_2$) spins as a function of primary mass and mass ratio for various values of $z_\text{\tiny \rm cut-off}$ (indicated by the same colors as in the right panel). The plot highlights that for unequal mass binaries, the PBH accretion model predicts the secondary spin component $\chi_2$ to be larger than the primary component $\chi_1$.
\textbf{Right:}
Predicted distribution of $\chi_\text{\tiny\rm eff}$ as a function of $m_1=m_2$ for three selected values of $z_\text{\tiny \rm cut-off}$.
}\label{mass_spin}
\end{figure*}

Another important property of a PBH is its spin $\chi$. Since extreme Gaussian perturbations tend to have nearly-spherical shape~\cite{bbks} and the collapse takes place in a radiation-dominated Universe, the initial dimensionless Kerr parameter $\chi \equiv J/M^2$ (where $J$ and $M$ are the angular momentum and mass of the BH) is expected to be below the percent level~\cite{DeLuca:2019buf,Mirbabayi:2019uph}. 
However, a nonvanishing spin can be acquired by PBHs forming binaries through an efficient phase of accretion~\cite{DeLuca:2020qqa,DeLuca:2020bjf} prior to the reionization epoch.

Accretion during the cosmic evolution was shown to be effective only for PBHs with masses above $m_\PBH\gtrsim  {\cal O}(10) M_\odot$.
Therefore, the PBH model predicts binaries with negligible spins in the ``light'' portion of the observable mass range of current ground-based detectors. 
At larger masses, a defining characteristic of the PBH model is the expected correlation between large binary total masses and large values of the spins of their PBH constituents, induced by accretion effects. 
In addition, the spin directions of PBHs in binaries are, at least following the modeling of accretion described in Ref.~\cite{DeLuca:2020qqa},  independent and randomly distributed on the sphere. 
We will consider this scenario in the remainder of the paper, but we warn the reader that details of the accretion dynamics are still rather uncertain, and exceptions to the prediction of random spin orientations are possible.
Overall, PBH accretion is still affected by large uncertainties, in particular coming from the impact of feedback effects~\cite{Ricotti:2007jk,Ali-Haimoud:2017rtz}, structure formation~\cite{Hasinger:2020ptw,Hutsi:2019hlw}, and early X-ray pre-heating (e.g.~\citep{Oh:2003pm}).
Therefore, in  recent years,  an additional hyperparameter (the cut-off redshift $z_\co \in [10,30]$) was introduced to  account for accretion model uncertainties~\cite{DeLuca:2020bjf}.
For each value of $z_\co$ there is a one-to-one correspondence between the initial and final masses, which can be computed according to the accretion model described in detail in Refs.~\citep{Ricotti:2007jk,Ricotti:2007au,DeLuca:2020qqa,DeLuca:2020bjf,DeLuca:2020fpg}. We highlight, for clarity, that a lower cut-off is associated to stronger accretion and vice-versa. Values above $z_\co \simeq 30$ effectively correspond to negligible accretion in the mass range of interest for LVK observations. 
For detections at high redshift $z>z_\co$ (as those potentially achievable with 3G detectors), one expects a characteristic $z-\chi$ correlation.

\begin{figure*}[th]
\centering
\includegraphics[width=0.501\textwidth]{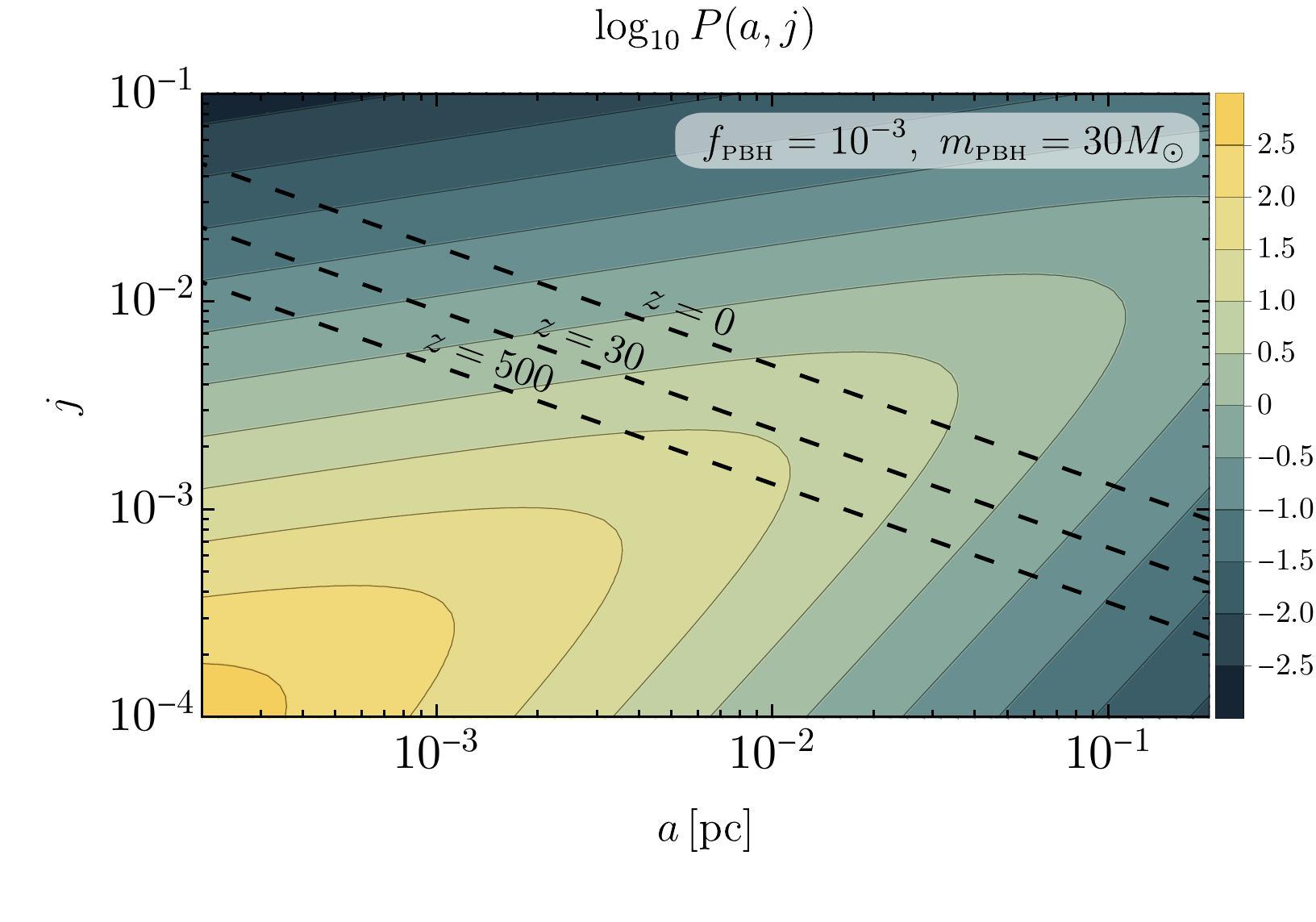}
\includegraphics[width=0.475\textwidth]{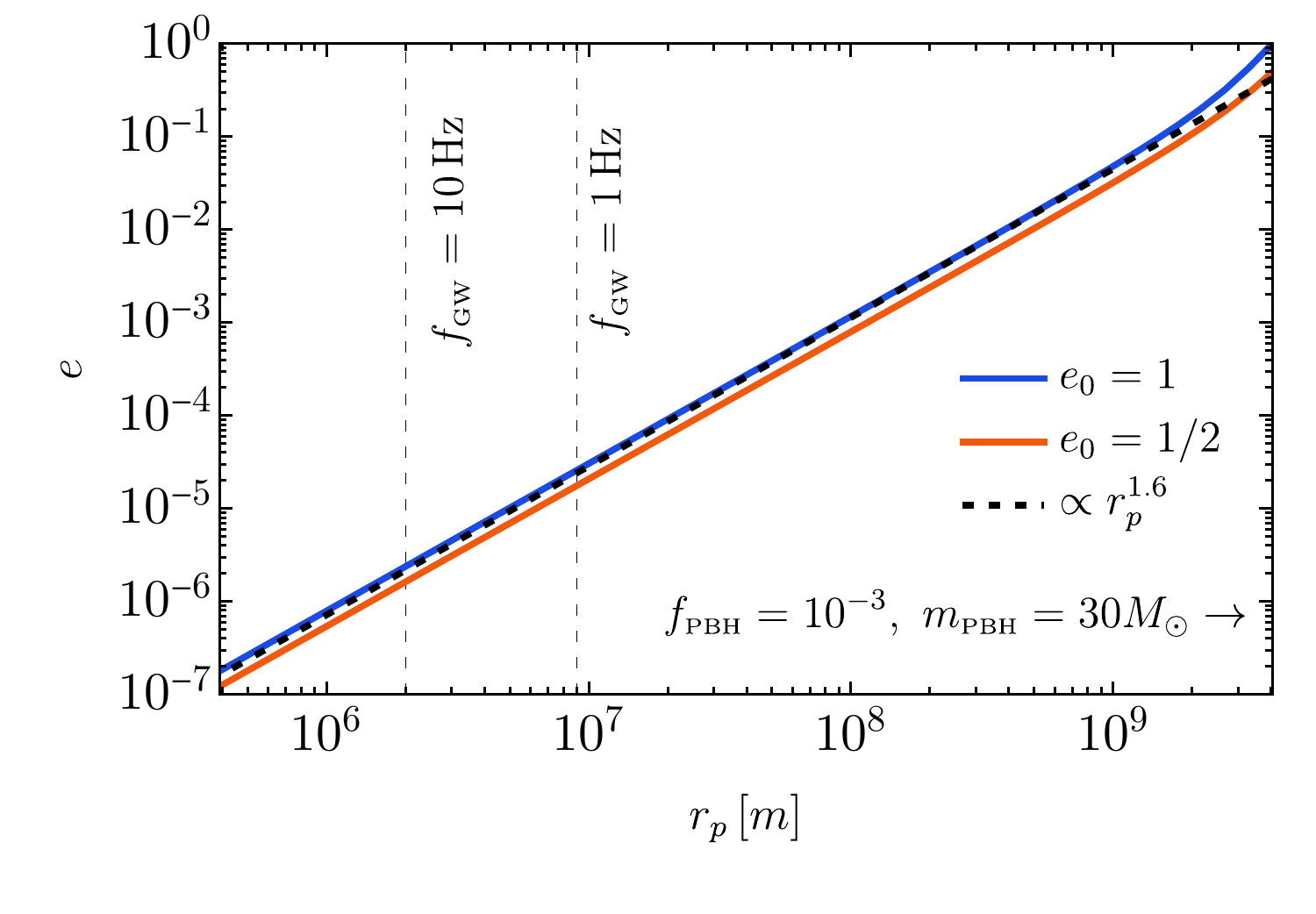}
\caption{
\textbf{Left:} Probability distribution of semimajor axis $a$ and rescaled angular momentum $j$ for PBH binaries of a population described by $f_\text{\tiny \rm PBH} =10^{-3}$ and $m_\text{\tiny \rm PBH}=30 M_\odot$. The black dashed lines indicate the combination of parameters giving a merger time equal to the age of the Universe at various redshift $z = [0,\, 30,\, 500]$.
\textbf{Right:} Eccentricity evolution for a characteristic binary merging at low redshift ($z\approx 0$) formed from a PBH population with $f_\text{\tiny \rm PBH} =10^{-3}$ and $m_\text{\tiny \rm PBH}=30 M_\odot$, whose initial pericenter distance $r_p$ at the binary formation epoch is indicated by the arrow. As an indication, we also show the power-law scaling relating the eccentricity $e$ to $r_p$.
}\label{fig1}
\end{figure*}

It is possible to derive an analytical fit of the relation between the masses and spins predicted at low redshift (that is, $z\lesssim z_\text{\tiny cut-off}$) as a function of $z_\co \supset[10,30]$. 
This fit describes the magnitude of both individual spins $\chi_i$ (where $i=1,2$) in PBH binaries as a function of the primary mass and mass ratio in the ranges $m_1 \lesssim 10^{2} M_\odot$ and $q \gtrsim 0.1$, respectively. 
The fit was derived using the results of the numerical integration of the equations describing accretion onto PBH in binaries as modelled in Refs.~\cite{DeLuca:2020bjf,DeLuca:2020qqa}. It could be useful when performing Bayesian parameter estimations assuming PBH motivated priors, or for searches in GW catalogs for a PBH-motivated mass-spin relation.
It can be written parametrically as
\begin{align}\label{spin-mass-relation-fit}
    \chi_i(m_1,q, z_\co)
    &\approx 10^{-2}
    + {\rm Min}\llp 
    0.988, 10^{f^b_{i}}
    \lp \frac{m_1}{M_\odot}-{f^ a_{i}}\rp^{f^c_{i}} \right. 
    \nonumber \\
    &\times \left .
    \Theta\lp\frac{m_1}{M_\odot} -{ f^a_{i} }\rp
    \rrp,
\end{align}
where $\Theta$ is the Heaviside theta function, and each coefficient $f^{a,b,c}_{i}$ depends on both $z_\co$  and $q$. Those functions are expanded as a polynomial series of the form
\begin{align}\label{fitmassspinrel}
f^{\alpha}_{i}(z_\co,q) &= 
\alpha_i^0+ 
\sum_{j=1}^3 \alpha_i^{z,j} \lp \frac{z_\co}{10}\rp^j
+
\sum_{j=1}^3\alpha_i^{q,j} q^j 
\nonumber \\
&+ \sum_{j,k=1}^2 \alpha_i^{j,k} \lp\frac{z_\co}{10}\rp^j q^k,
\end{align}
with $\alpha = [a,b,c]$. Note that the terms in the polynomial expansion involving the cut-off redshift are renormalized as a function of $(z_\co/10)$ for numerical convenience. The fit percentile error is below $10\%$ in the vast majority ($>98\%$) of the parameter space, while it degrades to around $30\%$ close to the boundaries of the $(m_1,\,q,\,z_\co)$ space. 
The coefficients in the analytical relation are reported in Appendix~\ref{appendix_fit_spins}.
In Fig.~\ref{mass_spin}, we show the expected distribution of $\chi_\text{\tiny eff}$ produced using Eq.~\eqref{spin-mass-relation-fit} and by averaging over the spin angles, as a function of PBH masses in binaries for various choices of $z_\text{\tiny cut-off}$. 

After this summary, we would like to stress once more that these predictions assume the standard PBH scenario, where PBHs are formed through the collapse of large overdensities during the radiation phase. There are other possible scenarios, such as formation from assembly of matterlike objects (particles, Q-balls, oscillons, etc.), domain walls and heavy quarks of a confining gauge theory, which may lead to different predictions for the PBH spin at formation~\cite{Harada:2017fjm,Flores:2021tmc,Dvali:2021byy,Eroshenko:2021sez}.
For instance, during an early matter-dominated phase, possibly following the end of inflation and preceding the reheating phase, PBHs may be formed in a pressureless environment and develop initial large, and possibly maximal,  spins~\cite{Harada:2017fjm,DeLuca:2021pls}. 
Such scenarios would require dedicated analyses, but we remark that the impact of accretion (when relevant) onto the mass-spin correlation and the properties of the remaining observables (i.e. redshift distribution, eccentricity and masses) remain consistent with the standard scenario.

\subsection{PBH eccentricity}\label{sec:th_ecc}

Another key prediction of the primordial model involves the eccentricity $e$ of PBH binaries. While formed with large eccentricity at high redshift, PBH binaries then have enough time to circularize before the GW signal can enter the observation band of current and future detectors.\footnote{Refs.~\cite{Cholis:2016kqi,Wang:2021qsu} analyzed the scenario where PBH binaries are formed dynamically in the late-time Universe and potentially retain large eccentricities. This channel was shown to provide a subdominant contribution to the overall merger rate in the standard scenario \cite{Ali-Haimoud:2017rtz}, as we discussed at the beginning of this section, and therefore we disregard it.} In this section we quantify this statement.

\subsubsection{Eccentricity distribution at formation}
We start by defining the mean PBH separation at matter-radiation equality as
\begin{equation}
\bar{x} \equiv \left( \frac{3 m_\PBH}{4 \pi f_\PBH \, \rho_\mathrm{eq}}  \right)^{1/3},
\end{equation}
where $\rho_\text{\tiny eq}$ is the average energy density at matter-radiation equality and $f_\PBH$ is the fraction of dark matter in PBHs. 
As predicted by the standard formation scenario in the absence of primordial nongaussianities, PBHs follow a Poisson spatial distribution at formation. One can show
that the differential probability distribution 
of the rescaled angular momentum $j\equiv \sqrt{1-e^2}$ reads~\cite{Ali-Haimoud:2017rtz,Kavanagh:2018ggo,Liu:2018ess}
\begin{align}
\label{eq:Pj}
j  P(j) &= \frac{y(j)^2}{\left(1+y(j)^2\right)^{3/2}},
\nonumber \\
y(j) &\equiv \frac{j}{0.5(1+\sigma_\text{\tiny eq}^2/f_\PBH^2)^{1/2}  \left({x}/{\bar{x}}\right)^3 },
\end{align}
where $\sigma_\text{\tiny eq} \approx 0.005$ indicates the variance of the Gaussian large-scale density perturbations at matter-radiation equality. This distribution is the result of both the surrounding PBHs and matter perturbations producing a torque on the PBH binary system during its formation.\footnote{Note that the description of the formation properties of PBH binaries was slightly improved in Ref.~\cite{Raidal:2018bbj}, accounting for the results of N-body simulations.
{ We neglect this small correction in our estimates, as it would not affect our conclusion that the eccentricity must be small when PBH binaries enter the sensitivity band of GW detectors}.}
Finally, the distribution describing both $j$ and the semimajor axis $a$ can be written as
\begin{equation}
\label{eq:Paj}
P (a, j) = \frac{3}{4 a^{1/4}} 
\left(\frac{f_\text{\tiny PBH}}{\alpha \bar{x}}\right)^{3/4}
P(j)
\exp \llp -\lp  \frac{x(a)}{\bar{x}}\rp^3\rrp,
\end{equation}
where 
\begin{equation}
x(a) \equiv \left(\frac{3\,  a\, m_\text{\tiny PBH}}{4 \pi \,\alpha \,\rho_\text{\tiny eq}} \right)^{1/4}
\end{equation}
and $\alpha \simeq 0.1$~\cite{Ali-Haimoud:2017rtz}. This distribution is shown in Fig.~\ref{fig1}.

We are interested in finding the probability distribution of the angular momentum of binaries constrained by the requirement of merging at redshift $z_\text{\tiny merge}$ (or time $t_\text{\tiny merge}$).
We compute, therefore, the merger time $t_\text{\tiny merge}$  of primordial binaries using Peter's formula~\cite{Peters:1963ux,Peters:1964zz} (see also~\cite{Mandel:2021fra}). For a binary of equal masses $m_1 = m_2 = m_\PBH $, initial eccentricity $e_{0}$ and semimajor axis $a_{0}$, and  in the limit of large initial eccentricity, one finds
\begin{align}
\tau = 
\frac{3}{170}\frac{a_{0}^{4}}{m_\PBH^{3} }\left(1-e_{0}^{2}\right)^{7/2}.
\label{eq:tau_m_2}
\end{align}
In the left panel of Fig.~\ref{fig1} we also show by dashed black lines the set of parameters $(a,j)$  giving rise to a merger at redshift $z=[0,30,100]$. Note that while the value of $f_\PBH$ has a major impact on the overall probability of forming a PBH binary (and the consequent overall merger rate), it affects only slightly the shape of the probability density function for the orbital parameters.

\subsubsection{Eccentricity evolution}\label{sec:orbital_evo}

The predictions for the initial binary parameters of PBHs should be used to forecast the final eccentricity when the GW enters the observability band of current ground-based detectors. As PBH binaries form at very high redshift, observable signals are coming from binaries which are initially wide enough so that the merger time is close to the current age of the Universe. 
As GW emission circularizes the orbit, one expects PBH binaries to lose any relevant eccentricity before detection. 

Let us show this explicitly. Rearranging the equations describing the orbital evolution under the effect of GW emission, and defining the pericenter distance 
\begin{equation}
r_p \equiv  a (1-e),    
\end{equation} 
one obtains~\cite{Maggiore:2007ulw}
\begin{align}
r_p \frac{{\rm d} e}{{\rm d} r_p} = e (1 + e) \frac{304 +121 e^{2}}{192 - 112 e + 168 e^{2} +47 e^{3}},
\label{eq:dedchi}
\end{align}
which, in the limit of quasicircular orbits $e \ll 1$, simplifies to 
\begin{equation}
 \frac{{\rm d ln } e}{{\rm d ln } r_p} \simeq  \frac{304 }{192 } \simeq 1.6.
\end{equation}
We show the evolution of the eccentricity as a function of $e_0$ in Fig.~\ref{fig1} (right panel). For a characteristic PBH binary formed by a narrow PBH population with $f_\PBH = 10^{-3}$ and $m_\PBH = 30 M_\odot$ and expected to merge at $z\simeq 0$, one finds an initial binary pericenter distance of the order of $r_p\simeq 4 \times 10^{9}\,{\rm m}$. 
In the figure we also indicate the eccentricity at which 
the binary would approximately enter the ET and LVK observable band with a frequency of $1$ and $10$ Hz, respectively.
Those frequencies corresponds to roughly $r_p \simeq 22 R_\text{\tiny Sch} \simeq 2\times 10^{6} {\rm m}$ 
($r_p \simeq 102 R_\text{\tiny Sch} \simeq 9\times 10^{6} {\rm m}$)
for $m_\PBH = 30 M_\odot$, where the eccentricity of the orbit has already been reduced to a value below $e\approx 10^{-5}$.

\begin{figure*}[t!]
\centering
\includegraphics[width=0.49\textwidth]{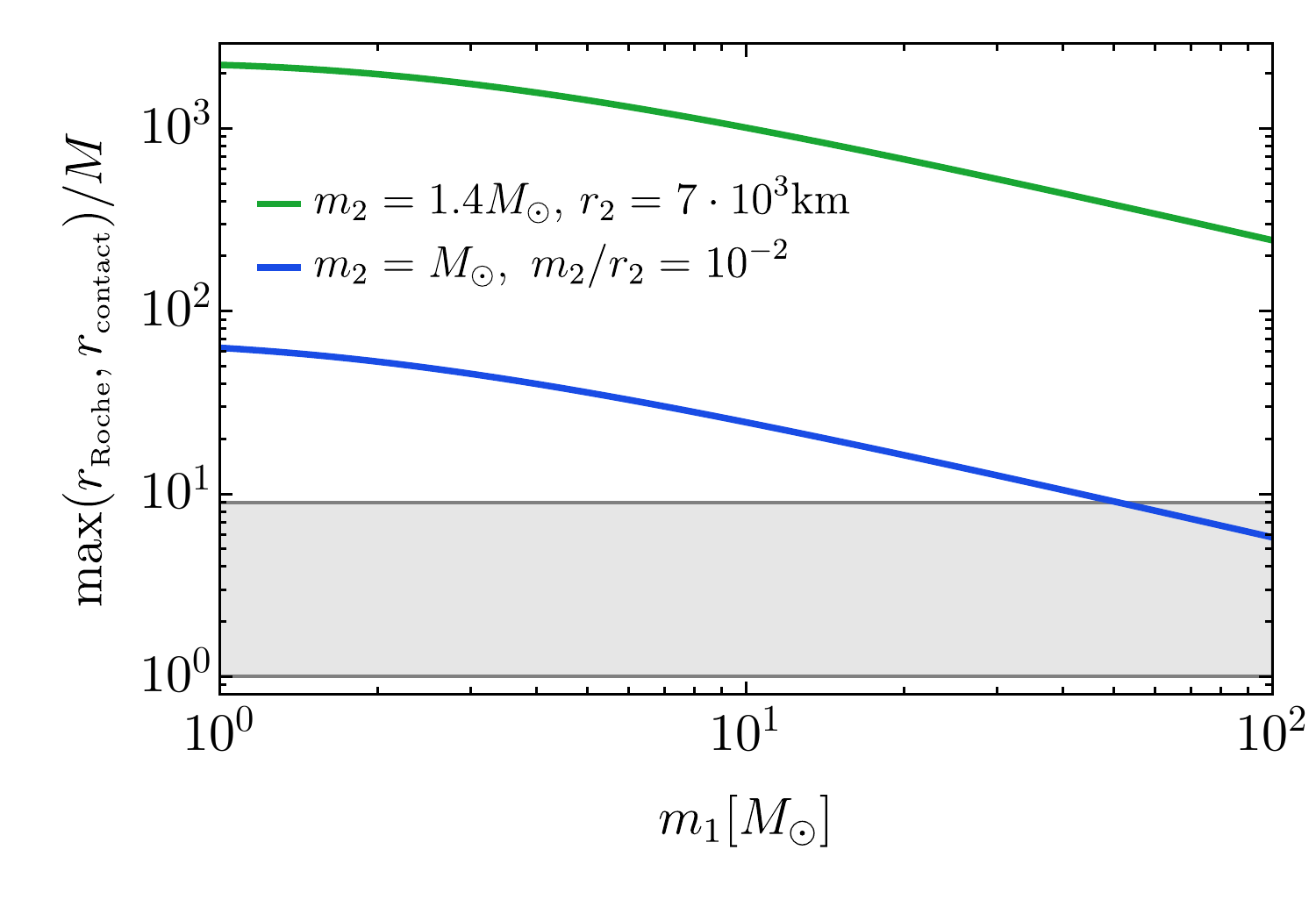}
\includegraphics[width=0.49\textwidth]{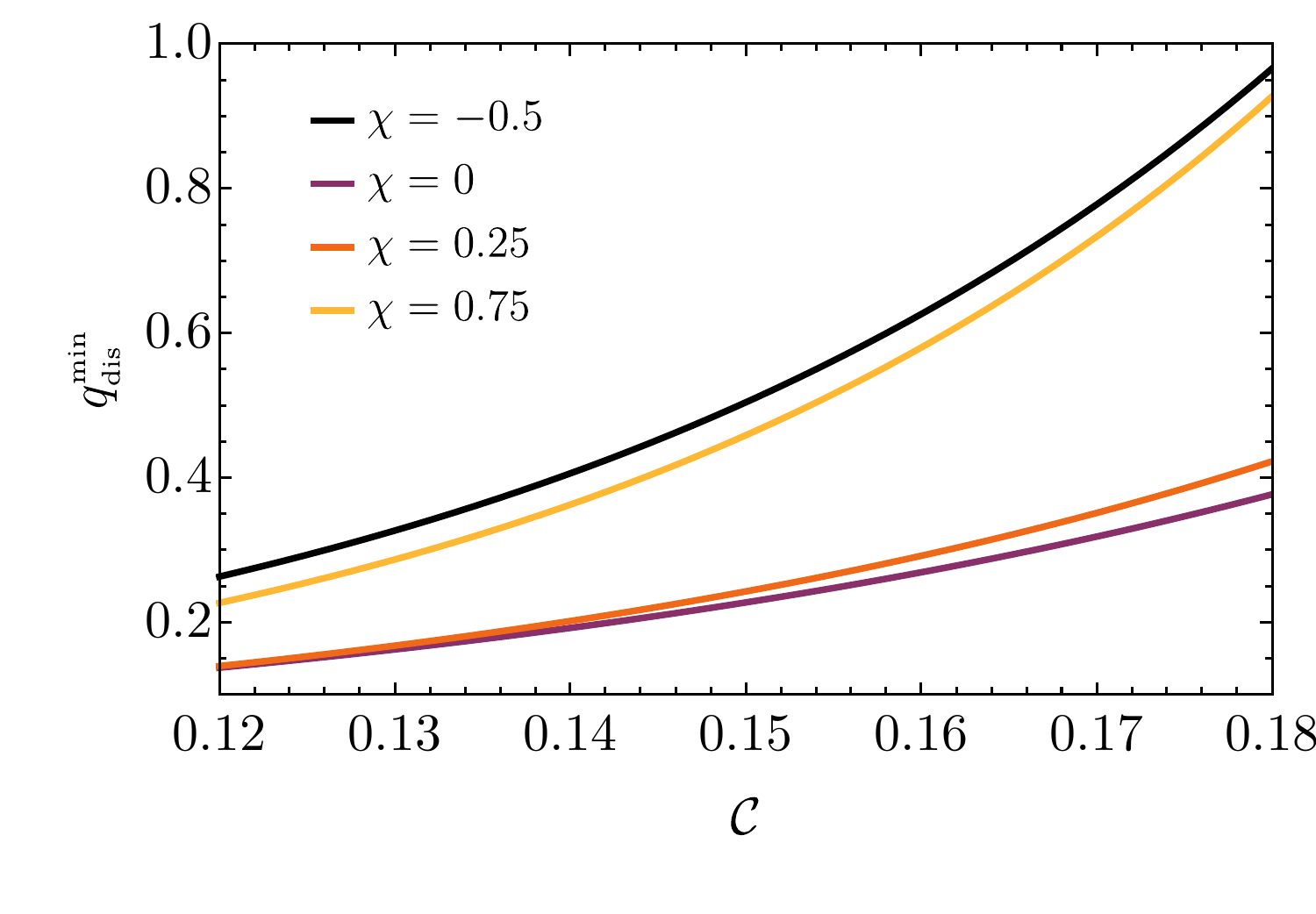}
\caption{\textbf{Left:} Maximum between the Roche radius and the contact radius, $r_2+2m_1$, normalized by the total binary mass $M$, as a function of the primary mass $m_1$ for two representative examples of secondary object: a white dwarf (blue) and a solar-mass mini boson star (green). The horizontal gray band brackets the ISCO of a spinning BH ($1<r_\text{\tiny \rm ISCO}/M<9$). 
This plot indicates that compact objects such as white dwarfs and mini boson stars are usually disrupted before reaching the ISCO. 
For both chosen examples, 
${\rm max}(r_\text{\tiny\rm  Roche}, r_\text{\tiny\rm  contact})=r_\text{\tiny\rm  Roche}$.
\textbf{Right:} Focusing on NS-BH binaries, the plot shows the minimum mass ratio above which the NSs get disrupted depending on the NS compactness ${\cal C}$ and BH spin $\chi$. We use the disruption criterion of Ref.~\cite{Pannarale:2015jia}, which is based on numerical NS-BH merger simulations. 
}\label{fig_Roche}
\end{figure*}

We can use the GW frequency evolution as a function of the eccentricity to estimate the eccentricity of the binary at the smaller frequencies accessible to 3G detectors. Since (see e.g.~\cite{Peters:1963ux,Peters:1964zz,2011A&A...527A..70K})  
\begin{equation}
\label{f(e)}
f_\text{\tiny GW}(e)=\frac{2}{P_0} 
\frac{(1-e^2)^{3/2}}{e^{18/19}}
\llp 1+\frac{121}{304}e^2\rrp^{-1305/2299} 
c_0^{3/2},
\end{equation}
where $c_0=[e_0^{12/19}(1+{121}e_0^2/{304})^{1305/2299}]/(1-e_0^2)$ and  $P_0$ is the initial orbital period,
we can infer that in the limit of small eccentricity 
\begin{equation}
    e \propto f_\text{\tiny GW}^{-19/18}.
\end{equation}
Since detectors such as ET are potentially sensitive down to frequencies $\approx 1 {\rm Hz}$, the above scaling shows that the eccentricity of the binary when it enters the 3G band is only a factor ${\cal O}(10)$ bigger than when it enters the LIGO band, 
and it is still negligible for PBH binaries.\footnote{For the mass range considered in this work, LISA will only be able to observe inspiralling binaries at frequencies $\gtrsim 2\times 10^{-2}$\,Hz (assuming an observation time $T_\text{\tiny obs}= 1$ yr)~\cite{Kyutoku:2016ppx}. The problem in this case is that large eccentricities are expected also in astrophysical dynamical formation channels, so eccentricity would not be a good way to discriminate { individual} PBH binaries in the LISA band~\cite{Samsing:2013kua,Nishizawa:2016jji,Breivik:2016ddj,Nishizawa:2016eza,Zevin:2021rtf}. {However, it may be possible to distinguish the PBH channel from other astrophysical models from the eccentricity distribution of the whole BBH population. } }

So far we have considered mergers happening at low redshift $z \lesssim 1$. In case of high-redshift mergers ($z\gtrsim 30$) predicted by the primordial scenario, the initial PBH binary semimajor axis is reduced by only a factor ${\cal O}$(2), as shown in Fig.~\ref{fig1}. This small change is due to the large sensitivity of the merger time to the initial semimajor axis: $\tau \propto a_0^4$, see Eq.~\eqref{eq:tau_m_2}. Therefore, when the binary enters the detectable frequency band of GW experiments, it is expected to have already circularized its orbit to an undetectable level.  This property allows us to distinguish primordial binaries from binaries produced by astrophysical dynamical formation channels, which may retain significant eccentricities (see e.g. Refs.~\cite{Samsing:2013kua,Nishizawa:2016jji,Breivik:2016ddj,Nishizawa:2016eza,Zevin:2021rtf}).

Let us recall one more time that our predictions are based on the assumption that PBH mergers are dominated by the binaries formed in the early Universe. If late-time Universe binaries contribute substantially to the observed events, the PBH binary eccentricity may be larger, and comparable to expectations from the astrophysical dynamical formation scenarios. This situation may be realized with strong PBH clustering suppressing the early-Universe binary merger rate, while enhancing the late-time Universe contribution~\cite{Jedamzik:2020ypm,Vaskonen:2019jpv,Trashorras:2020mwn,DeLuca:2020jug}. However, we stress that this scenario would require a large value of the PBH abundance ($f_\PBH\simeq 1$), which is in contrast with current PBH constraints in the LVK mass range~\cite{Carr:2020gox}.

\subsection{Tidal disruption and tidal deformability}\label{sec:tidal}
While PBHs below a few solar masses can be easily distinguished from a population of (heavier) stellar-origin BHs, they might be confused with other compact objects. For example, in standard astrophysical scenarios, white dwarfs and NSs are formed with masses above $\approx 0.2M_\odot$~\cite{Kilic:2006as} and $\approx 1 M_\odot$~\cite{Strobel:1999vn,Lattimer:2012nd,Silva:2016myw,Suwa:2018uni}, respectively.

A relevant discriminator in this case is provided by the Roche radius, ${r_\text{\tiny Roche}}$, below which the secondary object in a binary system gets tidally disrupted, if it is not a BH. The Roche radius is approximately
\begin{equation}
    {r_\text{\tiny Roche}}\sim 1.26 \, r_2 q^{-1/3} \,,\label{Roche}
\end{equation}
where $r_2$ is the radius of the secondary object.
If ${r_\text{\tiny Roche}}>r_\text{\tiny ISCO}\sim {\cal O}(M)$, the binary is tidally disrupted before merger, thus effectively cutting off the GW signal at the GW frequency corresponding to $r_\text{\tiny Roche}$.
Another relevant quantity to check is the contact radius which, assuming the primary is a BH, can be estimated as
\begin{equation}
    r_\text{\tiny contact}\sim 2 m_1+ r_2\,.
\end{equation}
If $r_\text{\tiny contact}>r_\text{\tiny ISCO}$ the contact frequency of the objects is lower than the ISCO frequency, and the point-particle approximation breaks down.

The left panel of Fig.~\ref{fig_Roche} shows that, for a typical white dwarf, ${\rm max}(r_\text{\tiny Roche}, r_\text{\tiny contact})=r_\text{\tiny Roche}$ is larger than the ISCO.  Therefore, the star is tidally disrupted well before the GW signal reaches the ISCO frequency. Less compact objects, such as brown dwarfs, are disrupted at even larger radii (smaller orbital frequencies).
Therefore, if $m_2< M_\odot$ (thus excluding NSs) the \emph{maximum} frequency of the coalescence can be used to detect a tidal disruption event and discriminate whether the secondary is a BH or a less compact star.
When the secondary is a NS, the possible outcomes are more complicated. We may still have nondisruptive mergers if the NS compactness (i.e. the ratio between the mass and the size of the NS, ${\cal C} \equiv m_\text{\tiny NS}/r_\text{\tiny NS}$) is large enough, or if the ratio between the secondary (NS) mass and the primary (BH) mass is sufficiently small: see the right panel of Fig.~\ref{fig_Roche}, based on the criterion in Ref.~\cite{Pannarale:2015jia}.
In this case, the absence of tidal disruption may not be used as a discriminator for the (primordial) BH nature of the secondary object.

Exotic compact objects~\cite{Cardoso:2019rvt} (e.g. boson stars~\cite{Liebling:2012fv}) would provide another possible explanation for a (sub)solar compact object. The compactness of a boson star depends strongly on its mass and on the scalar self interactions~\cite{Cardoso:2017cfl}. For the vanilla ``mini'' boson star model without self-interactions~\cite{Ruffini:1969qy}, the compactness is $m_2/r_2={\cal O}(0.01)$ near the maximum mass. The left panel of Fig.~\ref{fig_Roche} shows that also solar-mass mini boson stars would be tidally disrupted before the ISCO. In the presence of strong scalar self-interactions, boson stars can be as compact as a NS~\cite{Colpi:1986ye,Cardoso:2017cfl}, so in that case the tidal disruption is not a clear-cut discriminator.

Another key discriminator between PBHs and (sub)solar horizonless objects is the absence, in the former case, of tidal deformability contributions to the gravitational waveform. The tidal Love numbers are identically zero for a BH (see Refs.~\cite{Binnington:2009bb,Damour:2009vw,Damour:2009va,Pani:2015hfa,Pani:2015nua,Gurlebeck:2015xpa,Porto:2016zng,
LeTiec:2020spy, Chia:2020yla,LeTiec:2020bos,Hui:2020xxx,Charalambous:2021kcz,Charalambous:2021mea,Pereniguez:2021xcj} for literature on this topic) , whereas they are generically nonzero and model-dependent for any other compact object~\cite{Cardoso:2017cfl}.
The tidal Love numbers enter the GW phase in Eq.~\eqref{eq:Psiterms}
starting at 5 post-Newtonian (PN) order. We write
\begin{equation}\label{tidaldefwaf}
\Delta \Psi^\text{\tiny tidal}_\text{\tiny 6PN}
 = \Lambda  v^5 + \delta\Lambda v^6+{\cal O}(v^7)\,,
\end{equation}
where $v=(\pi M f)^{1/3}$ is the PN orbital velocity parameter, and the $5$PN and $6$PN terms are given by~\cite{Flanagan:2007ix,Vines:2011ud}
\begin{align}
 \Lambda &=\left(264 -\frac{288}{\eta_1}\right) \frac{\lambda_2^{(1)}}{M^5} +    (1\leftrightarrow 2)\,, \nonumber \\
 \delta\Lambda&=\left(  \frac{4595}{28}- \frac{15895}{28 \eta_1}  + \frac{5715 \eta_1}{14} -
 \frac{325 \eta_1^2}{7}   \right)\frac{\lambda_2^{(1)}}{M^5} +(1\leftrightarrow 2) \,,
\end{align}
in terms of the dominant (i.e., electric-type, quadrupolar) tidal Love number, $\lambda_2^{(i)} = 2m_{i}^5 k_2^{(i)}/3$, of the $i$-th body.
In the Newtonian approximation, the tidal Love number of an object is (see e.g.~\cite{PoissonWill})
\begin{equation}
    k_2^{(i)} \sim {\cal O}(0.01-0.1) \left(\frac{r_i}{m_i}\right)^5\,,
\end{equation}
where the precise value of the dimensionless prefactor depends on the nature of the object (for instance, on the equation of state in the case of a NS). Thus, less compact objects have the larger tidal deformability, and hence can be more easily discriminated from a BH.

Overall, any measurement of a nonzero tidal deformability in an object above a few solar masses would automatically imply that either the object is not a BH, or that the BH is surrounded by matter fields, in which case the total tidal Love number of the dressed BH is nonzero~\cite{Cardoso:2019upw,DeLuca:2021ite}.

Finally, a further discriminator would be the waveform corrections due to tidal heating terms in the case of BHs. This correction is due to dissipation at the event horizon~\cite{Alvi:2001mx} and is negligible for other compact objects~\cite{Cardoso:2019rvt,Maselli:2017cmm}. However, the contribution of the tidal heating is typically small. Unless the object is extremely compact so that $k_2\sim0$, tidal heating is subdominant with respect to the tidal deformability correction presented above.

\section{Measurement accuracy for key PBH binary discriminants} \label{sec:results}

In this section we quantify the measurability of the PBH discriminators presented above, following the flowchart of Fig.~\ref{fig:flowchart}. The statistical errors are computed using a Fisher information matrix approach, which provides an accurate estimate of the statistical errors in the high-SNR limit with Gaussian noise and in the absence of systematic biases in the waveform parameters~\cite{Vallisneri:2007ev}.  Our methodology is standard, and reviewed in Appendix~\ref{app:Fisher} for completeness.

We will present results for the planned future stage of the LIGO experiment (Ad.~LIGO), the 3G detector ET (in its ET-D configuration~\cite{Hild:2010id}) and LISA~\cite{Robson:2018ifk}. We do not explicitly report the analysis for the CE detector, because the CE noise power spectral density is qualitatively similar to ET. The noise power spectral densities used in our analysis are listed in Appendix~\ref{apppsd}.

As we discussed in the introduction, we will focus on binary mergers with individual component masses below ${\cal O}(10^2) M_\odot$. We leave the analysis of more massive events, up to the supermassive range of interest for LISA, for future work.

\begin{figure}[t!]
\centering
\includegraphics[width=0.49\textwidth]{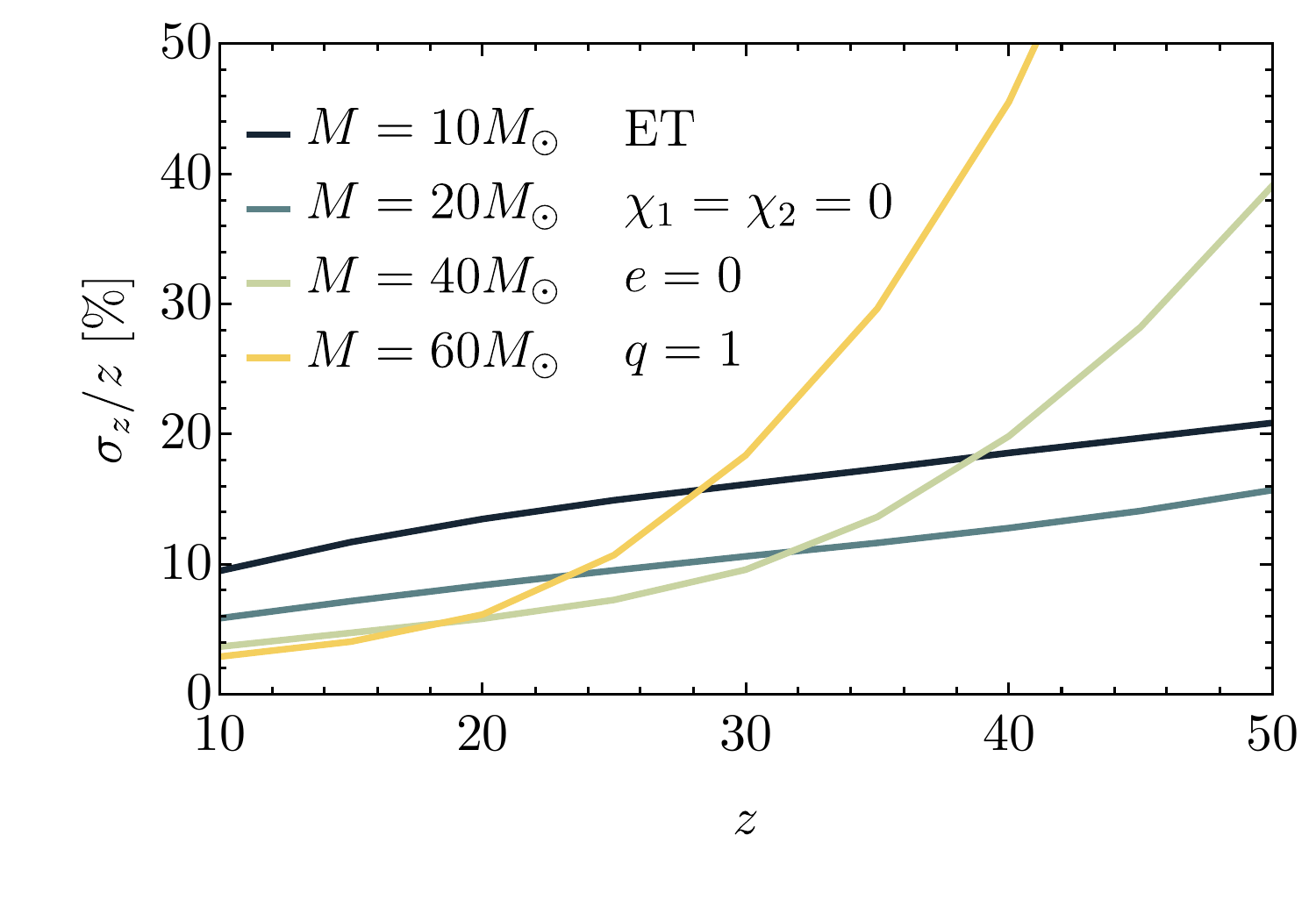}
\caption{ Relative percentage errors on redshift measurement for an optimally oriented source located at redshift $z$, detected
with ET.
The various colors correspond to different choices of total mass $M$. 
}\label{fig_highredshift}
\end{figure}

\begin{figure*}[th]
\centering
\includegraphics[width=0.520625\textwidth]{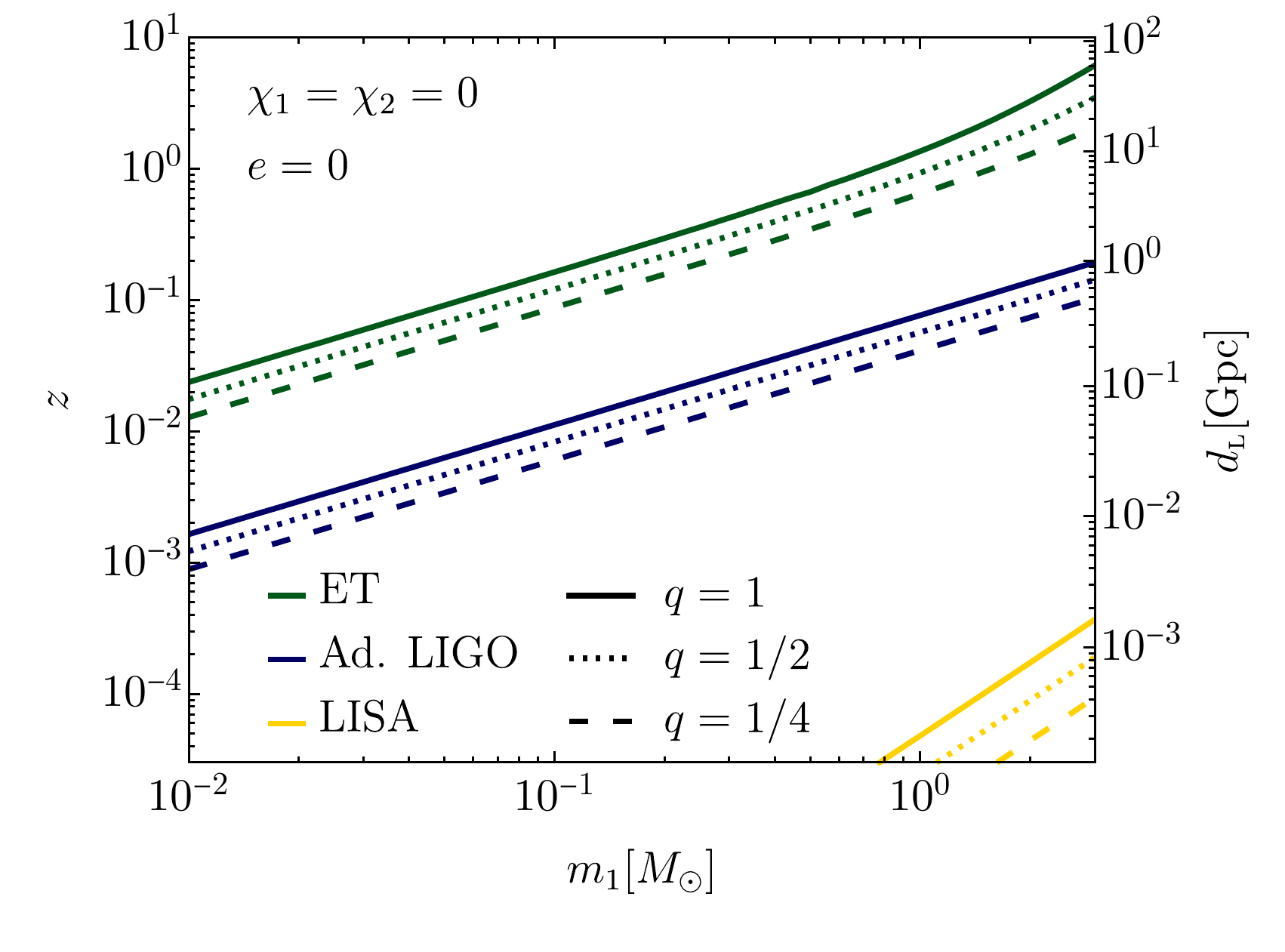}
\includegraphics[width=0.459375\textwidth]{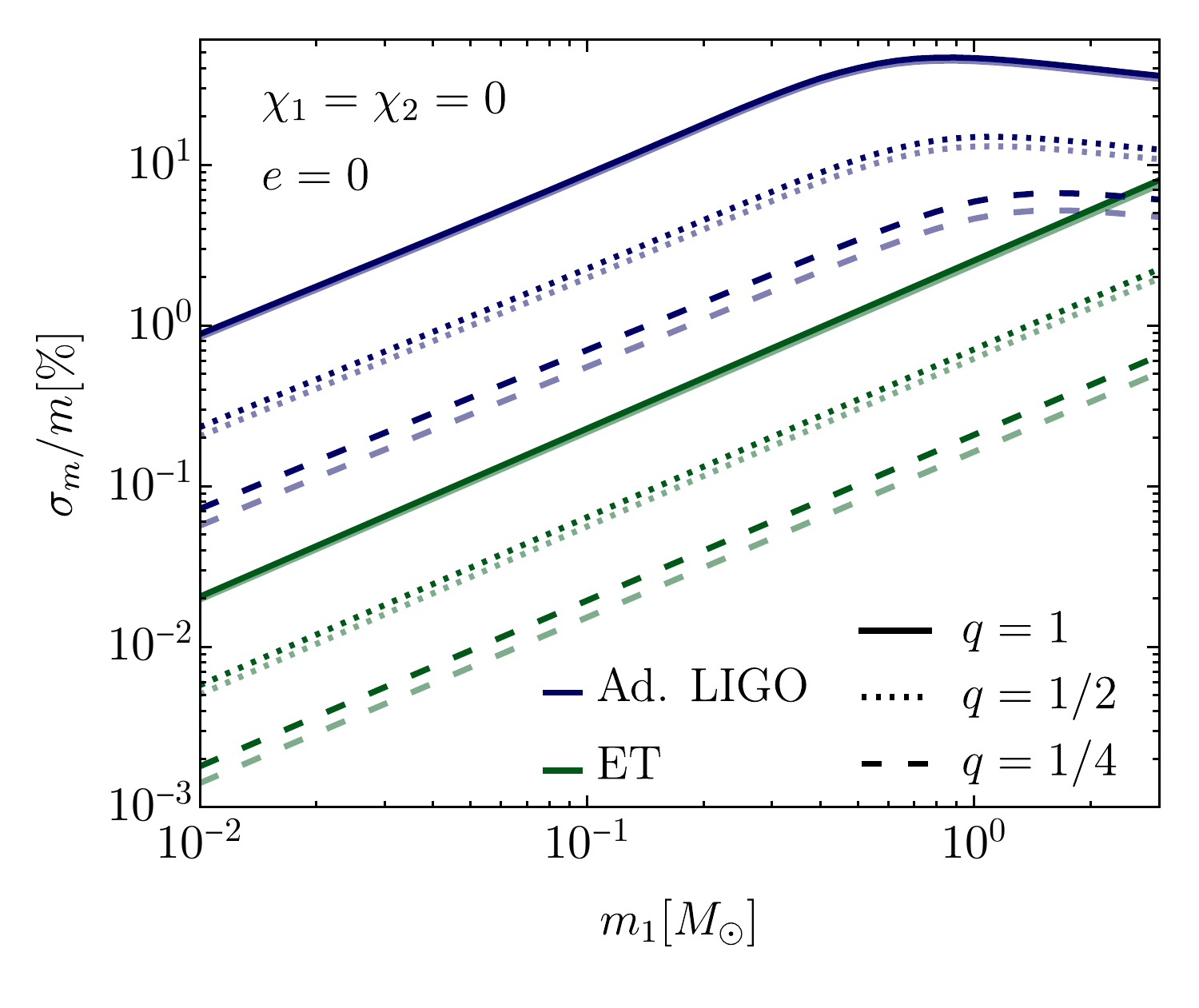}
\caption{
\textbf{Left:} Horizon redshift as a function of SNR for subsolar-mass mergers assuming noneccentric, spinless and optimally oriented binaries for different values of the mass ratio. In this mass range, due to the different range of frequencies probed by space-based detectors, LISA will have very limited reach and its horizon can only be seen in the bottom right corner.
\textbf{Right:}
Measurement errors of the binary's primary and secondary masses in the subsolar range.
The binaries are assumed to be located at the horizon redshift of Ad.~LIGO following the same assumptions as in the left panel. We assumed the same distance for ET to highlight the improved precision of the 3G detector. The dashed and dotted lines indicate smaller values of the mass ratio.
In the cases $q\neq 1$, the percentile uncertainties on the primary (secondary) mass is indicated with a darker (lighter) color, which corresponds to the higher (lower) line. 
 }\label{fig_subsolar}
\end{figure*}

\subsection{Redshift measurement accuracy}\label{error_redshift}
Next-generation interferometers such as CE and ET will be able to search for PBH mergers at redshift $z\gtrsim30$, where mergers of astrophysical origin should not occur~\cite{Koushiappas:2017kqm,DeLuca:2021wjr}. However, redshift measurements for such distant cosmological sources are typically inaccurate and prior-dependent~\cite{Ng:2021sqn}. 
In Fig.~\ref{fig_highredshift}, we show the measurement errors estimated using the Fisher matrix analysis for distant events with large source redshift and four selected values of the total mass. The measurement accuracy we obtain is consistent with the errors computed using a full Bayesian parameter estimation in Ref.~\cite{Ng:2021sqn}, but the Fisher formalism does not allow us to reproduce the bias towards smaller redshift observed in their results. This systematic bias is due to a combination of the assumed prior on the source redshift, and the asymmetric dependence of the errors on binary inclination angles ~\cite{Ng:2021sqn}, and it is partly responsible for the difficulty in confidently assessing the high-redshift ($z\gtrsim 30$) nature of distant binaries. Note also that we do not report results for total masses $M\gtrsim 60 M_{\odot}$ because, in that case, the SNR is dominated by the merger-ringdown portion of the GW signal, which we are not including in our simple estimates.

LISA will also be able to observe events with $z\gtrsim 30$ if the total mass of the binary is above $M \gtrsim 2 \times 10^3 M_\odot$ (see e.g. Refs.~\cite{Kaiser:2020tlg,DeLuca:2021hde}). Consistently with the rest of the paper, we defer a dedicated discussion of the high-mass region of the parameter space to future work.

\subsection{Horizon of subsolar detection and mass measurement accuracy}

In Fig.~\ref{fig_subsolar} (left panel), we show the horizon redshift for detecting subsolar binaries for Ad.~LIGO, ET, and LISA\footnote{See also~\cite{Barsanti:2021ydd} for a similar analysis in the context of extreme mass-ratio inspirals detectable by LISA and ET.}. We assume negligible spins and eccentricity, as expected for primordial BBHs. 

As one can infer from the plot, ET will extend the horizon redshift of Ad.~LIGO by more the one order of magnitude. In terms of maximum distance, in the subsolar mass range one obtains
\begin{equation}
    d_\text{\tiny L}^\text{\tiny hor}\approx
    \begin{cases}
      0.40 \, {\rm Gpc}\,  \lp \frac{{\cal M}}{M_\odot}\rp^{5/6}
      \quad {\rm for} \quad \text{Ad.~LIGO},\\
      7.1\, {\rm Gpc}\,  \lp \frac{{\cal M}}{M_\odot}\rp^{5/6}
      \, \, \, \quad {\rm for} \quad \text{ET},
    \end{cases}  
\end{equation}
where we introduced the chirp mass ${\cal M} \equiv \eta^{3/5} M$. This simple scaling is obtained thanks to the following simplifying conditions being met in this mass range: (i) the horizon falls below redshift $z\lesssim 1$;
(ii) the frequencies to which ground-based detectors are mostly sensitive practically always fall below the ISCO frequency for those light binaries; and
(iii) the amplitude of the GW signal scales like ${\cal A}\approx {\cal M}^{5/6}$, see Eq.~\eqref{eq:spawaveform2}.

On the other hand, due to the smaller frequencies probed by space-based detectors, LISA will have very limited reach in this mass range.
Therefore, the maximum distance that can be observed is greatly reduced, with an horizon falling much below the Mpc scale for the subsolar mass range and scaling as 
\begin{equation}
    d_\text{\tiny L}^\text{\tiny hor}\approx
    \begin{cases}
      0.030 \, {\rm Mpc}\,  \lp \frac{\cal M}{M_\odot}\rp^{5/6}
      \quad {\rm for} \quad {\cal M} \lesssim0.1 M_\odot,
      \\
      0.20 \, {\rm Mpc}\,  \lp \frac{\cal M}{M_\odot}\rp^{1.8}
      \,\,\,\,\, \quad {\rm for} \quad{\cal M}\gtrsim0.1 M_\odot,
    \end{cases}  
\end{equation}
 assuming an observation duration of $T_\text{\tiny obs} = 1 {\rm yr}$. The change in slope of the horizon luminosity distance as a function of mass can be explained by inspecting Eq.~\eqref{fconditionLISA}, which describes the frequencies spanned by the GW signal within the observation time $T_\text{\tiny obs}$. 
For ${\cal M}\lesssim 0.1 M_\odot$, the observed GW signal becomes effectively monochromatic and the SNR (or the horizon) is only affected by the GW amplitude, which in turn is controlled by the binary's chirp mass. 
For ${\cal M}\gtrsim 0.1 M_\odot$, LISA starts resolving part of the frequency evolution, and the SNR grows more steeply as a function of the binary mass due to the larger accessible frequency range. 

In Fig.~\ref{fig_subsolar} (right panel), we also show the error estimate on individual masses. To facilitate the comparison between the performance of the two experiments, for both Ad.~LIGO and ET we assume the binaries to be located at the same distance, chosen to be the Ad.~LIGO horizon. 
Errors decrease for smaller masses due to our choice of fixing the SNR of the source (${\rm SNR}=8$ for Ad.~LIGO by construction, much larger but almost constant around ${\rm SNR}\approx 110$ for ET), and to the corresponding larger number of cycles spanned by the GW signal in the detector band. 
For ET (green lines), we observe a similar dependence of the error on the primary mass, but with an improved overall measurement accuracy, 
which scales faster then the linear dependence on the SNR because 3G detectors have a smaller frequency cut-off than Ad.~LIGO, and therefore a larger number of observable cycles. 
At fixed SNR, more asymmetric binaries yield smaller relative errors on the reconstructed mass parameters. 

Overall, our results indicate that 3G detectors will be able to measure the mass of subsolar events with an extremely high precision, below the percent level.

\begin{figure}[t!]
\centering
\includegraphics[width=0.49\textwidth]{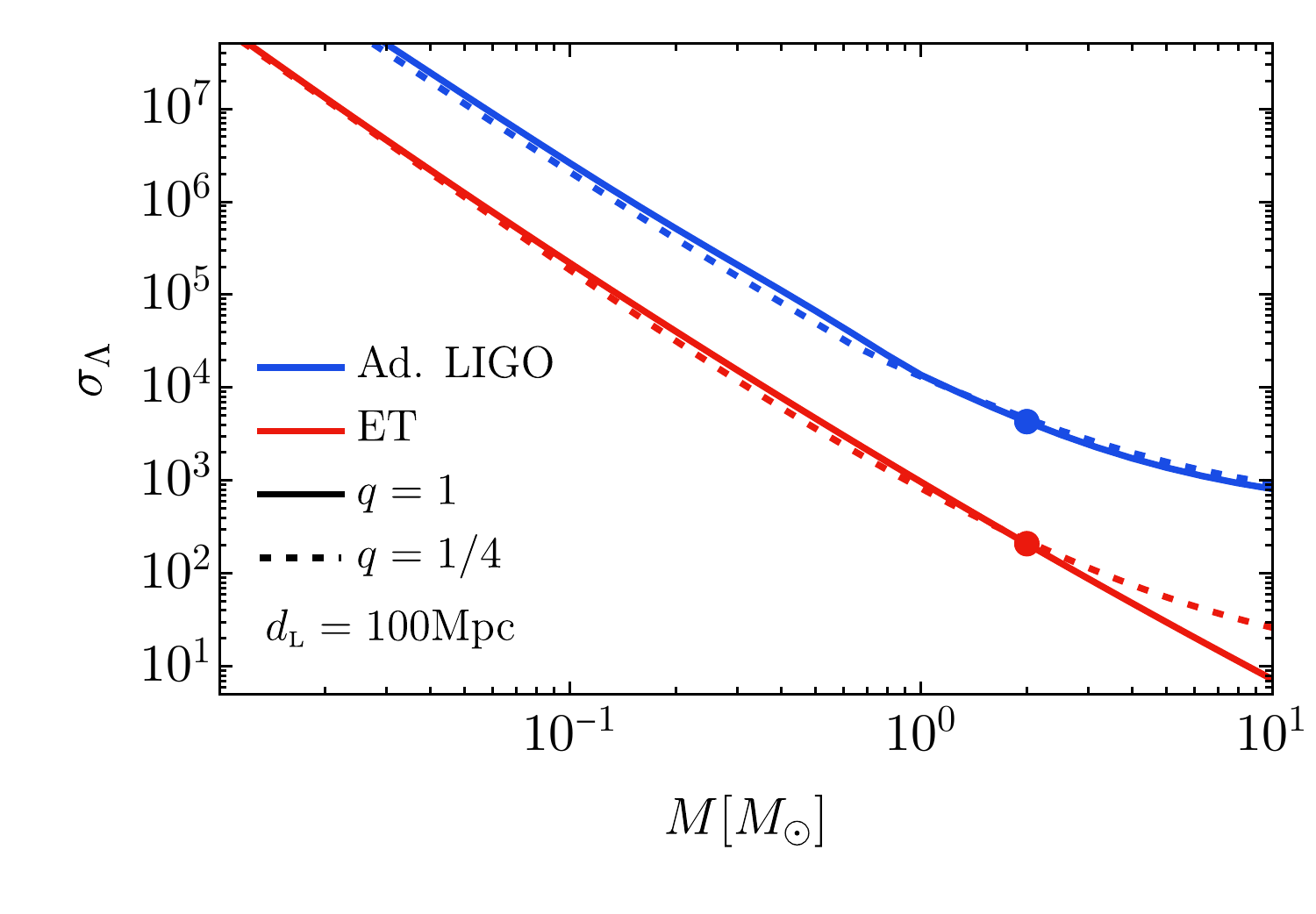}
\caption{
Measurement precision of the deformability parameter $\Lambda$ at both Ad.~LIGO and ET.
The binary is assumed to have spinless components and negligible eccentricity and deformability, as predicted by the PBH scenario. The solid (dashed) line indicates the result for $q=1$ ($q=1/4$).
}\label{fig_Lambda}
\end{figure}

\begin{figure*}[th]
\centering
\includegraphics[width=0.32\textwidth]{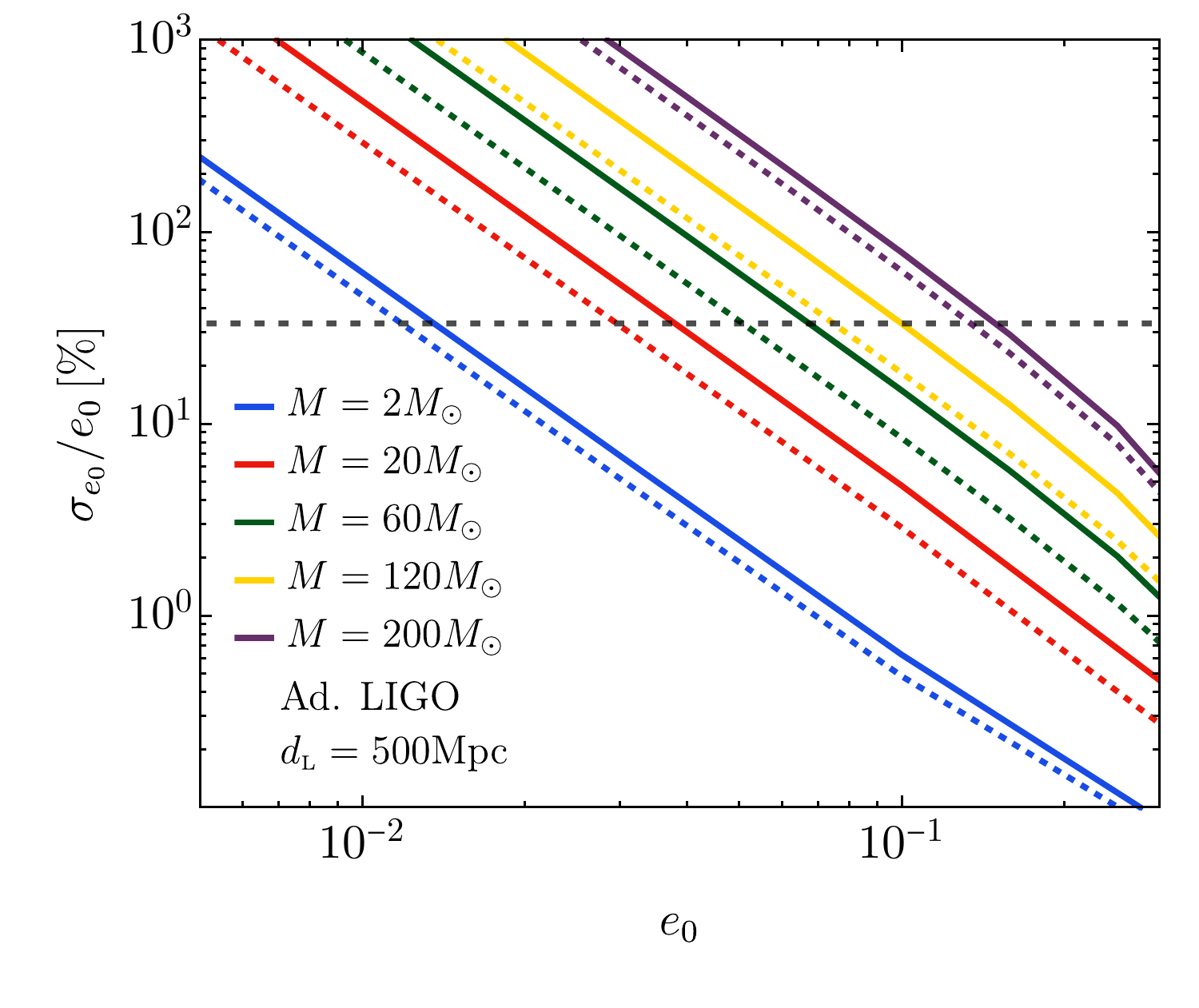}
\includegraphics[width=0.32\textwidth]{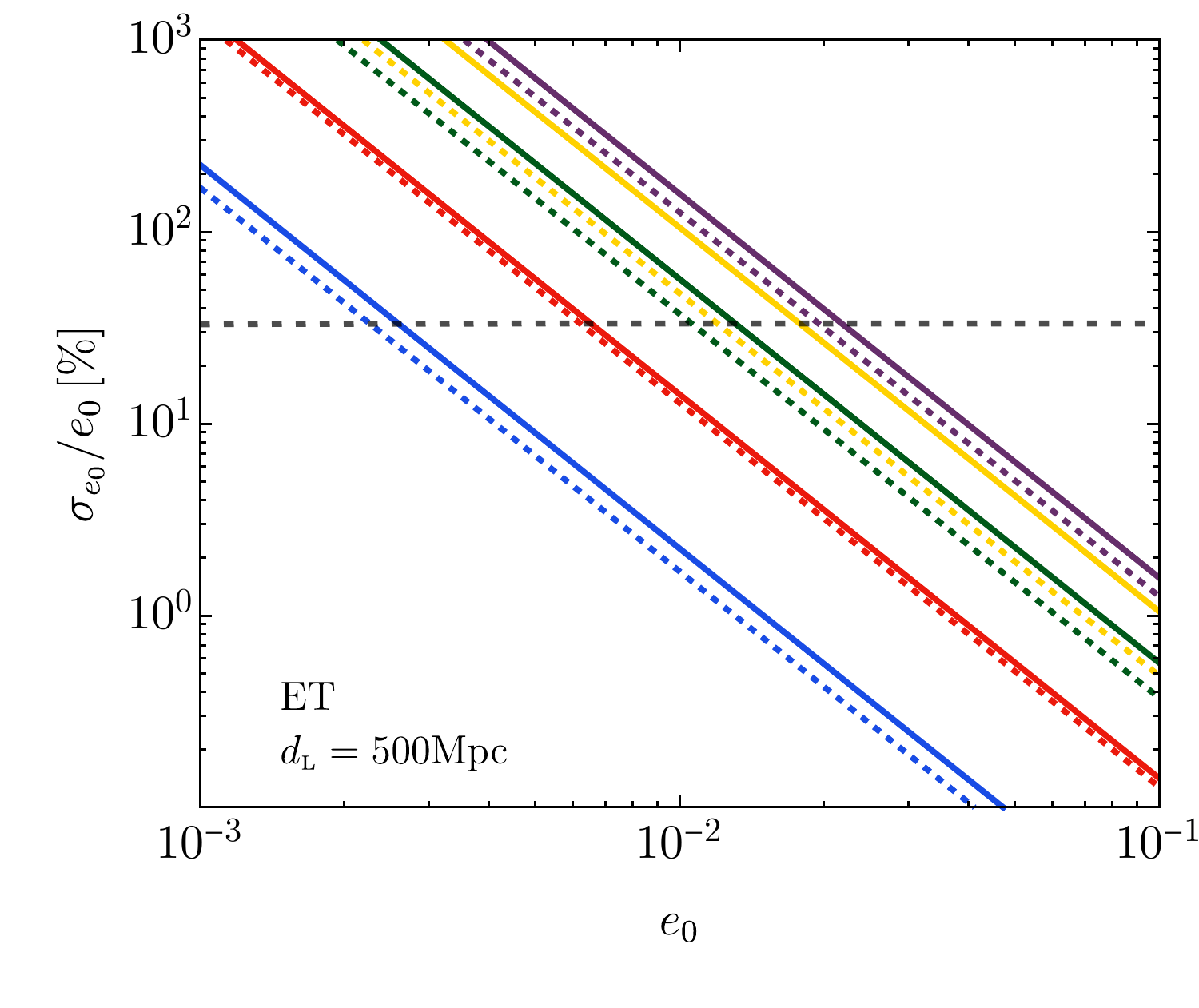}
\includegraphics[width=0.32\textwidth]{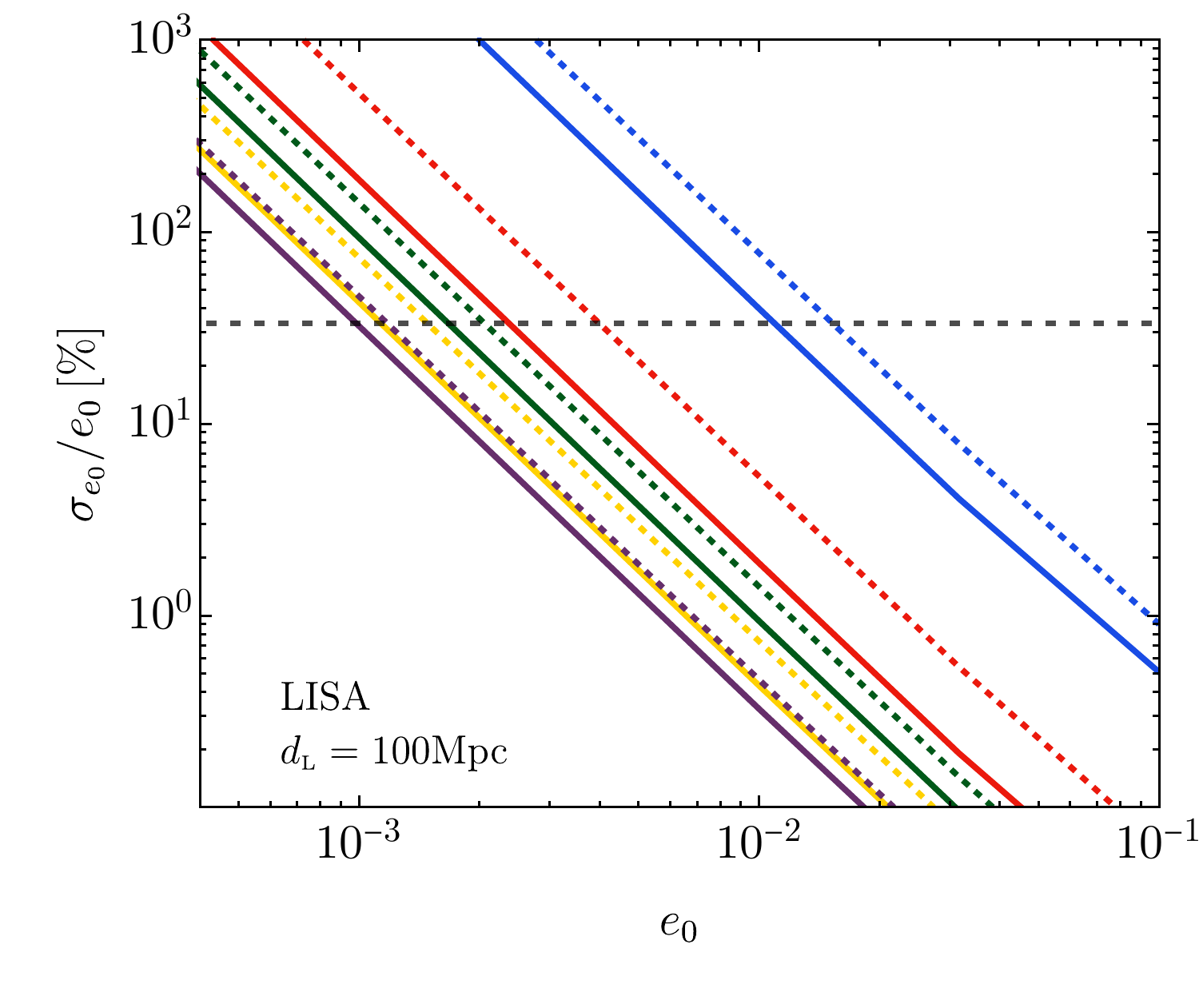}
\caption{
Percentage measurement 
uncertainty on eccentricity for various values of BBH total mass $M$.
The solid, dashed lines correspond to $q=1$ and $q=1/3$, respectively. 
We can confidently exclude vanishing eccentricity (at 3-$\sigma$ level) below the horizontal gray dashed line. 
The binaries are assumed to be at a distance of $d_\text{\tiny\rm L} = 500 {\rm Mpc}$ ($ 100 {\rm Mpc}$) for the ground-based (space-based) experiment, respectively. We have checked that varying the BH spins does not significantly affect the eccentricity measurements.
\textbf{Left:} Ad.~LIGO. 
\textbf{Center:} ET.
\textbf{Right:} LISA.
}\label{fig_eccent}
\end{figure*}

\subsection{Tidal deformability and tidal disruption}\label{error_tidal}

In Fig.~\ref{fig_Lambda}, we show the measurement errors on the binary tidal deformability $\Lambda$ as a function of the total mass $M<10 M_\odot$, assuming negligible spins, eccentricity, and deformability, as predicted by the PBH scenario.
We also report two representative values of the mass ratio ($q=1$ and $q=1/4$), to show its effects of the measurement errors on $\Lambda$.
In this case, we show the errors as a function of the binary total mass $M$ and not of the primary mass $m_1$, as done in the previous plots. Indeed, we find that $\sigma_\Lambda$ depends mostly on $M$.

The typical deformability expected for a BH-NS binary is approximately in the range $\Lambda\in (100,5000)$, depending on the NS equation of state~\cite{LIGOScientific:2019eut} and on the mass ratio. 
Fig.~\ref{fig_Lambda} shows that Ad.~LIGO (ET) would be able to exclude $\Lambda=0$ at $3\sigma$ (i.e., $\Lambda - 3 \sigma_\Lambda > 0$) for a symmetric subsolar-mass binary at $d_\text{\tiny L}=100\,{\rm Mpc}$ only if $\Lambda>1.3 \times 10^4$ ($\Lambda>6.27 \times 10^2$).\footnote{The errors estimated with the present analysis are consistent with the one reported in  Ref.~\cite{Pacilio:2021jmq} (found by a full Bayesian parameter estimation) once translated in terms of the reduced deformability parameter $\tilde \Lambda \equiv -2\Lambda / 39$ (see also Ref.~\cite{Favata:2013rwa}).}
Therefore, the constraining power of Ad.~LIGO is limited for this discriminator, while ET could exclude the primordial origin of a subsolar-mass binary, based on the tidal deformability measurements, only for the least compact NSs, which are already marginally in tension with GW170817~\cite{LIGOScientific:2017vwq}.

Less compact objects like white dwarfs (or hypothetical mini boson stars) have much larger tidal deformability, which can be therefore measured accurately given the estimates in Fig.~\ref{fig_Lambda}. However, as previously discussed, these objects are tidally disrupted well before the ISCO frequency. In this case the GW signal is abruptly suppressed at the frequency corresponding to the Roche or contact radius, so it can presumably be distinguished more easily from the ``smooth'' inspiral signal of a BH binary.

For this range of masses the measurement accuracy on $\Lambda$ with LISA is very low, since LISA can only observe the early inspiral and tidal effects enter at high PN order. Therefore, we do not show LISA results in this case.

Finally, note that our Fisher analysis for the errors on $\Lambda$ include the eccentricity in the waveform parameters. We explicitly checked that removing $e$ (i.e. assuming $e=0$ or that it is known a priori) does not affect the error estimates, even though one expects  that reducing the dimensionality of the problem would result in better constraining power. This is because the eccentricity and deformability mostly impact separate phases of the inspiral: eccentricity is larger at small frequencies, while the tidal deformability (5PN+) effects become relevant close to the ISCO frequency. Therefore, $e$ and $\Lambda$ are effectively uncorrelated in the Fisher matrix, and removing one of the two parameters does not reduce uncertainties on the other one.

\subsection{Eccentricity measurement accuracy}

A firm  prediction of the scenario involving PBH binaries formed in the early Universe is that their orbit circularizes before entering the observability band of ground-based detectors (see Sec.~\ref{sec:th_ecc}). 
In Fig.~\ref{fig_eccent}, we show the orbital eccentricity measurement accuracy in Ad.~LIGO, ET and, LISA as a function of the binary eccentricity for selected values of the binary masses. 
Consistently with the results of Ref.~\cite{Favata:2021vhw} for the case of Ad.~LIGO, we find that vanishing eccentricity can be ruled out at 3$\sigma$ level, for a binary with total mass $M$ located at a distance $d_\text{\tiny L} = 500\,{\rm Mpc}$, if $e_0$ is larger than (see also Ref.~\cite{Lower:2018seu})
\begin{equation}
e_0^\text{\tiny Ad.~LIGO} \gtrsim 5.5 \times 10^{-3} \lp  \frac{M}{M_\odot}    \rp^{0.62},
\end{equation}
with only a negligible dependence on the individual spins of the binary components and a minor dependence on the mass ratio. ET will be able to constrain the eccentricity down to lower values, with a minimum resolvable eccentricity scaling with the binary total mass as
\begin{equation}
e_0^\text{\tiny ET} \gtrsim 1.8 \times 10^{-3} \lp  \frac{M}{M_\odot}    \rp^{0.48}.
\end{equation}
Finally, assuming a binary located at $d_\text{\tiny L} =100$Mpc distance, for LISA one obtains (see also Refs.~\cite{Nishizawa:2016jji,Nishizawa:2016eza})
\begin{equation}
e_0^\text{\tiny LISA} \gtrsim 1.6 \times 10^{-2} \lp  \frac{M}{M_\odot}    \rp^{-0.56}.
\end{equation}
It is interesting to stress the trend observed in the relative accuracy as a function of $M$. As the eccentricity decreases during the binary evolution, most of the constraining power comes from low frequencies (see the discussion in Sec.~\ref{sec:orbital_evo}). In both the Ad.~LIGO and ET cases, a heavier binary enters in the observable frequency band closer to the merger time. For this reason, a larger mass implies larger errors on the eccentricity. 
On the other hand, LISA is mostly sensitive to smaller frequencies, and larger masses imply smaller errors $\sigma_{e_0}$ due both to the wider frequencies observable at fixed observation time, and to the larger SNR.
This trend can be observed in the right panel of Fig.~\ref{fig_eccent}.

\begin{figure*}[th]
\centering
\includegraphics[width=0.32\textwidth]{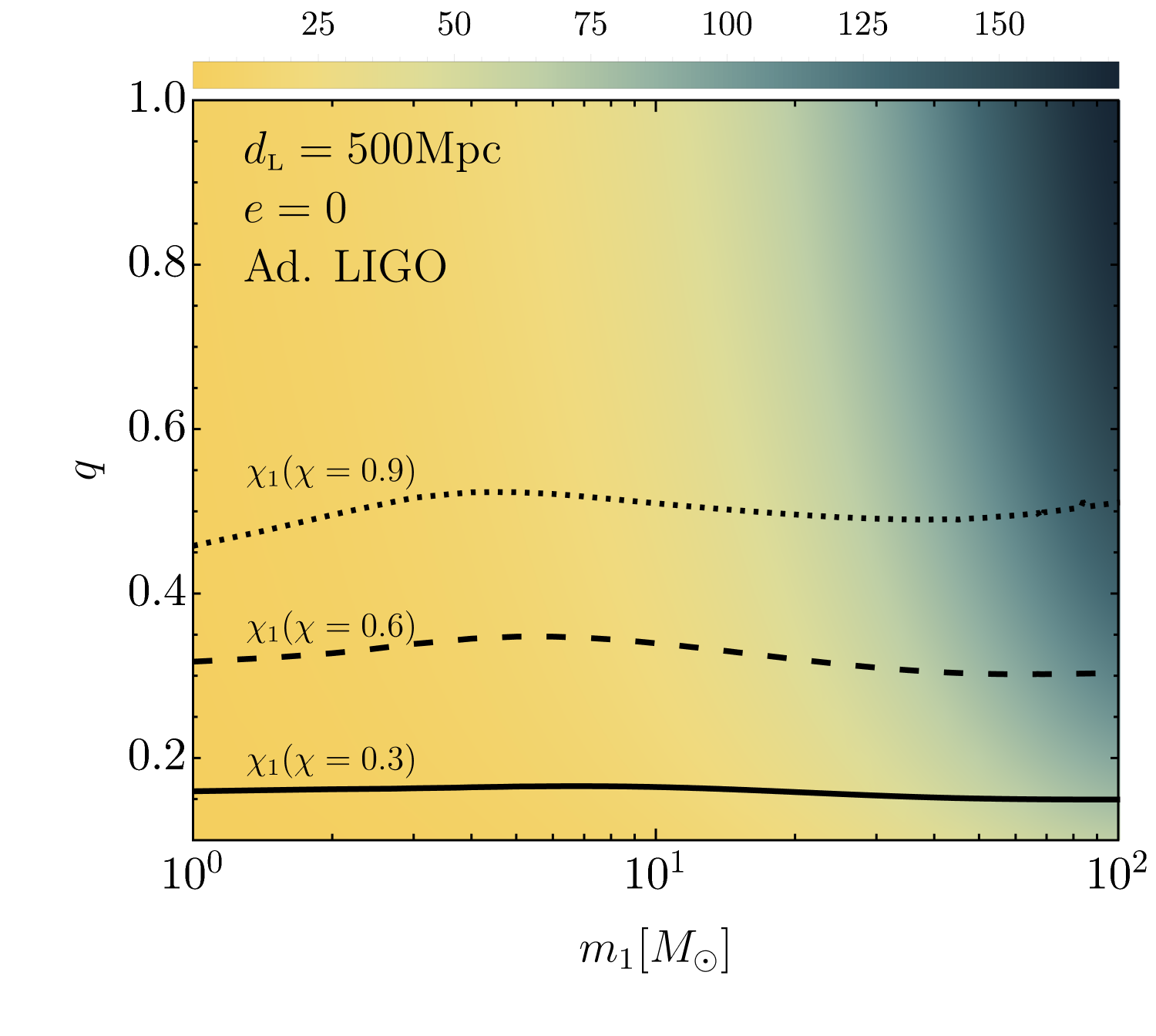}
\includegraphics[width=0.32\textwidth]{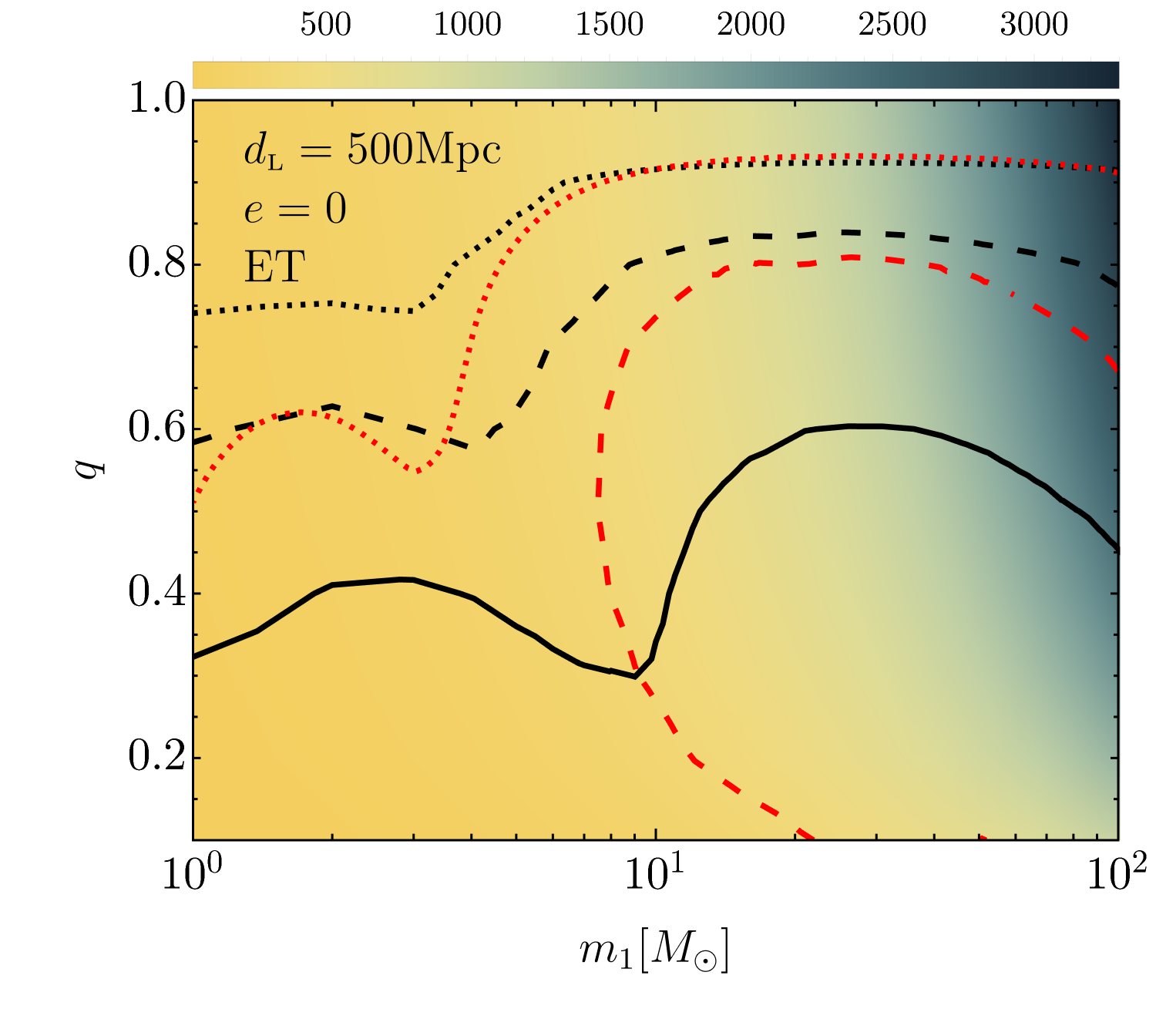}
\includegraphics[width=0.32\textwidth]{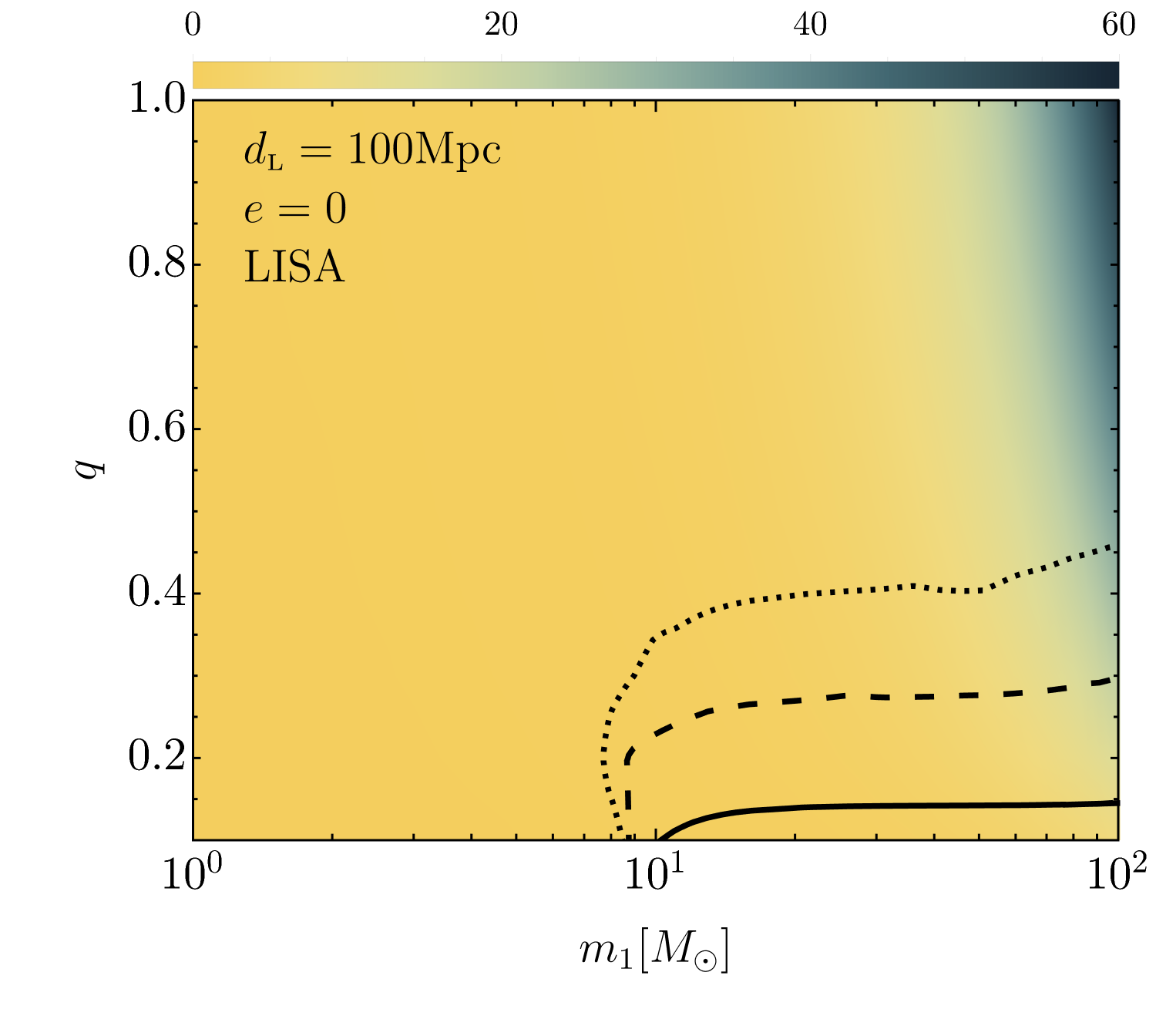}
\caption{
The region below black/red lines indicates the combinations of $(m_1,q)$ for which one can confidently exclude vanishing spins (at 3-$\sigma$ level) assuming different values of the component BH spins ($\chi_1=\chi_2 = \chi$). 
The solid, dashed and dotted lines correspond to $\chi=[0.3, 0.6,0.9]$, respectively.  Black (red) lines correspond to the primary (secondary) BH spins. The color code indicates the SNR, and the source is assumed to be at a distance $d_\text{\tiny \rm L}=500${\rm Mpc} ($d_\text{\tiny \rm L}=100${\rm Mpc}) for ground-based (spaced-based) detectors, respectively.
\textbf{Left:} Ad.~LIGO. 
\textbf{Center:} ET. 
\textbf{Right:} LISA.}\label{fig_light-spin}
\end{figure*}

\begin{figure*}[th]
\centering
\includegraphics[width=0.32\textwidth]{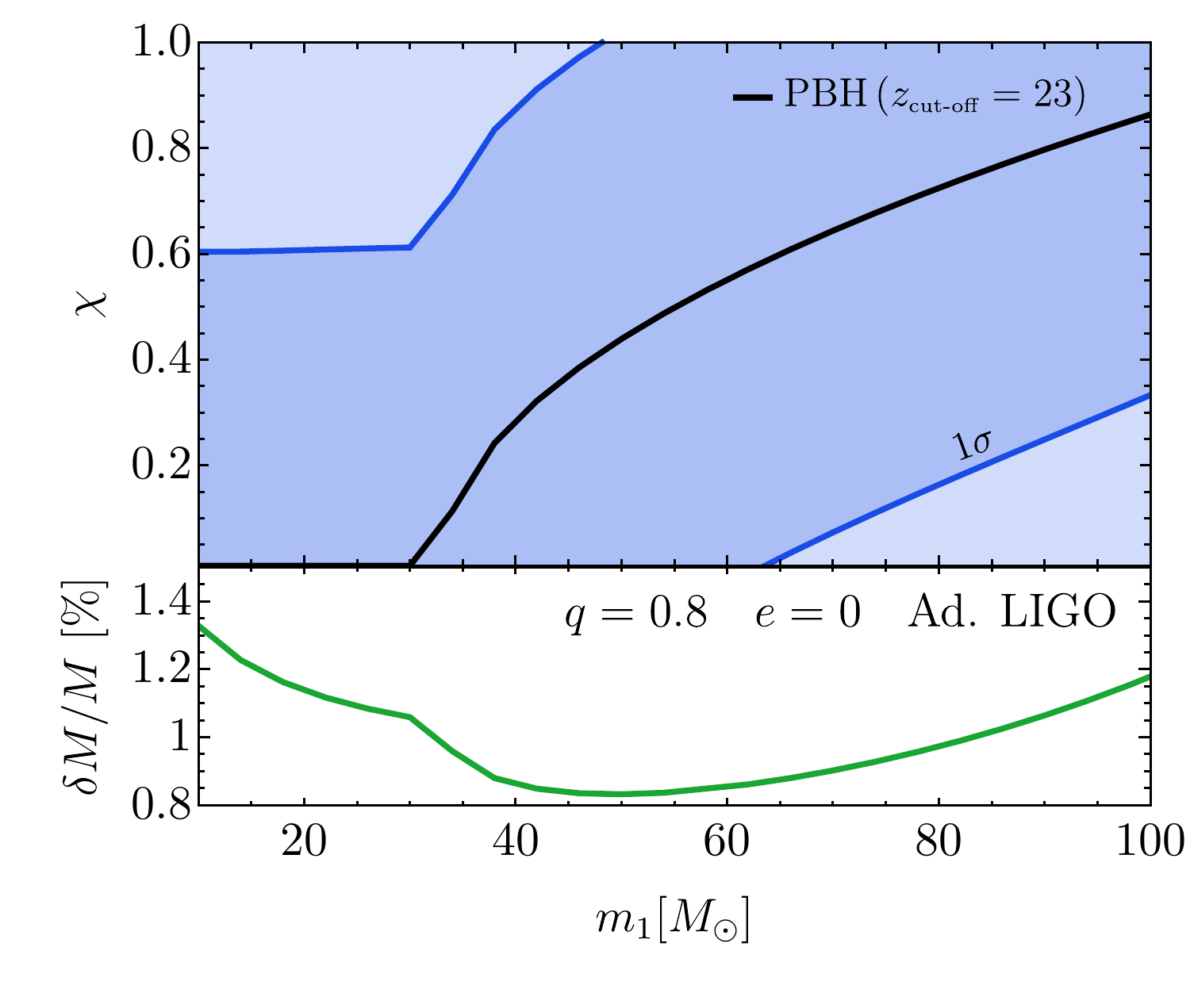}
\includegraphics[width=0.32\textwidth]{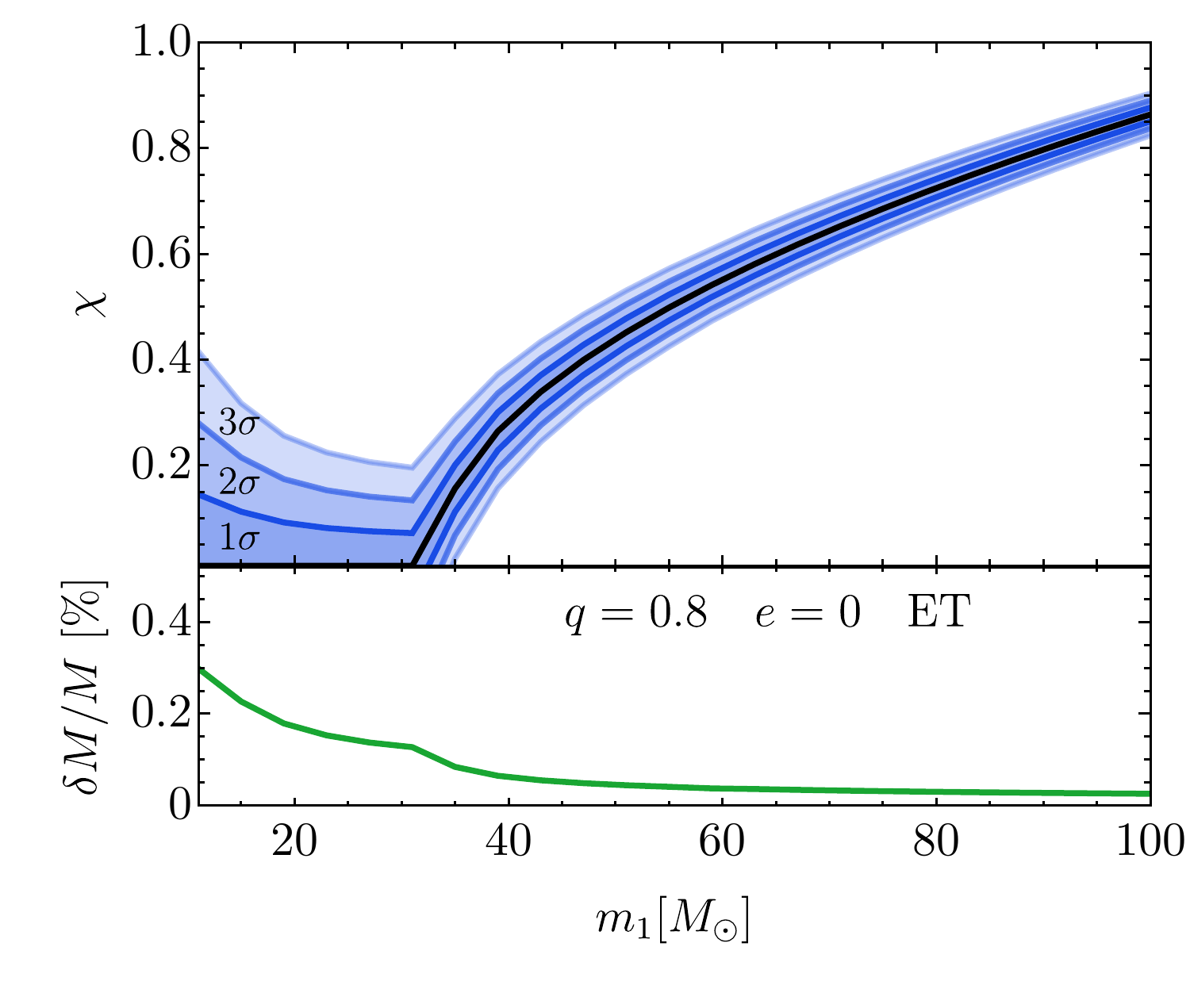}
\includegraphics[width=0.32\textwidth]{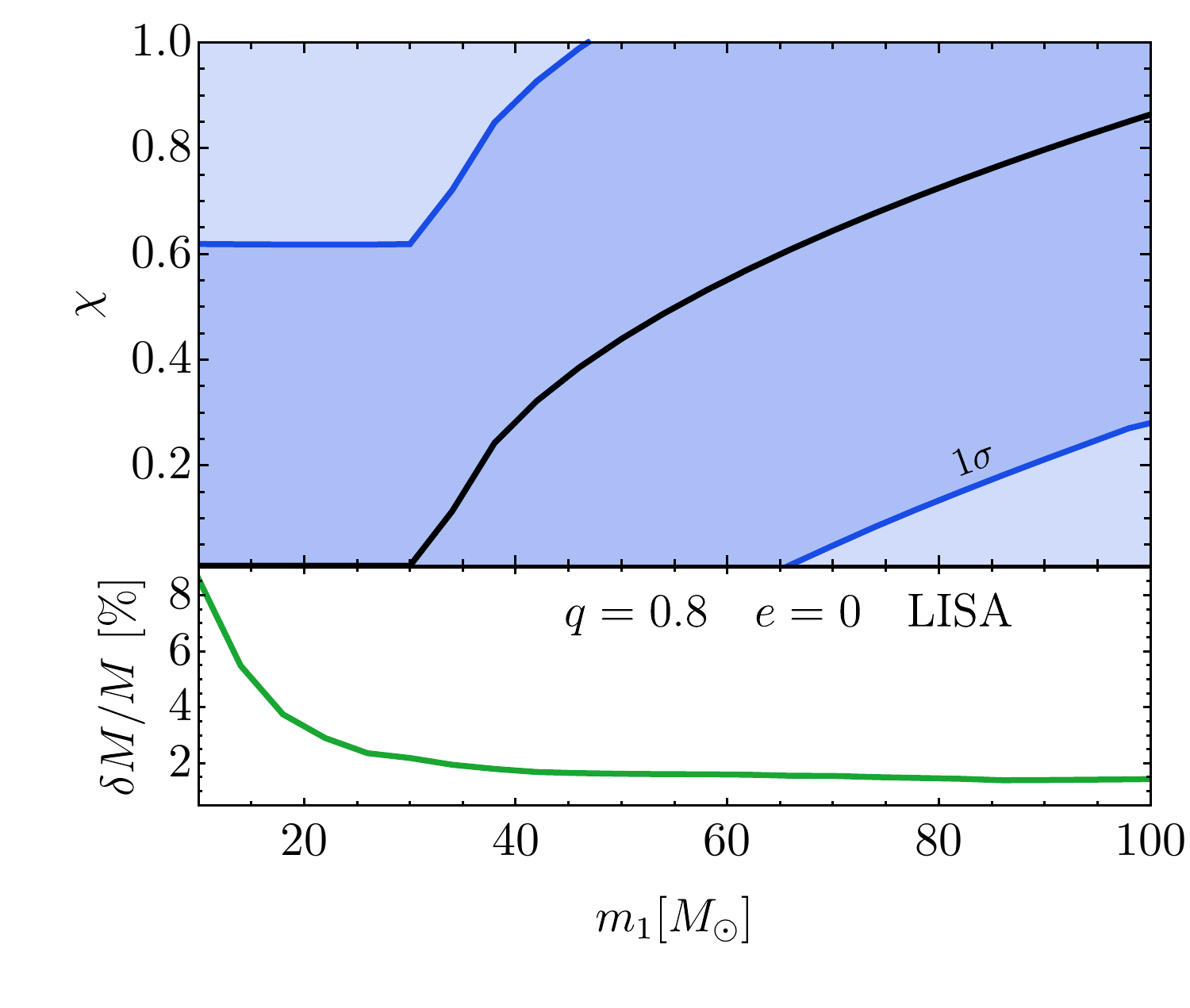}
\caption{Constraints (at 1, 2, and 3$\sigma$) around the PBH mass-spin correlation of Eq.~\eqref{spin-mass-relation-fit} with $z_\text{\tiny \rm cut-off} = 23$. For illustration, we consider circular binaries with $q=0.8$ at $d_\text{\tiny L}=100{\rm Mpc}$.
\textbf{Left:} Ad.~LIGO. 
\textbf{Center:} ET.
\textbf{Right:} LISA.
}\label{fig_spinmass}
\end{figure*}

\begin{figure*}[th]
\centering
\includegraphics[width=0.32\textwidth]{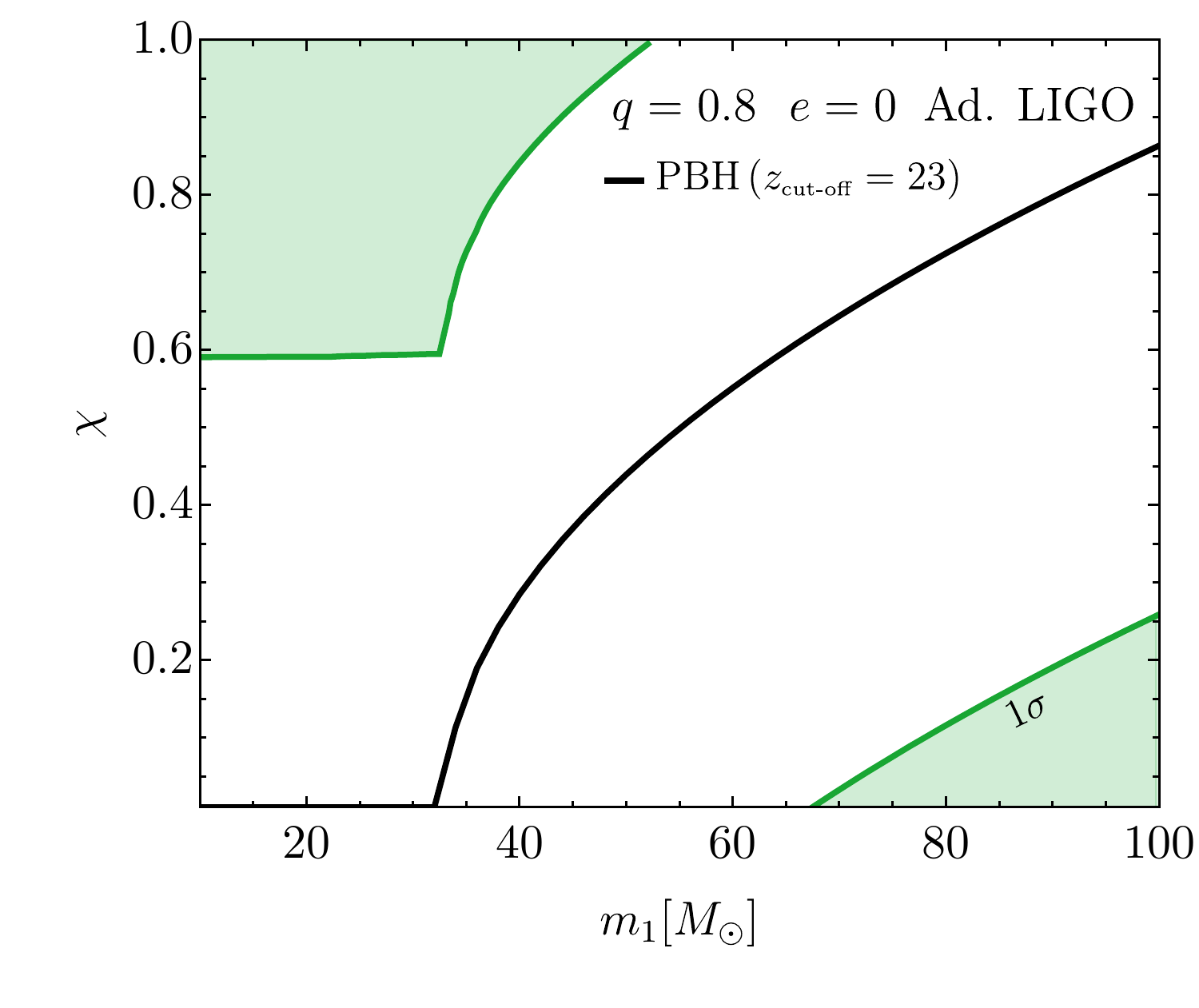}
\includegraphics[width=0.32\textwidth]{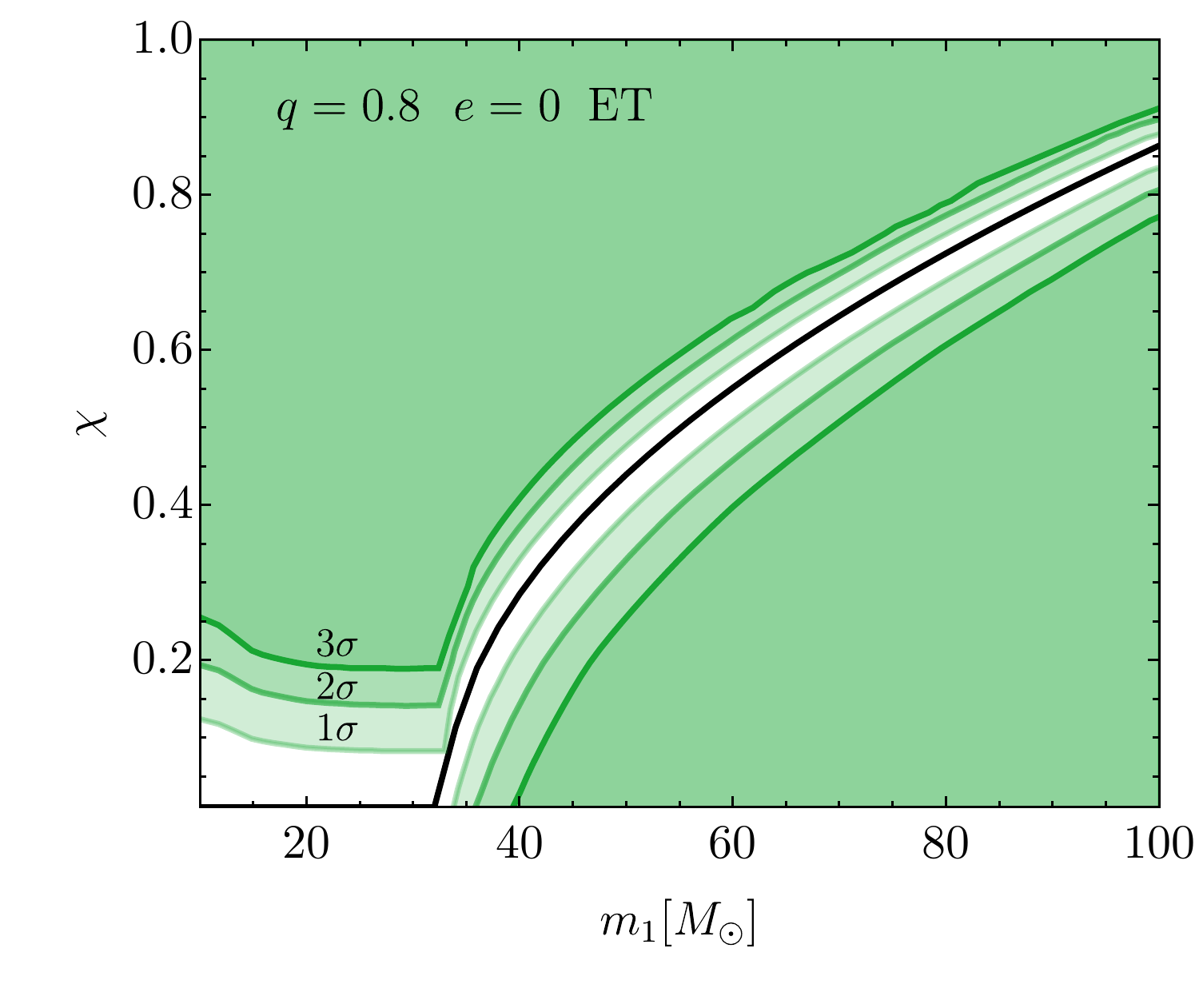}
\includegraphics[width=0.32\textwidth]{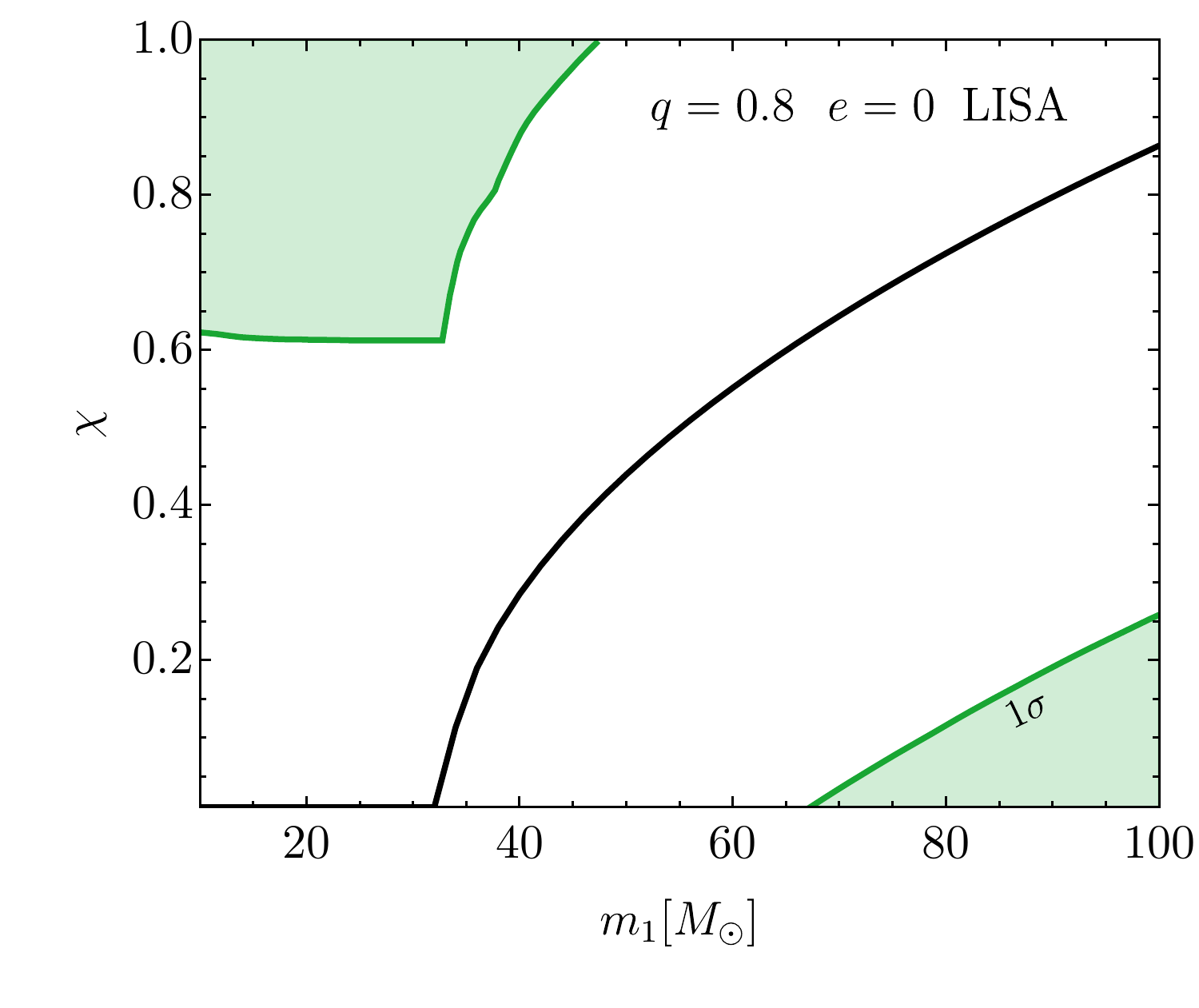}
\caption{Same as Fig.~\ref{fig_spinmass}, but we now show in green the region in the $(m_1,\,\chi)$ parameter space in which we can conclude at 3$\sigma$ C.L. that the event is {\em incompatible} with the primordial scenario with $z_\text{\tiny\rm cut-off} =23$ (the predicted correlation is shown with a black line).
\textbf{Left:} Ad.~LIGO. 
\textbf{Center:} ET.
\textbf{Right:} LISA.
}\label{fig_spinmass2}
\end{figure*}

\subsection{Spin measurement accuracy}

In the standard PBH formation scenario, binaries composed of individual PBHs lighter than $m_\text{\tiny PBH}\simeq 10 M_\odot$ retain small spins, as accretion is always ineffective in spinning up individual components (see Sec.~\ref{sec:th_mspins}). Therefore, measuring a nonzero spin for a sub-10$M_\odot$ object would be in tension with a primordial origin (unless we allow for other PBH formation scenarios). 
At larger masses, the prediction for the spins of primordial binaries becomes uncertain. In particular, binary component spins may still remain negligible up to masses above $m_\text{\tiny PBH} \gtrsim 10^2 M_\odot$, provided accretion is inefficient (i.e., with the accretion hyperparameter $z_\co \approx 30$, see Sec.~\ref{sec:th_mspins}).
Therefore, for completeness, we also report whether the spin measurement accuracy is enough to exclude negligible spins 
in the range of masses $[10,10^2]M_\odot$.
In Sec.~\ref{subsecmassspin} we will address the case of more efficient accretion, and tests of the resulting mass-spin correlation.

In Fig.~\ref{fig_light-spin} (left and center panels), we show the parameter space in which Ad.~LIGO and ET can confidently exclude negligible spins: i.e. we impose $\chi - 3 \sigma_\chi > 0$, so that we can rule out the primordial origin of an event. We place the source at a distance $d_\text{\tiny L}=500$\,Mpc. The performance of the detectors would, of course, improve for closer sources.  Under the assumption of aligned spins, in the limit $q=1$ it is not possible to make independent measurements of the individual component spins. This is because, when setting $q=1$, our waveform model described in App.~\ref{appwfmodel} is completely determined by $\chi_{s} \equiv (1/2) (\chi_{1}+\chi_{2})$, and the derivative with respect to the antisymmetric spin $\chi_{a} \equiv (1/2)(\chi_{1}-\chi_{2})$ in the Fisher information matrix vanishes identically.  Therefore, in this limit, the results of the Fisher analysis only provide the uncertainty on $\chi_{s}$, and there is a complete degeneracy between $\chi_{1}$ and $\chi_{2}$.

In the Ad.~LIGO case (left panel), the primary spin can be distinguished from zero for fairly asymmetric sources ($q\lesssim 0.5$) and large primary spin $\chi_1 \gtrsim 0.3$. On the other hand, the secondary spin is never distinguishable from zero within the 3$\sigma$ confidence limit (C.L.).
We can also explain the nearly horizontal behavior of the bound in the $(m_1,\,q)$ plane. While for a binary located at a fixed distance, larger BBH masses imply larger SNR, they also lead to a smaller number of cycles in the detector band. The two effects compensate each other, giving rise to a comparable spin measurement accuracy in the mass range $m_\text{\tiny PBH} \subset [10,10^2] M_\odot$.

In the ET case (central panel), when $m_1 \gtrsim 10 M_\odot$ the larger SNR reduces the error on both spins. We can now rule out negligible primary or secondary spins if the mass ratio is $q\lesssim 0.8$ for $\chi_{1,2}\gtrsim 0.6$.  When instead $m_1\lesssim 10M_\odot$, a primary spin of magnitude $\chi_1 =[0.3,0.6,0.9]$ can be constrained away from zero if $q\lesssim[0.35,0.6,0.75]$, and the secondary spin is only resolved if $\chi_2\gtrsim 0.9$ and $q\lesssim 0.6$.

LISA (right panel) has a smaller reach in this mass range, so we report results for binaries located at a distance $d_\text{\tiny L} =100$Mpc. Due to the small SNR, it is not possible to place bounds on the individual spins for primary masses below ${\cal O}(10)M_\odot$. For heavier masses, we can only constrained away from zero large primary spins, and only as long as the mass ratio $q\lesssim 0.4$.

\subsection{Testing the predicted mass-spin correlations}\label{subsecmassspin}

As discussed in Sec.~\ref{sec:th_mspins}, accretion effects would imprint a characteristic correlation between masses and spins of PBHs in binaries. If the modelling of PBH accretion is accurate enough, the mass-spin relation can (at least in principle) be compared with a single event to test its consistency with a primordial origin. 
In this section we forecast the accuracy with which current and future experiments could measure both masses and spins in the range above $m_1 \gtrsim 10 M_\odot $, where accretion effects may become relevant. For concreteness, we will fix the hyperparameter $z_\co = 23$, motivated by recent comparisons between the PBH scenario and current data~\cite{Wong:2020yig,DeLuca:2021wjr,Franciolini:2021tla}, even though different values are still possible. This should only be regarded as illustrative. A more precise determination of $z_\co$ is necessary before the test proposed here can be applied to GW events. 

In Fig.~\ref{fig_spinmass} we show the accuracy with which various GW detectors could constrain the mass-spin correlation. We assume that we are observing a binary with $q=0.8$, and that the mass-spin correlation is consistent with the predictions of the primordial scenario with $z_\co=23$. The top (bottom) panel shows measurement errors for the individual spins (total mass).
For both Ad.~LIGO and LISA, the SNR for a source located at $d\text{\tiny L} = 100$Mpc is too low to allow for a significant measurement of the individual PBH spins, even though the measurement accuracy on the total mass is rather high. On the contrary, ET will be able to measure the total mass with subpercent accuracy and also to constrain the spin of both individual components, as long as $m_1\gtrsim 40 M_\odot$.

In Fig.~\ref{fig_spinmass2} we identify the parameter space in the mass-spin plane that is {\em incompatible} with the PBH prediction within the GW measurement errors. We inject GW signals in the entire $(m,\,\chi)$ plane, and determine which region can be deemed incompatible with the primordial hypothesis at $3\sigma$. In agreement with the qualitative results in Fig.~\ref{fig_spinmass}, we find that both Ad.~LIGO and LISA will not be able to test the primordial hypothesis on a single-event basis, while ET can place good constraints in most of the parameter space. 
We conclude that 3G detectors will have large enough SNR to test the primordial hypothesis based on the mass-spin relation, as long as systematic uncertainties in the accretion model are small enough by the time the detectors are taking data.

\section{A case study: the GWTC-3 catalog as observed now and by 3G detectors} \label{sec:nGWe}

In this section we apply the algorithm to assess the primordial nature of individual GW sources developed above, and summarized in Fig.~\ref{fig:flowchart}, to the events reported in the GWTC-3 catalog~\cite{LIGOScientific:2021djp,LIGOScientific:2021psn}. In Sec.~\ref{sec:nGWe_AdLIGO} we ask whether we can draw any conclusion from the observed properties of the GWTC-3 events. Then, in Sec.~\ref{sec:nGWe_3G}, we extrapolate current observations to the estimated measurement accuracy achievable with 3G detectors to understand if any of the current may be confidently classified as primordial (or not) in the near future.

\begin{figure*}[t!]
\centering
\includegraphics[width=0.459\textwidth]{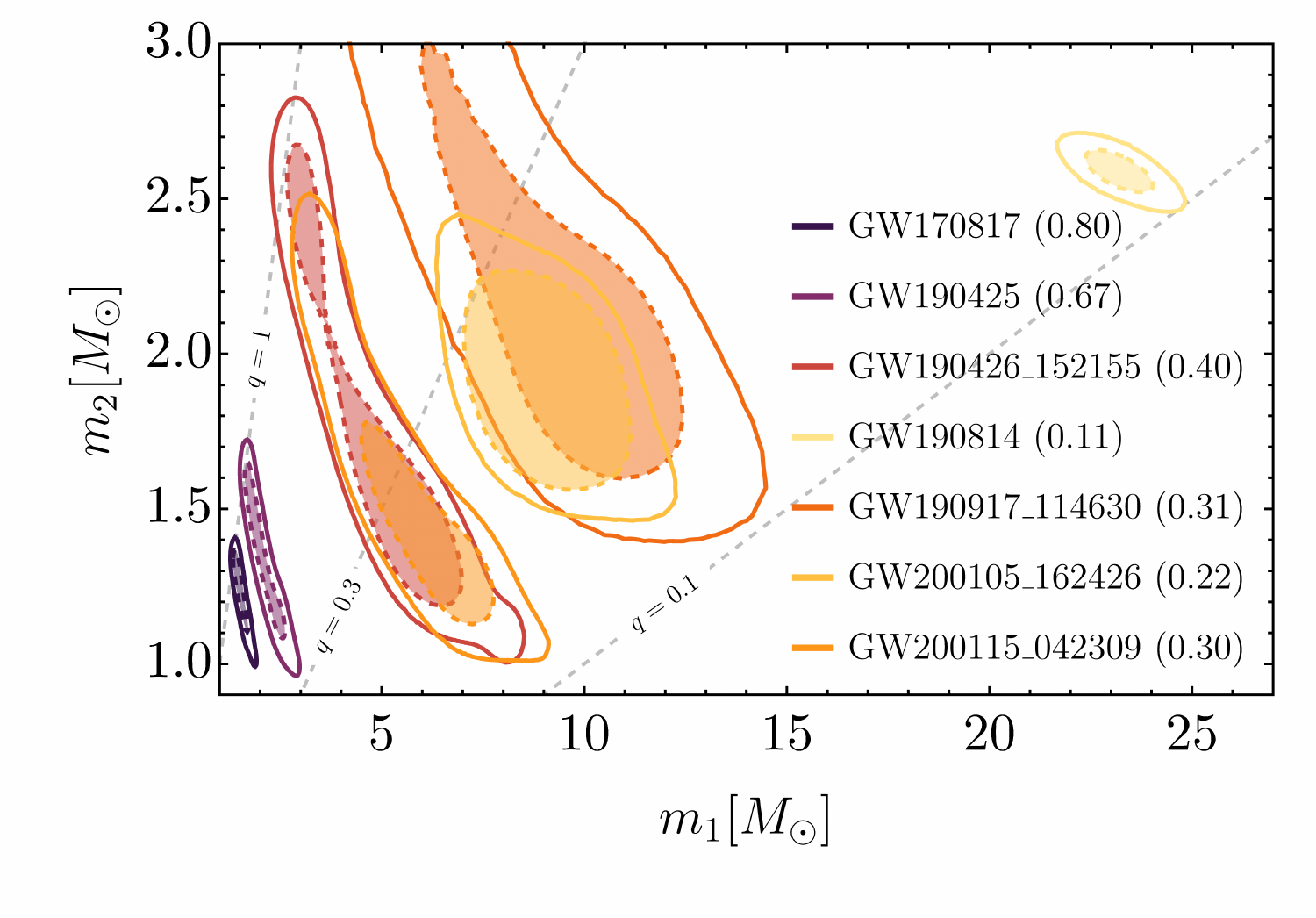}
\includegraphics[width=0.520\textwidth]{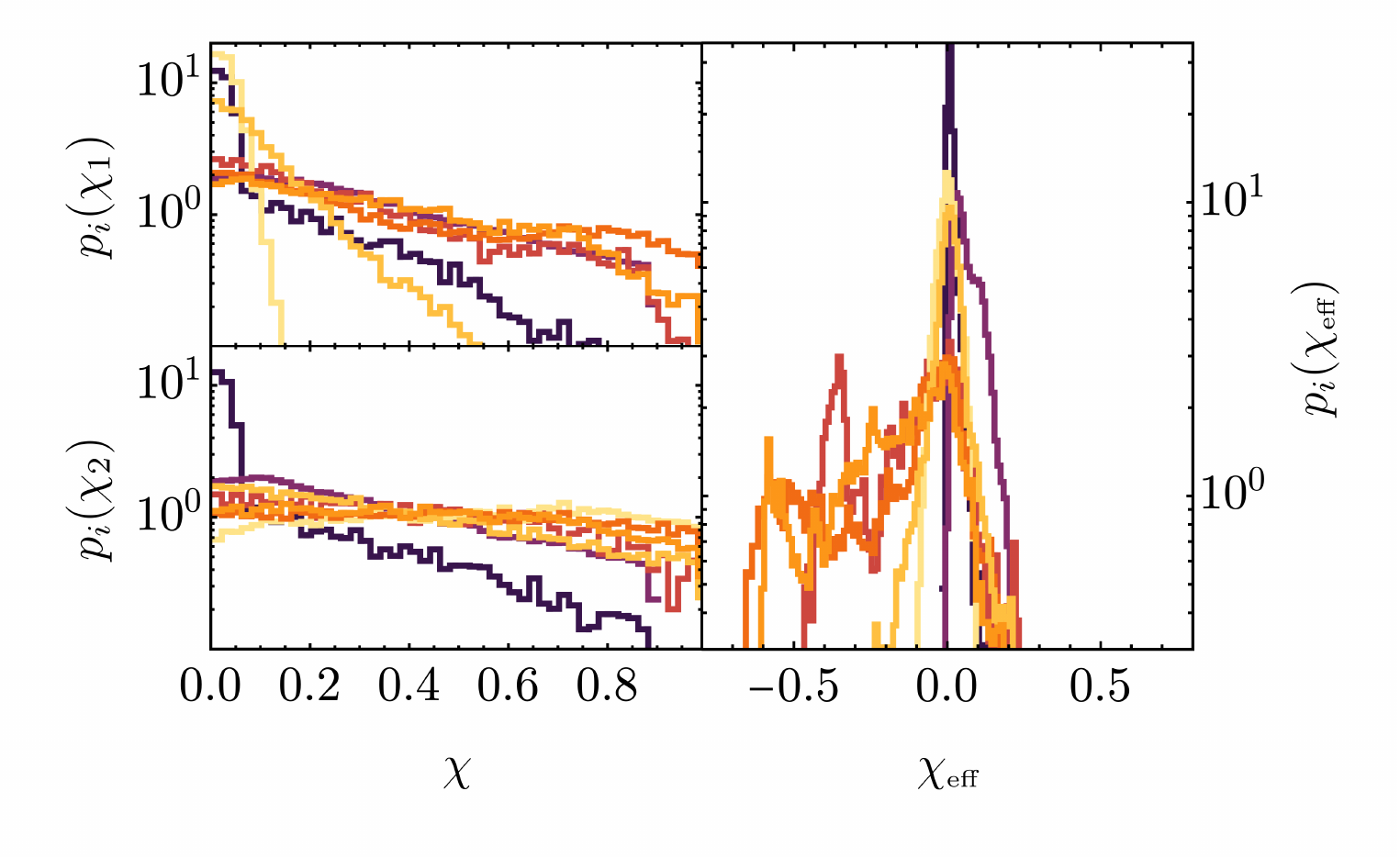}
\caption{  
Summary plot showing all 7 events reported in the GWTC-3 catalog having ${\rm FAR}<1/{\rm yr}$ and at least one of the masses below $3 M_\odot$~\cite{LIGOScientific:2021djp,LIGOScientific:2021psn}. 
\textbf{Left:} Posterior distributions in the $(m_1,m_2)$ plane. The shaded region corresponds to $50$\% C.L., while the outer contour to  $90$\% C.L. The posterior shapes roughly follow lines of constant chirp mass ${\cal M}$, as that is the best measured parameter from the waveform. To facilitate the discussion in the text, we report the mean value of mass ratio $q$ within brackets on the side of each event's name.
\textbf{Right:} Posterior distributions for both the individual spins $\chi_i$ and the effective inspiral spin parameter $\chi_\text{\tiny \rm eff}$, following the same color code as in the left panel.
}\label{fig_GWTC-3_2}
\end{figure*}

\begin{figure*}[t!]
\centering
\includegraphics[width=0.9\textwidth]{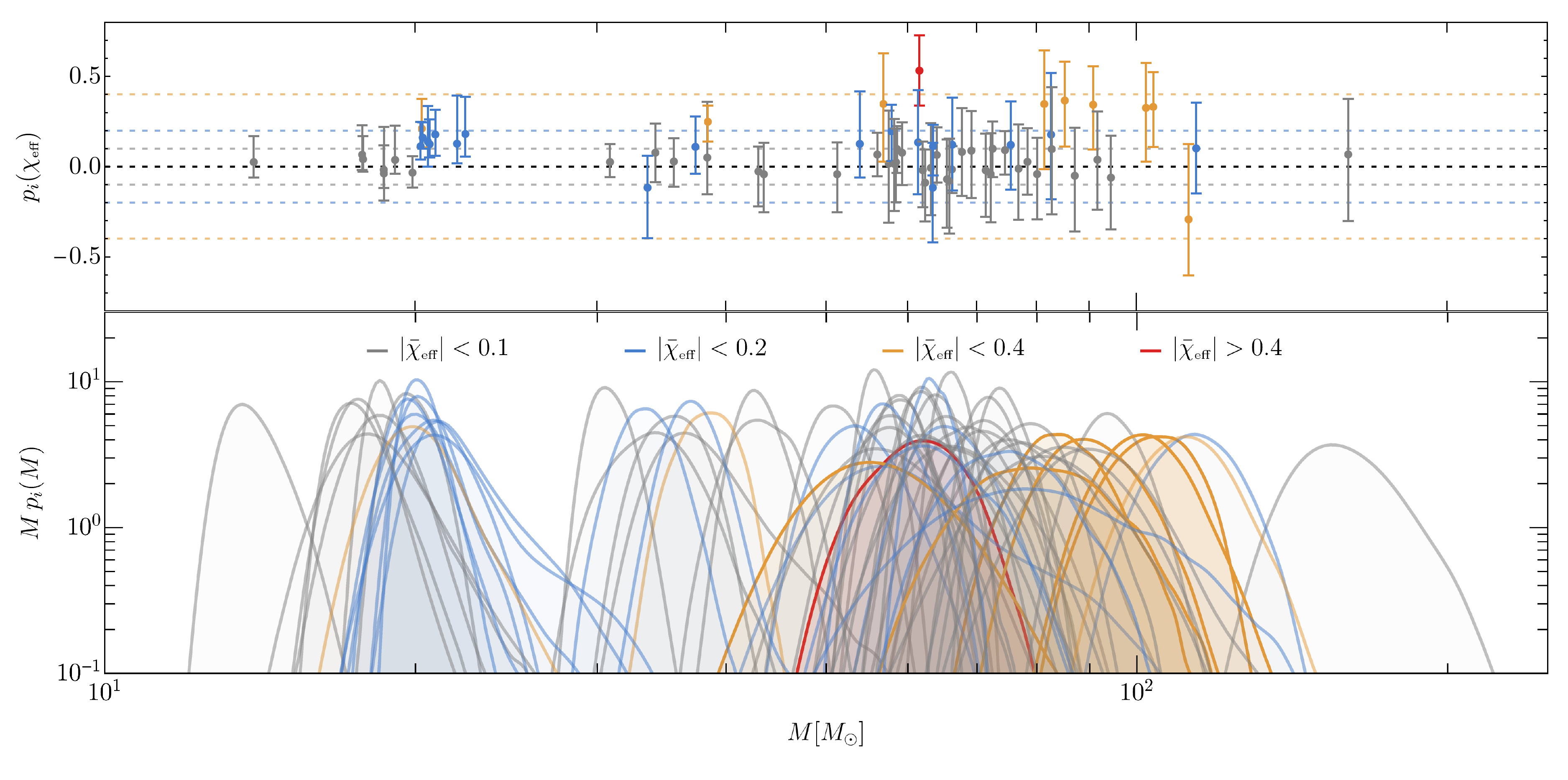}
\caption{  
Summary plot showing all 69 events reported in the GWTC-3 catalog having ${\rm FAR}<1/{\rm yr}$ and both masses $\gtrsim 3 M_\odot$ which are confidently regarded as BBH events~\cite{LIGOScientific:2021djp,LIGOScientific:2021psn}. We show the posterior distribution for the total mass of the binaries (bottom panel), while the color code indicates the mean effective spin parameter. The top panel shows the 90\% C.L. for $\chi_\text{\tiny \rm eff}$ of each event, following the same color code (the dashed horizontal lines bracket each color group). 
}\label{fig_GWTC-3}
\end{figure*}

\subsection{GWTC-3 events}\label{sec:nGWe_AdLIGO}

As the current GW detection horizon is within $z\lesssim 1$, no indication of the primordial nature of the single events can be drawn from current redshift observations.

Additionally, in the GWTC-3 LVK catalog, the eccentricity of the binaries was not measured, as the waveform models used for this analysis work under the assumption of zero eccentricity~\cite{LIGOScientific:2021djp}.
However, a reanalysis of the events from the O1/O2/O3a runs~\cite{Romero-Shaw:2021ual} suggested that GW190521 and GW190620 may present hints of a nonzero eccentricity  (see also~\cite{10.1093/mnras/stz2996,OShea:2021ugg,Romero-Shaw:2020thy,Gayathri:2020coq,Gamba:2021gap}). Most of these analyses use waveform models that neglect higher harmonics and spin precession. Since both of these effects are known to be important for an unbiased estimation of the parameters of the binary~\cite{Romero-Shaw:2020thy,Hoy:2021dqg}, a nonzero eccentricity measurement may still be driven by the inaccuracy of the waveform models. Even if these events are confirmed as having nonzero eccentricity, this would only exclude the primordial origin of two events. In summary, the large majority of the events detected so far has an eccentricity compatible with zero, and therefore this discriminant of their PBH nature is still inconclusive.

As for the tidal deformability, the only LVK event having tidal deformability signatures is GW170817, whose posterior is anyway compatible with $\Lambda \approx 0$.
Had the electromagnetic counterpart~\cite{LIGOScientific:2017ync} of this event not been observed, it would have been impossible to confidently rule out the possibility that GW170817 may be a PBH binary, rather than a NS binary. 

Short of constraints coming from redshift, eccentricity, and tidal deformabilities, we are left with the masses and spins to test whether the mergers detected so far are of primordial origin.
No events with masses below the threshold of $\approx 1 M_\odot$ have been confidently detected so far, implying that no smoking-gun detection based on light PBH binaries is available~~\cite{LIGOScientific:2019kan,Nitz:2021mzz,Nitz:2021vqh,LIGOScientific:2021job}. 

In Figs.~\ref{fig_GWTC-3_2} and \ref{fig_GWTC-3} we show summary plots of the events reported in the GWTC-3 catalog having false-alarm-rate~(FAR) above the threshold of $1/{\rm yr}$. We divide the events in the catalog following the analysis in Ref.~\cite{LIGOScientific:2021psn}. The first class includes events having at least one object with mass below $\approx 3 M_\odot$ (potentially consistent with binaries comprising NSs). The second class includes events where both component masses are above $\approx 3 M_\odot$. 

Let us first focus on the distribution of $\chi_\text{\tiny \rm eff}$ for the first class.
In Fig.~\ref{fig_GWTC-3_2} (right panel) we clearly see that the distribution of $\chi_\text{\tiny \rm eff}$ is mostly peaked around zero. 
The legend of the left panel of Fig.~\ref{fig_GWTC-3_2} shows their inferred mass ratio (in round parentheses). At smaller mass ratios the primary spin becomes better constrained: for example, the posterior distribution of GW190814 falls almost entirely within $\chi_{\rm eff} \lesssim 0.1$. Given the small mass ratio and the relatively small total mass, the observed smallness of the primary spin of GW190814 would not be in tension with the primordial scenario even in the presence of strong accretion (say, with $z_\co\approx 10$). 

The secondary spin is always mostly unconstrained. A potential exception is GW170817~\cite{LIGOScientific:2017vwq}, for which we can infer $\chi_2<0.5$ at $90\%$~C.L., mainly because the mass ratio $q$ is close to unity ($m_1 = 1.46^{+0.12}_{-0.10} M_\odot$ and $m_2 = 1.27^{+0.09}_{-0.09} M_\odot$). Overall, we conclude that it is not possible to rule out the primordial origin of GWTC-3 in the ``first class'' based only on their mass and spin measurements.
The one obvious exception is GW170817, where the observed electromagnetic counterpart~\cite{LIGOScientific:2017ync}  (not expected from a BBH merger\footnote{A possible association between the BBH event GW190521 and an EM flare was suggested in Ref.~\cite{Graham:2020gwr}, assuming that the merger took place in an AGN disk.}) allows us to confidently identify the event as a binary NS merger.\footnote{We do not address the possibility of mixed astrophysical-PBH mergers (i.e. binaries formed dynamically with a compact object coming from each population) because,
generally, the merger rates produced by dynamical capture are insufficient to explain the observed events under reasonable assumptions on the dark matter overdensity in star clusters~\cite{Tsai:2020hpi,Kritos:2020wcl,Sasaki:2021iuc}. We note, however, that multiple exchange interactions may boost the formation rate for those mixed objects, as suggested in Refs.~\cite{Kritos:2020wcl,Kritos:2021yty}. }

Consider now the second class of GWTC-3 events, those for which both masses are above $\approx 3M_\odot$ (see Fig.~\ref{fig_GWTC-3}).
We can further divide these events in two broad categories: one containing events with mean total mass below $\approx 30 M_\odot$ (corresponding to the first peak in the BBH population distribution identified by the latest LVK population analysis~\cite{LIGOScientific:2021psn}) and the remaining, more massive binaries. 

Within the first category, even though no precise measurement of individual spins was performed so far, we can check whether individual events are consistent with the primordial hypothesis by assessing whether $\chi_\text{\tiny eff}$ can be compatible with $|\chi_\text{\tiny eff}|\lesssim 10^{-2}$, which is the prediction for PBH binaries in this mass range.  
Out of the 14 events in the first class, five 
(GW151226, GW190720\_000836, GW190728\_064510, GW191103\_012549, GW191204\_171526, and GW191216\_213338) have $\chi_\text{\tiny eff}>10^{-2}$ at 99\% C.L., so they are in tension with the standard PBH scenario. 

At larger masses, only three events are found to be spinning (that is, in our context, $|\chi_\text{\tiny eff}|\gtrsim 10^{-2}$ at 99\%~C.L.). 
The fastest spinning one, GW190517\_055101,
has $\chi_\text{\tiny eff} = 0.52_{-0.19}^{+0.19}$, and it could only be compatible with the PBH scenario in the presence of some accretion ($z_\co\lesssim 20$). 
In addition, due to its smaller total mass, GW190412 would be 
compatible with the primordial scenario only if  $z_\co \lesssim  16$. Finally, the large $|\chi_\text{\tiny eff}|$ of GW190519\_153544 cannot be used to provide information on its primordial origin because the event has large mass, which would allow for any values of $z_\co$ within the posterior.

Finally, let us focus on the individual spins. Only two out of the 69 massive binaries in the GWTC-3 catalog have primary BH spins incompatible with zero at more than $99\%$~C.L.: these are the aforementioned GW190517\_055101 and GW191109\_010717. The latter event is ``special'' also because the posterior for the effective spin $\chi_\text{\tiny eff}$ has the largest support at negative values. Given their masses, we can only conclude that these two events would be incompatible with large values of cut-off $z_\co$ (close to $\approx 30$), i.e. with negligible accretion. 

We remark that the qualitative analysis of the GWTC-3 events presented in this section is based on the LVK parameter estimation analysis, which assumes uniformative priors. 
As current detections are still characterized by relatively small SNR values, the interpretation of some events can be sensitive to the choice of priors~\cite{Pankow:2016udj,Vitale:2017cfs,Mandel:2020lhv,Gerosa:2020bjb,Zevin:2020gxf} (see Ref.~\cite{Bhagwat:2020bzh} for an interpretation of the events assuming PBH-motivated priors). 
{
Additionally, as pointed out by Ref.~\cite{Romero-Shaw:2021ual}, there is a correlation between the aligned spins and eccentricity obtained from GW parameter estimation. Therefore, GW data analyzed with waveforms neglecting eccentricity may be affected by systematic errors, so that eccentric systems may be misinterpreted as quasicircular systems with nonzero aligned spin.}

{It was first suggested in Ref.~\cite{Callister:2021fpo} (and then confirmed in Ref.~\cite{LIGOScientific:2021psn}) that a fraction of the GWTC-3 events shows a peculiar correlation between $q$ and $\chi_\text{\tiny eff}$: more asymmetric binaries tend to have larger positive values of $\chi_\text{\tiny eff}$. This property of the population cannot conclusively determine the PBH origin of individual events, but the observed correlation would be in contrast with the primordial scenario, which predicts a wider distribution of $\chi_\text{\tiny eff}$ for large total masses, and small values of $|\chi_\text{\tiny eff}|$ for small $q$ at fixed primary mass. As pointed out in Ref.~\cite{Franciolini:2022iaa}, GWTC-3 data also support the hypothesis that a fraction of events may be characterized by a mass-spin correlation which closely resembles the one expected in the PBH scenario discussed here. However, the PBH origin of the events is still indistinguishable from astrophysical formation in the dynamical channel with current statistics.}

\subsection{GWTC-3-like events as observed by 3G detectors} \label{sec:nGWe_3G}

It is interesting to test whether future experiments would be able to provide enough information to exclude or confirm the primordial origin of some of the GWTC-3 events. We can consider explicit examples and use their inferred parameters in a Fisher matrix analysis to forecast how accurately ET could measure the two individual spins (we neglect $z$, $e$, and $\Lambda$ since, as discussed above, the measured valued for these quantities in the GWTC-3 catalog are uninteresting for ruling out PBHs.)

We consider a representative candidate from each of the various groups of events, including low-mass events, events belonging to the two peaks in the mass distributionidentified in GWTC-3~\cite{LIGOScientific:2021psn}, and upper mass gap events.

\noindent {\bf \em Low-mass events (e.g. GW190814).} This event is regarded as a potential outlier of both the astrophysical BH and NS populations~\cite{LIGOScientific:2021psn}. A similar event detected by ET would have $\mathrm{SNR} \approx 820$, so we would measure the mass parameters with percent precision and the primary spin with $\sigma_{\chi_1} \approx 0.04$. This implies that, given the current median value of $\chi_1 \approx 0.034$, one would not be able to rule out the primordial origin for the primary component of the binary.  Because of the small mass ratio, the secondary spin will be poorly measured, with $\sigma_{\chi_2} \approx 0.46$.  Therefore, it would not be possible to rule out the possibility that the second object may be either a BH formed from stellar collapse, or a second-generation BH resulting from a previous binary NS merger. Similar conclusions apply to GW190917\_114630, despite the relatively smaller difference between the individual masses.
    
\noindent {\bf \em First peak in the mass distribution (e.g. GW191204\_171526).} This is a representative event for the first peak in the BBH mass distribution identified by the LVK population analysis~\cite{LIGOScientific:2021psn}, with individual masses $m_1= 11.9 ^{+3.3}_{-1.8} M_\odot$ and $m_2= 8.2 ^{+1.4}_{-1.6} M_\odot$. As previously discussed, PBHs with such masses are predicted to retain small spins in the standard scenario. A detection of such an event by ET would have $\mathrm{SNR} \approx 455$ and subpercent precision in measuring mass parameters. The relative errors on the spins would be around $\sigma_{\chi_1}\approx 0.16$ and $\sigma_{\chi_2}\approx 0.25$, thus allowing us to rule out the primordial origin at $3\sigma$ for similar events with spins larger than $\chi_1\gtrsim 0.5$ and $\chi_2\gtrsim 0.75$.  Ruling out of the primordial origin for events in this region of the parameter space would not be possible with Ad.~LIGO.  Given the large number of events falling in this mass range, which is currently expected to dominate the BBH population, these findings confirm the importance of 3G detectors for identifying PBH binaries.
    
\noindent {\bf \em Second peak in the mass distribution (e.g. GW200129\_ 065458).} Similarly to the previous case, we pick this event as representative of the second peak in the BBH mass distribution.  In ET this event would have $\mathrm{SNR}\approx 860$, but owing to the larger masses, the spin measurement accuracy is somewhat reduced, with $\sigma_{\chi_1}\approx 0.24$ and $\sigma_{\chi_2}\approx 0.33$. The reduced spin measurement accuracy and the uncertainties in the accretion model make it challenging to probe the PBH nature of events in this mass range even for ET, unless less distant events are observed (as assumed in Figs.~\ref{fig_spinmass} and \ref{fig_spinmass2}).

\noindent {\bf \em Upper mass gap events (e.g. GW190521).} Our conclusions on the massive events of the GWTC-3 catalog apply also to GW190521-like objects.

\section{Conclusions and future work}\label{sec:conclusions}

In previous work we carried out a population analysis and asked whether a subpopulations of PBHs can be compatible with the observed catalog of GW events, given current uncertainties in astrophysical formation scenarios~\cite{Franciolini:2021tla}.
In this work we asked a complementary question: what are the key observables which may allow us to assess the primordial origin of BHs at the {\em single-event} level? 

We have taken a conservative point of view: we first identified the crucial combinations of binary parameters that would allow us to draw conclusions on the primordial origin of the events, and then we quantified how accurately present and planned experiments (including Ad.~LIGO, ET, and LISA) could measure those key observables. 

Our findings can be summarized as follows (see Fig.~\ref{fig:flowchart}):
%

\noindent
{\bf \em Large-redshift observations.} A smoking-gun signal of PBHs would be the detection of a GW signal at redshift larger than $z\gtrsim 30$, as long as we can confidently set lower bounds on the source redshift. We have estimated uncertainties in the source redshift measurements as a function of the PBH mass, essentially confirming the findings of Ref.~\cite{Ng:2021sqn}.  For events at $z>z_\co$ (which might be smaller than $z\approx30$), another possibility is to use the characteristic $z-\chi$ correlation predicted in the standard PBH formation scenario.

\noindent
{\bf \em Eccentricity.} In the standard formation scenario considered here, primordial binaries do not retain any relevant eccentricity at observable redshifts. We investigated the lower bounds on $e$ above which the primordial nature of the mergers may be excluded. This signature will be useful to rule out the primordial origin of an event only when it retains some significant eccentricity, and even then we should allow for the possibility of an astrophysical origin in dynamical formation scenarios.

\noindent
{\bf \em Subsolar-mass events and tidal deformability.}
Even subsolar, zero-eccentricity events may not be PBHs if their tidal deformability is nonzero. Future 3G detectors will be able to measure the mass of BHs in binaries with subpercent accuracy. This is often sufficient to confidently claim the primordial nature of the compact object. Possible alternatives, such as white dwarfs and mini-boson stars, can be distinguished from PBHs by using the characteristic pre-ISCO cut-off in the GW signal caused by tidal disruption.

\noindent
{\bf \em Masses and spins.} We have quantified the mass and spin measurement accuracy achievable by 3G detectors in the solar mass range, showing that ET will be able to test the mass-spin correlation predicted in the standard PBH formation scenario (see Figs.~\ref{fig_spinmass} and \ref{fig_spinmass2}). This test could only be performed on single events in the future if systematic theoretical uncertainties on PBH accretion are significantly reduced.


As a proof of principle, we have applied this strategy to the events in the recently released GWTC-3 catalog. Due to the relatively low SNR of the binary mergers observed in current detectors, there are very few events that can be deemed incompatible with a primordial origin, and there are no smoking-gun signatures of PBHs in the current catalog.
We then quantified how 3G detectors will ameliorate the current state of affairs by estimating what could be learned at higher SNRs from the same GW events already present in the GWTC-3 catalog. This is, of course, a very conservative scenario, because the most informative events are likely to be just those that are not observable with current interferometers.

This work can be extended in various directions. It will be interesting to study what can be learned from LISA observations of massive BHs, for which PBH formation predictions are still under investigation, mainly because it is difficult to quantify the effect of accretion.  The simple inspiral Fisher matrix analysis we performed can and should be improved through a full Bayesian parameter estimation framework of the complete inspiral-merger-ringdown waveforms (see e.g. \cite{Vitale:2016icu,Moore:2019vjj,Favata:2021vhw,OShea:2021ugg}). This is especially relevant for low-SNR events. In addition, it is possible that the correlations between certain observable parameters (such as chirp mass and eccentricity) may differ between primordial and astrophysical formation models (see e.g.~\cite{Breivik:2016ddj}). These correlations may either enhance or reduce the constraining power of future detectors.  Finally, it will be interesting to understand how to best optimize the 3G detector network and to investigate the potential of multiband events~\cite{Sesana:2016ljz,Wong:2018uwb,Cutler:2019krq,Gerosa:2019dbe,Moore:2019pke,Ewing:2020brd,Buscicchio:2021dph} to better assess the (primordial or astrophysical) nature of the observed merging events.

\begin{table*}[t!]
\footnotesize
\renewcommand{\arraystretch}{1.2}
\caption{Mass-spin relation for PBH binaries: numerical coefficients in the analytical fit described in Eq.~\eqref{fitmassspinrel}.}\label{tabcoeff}
\vspace{0cm}
\begin{tabularx}{2. \columnwidth}{
>{\centering}X>{\centering}X>{\centering}X
>{\centering}X>{\centering}X>{\centering}X
>{\centering}X>{\centering}X>{\centering}X
>{\centering}X>{\centering \arraybackslash}X
}
\midrule
\midrule
$a^0_1$ &  $a^{z,1}_1$ &$a^{z,2}_1$&$a^{z,3}_1$&
$a^{q,1}_1$ &$a^{q,2}_1$& $a^{q,3}_1$& $a^{1,1}_1$& $a^{2,1}_1$& $a^{1,2}_1$ &$a^{2,2}_1$
\\
$57.8531$ & $-66.8879$ &  $43.9529$ & $-5.46522$ &
$-56.4905$ & $39.4605$ & $-10.5127$ &
$37.4532$ &  $-17.5600$ & $-17.5899$ & $7.32670$ 
\\
\midrule 
$b^0_1$ & $b^{z,1}_1$ &$b^{z,2}_1$&$b^{z,3}_1$&
$b^{q,1}_1$ &$b^{q,2}_1$& $b^{q,3}_1$& $b^{1,1}_1$& $b^{2,1}_1$& $b^{1,2}_1$ &$b^{2,2}_1$
\\
$2.14680 $&$ -3.65483 $&$ 1.23732 $&$ -0.185276$&$ -1.59262$&$ -1.33445$&$ 0.940219$&$ 2.48367$&$ 0.0136971$&$ -0.313974$&$ -0.218091$
\\ 
\midrule
$c^0_1$ & $c^{z,1}_1$ &$c^{z,2}_1$&$c^{z,3}_1$&
$c^{q,1}_1$ &$c^{q,2}_1$& $c^{q,3}_1$& $c^{1,1}_1$& $c^{2,1}_1$& $c^{1,2}_1$ &$c^{2,2}_1$
\\
$0.441418$ & $-0.738179$ & $0.834177$ & $-0.175491$ & $-0.231674$ & $2.12451$ & $-0.787300$ & $-0.0461876$ & $-1.20687$ & $-0.234563$ & $ 0.477210$
\\ 
\midrule
$a^0_2$ & $a^{z,1}_2$ &$a^{z,2}_2$&$a^{z,3}_2$&
$a^{q,1}_2$ &$a^{q,2}_2$& $a^{q,3}_2$& $a^{1,1}_2$& $a^{2,1}_2$& $a^{1,2}_2$ &$a^{2,2}_2$
\\
$44.3220 $&$-72.7617$&$ 50.9837$&$ -8.27027$&$ 19.8378$&$ -33.8142$&$18.3605 $&$ -6.80676 $&$ 1.95003 $&$ 0.0581762$&$ -0.957243$
\\
\midrule
$b^0_2$ & $b^{z,1}_2$ &$b^{z,2}_2$&$b^{z,3}_2$&
$b^{q,1}_2$ &$b^{q,2}_2$& $b^{q,3}_2$& $b^{1,1}_2$& $b^{2,1}_2$& $b^{1,2}_2$ &$b^{2,2}_2$
\\
$3.65282$&$-6.94442$&$ 3.55860$&$-0.630911$&$ -0.474109$&$-0.199862$&$ 0.0523957$&$ 0.737077$&$ 0.0855668$&$ -0.178022$&$ -0.0212303$
\\ 
\midrule
$c^0_2$ & $c^{z,1}_2$ &$c^{z,2}_2$&$c^{z,3}_2$&
$c^{q,1}_2$ &$c^{q,2}_2$& $c^{q,3}_2$& $c^{1,1}_2$& $c^{2,1}_2$& $c^{1,2}_2$ &$c^{2,2}_2$
\\
$ -0.189439 $&$ 1.28502$&$ -0.587638$&$0.0864602
$&$ -0.905386 $&$ 1.25085
$&$ -0.346207 $&$ 0.158765$&$ -0.447706$&$ -0.0109165 $&$  0.0945026$
\\
\midrule
\midrule
\end{tabularx}
\label{tabfit}
\end{table*}

\begin{acknowledgments}
We thank V. De Luca, K.~Kritos and C. Pacilio for useful discussions and E. Kovetz for comments on the manuscript.
We acknowledge financial support provided under the European Union's H2020 ERC, Starting Grant agreement no.~DarkGRA--757480, and under the MIUR PRIN and FARE programmes (GW-NEXT, CUP:~B84I20000100001), and support from the Amaldi Research Center funded by the MIUR program ``Dipartimento di Eccellenza" (CUP:~B81I18001170001).
R.C. and E.B. are supported by NSF Grants No. PHY-1912550, AST-2006538, PHY-090003 and PHY-20043, as well as NASA Grants No. 17-ATP17-0225, 19-ATP19-0051 and 20-LPS20-0011.
A.R. is supported by the Swiss National Science Foundation 
(SNSF), project {\sl The nonGaussian Universe and Cosmological Symmetries}, project number: 200020-178787.
\end{acknowledgments}

\appendix

\section{Fits of mass-spin relation for PBH binaries as a result of accretion}\label{appendix_fit_spins}

In this appendix we provide the numerical coefficients specifying the analytical relation between masses and spins for PBH binaries at redshift smaller than $z_\co \in [10,30]$, see Eq.~\eqref{spin-mass-relation-fit}.
These coefficient are reported in Table~\ref{tabcoeff}. \texttt{Mathematica} and \texttt{Python} codes with the relevant tabulated functions are publicly available at the GIT repository linked in~\cite{webpage}.
The analytical fit may be useful when performing Bayesian parameter estimations assuming PBH motivated priors, or for searches in the GW catalog for a PBH motivated mass-spin relation.

\section{Methodology}\label{app:Fisher}

In this appendix we review the methodology adopted to derive the main results contained in the paper. We start by reviewing the waveform model we use, mainly following Ref.~\cite{Favata:2021vhw} and references therein. Then we review the Fisher matrix method and we list our chosen power spectral density curves for the GW experiments discussed in this work, as well as the frequency range used in the integrations.

\subsection{Waveform model}\label{appwfmodel}

We define the GW signal in Fourier space adopting the stationary phase approximation~(SPA). We can write  
\be\label{eq:spawaveform}
\tilde{h}(f) = {\mathcal A} e^{i \Psi}, 
\ee
where
\begin{multline}\label{eq:spawaveform2}
{\mathcal A} = -
\lp\frac{5\pi\eta}{96}\rp^{1/2} \left(\frac{M^2}{D}\right)  (\pi M f)^{-7/6} \\
\times \left[(1+\cos^2(\iota))^2 F_{+}^2 + 4 \cos^2(\iota) F_{\times}^2 \right]^{1/2}.
\end{multline}
Eccentric corrections are only introduced in the phase evolution $\Psi$ via a ``post-circular''~\cite{Yunes:2008tw} low-eccentricity expansion accurate to $O(e_0^2)$, presented below. This waveform is an extension of the one presented in Refs.~\cite{Moore:2016qxz}
with the inclusion of spin effects performed in~\cite{Favata:2021vhw}.
Also, in the previous formula, we introduced the binary inclination angle $\iota$ relative to the line-of-sight, the distance to the detector $D$, the antenna pattern functions $F_{+,\times}$ and the symmetric mass ratio $\eta=m_1 m_2/M^2$. 
The SPA phase can be written as a sum of PN corrections:
\begin{multline}
\label{eq:Psiterms}
\Psi(f) = \phi_c + 2\pi f t_c + \frac{3}{128 \eta v^5} \Big(1 + 
\Delta \Psi^\text{\tiny tidal}_\text{\tiny 6PN}
\\
+\Delta \Psi_\text{\tiny 3.5PN}^\text{\tiny circ.}+ \Delta \Psi_\text{\tiny 4PN}^\text{\tiny spin,circ.} + \Delta \Psi_\text{\tiny 3PN}^\text{\tiny ecc.}  \Big),
\end{multline}
where $t_c$ and $\phi_c$ are the coalescence time and phase, and $v\equiv (\pi M f)^{1/3}$ is the PN orbital velocity parameter.
The tidal deformability terms we include in the waveform, starting at 5PN order, are defined in Eq.~\eqref{tidaldefwaf}.

The standard 3.5PN circular contribution is 
\begin{equation}
    \Delta \Psi_\text{\tiny 3.5PN}^\text{\tiny circ.} = \sum_{n=2}^{7} c_n(\eta) v^n,
\end{equation}
 where the coefficients $c_n(\eta)$ are found in Eq.~(3.18) of~\cite{PhysRevD.80.084043}, and the 2.5PN and 3PN coefficients depend also on $\ln v$.

Spin effects up to 4PN order add a contribution
\begin{multline}
\Delta \Psi_\text{\tiny 4PN}^\text{\tiny spin,circ.} = 4 \beta_{1.5} v^3 -10 \sigma v^4 + v^5 \ln v^3 \left[ \frac{40}{9} \beta_{2.5}
\right. \\ \left.
- \beta_{1.5} \left( \frac{3715}{189} + \frac{220}{9}\eta \right) \right] + {\mathcal P}_6 v^6 + {\mathcal P}_7 v^7 +{\mathcal P}_8 v^8,
\end{multline}
where $\beta_{1.5}$ is the 1.5PN spin-orbit term~\cite{Kidder:1992fr,PhysRevD.48.1860,Kidder:1995zr}
\be
\label{eq:beta15}
\beta_{1.5} = \sum_{i=1,2} \chi_i \kappa_i \left(\frac{113}{12} \frac{m_i^2}{M^2} + \frac{25}{4} \eta \right).
\ee 
The 2PN spin-spin term includes three distinct contributions  $\sigma = \sigma_{\text{\tiny S}_1 \text{\tiny S}_2} + \sigma_\text{\tiny QM}+ \sigma_\text{\tiny self spin}$~\cite{Mikoczi:2005dn}:

\noindent
{\it i)} the standard spin-spin interaction~\cite{Kidder:1992fr,Kidder:1995zr}
\be
\sigma_{\text{\tiny S}_1 \text{\tiny S}_2} = \frac{1}{48} \eta \chi_1 \chi_2 (721 \kappa_1 \kappa_2 - 247 \gamma_{12});
\ee
{\it ii)} the quadrupole-monopole term~\cite{Poisson:1997ha}
\begin{align}
\sigma_\text{\tiny QM} 
&= \frac{5}{2} \sum_{i=1,2} \chi_i^2 \left( \frac{m_i}{M} \right)^2 (3\kappa_i^2 -1);
\end{align}
{\it iii)} the self-spin interaction ~\cite{Mikoczi:2005dn,Gergely:1999pd}
\be
\sigma_\text{\tiny SS-self} = \frac{1}{96} \sum_{i=1,2} \chi_i^2 \left(\frac{m_i}{M} \right)^2 (7-\kappa_i^2).
\ee
In the previous equations $\chi_i$ denotes the dimensionless spin parameter, $\kappa_i = \hat{{\bm s}}_i \cdot \hat{{\bm L}}_N$ is the cosine of the angle between the $i^\text{\tiny th}$ spin direction $\hat{{\bm s}}_i$ and the Newtonian orbital angular momentum direction $\hat {\bm L}_N$,
and $\gamma_{12} = {\hat{\bm s}}_1 \cdot {\hat{\bm s}}_2$. 

The 2.5PN spin-orbit term $\beta_{2.5}$  is ~\cite{PhysRevD.74.104034}
\begin{align}
\label{eq:beta25SO}
\beta^\text{\tiny SO}_{2.5} = \sum_{i=1,2} \chi_i \kappa_i \left[ \frac{m_i^2}{M^2} \left( -\frac{31319}{1008} + \frac{1159}{24}\eta \right)
\right. \nonumber \\ \left.
+ \eta \left( -\frac{809}{84} + \frac{281}{8}\eta  \right) \right] \;,
\end{align}
where BH absorption terms (i.e., tidal heating) were neglected~\cite{Alvi:2001mx}. 
The subsequent 3PN, 3.5PN, and 4PN terms ${\mathcal P}_6$, ${\mathcal P}_7$, and ${\mathcal P}_8$ can be found in Ref.~\cite{Mishra:2016whh}.
This analysis assumes nonprecessing (aligned) spins, and therefore the parameters $\beta_{(\cdots)}$ and $\sigma_{(\cdots)}$ are constant in time and  functions of $\chi_i$. 

Leading-order in eccentricity corrections to the SPA phase were derived up to to 3PN order in Ref.~\cite{Moore:2016qxz}, building upon previous results on eccentric binaries~\cite{PhysRev.131.435,1992MNRAS.254..146J,Gopakumar:1997ng,Arun:2007rg,Arun:2007sg,Arun:2009mc}. Following Ref.~\cite{Favata:2021vhw}, we use the full 3PN expression in our calculations, whose structure is of the form
\begin{align}
\label{eq:Psiecc}
\Delta \Psi_\text{\tiny 3PN}^\text{\tiny ecc.} =
-\frac{2355}{1462} e_0^2 \left( \frac{v_0}{v} \right)^{19/3} \left[
1 + v^2 \left( \frac{299\,076\,223}{81\,976\,608}  \right. \right.
\nonumber \\
\left. \left. + \frac{18\,766\,963}{2\,927\,736}\eta \right)
+ v_0^2 \left( \frac{2833}{1008} - \frac{197}{36} \eta \right) + \cdots + O(v^6)
\right].
\end{align}
Here, $e_0$ is the eccentricity at a reference frequency $f_0$, and $v_0 \equiv (\pi M f_0)^{1/3}$. The choice of $f_0$ is arbitrary, and throughout this paper we set $f_0= 10$\,Hz, following Ref.~\cite{Favata:2021vhw}.

We additionally introduce the effect of cosmological redshift by replacing, in Eq.~\eqref{eq:spawaveform}, the distance $D$ by the luminosity distance $d_\text{\tiny L}$, defined as
\be
d_\text{\tiny L}(z) = \frac{c}{H_0} (1+z) \int_0^z \frac{dz'}{\sqrt{\Omega_M (1+z')^3 + \Omega_{\Lambda}}},
\ee
where $H_0=100 h \, {\rm (km/s)/Mpc}$, $h=0.6790$, $\Omega_M=0.3065$, and $\Omega_{\Lambda}=0.6935$ ~\cite{Planck:2015fie}.
Also, the redshift of the GW frequency can be accounted for in Eq.~\eqref{eq:spawaveform} by replacing the total mass with the observer-frame total mass $M \rightarrow M_\text{\tiny obs}=(1+z) M$.
Throughout this work, $M$ refers to the source-frame total mass.

\subsection{Fisher matrix analysis}

The Fisher information matrix is  often used to assess the parameter estimation capabilities of GW detectors (see, for example, Refs.~\cite{PhysRevD.46.5236,PhysRevD.47.2198,Cutler:1994ys,Poisson:1995ef,Berti:2004bd,Ajith:2009fz,Cardoso:2017cfl}, as well as Refs.~\cite{Vallisneri:2007ev,Rodriguez:2013mla} for discussions of the limitations of this approach).

The output $s(t)$ of a general GW interferometer can be written as the sum of the GW signal $h (t,\vec \xi)$ and the stationary detector noise $n(t)$. The posterior distribution for the hyperparameters $\vec \theta$ can be approximated by
\be\label{pos_dist_F}
p (\vec \theta| s) \propto \pi (\vec \theta) e^{-\frac{1}{2}(h (\vec \theta) - s|h (\vec \theta) - s)}
\ee
in terms of the prior distribution $\pi (\vec \theta)$. Here we ahve introduced the inner product
\be\label{innprod}
(g|h) = 2\int_{f_\text{\tiny min}}^{f_\text{\tiny max}} \d f \frac{h (f) g^* (f) + h^*(f) g(f)}{S_n(f)}.
\ee
In Eq.~\eqref{innprod},
 $S_n(f)$ is the detector noise power spectral density, and $f_\text{\tiny min}$ ($f_\text{\tiny max}$) is the characteristic minimum (maximum) frequency of integration. The frequency band of interest for each GW experiment will be discussed in Sec.~\ref{freqrange} below.

Following the principle of the maximum-likelihood estimator, the central values of the hyperparameters are approximated by the point $\vec\theta \equiv \vec \theta_\text{\tiny p}$ where the likelihood peaks. In the limit of large signal-to-noise ratio~(SNR$\equiv\sqrt{(h|h)}$), one can perform a Taylor expansion of Eq.~\eqref{pos_dist_F} and get
\be
p (\vec \theta| s) \propto \pi (\vec \theta) e^{-\frac{1}{2}\Gamma_{ab} \Delta \theta^a \Delta \theta^b},
\ee
where $\Delta \vec \theta = \vec \theta_\text{\tiny p} - \vec \theta$ and we have introduced the Fisher matrix 
\be
\Gamma_{ab} = \lp \frac{\partial h}{\partial \theta^a} \bigg | \frac{\partial h}{\partial \theta^b} \rp_{\vec \theta = \vec \theta_\text{\tiny p}}\,.
\ee
The errors on the hyperparameters are, therefore, given by $\sigma_a = \sqrt{\Sigma^{aa}}$, where $\Sigma^{ab} = \lp \Gamma^{-1}\rp^{ab}$ is the covariance matrix.

Our parameter set is the following:
\begin{equation}
\label{eq:theta}
\theta_a = (t_c, \phi_c, \ln M, \ln \eta, \chi_1, \chi_2, \ln e_0),
\end{equation}
with the addition of the redshift $z$ in Sec.~\ref{error_redshift} and of the tidal deformability $\Lambda$ in Sec.~\ref{error_tidal}.
Following Ref.~\cite{Favata:2021vhw},
we use Gaussian priors on the parameters  $\phi_c \in [-\pi, \pi]$, $\chi_{1,2} \in [-1,1]$, corresponding to
\be
\label{eq:priors}
\delta \phi_c = \pi, \qquad \delta \chi_{1,2} = 1,
\ee
by adding to the diagonal elements of our Fisher matrix terms of the form $\Gamma^0_{aa} = 1/(\delta \theta_a)^2$.

Throughout this work, we always consider sources which are optimally oriented with respect to the detector. This means that orientation-dependent terms in the amplitude take the  value
 \begin{equation}
     (1+\cos^2(\iota))^2 F_+^2 + 4 \cos^2(\iota) F_{\times}^2=4
 \end{equation} 
and the optimal SNR $\rho_\text{\tiny opt}$ can be computed using
 \be
\rho_\text{\tiny opt}^2 = \frac{5}{6 \pi^{4/3}} (1+z)^{5/3} \frac{\eta M^{5/3}}{d_\text{\tiny L}^2(z)} \int_{f_\text{\tiny low}}^{f_\text{\tiny high}} \frac{f^{-7/3}}{S_n(f)} \d f  .
\ee

\begin{figure}[t!]
\centering
\includegraphics[width=.5
\textwidth]{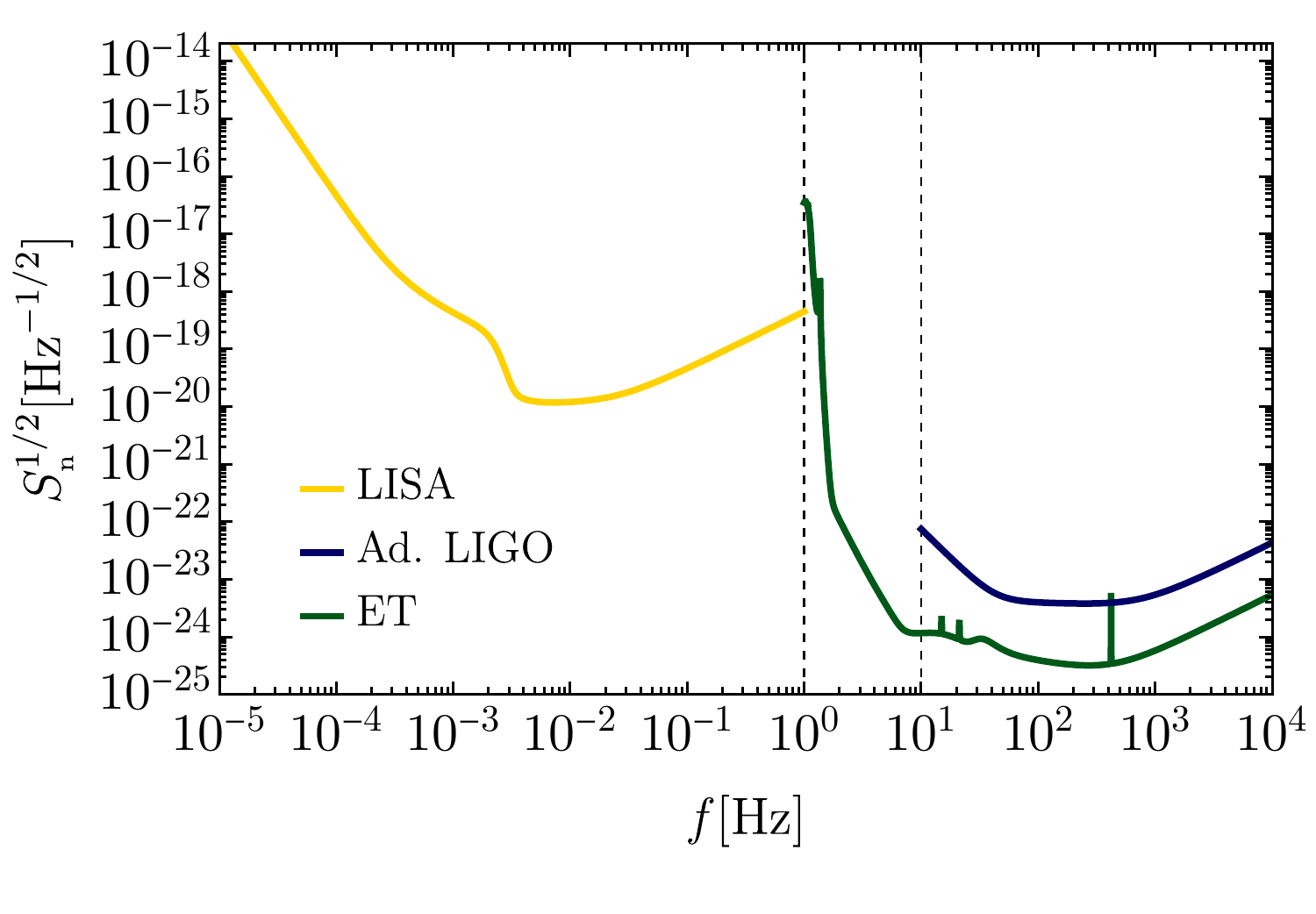}
\caption{ 
Noise power spectral densities for LISA, ET and Ad.~LIGO. The dashed vertical lines indicate the two minimum frequencies for ground-based detectors, i.e. $f_\text{\tiny\rm low}^{\text{\tiny\rm Ad. ET}} = 1{\rm Hz}$ and $f_\text{\tiny\rm low}^{\text{\tiny\rm Ad.~LIGO}} = 10{\rm Hz}$.
}\label{fig_psd}
\end{figure}

\subsubsection{Power spectral density curves}\label{apppsd}
For Ad.~LIGO, we consider the expected power spectral density (PSD) of the ``zero-detuning, high power'' configuration~\cite{Harry_2010}:
\begin{align}
S_n(f) & =  10^{-48} {\rm Hz}^{-1}\left(0.0152\,x^{-4} + 0.2935\,x^{9/4}  \right. 
\nonumber \\
& 
\left.
+ 2.7951\,x^{3/2}
-   6.5080\,x^{3/4} + 17.7622\right),
\end{align}
where $x = f/(245.4\,{\rm Hz})$ (see also Eq.~(4.7) of Ref.~\cite{Ajith:2011ec}). 
We adopt the ET-D sensitivity curves from Ref.~\cite{Hild:2010id}.
Finally, we consider the LISA PSD of Ref.~\cite{Robson:2018ifk} (see also~\cite{Maselli:2021men}),
that provides an analytic fit for the detector noise. The PSD consists of two parts: the instrumental noise and the confusion noise produced by unresolved galactic binaries, i.e. 
\begin{equation}
S_{n}(f)=S^\tn{Ins}_{n}(f)+S^\tn{WDN}_{n}(f) \ ,
\end{equation}
where 
\begin{align}
S^\tn{Ins}_{n}(f)&=A_1\left(P_\tn{OMS}+2[1+\cos^2(f/f_\star)]
\frac{P_\tn{acc}}{(2\pi f)^4}\right)
\nonumber\\
&\times \left(1+\frac{6}{10}\frac{f^2}{f_\star^2}\right)\ ,
\end{align}
$A_1={10}/{3L^2}$, $L=2.5$Gm, $f_\star=19.09$mHz, while 
\begin{align}
P_\tn{OMS}&=(1.5\times 10^{-11}\tn{m})^2\left[1+\left(\frac{2\tn{mHz}}{f}\right)^4\right]\ \tn{Hz}^{-1}\nonumber\ ,\\ 
P_\tn{ACC}&=(3\times 10^{-15}\tn{m s}^{-2})^2\left[1+\left(\frac{0.4\tn{mHz}}{f}\right)^2\right]
\nonumber \\
&\times \left[1+\left(\frac{f}{8\tn{mHz}}\right)^4\right]\ \tn{Hz}^{-1} .
\end{align}
For the white dwarf contribution, we use
\begin{equation}
S^\tn{WDN}_{n}=A_2 f^{-7/3}e^{-f^\alpha+\beta f\sin(\kappa f)}[1+\tanh(\gamma(f_k-f))]\ ,
\end{equation}
with the amplitude $A_2=9\times 10^{-45} \tn{Hz}^{-1}$, and the coefficients 
$(\alpha,\beta,\kappa,\gamma,f_k)=(0.171,292,1020,1680,0.00215)$.
The noise spectral densities are shown in Fig.~\ref{fig_psd}.

\subsubsection{Frequency range}\label{freqrange}

We set $f_\text{\tiny min}$ to the minimum frequency detectable by the interferometer.
In particular, we adopt as minimum frequencies $f_\text{\tiny low}^\text{\tiny Ad.~LIGO} =10\, {\rm Hz}$
($f_\text{\tiny low}^\text{\tiny ET} =1\, {\rm Hz}$)
for the Ad.~LIGO (ET) case. 
 For LISA we take  $f^\text{\tiny LISA}_\text{\tiny low} = {\rm max}[10^{-5} \, {\rm Hz}, f^\text{\tiny LISA}_\text{\tiny obs}$], i.e. the maximum frequency between the cutoff frequency below which the LISA noise curve is not well characterized ($10^{-5} \, {\rm Hz}$) and the frequency corresponding to a binary that spends $T_\text{\tiny obs}=1\,{\rm yr}$ to span the frequency band up to $f_\text{\tiny max}^\text{\tiny LISA}$. We use a LISA maximum frequency $f_\text{\tiny max}^\text{\tiny LISA}=1\,{\rm Hz}$.
Therefore, we find~\cite{Berti:2004bd}
\be\label{fconditionLISA}
f^\text{\tiny LISA}_\text{\tiny obs} = 
\llp
\lp \frac{f_\text{\tiny max}^\text{\tiny LISA}}{\rm Hz}\rp^{-8/3}
+
49
\lp \frac{\mathcal{M}}{ M_\odot} \rp^{5/3} \lp \frac{T_\text{\tiny obs}}{ \rm yr} \rp
\rrp^{-3/8}
\,{\rm Hz}.
\ee
On the other hand,
the maximum frequency is set 
 by the smallest value between either the maximum frequency reached by the detector 
 \begin{equation}
     f_\text{\tiny max}^{\text{(Ad.~LIGO,ET,LISA)}} = (10^4,10^4,1) \, {\rm Hz}
 \end{equation} 
or
the ISCO frequency of the binary system, defined as
\be
\label{eq:fisco}
f_{\text{\tiny ISCO}, z} = \frac{1}{1+z} \frac{\hat{\Omega}_\text{\tiny ISCO}(\chi_f)}{\pi M_f}\,,
\ee
where $\hat{\Omega}_\text{\tiny ISCO}(\chi) \equiv M_\text{\tiny Kerr} \Omega_\text{\tiny ISCO} $ is the dimensionless angular frequency for a circular, equatorial orbit around a Kerr BH with mass $M_\text{\tiny kerr}$ and spin parameter $\chi$~\cite{1972ApJ...178..347B}, while $\chi_f$ and $M_f$ are the final spin and mass of the BH merger remnant, whose full expressions (based on fits to numerical relativity simulations~\cite{Husa:2015iqa,Hofmann:2016yih}) can be found in Appendix~B of Ref.~\cite{Favata:2021vhw}.

\bibliography{main}

\begin{thebibliography}{226}%
\makeatletter
\providecommand \@ifxundefined [1]{%
 \@ifx{#1\undefined}
}%
\providecommand \@ifnum [1]{%
 \ifnum #1\expandafter \@firstoftwo
 \else \expandafter \@secondoftwo
 \fi
}%
\providecommand \@ifx [1]{%
 \ifx #1\expandafter \@firstoftwo
 \else \expandafter \@secondoftwo
 \fi
}%
\providecommand \natexlab [1]{#1}%
\providecommand \enquote  [1]{``#1''}%
\providecommand \bibnamefont  [1]{#1}%
\providecommand \bibfnamefont [1]{#1}%
\providecommand \citenamefont [1]{#1}%
\providecommand \href@noop [0]{\@secondoftwo}%
\providecommand \href [0]{\begingroup \@sanitize@url \@href}%
\providecommand \@href[1]{\@@startlink{#1}\@@href}%
\providecommand \@@href[1]{\endgroup#1\@@endlink}%
\providecommand \@sanitize@url [0]{\catcode `\\12\catcode `\$12\catcode
  `\&12\catcode `\#12\catcode `\^12\catcode `\_12\catcode `\%12\relax}%
\providecommand \@@startlink[1]{}%
\providecommand \@@endlink[0]{}%
\providecommand \url  [0]{\begingroup\@sanitize@url \@url }%
\providecommand \@url [1]{\endgroup\@href {#1}{\urlprefix }}%
\providecommand \urlprefix  [0]{URL }%
\providecommand \Eprint [0]{\href }%
\providecommand \doibase [0]{http://dx.doi.org/}%
\providecommand \selectlanguage [0]{\@gobble}%
\providecommand \bibinfo  [0]{\@secondoftwo}%
\providecommand \bibfield  [0]{\@secondoftwo}%
\providecommand \translation [1]{[#1]}%
\providecommand \BibitemOpen [0]{}%
\providecommand \bibitemStop [0]{}%
\providecommand \bibitemNoStop [0]{.\EOS\space}%
\providecommand \EOS [0]{\spacefactor3000\relax}%
\providecommand \BibitemShut  [1]{\csname bibitem#1\endcsname}%
\let\auto@bib@innerbib\@empty
\bibitem [{\citenamefont {Zel'dovich}\ and\ \citenamefont
  {Novikov}(1967)}]{Zeldovich:1967lct}%
  \BibitemOpen
  \bibfield  {author} {\bibinfo {author} {\bibfnamefont {Y.~B.}\ \bibnamefont
  {Zel'dovich}}\ and\ \bibinfo {author} {\bibfnamefont {I.~D.}\ \bibnamefont
  {Novikov}},\ }\href@noop {} {\bibfield  {journal} {\bibinfo  {journal}
  {Soviet Astron. AJ (Engl. Transl. ),}\ }\textbf {\bibinfo {volume} {10}},\
  \bibinfo {pages} {602} (\bibinfo {year} {1967})}\BibitemShut {NoStop}%
\bibitem [{\citenamefont {Hawking}(1974)}]{Hawking:1974rv}%
  \BibitemOpen
  \bibfield  {author} {\bibinfo {author} {\bibfnamefont {S.~W.}\ \bibnamefont
  {Hawking}},\ }\href {\doibase 10.1038/248030a0} {\bibfield  {journal}
  {\bibinfo  {journal} {Nature}\ }\textbf {\bibinfo {volume} {248}},\ \bibinfo
  {pages} {30} (\bibinfo {year} {1974})}\BibitemShut {NoStop}%
\bibitem [{\citenamefont {Chapline}(1975)}]{Chapline:1975ojl}%
  \BibitemOpen
  \bibfield  {author} {\bibinfo {author} {\bibfnamefont {G.~F.}\ \bibnamefont
  {Chapline}},\ }\href {\doibase 10.1038/253251a0} {\bibfield  {journal}
  {\bibinfo  {journal} {Nature}\ }\textbf {\bibinfo {volume} {253}},\ \bibinfo
  {pages} {251} (\bibinfo {year} {1975})}\BibitemShut {NoStop}%
\bibitem [{\citenamefont {Carr}(1975)}]{Carr:1975qj}%
  \BibitemOpen
  \bibfield  {author} {\bibinfo {author} {\bibfnamefont {B.~J.}\ \bibnamefont
  {Carr}},\ }\href {\doibase 10.1086/153853} {\bibfield  {journal} {\bibinfo
  {journal} {Astrophys. J.}\ }\textbf {\bibinfo {volume} {201}},\ \bibinfo
  {pages} {1} (\bibinfo {year} {1975})}\BibitemShut {NoStop}%
\bibitem [{\citenamefont {Ivanov}\ \emph {et~al.}(1994)\citenamefont {Ivanov},
  \citenamefont {Naselsky},\ and\ \citenamefont {Novikov}}]{Ivanov:1994pa}%
  \BibitemOpen
  \bibfield  {author} {\bibinfo {author} {\bibfnamefont {P.}~\bibnamefont
  {Ivanov}}, \bibinfo {author} {\bibfnamefont {P.}~\bibnamefont {Naselsky}}, \
  and\ \bibinfo {author} {\bibfnamefont {I.}~\bibnamefont {Novikov}},\ }\href
  {\doibase 10.1103/PhysRevD.50.7173} {\bibfield  {journal} {\bibinfo
  {journal} {Phys. Rev. D}\ }\textbf {\bibinfo {volume} {50}},\ \bibinfo
  {pages} {7173} (\bibinfo {year} {1994})}\BibitemShut {NoStop}%
\bibitem [{\citenamefont {Garcia-Bellido}\ \emph {et~al.}(1996)\citenamefont
  {Garcia-Bellido}, \citenamefont {Linde},\ and\ \citenamefont
  {Wands}}]{GarciaBellido:1996qt}%
  \BibitemOpen
  \bibfield  {author} {\bibinfo {author} {\bibfnamefont {J.}~\bibnamefont
  {Garcia-Bellido}}, \bibinfo {author} {\bibfnamefont {A.~D.}\ \bibnamefont
  {Linde}}, \ and\ \bibinfo {author} {\bibfnamefont {D.}~\bibnamefont
  {Wands}},\ }\href {\doibase 10.1103/PhysRevD.54.6040} {\bibfield  {journal}
  {\bibinfo  {journal} {Phys. Rev. D}\ }\textbf {\bibinfo {volume} {54}},\
  \bibinfo {pages} {6040} (\bibinfo {year} {1996})},\ \Eprint
  {http://arxiv.org/abs/astro-ph/9605094} {arXiv:astro-ph/9605094} \BibitemShut
  {NoStop}%
\bibitem [{\citenamefont {Ivanov}(1998)}]{Ivanov:1997ia}%
  \BibitemOpen
  \bibfield  {author} {\bibinfo {author} {\bibfnamefont {P.}~\bibnamefont
  {Ivanov}},\ }\href {\doibase 10.1103/PhysRevD.57.7145} {\bibfield  {journal}
  {\bibinfo  {journal} {Phys. Rev. D}\ }\textbf {\bibinfo {volume} {57}},\
  \bibinfo {pages} {7145} (\bibinfo {year} {1998})},\ \Eprint
  {http://arxiv.org/abs/astro-ph/9708224} {arXiv:astro-ph/9708224} \BibitemShut
  {NoStop}%
\bibitem [{\citenamefont {Blinnikov}\ \emph {et~al.}(2016)\citenamefont
  {Blinnikov}, \citenamefont {Dolgov}, \citenamefont {Porayko},\ and\
  \citenamefont {Postnov}}]{Blinnikov:2016bxu}%
  \BibitemOpen
  \bibfield  {author} {\bibinfo {author} {\bibfnamefont {S.}~\bibnamefont
  {Blinnikov}}, \bibinfo {author} {\bibfnamefont {A.}~\bibnamefont {Dolgov}},
  \bibinfo {author} {\bibfnamefont {N.~K.}\ \bibnamefont {Porayko}}, \ and\
  \bibinfo {author} {\bibfnamefont {K.}~\bibnamefont {Postnov}},\ }\href
  {\doibase 10.1088/1475-7516/2016/11/036} {\bibfield  {journal} {\bibinfo
  {journal} {JCAP}\ }\textbf {\bibinfo {volume} {1611}},\ \bibinfo {pages}
  {036} (\bibinfo {year} {2016})},\ \Eprint {http://arxiv.org/abs/1611.00541}
  {arXiv:1611.00541 [astro-ph.HE]} \BibitemShut {NoStop}%
\bibitem [{\citenamefont {Carr}\ \emph {et~al.}(2020)\citenamefont {Carr},
  \citenamefont {Kohri}, \citenamefont {Sendouda},\ and\ \citenamefont
  {Yokoyama}}]{Carr:2020gox}%
  \BibitemOpen
  \bibfield  {author} {\bibinfo {author} {\bibfnamefont {B.}~\bibnamefont
  {Carr}}, \bibinfo {author} {\bibfnamefont {K.}~\bibnamefont {Kohri}},
  \bibinfo {author} {\bibfnamefont {Y.}~\bibnamefont {Sendouda}}, \ and\
  \bibinfo {author} {\bibfnamefont {J.}~\bibnamefont {Yokoyama}},\ }\href@noop
  {} {\  (\bibinfo {year} {2020})},\ \Eprint {http://arxiv.org/abs/2002.12778}
  {arXiv:2002.12778 [astro-ph.CO]} \BibitemShut {NoStop}%
\bibitem [{\citenamefont {{Volonteri}}(2010)}]{2010A&ARv..18..279V}%
  \BibitemOpen
  \bibfield  {author} {\bibinfo {author} {\bibfnamefont {M.}~\bibnamefont
  {{Volonteri}}},\ }\href {\doibase 10.1007/s00159-010-0029-x} {\bibfield
  {journal} {\bibinfo  {journal} {\aapr}\ }\textbf {\bibinfo {volume} {18}},\
  \bibinfo {pages} {279} (\bibinfo {year} {2010})},\ \Eprint
  {http://arxiv.org/abs/1003.4404} {arXiv:1003.4404 [astro-ph.CO]} \BibitemShut
  {NoStop}%
\bibitem [{\citenamefont {Clesse}\ and\ \citenamefont
  {Garc\'\i{}a-Bellido}(2015)}]{Clesse:2015wea}%
  \BibitemOpen
  \bibfield  {author} {\bibinfo {author} {\bibfnamefont {S.}~\bibnamefont
  {Clesse}}\ and\ \bibinfo {author} {\bibfnamefont {J.}~\bibnamefont
  {Garc\'\i{}a-Bellido}},\ }\href {\doibase 10.1103/PhysRevD.92.023524}
  {\bibfield  {journal} {\bibinfo  {journal} {Phys. Rev. D}\ }\textbf {\bibinfo
  {volume} {92}},\ \bibinfo {pages} {023524} (\bibinfo {year} {2015})},\
  \Eprint {http://arxiv.org/abs/1501.07565} {arXiv:1501.07565 [astro-ph.CO]}
  \BibitemShut {NoStop}%
\bibitem [{\citenamefont {Serpico}\ \emph {et~al.}(2020)\citenamefont
  {Serpico}, \citenamefont {Poulin}, \citenamefont {Inman},\ and\ \citenamefont
  {Kohri}}]{Serpico:2020ehh}%
  \BibitemOpen
  \bibfield  {author} {\bibinfo {author} {\bibfnamefont {P.~D.}\ \bibnamefont
  {Serpico}}, \bibinfo {author} {\bibfnamefont {V.}~\bibnamefont {Poulin}},
  \bibinfo {author} {\bibfnamefont {D.}~\bibnamefont {Inman}}, \ and\ \bibinfo
  {author} {\bibfnamefont {K.}~\bibnamefont {Kohri}},\ }\href {\doibase
  10.1103/PhysRevResearch.2.023204} {\bibfield  {journal} {\bibinfo  {journal}
  {Phys. Rev. Res.}\ }\textbf {\bibinfo {volume} {2}},\ \bibinfo {pages}
  {023204} (\bibinfo {year} {2020})},\ \Eprint
  {http://arxiv.org/abs/2002.10771} {arXiv:2002.10771 [astro-ph.CO]}
  \BibitemShut {NoStop}%
\bibitem [{\citenamefont {Abbott}\ \emph
  {et~al.}(2019{\natexlab{a}})\citenamefont {Abbott} \emph
  {et~al.}}]{LIGOScientific:2018mvr}%
  \BibitemOpen
  \bibfield  {author} {\bibinfo {author} {\bibfnamefont {B.~P.}\ \bibnamefont
  {Abbott}} \emph {et~al.} (\bibinfo {collaboration} {LIGO Scientific,
  Virgo}),\ }\href {\doibase 10.1103/PhysRevX.9.031040} {\bibfield  {journal}
  {\bibinfo  {journal} {Phys. Rev. X}\ }\textbf {\bibinfo {volume} {9}},\
  \bibinfo {pages} {031040} (\bibinfo {year} {2019}{\natexlab{a}})},\ \Eprint
  {http://arxiv.org/abs/1811.12907} {arXiv:1811.12907 [astro-ph.HE]}
  \BibitemShut {NoStop}%
\bibitem [{\citenamefont {Abbott}\ \emph
  {et~al.}(2021{\natexlab{a}})\citenamefont {Abbott} \emph
  {et~al.}}]{LIGOScientific:2020ibl}%
  \BibitemOpen
  \bibfield  {author} {\bibinfo {author} {\bibfnamefont {R.}~\bibnamefont
  {Abbott}} \emph {et~al.} (\bibinfo {collaboration} {LIGO Scientific,
  Virgo}),\ }\href {\doibase 10.1103/PhysRevX.11.021053} {\bibfield  {journal}
  {\bibinfo  {journal} {Phys. Rev. X}\ }\textbf {\bibinfo {volume} {11}},\
  \bibinfo {pages} {021053} (\bibinfo {year} {2021}{\natexlab{a}})},\ \Eprint
  {http://arxiv.org/abs/2010.14527} {arXiv:2010.14527 [gr-qc]} \BibitemShut
  {NoStop}%
\bibitem [{\citenamefont {Abbott}\ \emph
  {et~al.}(2021{\natexlab{b}})\citenamefont {Abbott} \emph
  {et~al.}}]{LIGOScientific:2021djp}%
  \BibitemOpen
  \bibfield  {author} {\bibinfo {author} {\bibfnamefont {R.}~\bibnamefont
  {Abbott}} \emph {et~al.} (\bibinfo {collaboration} {LIGO Scientific, VIRGO,
  KAGRA}),\ }\href@noop {} {\  (\bibinfo {year} {2021}{\natexlab{b}})},\
  \Eprint {http://arxiv.org/abs/2111.03606} {arXiv:2111.03606 [gr-qc]}
  \BibitemShut {NoStop}%
\bibitem [{\citenamefont {Bird}\ \emph {et~al.}(2016)\citenamefont {Bird},
  \citenamefont {Cholis}, \citenamefont {Muñoz}, \citenamefont {Ali-Haïmoud},
  \citenamefont {Kamionkowski}, \citenamefont {Kovetz}, \citenamefont
  {Raccanelli},\ and\ \citenamefont {Riess}}]{Bird:2016dcv}%
  \BibitemOpen
  \bibfield  {author} {\bibinfo {author} {\bibfnamefont {S.}~\bibnamefont
  {Bird}}, \bibinfo {author} {\bibfnamefont {I.}~\bibnamefont {Cholis}},
  \bibinfo {author} {\bibfnamefont {J.~B.}\ \bibnamefont {Muñoz}}, \bibinfo
  {author} {\bibfnamefont {Y.}~\bibnamefont {Ali-Haïmoud}}, \bibinfo {author}
  {\bibfnamefont {M.}~\bibnamefont {Kamionkowski}}, \bibinfo {author}
  {\bibfnamefont {E.~D.}\ \bibnamefont {Kovetz}}, \bibinfo {author}
  {\bibfnamefont {A.}~\bibnamefont {Raccanelli}}, \ and\ \bibinfo {author}
  {\bibfnamefont {A.~G.}\ \bibnamefont {Riess}},\ }\href {\doibase
  10.1103/PhysRevLett.116.201301} {\bibfield  {journal} {\bibinfo  {journal}
  {Phys. Rev. Lett.}\ }\textbf {\bibinfo {volume} {116}},\ \bibinfo {pages}
  {201301} (\bibinfo {year} {2016})},\ \Eprint
  {http://arxiv.org/abs/1603.00464} {arXiv:1603.00464 [astro-ph.CO]}
  \BibitemShut {NoStop}%
\bibitem [{\citenamefont {Sasaki}\ \emph {et~al.}(2016)\citenamefont {Sasaki},
  \citenamefont {Suyama}, \citenamefont {Tanaka},\ and\ \citenamefont
  {Yokoyama}}]{Sasaki:2016jop}%
  \BibitemOpen
  \bibfield  {author} {\bibinfo {author} {\bibfnamefont {M.}~\bibnamefont
  {Sasaki}}, \bibinfo {author} {\bibfnamefont {T.}~\bibnamefont {Suyama}},
  \bibinfo {author} {\bibfnamefont {T.}~\bibnamefont {Tanaka}}, \ and\ \bibinfo
  {author} {\bibfnamefont {S.}~\bibnamefont {Yokoyama}},\ }\href {\doibase
  10.1103/PhysRevLett.121.059901, 10.1103/PhysRevLett.117.061101} {\bibfield
  {journal} {\bibinfo  {journal} {Phys. Rev. Lett.}\ }\textbf {\bibinfo
  {volume} {117}},\ \bibinfo {pages} {061101} (\bibinfo {year} {2016})},\
  \bibinfo {note} {[erratum: Phys. Rev. Lett.121,no.5,059901(2018)]},\ \Eprint
  {http://arxiv.org/abs/1603.08338} {arXiv:1603.08338 [astro-ph.CO]}
  \BibitemShut {NoStop}%
\bibitem [{\citenamefont {Eroshenko}(2018)}]{Eroshenko:2016hmn}%
  \BibitemOpen
  \bibfield  {author} {\bibinfo {author} {\bibfnamefont {Y.~N.}\ \bibnamefont
  {Eroshenko}},\ }\href {\doibase 10.1088/1742-6596/1051/1/012010} {\bibfield
  {journal} {\bibinfo  {journal} {J. Phys. Conf. Ser.}\ }\textbf {\bibinfo
  {volume} {1051}},\ \bibinfo {pages} {012010} (\bibinfo {year} {2018})},\
  \Eprint {http://arxiv.org/abs/1604.04932} {arXiv:1604.04932 [astro-ph.CO]}
  \BibitemShut {NoStop}%
\bibitem [{\citenamefont {Wang}\ \emph {et~al.}(2018)\citenamefont {Wang},
  \citenamefont {Wang}, \citenamefont {Huang},\ and\ \citenamefont
  {Li}}]{Wang:2016ana}%
  \BibitemOpen
  \bibfield  {author} {\bibinfo {author} {\bibfnamefont {S.}~\bibnamefont
  {Wang}}, \bibinfo {author} {\bibfnamefont {Y.-F.}\ \bibnamefont {Wang}},
  \bibinfo {author} {\bibfnamefont {Q.-G.}\ \bibnamefont {Huang}}, \ and\
  \bibinfo {author} {\bibfnamefont {T.~G.~F.}\ \bibnamefont {Li}},\ }\href
  {\doibase 10.1103/PhysRevLett.120.191102} {\bibfield  {journal} {\bibinfo
  {journal} {Phys. Rev. Lett.}\ }\textbf {\bibinfo {volume} {120}},\ \bibinfo
  {pages} {191102} (\bibinfo {year} {2018})},\ \Eprint
  {http://arxiv.org/abs/1610.08725} {arXiv:1610.08725 [astro-ph.CO]}
  \BibitemShut {NoStop}%
\bibitem [{\citenamefont {Ali-Haïmoud}\ \emph {et~al.}(2017)\citenamefont
  {Ali-Haïmoud}, \citenamefont {Kovetz},\ and\ \citenamefont
  {Kamionkowski}}]{Ali-Haimoud:2017rtz}%
  \BibitemOpen
  \bibfield  {author} {\bibinfo {author} {\bibfnamefont {Y.}~\bibnamefont
  {Ali-Haïmoud}}, \bibinfo {author} {\bibfnamefont {E.~D.}\ \bibnamefont
  {Kovetz}}, \ and\ \bibinfo {author} {\bibfnamefont {M.}~\bibnamefont
  {Kamionkowski}},\ }\href {\doibase 10.1103/PhysRevD.96.123523} {\bibfield
  {journal} {\bibinfo  {journal} {Phys. Rev.}\ }\textbf {\bibinfo {volume}
  {D96}},\ \bibinfo {pages} {123523} (\bibinfo {year} {2017})},\ \Eprint
  {http://arxiv.org/abs/1709.06576} {arXiv:1709.06576 [astro-ph.CO]}
  \BibitemShut {NoStop}%
\bibitem [{\citenamefont {Chen}\ and\ \citenamefont
  {Huang}(2018)}]{Chen:2018czv}%
  \BibitemOpen
  \bibfield  {author} {\bibinfo {author} {\bibfnamefont {Z.-C.}\ \bibnamefont
  {Chen}}\ and\ \bibinfo {author} {\bibfnamefont {Q.-G.}\ \bibnamefont
  {Huang}},\ }\href {\doibase 10.3847/1538-4357/aad6e2} {\bibfield  {journal}
  {\bibinfo  {journal} {Astrophys. J.}\ }\textbf {\bibinfo {volume} {864}},\
  \bibinfo {pages} {61} (\bibinfo {year} {2018})},\ \Eprint
  {http://arxiv.org/abs/1801.10327} {arXiv:1801.10327 [astro-ph.CO]}
  \BibitemShut {NoStop}%
\bibitem [{\citenamefont {Raidal}\ \emph {et~al.}(2019)\citenamefont {Raidal},
  \citenamefont {Spethmann}, \citenamefont {Vaskonen},\ and\ \citenamefont
  {Veermäe}}]{Raidal:2018bbj}%
  \BibitemOpen
  \bibfield  {author} {\bibinfo {author} {\bibfnamefont {M.}~\bibnamefont
  {Raidal}}, \bibinfo {author} {\bibfnamefont {C.}~\bibnamefont {Spethmann}},
  \bibinfo {author} {\bibfnamefont {V.}~\bibnamefont {Vaskonen}}, \ and\
  \bibinfo {author} {\bibfnamefont {H.}~\bibnamefont {Veermäe}},\ }\href
  {\doibase 10.1088/1475-7516/2019/02/018} {\bibfield  {journal} {\bibinfo
  {journal} {JCAP}\ }\textbf {\bibinfo {volume} {02}},\ \bibinfo {pages} {018}
  (\bibinfo {year} {2019})},\ \Eprint {http://arxiv.org/abs/1812.01930}
  {arXiv:1812.01930 [astro-ph.CO]} \BibitemShut {NoStop}%
\bibitem [{\citenamefont {Liu}\ \emph {et~al.}(2019{\natexlab{a}})\citenamefont
  {Liu}, \citenamefont {Guo},\ and\ \citenamefont {Cai}}]{Liu:2019rnx}%
  \BibitemOpen
  \bibfield  {author} {\bibinfo {author} {\bibfnamefont {L.}~\bibnamefont
  {Liu}}, \bibinfo {author} {\bibfnamefont {Z.-K.}\ \bibnamefont {Guo}}, \ and\
  \bibinfo {author} {\bibfnamefont {R.-G.}\ \bibnamefont {Cai}},\ }\href
  {\doibase 10.1140/epjc/s10052-019-7227-0} {\bibfield  {journal} {\bibinfo
  {journal} {Eur. Phys. J. C}\ }\textbf {\bibinfo {volume} {79}},\ \bibinfo
  {pages} {717} (\bibinfo {year} {2019}{\natexlab{a}})},\ \Eprint
  {http://arxiv.org/abs/1901.07672} {arXiv:1901.07672 [astro-ph.CO]}
  \BibitemShut {NoStop}%
\bibitem [{\citenamefont {H\"utsi}\ \emph {et~al.}(2019)\citenamefont
  {H\"utsi}, \citenamefont {Raidal},\ and\ \citenamefont
  {Veerm\"ae}}]{Hutsi:2019hlw}%
  \BibitemOpen
  \bibfield  {author} {\bibinfo {author} {\bibfnamefont {G.}~\bibnamefont
  {H\"utsi}}, \bibinfo {author} {\bibfnamefont {M.}~\bibnamefont {Raidal}}, \
  and\ \bibinfo {author} {\bibfnamefont {H.}~\bibnamefont {Veerm\"ae}},\ }\href
  {\doibase 10.1103/PhysRevD.100.083016} {\bibfield  {journal} {\bibinfo
  {journal} {Phys. Rev. D}\ }\textbf {\bibinfo {volume} {100}},\ \bibinfo
  {pages} {083016} (\bibinfo {year} {2019})},\ \Eprint
  {http://arxiv.org/abs/1907.06533} {arXiv:1907.06533 [astro-ph.CO]}
  \BibitemShut {NoStop}%
\bibitem [{\citenamefont {Vaskonen}\ and\ \citenamefont
  {Veerm\"ae}(2020)}]{Vaskonen:2019jpv}%
  \BibitemOpen
  \bibfield  {author} {\bibinfo {author} {\bibfnamefont {V.}~\bibnamefont
  {Vaskonen}}\ and\ \bibinfo {author} {\bibfnamefont {H.}~\bibnamefont
  {Veerm\"ae}},\ }\href {\doibase 10.1103/PhysRevD.101.043015} {\bibfield
  {journal} {\bibinfo  {journal} {Phys. Rev. D}\ }\textbf {\bibinfo {volume}
  {101}},\ \bibinfo {pages} {043015} (\bibinfo {year} {2020})},\ \Eprint
  {http://arxiv.org/abs/1908.09752} {arXiv:1908.09752 [astro-ph.CO]}
  \BibitemShut {NoStop}%
\bibitem [{\citenamefont {Gow}\ \emph {et~al.}(2020)\citenamefont {Gow},
  \citenamefont {Byrnes}, \citenamefont {Hall},\ and\ \citenamefont
  {Peacock}}]{Gow:2019pok}%
  \BibitemOpen
  \bibfield  {author} {\bibinfo {author} {\bibfnamefont {A.~D.}\ \bibnamefont
  {Gow}}, \bibinfo {author} {\bibfnamefont {C.~T.}\ \bibnamefont {Byrnes}},
  \bibinfo {author} {\bibfnamefont {A.}~\bibnamefont {Hall}}, \ and\ \bibinfo
  {author} {\bibfnamefont {J.~A.}\ \bibnamefont {Peacock}},\ }\href {\doibase
  10.1088/1475-7516/2020/01/031} {\bibfield  {journal} {\bibinfo  {journal}
  {JCAP}\ }\textbf {\bibinfo {volume} {01}},\ \bibinfo {pages} {031} (\bibinfo
  {year} {2020})},\ \Eprint {http://arxiv.org/abs/1911.12685} {arXiv:1911.12685
  [astro-ph.CO]} \BibitemShut {NoStop}%
\bibitem [{\citenamefont {Wu}(2020)}]{Wu:2020drm}%
  \BibitemOpen
  \bibfield  {author} {\bibinfo {author} {\bibfnamefont {Y.}~\bibnamefont
  {Wu}},\ }\href {\doibase 10.1103/PhysRevD.101.083008} {\bibfield  {journal}
  {\bibinfo  {journal} {Phys. Rev.}\ }\textbf {\bibinfo {volume} {D101}},\
  \bibinfo {pages} {083008} (\bibinfo {year} {2020})},\ \Eprint
  {http://arxiv.org/abs/2001.03833} {arXiv:2001.03833 [astro-ph.CO]}
  \BibitemShut {NoStop}%
\bibitem [{\citenamefont {De~Luca}\ \emph
  {et~al.}(2020{\natexlab{a}})\citenamefont {De~Luca}, \citenamefont
  {Franciolini}, \citenamefont {Pani},\ and\ \citenamefont
  {Riotto}}]{DeLuca:2020qqa}%
  \BibitemOpen
  \bibfield  {author} {\bibinfo {author} {\bibfnamefont {V.}~\bibnamefont
  {De~Luca}}, \bibinfo {author} {\bibfnamefont {G.}~\bibnamefont
  {Franciolini}}, \bibinfo {author} {\bibfnamefont {P.}~\bibnamefont {Pani}}, \
  and\ \bibinfo {author} {\bibfnamefont {A.}~\bibnamefont {Riotto}},\ }\href
  {\doibase 10.1088/1475-7516/2020/06/044} {\bibfield  {journal} {\bibinfo
  {journal} {JCAP}\ }\textbf {\bibinfo {volume} {06}},\ \bibinfo {pages} {044}
  (\bibinfo {year} {2020}{\natexlab{a}})},\ \Eprint
  {http://arxiv.org/abs/2005.05641} {arXiv:2005.05641 [astro-ph.CO]}
  \BibitemShut {NoStop}%
\bibitem [{\citenamefont {Hall}\ \emph {et~al.}(2020)\citenamefont {Hall},
  \citenamefont {Gow},\ and\ \citenamefont {Byrnes}}]{Hall:2020daa}%
  \BibitemOpen
  \bibfield  {author} {\bibinfo {author} {\bibfnamefont {A.}~\bibnamefont
  {Hall}}, \bibinfo {author} {\bibfnamefont {A.~D.}\ \bibnamefont {Gow}}, \
  and\ \bibinfo {author} {\bibfnamefont {C.~T.}\ \bibnamefont {Byrnes}},\
  }\href {\doibase 10.1103/PhysRevD.102.123524} {\bibfield  {journal} {\bibinfo
   {journal} {Phys. Rev. D}\ }\textbf {\bibinfo {volume} {102}},\ \bibinfo
  {pages} {123524} (\bibinfo {year} {2020})},\ \Eprint
  {http://arxiv.org/abs/2008.13704} {arXiv:2008.13704 [astro-ph.CO]}
  \BibitemShut {NoStop}%
\bibitem [{\citenamefont {Wong}\ \emph {et~al.}(2021)\citenamefont {Wong},
  \citenamefont {Franciolini}, \citenamefont {De~Luca}, \citenamefont
  {Baibhav}, \citenamefont {Berti}, \citenamefont {Pani},\ and\ \citenamefont
  {Riotto}}]{Wong:2020yig}%
  \BibitemOpen
  \bibfield  {author} {\bibinfo {author} {\bibfnamefont {K.~W.~K.}\
  \bibnamefont {Wong}}, \bibinfo {author} {\bibfnamefont {G.}~\bibnamefont
  {Franciolini}}, \bibinfo {author} {\bibfnamefont {V.}~\bibnamefont
  {De~Luca}}, \bibinfo {author} {\bibfnamefont {V.}~\bibnamefont {Baibhav}},
  \bibinfo {author} {\bibfnamefont {E.}~\bibnamefont {Berti}}, \bibinfo
  {author} {\bibfnamefont {P.}~\bibnamefont {Pani}}, \ and\ \bibinfo {author}
  {\bibfnamefont {A.}~\bibnamefont {Riotto}},\ }\href {\doibase
  10.1103/PhysRevD.103.023026} {\bibfield  {journal} {\bibinfo  {journal}
  {Phys. Rev.}\ }\textbf {\bibinfo {volume} {D103}},\ \bibinfo {pages} {023026}
  (\bibinfo {year} {2021})},\ \Eprint {http://arxiv.org/abs/2011.01865}
  {arXiv:2011.01865 [gr-qc]} \BibitemShut {NoStop}%
\bibitem [{\citenamefont {Hütsi}\ \emph {et~al.}(2021)\citenamefont {Hütsi},
  \citenamefont {Raidal}, \citenamefont {Vaskonen},\ and\ \citenamefont
  {Veermäe}}]{Hutsi:2020sol}%
  \BibitemOpen
  \bibfield  {author} {\bibinfo {author} {\bibfnamefont {G.}~\bibnamefont
  {Hütsi}}, \bibinfo {author} {\bibfnamefont {M.}~\bibnamefont {Raidal}},
  \bibinfo {author} {\bibfnamefont {V.}~\bibnamefont {Vaskonen}}, \ and\
  \bibinfo {author} {\bibfnamefont {H.}~\bibnamefont {Veermäe}},\ }\href
  {\doibase 10.1088/1475-7516/2021/03/068} {\bibfield  {journal} {\bibinfo
  {journal} {JCAP}\ }\textbf {\bibinfo {volume} {2103}},\ \bibinfo {pages}
  {068} (\bibinfo {year} {2021})},\ \Eprint {http://arxiv.org/abs/2012.02786}
  {arXiv:2012.02786 [astro-ph.CO]} \BibitemShut {NoStop}%
\bibitem [{\citenamefont {Kritos}\ \emph {et~al.}(2021)\citenamefont {Kritos},
  \citenamefont {De~Luca}, \citenamefont {Franciolini}, \citenamefont
  {Kehagias},\ and\ \citenamefont {Riotto}}]{Kritos:2020wcl}%
  \BibitemOpen
  \bibfield  {author} {\bibinfo {author} {\bibfnamefont {K.}~\bibnamefont
  {Kritos}}, \bibinfo {author} {\bibfnamefont {V.}~\bibnamefont {De~Luca}},
  \bibinfo {author} {\bibfnamefont {G.}~\bibnamefont {Franciolini}}, \bibinfo
  {author} {\bibfnamefont {A.}~\bibnamefont {Kehagias}}, \ and\ \bibinfo
  {author} {\bibfnamefont {A.}~\bibnamefont {Riotto}},\ }\href {\doibase
  10.1088/1475-7516/2021/05/039} {\bibfield  {journal} {\bibinfo  {journal}
  {JCAP}\ }\textbf {\bibinfo {volume} {05}},\ \bibinfo {pages} {039} (\bibinfo
  {year} {2021})},\ \Eprint {http://arxiv.org/abs/2012.03585} {arXiv:2012.03585
  [gr-qc]} \BibitemShut {NoStop}%
\bibitem [{\citenamefont {De~Luca}\ \emph
  {et~al.}(2021{\natexlab{a}})\citenamefont {De~Luca}, \citenamefont
  {Franciolini}, \citenamefont {Pani},\ and\ \citenamefont
  {Riotto}}]{DeLuca:2021wjr}%
  \BibitemOpen
  \bibfield  {author} {\bibinfo {author} {\bibfnamefont {V.}~\bibnamefont
  {De~Luca}}, \bibinfo {author} {\bibfnamefont {G.}~\bibnamefont
  {Franciolini}}, \bibinfo {author} {\bibfnamefont {P.}~\bibnamefont {Pani}}, \
  and\ \bibinfo {author} {\bibfnamefont {A.}~\bibnamefont {Riotto}},\ }\href
  {\doibase 10.1088/1475-7516/2021/05/003} {\bibfield  {journal} {\bibinfo
  {journal} {JCAP}\ }\textbf {\bibinfo {volume} {05}},\ \bibinfo {pages} {003}
  (\bibinfo {year} {2021}{\natexlab{a}})},\ \Eprint
  {http://arxiv.org/abs/2102.03809} {arXiv:2102.03809 [astro-ph.CO]}
  \BibitemShut {NoStop}%
\bibitem [{\citenamefont {Deng}(2021)}]{Deng:2021ezy}%
  \BibitemOpen
  \bibfield  {author} {\bibinfo {author} {\bibfnamefont {H.}~\bibnamefont
  {Deng}},\ }\href {\doibase 10.1088/1475-7516/2021/04/058} {\bibfield
  {journal} {\bibinfo  {journal} {JCAP}\ }\textbf {\bibinfo {volume} {04}},\
  \bibinfo {pages} {058} (\bibinfo {year} {2021})},\ \Eprint
  {http://arxiv.org/abs/2101.11098} {arXiv:2101.11098 [astro-ph.CO]}
  \BibitemShut {NoStop}%
\bibitem [{\citenamefont {Kimura}\ \emph {et~al.}(2021)\citenamefont {Kimura},
  \citenamefont {Suyama}, \citenamefont {Yamaguchi},\ and\ \citenamefont
  {Zhang}}]{Kimura:2021sqz}%
  \BibitemOpen
  \bibfield  {author} {\bibinfo {author} {\bibfnamefont {R.}~\bibnamefont
  {Kimura}}, \bibinfo {author} {\bibfnamefont {T.}~\bibnamefont {Suyama}},
  \bibinfo {author} {\bibfnamefont {M.}~\bibnamefont {Yamaguchi}}, \ and\
  \bibinfo {author} {\bibfnamefont {Y.-L.}\ \bibnamefont {Zhang}},\ }\href
  {\doibase 10.1088/1475-7516/2021/04/031} {\bibfield  {journal} {\bibinfo
  {journal} {JCAP}\ }\textbf {\bibinfo {volume} {04}},\ \bibinfo {pages} {031}
  (\bibinfo {year} {2021})},\ \Eprint {http://arxiv.org/abs/2102.05280}
  {arXiv:2102.05280 [astro-ph.CO]} \BibitemShut {NoStop}%
\bibitem [{\citenamefont {Franciolini}\ \emph {et~al.}(2021)\citenamefont
  {Franciolini}, \citenamefont {Baibhav}, \citenamefont {De~Luca},
  \citenamefont {Ng}, \citenamefont {Wong}, \citenamefont {Berti},
  \citenamefont {Pani}, \citenamefont {Riotto},\ and\ \citenamefont
  {Vitale}}]{Franciolini:2021tla}%
  \BibitemOpen
  \bibfield  {author} {\bibinfo {author} {\bibfnamefont {G.}~\bibnamefont
  {Franciolini}}, \bibinfo {author} {\bibfnamefont {V.}~\bibnamefont
  {Baibhav}}, \bibinfo {author} {\bibfnamefont {V.}~\bibnamefont {De~Luca}},
  \bibinfo {author} {\bibfnamefont {K.~K.~Y.}\ \bibnamefont {Ng}}, \bibinfo
  {author} {\bibfnamefont {K.~W.~K.}\ \bibnamefont {Wong}}, \bibinfo {author}
  {\bibfnamefont {E.}~\bibnamefont {Berti}}, \bibinfo {author} {\bibfnamefont
  {P.}~\bibnamefont {Pani}}, \bibinfo {author} {\bibfnamefont {A.}~\bibnamefont
  {Riotto}}, \ and\ \bibinfo {author} {\bibfnamefont {S.}~\bibnamefont
  {Vitale}},\ }\href@noop {} {\  (\bibinfo {year} {2021})},\ \Eprint
  {http://arxiv.org/abs/2105.03349} {arXiv:2105.03349 [gr-qc]} \BibitemShut
  {NoStop}%
\bibitem [{\citenamefont {Bavera}\ \emph {et~al.}(2021)\citenamefont {Bavera},
  \citenamefont {Franciolini}, \citenamefont {Cusin}, \citenamefont {Riotto},
  \citenamefont {Zevin},\ and\ \citenamefont {Fragos}}]{Bavera:2021wmw}%
  \BibitemOpen
  \bibfield  {author} {\bibinfo {author} {\bibfnamefont {S.~S.}\ \bibnamefont
  {Bavera}}, \bibinfo {author} {\bibfnamefont {G.}~\bibnamefont {Franciolini}},
  \bibinfo {author} {\bibfnamefont {G.}~\bibnamefont {Cusin}}, \bibinfo
  {author} {\bibfnamefont {A.}~\bibnamefont {Riotto}}, \bibinfo {author}
  {\bibfnamefont {M.}~\bibnamefont {Zevin}}, \ and\ \bibinfo {author}
  {\bibfnamefont {T.}~\bibnamefont {Fragos}},\ }\href@noop {} {\  (\bibinfo
  {year} {2021})},\ \Eprint {http://arxiv.org/abs/2109.05836} {arXiv:2109.05836
  [astro-ph.CO]} \BibitemShut {NoStop}%
\bibitem [{\citenamefont {Liu}\ \emph {et~al.}(2021)\citenamefont {Liu},
  \citenamefont {Yang}, \citenamefont {Guo},\ and\ \citenamefont
  {Cai}}]{Liu:2021jnw}%
  \BibitemOpen
  \bibfield  {author} {\bibinfo {author} {\bibfnamefont {L.}~\bibnamefont
  {Liu}}, \bibinfo {author} {\bibfnamefont {X.-Y.}\ \bibnamefont {Yang}},
  \bibinfo {author} {\bibfnamefont {Z.-K.}\ \bibnamefont {Guo}}, \ and\
  \bibinfo {author} {\bibfnamefont {R.-G.}\ \bibnamefont {Cai}},\ }\href@noop
  {} {\  (\bibinfo {year} {2021})},\ \Eprint {http://arxiv.org/abs/2112.05473}
  {arXiv:2112.05473 [astro-ph.CO]} \BibitemShut {NoStop}%
\bibitem [{\citenamefont {De~Luca}\ \emph
  {et~al.}(2021{\natexlab{b}})\citenamefont {De~Luca}, \citenamefont
  {Franciolini}, \citenamefont {Pani},\ and\ \citenamefont
  {Riotto}}]{DeLuca:2021hde}%
  \BibitemOpen
  \bibfield  {author} {\bibinfo {author} {\bibfnamefont {V.}~\bibnamefont
  {De~Luca}}, \bibinfo {author} {\bibfnamefont {G.}~\bibnamefont
  {Franciolini}}, \bibinfo {author} {\bibfnamefont {P.}~\bibnamefont {Pani}}, \
  and\ \bibinfo {author} {\bibfnamefont {A.}~\bibnamefont {Riotto}},\ }\href
  {\doibase 10.1088/1475-7516/2021/11/039} {\bibfield  {journal} {\bibinfo
  {journal} {JCAP}\ }\textbf {\bibinfo {volume} {11}},\ \bibinfo {pages} {039}
  (\bibinfo {year} {2021}{\natexlab{b}})},\ \Eprint
  {http://arxiv.org/abs/2106.13769} {arXiv:2106.13769 [astro-ph.CO]}
  \BibitemShut {NoStop}%
\bibitem [{\citenamefont {Pujolas}\ \emph {et~al.}(2021)\citenamefont
  {Pujolas}, \citenamefont {Vaskonen},\ and\ \citenamefont
  {Veerm\"ae}}]{Pujolas:2021yaw}%
  \BibitemOpen
  \bibfield  {author} {\bibinfo {author} {\bibfnamefont {O.}~\bibnamefont
  {Pujolas}}, \bibinfo {author} {\bibfnamefont {V.}~\bibnamefont {Vaskonen}}, \
  and\ \bibinfo {author} {\bibfnamefont {H.}~\bibnamefont {Veerm\"ae}},\ }\href
  {\doibase 10.1103/PhysRevD.104.083521} {\bibfield  {journal} {\bibinfo
  {journal} {Phys. Rev. D}\ }\textbf {\bibinfo {volume} {104}},\ \bibinfo
  {pages} {083521} (\bibinfo {year} {2021})},\ \Eprint
  {http://arxiv.org/abs/2107.03379} {arXiv:2107.03379 [astro-ph.CO]}
  \BibitemShut {NoStop}%
\bibitem [{\citenamefont {Abbott}\ \emph
  {et~al.}(2021{\natexlab{c}})\citenamefont {Abbott} \emph
  {et~al.}}]{LIGOScientific:2021psn}%
  \BibitemOpen
  \bibfield  {author} {\bibinfo {author} {\bibfnamefont {R.}~\bibnamefont
  {Abbott}} \emph {et~al.} (\bibinfo {collaboration} {LIGO Scientific, VIRGO,
  KAGRA}),\ }\href@noop {} {\  (\bibinfo {year} {2021}{\natexlab{c}})},\
  \Eprint {http://arxiv.org/abs/2111.03634} {arXiv:2111.03634 [astro-ph.HE]}
  \BibitemShut {NoStop}%
\bibitem [{\citenamefont {Sasaki}\ \emph {et~al.}(2018)\citenamefont {Sasaki},
  \citenamefont {Suyama}, \citenamefont {Tanaka},\ and\ \citenamefont
  {Yokoyama}}]{Sasaki:2018dmp}%
  \BibitemOpen
  \bibfield  {author} {\bibinfo {author} {\bibfnamefont {M.}~\bibnamefont
  {Sasaki}}, \bibinfo {author} {\bibfnamefont {T.}~\bibnamefont {Suyama}},
  \bibinfo {author} {\bibfnamefont {T.}~\bibnamefont {Tanaka}}, \ and\ \bibinfo
  {author} {\bibfnamefont {S.}~\bibnamefont {Yokoyama}},\ }\href {\doibase
  10.1088/1361-6382/aaa7b4} {\bibfield  {journal} {\bibinfo  {journal} {Class.
  Quant. Grav.}\ }\textbf {\bibinfo {volume} {35}},\ \bibinfo {pages} {063001}
  (\bibinfo {year} {2018})},\ \Eprint {http://arxiv.org/abs/1801.05235}
  {arXiv:1801.05235 [astro-ph.CO]} \BibitemShut {NoStop}%
\bibitem [{\citenamefont {Green}\ and\ \citenamefont
  {Kavanagh}(2021)}]{Green:2020jor}%
  \BibitemOpen
  \bibfield  {author} {\bibinfo {author} {\bibfnamefont {A.~M.}\ \bibnamefont
  {Green}}\ and\ \bibinfo {author} {\bibfnamefont {B.~J.}\ \bibnamefont
  {Kavanagh}},\ }\href {\doibase 10.1088/1361-6471/abc534} {\bibfield
  {journal} {\bibinfo  {journal} {J. Phys. G}\ }\textbf {\bibinfo {volume}
  {48}},\ \bibinfo {pages} {4} (\bibinfo {year} {2021})},\ \Eprint
  {http://arxiv.org/abs/2007.10722} {arXiv:2007.10722 [astro-ph.CO]}
  \BibitemShut {NoStop}%
\bibitem [{\citenamefont {Franciolini}(2021)}]{Franciolini:2021nvv}%
  \BibitemOpen
  \bibfield  {author} {\bibinfo {author} {\bibfnamefont {G.}~\bibnamefont
  {Franciolini}},\ }\emph {\bibinfo {title} {{Primordial Black Holes: from
  Theory to Gravitational Wave Observations}}},\ \href@noop {} {\bibinfo {type}
  {Other thesis}} (\bibinfo {year} {2021}),\ \Eprint
  {http://arxiv.org/abs/2110.06815} {arXiv:2110.06815 [astro-ph.CO]}
  \BibitemShut {NoStop}%
\bibitem [{\citenamefont {Clesse}\ and\ \citenamefont
  {Garcia-Bellido}(2020)}]{Clesse:2020ghq}%
  \BibitemOpen
  \bibfield  {author} {\bibinfo {author} {\bibfnamefont {S.}~\bibnamefont
  {Clesse}}\ and\ \bibinfo {author} {\bibfnamefont {J.}~\bibnamefont
  {Garcia-Bellido}},\ }\href@noop {} {\  (\bibinfo {year} {2020})},\ \Eprint
  {http://arxiv.org/abs/2007.06481} {arXiv:2007.06481 [astro-ph.CO]}
  \BibitemShut {NoStop}%
\bibitem [{\citenamefont {De~Luca}\ \emph
  {et~al.}(2021{\natexlab{c}})\citenamefont {De~Luca}, \citenamefont
  {Desjacques}, \citenamefont {Franciolini}, \citenamefont {Pani},\ and\
  \citenamefont {Riotto}}]{DeLuca:2020sae}%
  \BibitemOpen
  \bibfield  {author} {\bibinfo {author} {\bibfnamefont {V.}~\bibnamefont
  {De~Luca}}, \bibinfo {author} {\bibfnamefont {V.}~\bibnamefont {Desjacques}},
  \bibinfo {author} {\bibfnamefont {G.}~\bibnamefont {Franciolini}}, \bibinfo
  {author} {\bibfnamefont {P.}~\bibnamefont {Pani}}, \ and\ \bibinfo {author}
  {\bibfnamefont {A.}~\bibnamefont {Riotto}},\ }\href {\doibase
  10.1103/PhysRevLett.126.051101} {\bibfield  {journal} {\bibinfo  {journal}
  {Phys. Rev. Lett.}\ }\textbf {\bibinfo {volume} {126}},\ \bibinfo {pages}
  {051101} (\bibinfo {year} {2021}{\natexlab{c}})},\ \Eprint
  {http://arxiv.org/abs/2009.01728} {arXiv:2009.01728 [astro-ph.CO]}
  \BibitemShut {NoStop}%
\bibitem [{\citenamefont {Bhagwat}\ \emph {et~al.}(2021)\citenamefont
  {Bhagwat}, \citenamefont {De~Luca}, \citenamefont {Franciolini},
  \citenamefont {Pani},\ and\ \citenamefont {Riotto}}]{Bhagwat:2020bzh}%
  \BibitemOpen
  \bibfield  {author} {\bibinfo {author} {\bibfnamefont {S.}~\bibnamefont
  {Bhagwat}}, \bibinfo {author} {\bibfnamefont {V.}~\bibnamefont {De~Luca}},
  \bibinfo {author} {\bibfnamefont {G.}~\bibnamefont {Franciolini}}, \bibinfo
  {author} {\bibfnamefont {P.}~\bibnamefont {Pani}}, \ and\ \bibinfo {author}
  {\bibfnamefont {A.}~\bibnamefont {Riotto}},\ }\href {\doibase
  10.1088/1475-7516/2021/01/037} {\bibfield  {journal} {\bibinfo  {journal}
  {JCAP}\ }\textbf {\bibinfo {volume} {01}},\ \bibinfo {pages} {037} (\bibinfo
  {year} {2021})},\ \Eprint {http://arxiv.org/abs/2008.12320} {arXiv:2008.12320
  [astro-ph.CO]} \BibitemShut {NoStop}%
\bibitem [{\citenamefont {Reitze}\ \emph {et~al.}(2019)\citenamefont {Reitze}
  \emph {et~al.}}]{Reitze:2019iox}%
  \BibitemOpen
  \bibfield  {author} {\bibinfo {author} {\bibfnamefont {D.}~\bibnamefont
  {Reitze}} \emph {et~al.},\ }\href@noop {} {\bibfield  {journal} {\bibinfo
  {journal} {Bull. Am. Astron. Soc.}\ }\textbf {\bibinfo {volume} {51}},\
  \bibinfo {pages} {035} (\bibinfo {year} {2019})},\ \Eprint
  {http://arxiv.org/abs/1907.04833} {arXiv:1907.04833 [astro-ph.IM]}
  \BibitemShut {NoStop}%
\bibitem [{\citenamefont {Hild}\ \emph {et~al.}(2011)\citenamefont {Hild} \emph
  {et~al.}}]{Hild:2010id}%
  \BibitemOpen
  \bibfield  {author} {\bibinfo {author} {\bibfnamefont {S.}~\bibnamefont
  {Hild}} \emph {et~al.},\ }\href {\doibase 10.1088/0264-9381/28/9/094013}
  {\bibfield  {journal} {\bibinfo  {journal} {Class. Quant. Grav.}\ }\textbf
  {\bibinfo {volume} {28}},\ \bibinfo {pages} {094013} (\bibinfo {year}
  {2011})},\ \Eprint {http://arxiv.org/abs/1012.0908} {arXiv:1012.0908 [gr-qc]}
  \BibitemShut {NoStop}%
\bibitem [{\citenamefont {{Amaro-Seoane}}\ \emph {et~al.}(2017)\citenamefont
  {{Amaro-Seoane}}, \citenamefont {{Audley}}, \citenamefont {{Babak}},
  \citenamefont {{Baker}}, \citenamefont {{Barausse}}, \citenamefont
  {{Bender}}, \citenamefont {{Berti}}, \citenamefont {{Binetruy}},
  \citenamefont {{Born}}, \citenamefont {{Bortoluzzi}}, \citenamefont {{Camp}},
  \citenamefont {{Caprini}}, \citenamefont {{Cardoso}}, \citenamefont
  {{Colpi}}, \citenamefont {{Conklin}}, \citenamefont {{Cornish}},
  \citenamefont {{Cutler}}, \citenamefont {{Danzmann}}, \citenamefont
  {{Dolesi}}, \citenamefont {{Ferraioli}}, \citenamefont {{Ferroni}},
  \citenamefont {{Fitzsimons}}, \citenamefont {{Gair}}, \citenamefont {{Gesa
  Bote}}, \citenamefont {{Giardini}}, \citenamefont {{Gibert}}, \citenamefont
  {{Grimani}}, \citenamefont {{Halloin}}, \citenamefont {{Heinzel}},
  \citenamefont {{Hertog}}, \citenamefont {{Hewitson}}, \citenamefont
  {{Holley-Bockelmann}}, \citenamefont {{Hollington}}, \citenamefont
  {{Hueller}}, \citenamefont {{Inchauspe}}, \citenamefont {{Jetzer}},
  \citenamefont {{Karnesis}}, \citenamefont {{Killow}}, \citenamefont
  {{Klein}}, \citenamefont {{Klipstein}}, \citenamefont {{Korsakova}},
  \citenamefont {{Larson}}, \citenamefont {{Livas}}, \citenamefont {{Lloro}},
  \citenamefont {{Man}}, \citenamefont {{Mance}}, \citenamefont {{Martino}},
  \citenamefont {{Mateos}}, \citenamefont {{McKenzie}}, \citenamefont
  {{McWilliams}}, \citenamefont {{Miller}}, \citenamefont {{Mueller}},
  \citenamefont {{Nardini}}, \citenamefont {{Nelemans}}, \citenamefont
  {{Nofrarias}}, \citenamefont {{Petiteau}}, \citenamefont {{Pivato}},
  \citenamefont {{Plagnol}}, \citenamefont {{Porter}}, \citenamefont
  {{Reiche}}, \citenamefont {{Robertson}}, \citenamefont {{Robertson}},
  \citenamefont {{Rossi}}, \citenamefont {{Russano}}, \citenamefont {{Schutz}},
  \citenamefont {{Sesana}}, \citenamefont {{Shoemaker}}, \citenamefont
  {{Slutsky}}, \citenamefont {{Sopuerta}}, \citenamefont {{Sumner}},
  \citenamefont {{Tamanini}}, \citenamefont {{Thorpe}}, \citenamefont
  {{Troebs}}, \citenamefont {{Vallisneri}}, \citenamefont {{Vecchio}},
  \citenamefont {{Vetrugno}}, \citenamefont {{Vitale}}, \citenamefont
  {{Volonteri}}, \citenamefont {{Wanner}}, \citenamefont {{Ward}},
  \citenamefont {{Wass}}, \citenamefont {{Weber}}, \citenamefont {{Ziemer}},\
  and\ \citenamefont {{Zweifel}}}]{2017arXiv170200786A}%
  \BibitemOpen
  \bibfield  {author} {\bibinfo {author} {\bibfnamefont {P.}~\bibnamefont
  {{Amaro-Seoane}}}, \bibinfo {author} {\bibfnamefont {H.}~\bibnamefont
  {{Audley}}}, \bibinfo {author} {\bibfnamefont {S.}~\bibnamefont {{Babak}}},
  \bibinfo {author} {\bibfnamefont {J.}~\bibnamefont {{Baker}}}, \bibinfo
  {author} {\bibfnamefont {E.}~\bibnamefont {{Barausse}}}, \bibinfo {author}
  {\bibfnamefont {P.}~\bibnamefont {{Bender}}}, \bibinfo {author}
  {\bibfnamefont {E.}~\bibnamefont {{Berti}}}, \bibinfo {author} {\bibfnamefont
  {P.}~\bibnamefont {{Binetruy}}}, \bibinfo {author} {\bibfnamefont
  {M.}~\bibnamefont {{Born}}}, \bibinfo {author} {\bibfnamefont
  {D.}~\bibnamefont {{Bortoluzzi}}}, \bibinfo {author} {\bibfnamefont
  {J.}~\bibnamefont {{Camp}}}, \bibinfo {author} {\bibfnamefont
  {C.}~\bibnamefont {{Caprini}}}, \bibinfo {author} {\bibfnamefont
  {V.}~\bibnamefont {{Cardoso}}}, \bibinfo {author} {\bibfnamefont
  {M.}~\bibnamefont {{Colpi}}}, \bibinfo {author} {\bibfnamefont
  {J.}~\bibnamefont {{Conklin}}}, \bibinfo {author} {\bibfnamefont
  {N.}~\bibnamefont {{Cornish}}}, \bibinfo {author} {\bibfnamefont
  {C.}~\bibnamefont {{Cutler}}}, \bibinfo {author} {\bibfnamefont
  {K.}~\bibnamefont {{Danzmann}}}, \bibinfo {author} {\bibfnamefont
  {R.}~\bibnamefont {{Dolesi}}}, \bibinfo {author} {\bibfnamefont
  {L.}~\bibnamefont {{Ferraioli}}}, \bibinfo {author} {\bibfnamefont
  {V.}~\bibnamefont {{Ferroni}}}, \bibinfo {author} {\bibfnamefont
  {E.}~\bibnamefont {{Fitzsimons}}}, \bibinfo {author} {\bibfnamefont
  {J.}~\bibnamefont {{Gair}}}, \bibinfo {author} {\bibfnamefont
  {L.}~\bibnamefont {{Gesa Bote}}}, \bibinfo {author} {\bibfnamefont
  {D.}~\bibnamefont {{Giardini}}}, \bibinfo {author} {\bibfnamefont
  {F.}~\bibnamefont {{Gibert}}}, \bibinfo {author} {\bibfnamefont
  {C.}~\bibnamefont {{Grimani}}}, \bibinfo {author} {\bibfnamefont
  {H.}~\bibnamefont {{Halloin}}}, \bibinfo {author} {\bibfnamefont
  {G.}~\bibnamefont {{Heinzel}}}, \bibinfo {author} {\bibfnamefont
  {T.}~\bibnamefont {{Hertog}}}, \bibinfo {author} {\bibfnamefont
  {M.}~\bibnamefont {{Hewitson}}}, \bibinfo {author} {\bibfnamefont
  {K.}~\bibnamefont {{Holley-Bockelmann}}}, \bibinfo {author} {\bibfnamefont
  {D.}~\bibnamefont {{Hollington}}}, \bibinfo {author} {\bibfnamefont
  {M.}~\bibnamefont {{Hueller}}}, \bibinfo {author} {\bibfnamefont
  {H.}~\bibnamefont {{Inchauspe}}}, \bibinfo {author} {\bibfnamefont
  {P.}~\bibnamefont {{Jetzer}}}, \bibinfo {author} {\bibfnamefont
  {N.}~\bibnamefont {{Karnesis}}}, \bibinfo {author} {\bibfnamefont
  {C.}~\bibnamefont {{Killow}}}, \bibinfo {author} {\bibfnamefont
  {A.}~\bibnamefont {{Klein}}}, \bibinfo {author} {\bibfnamefont
  {B.}~\bibnamefont {{Klipstein}}}, \bibinfo {author} {\bibfnamefont
  {N.}~\bibnamefont {{Korsakova}}}, \bibinfo {author} {\bibfnamefont {S.~L.}\
  \bibnamefont {{Larson}}}, \bibinfo {author} {\bibfnamefont {J.}~\bibnamefont
  {{Livas}}}, \bibinfo {author} {\bibfnamefont {I.}~\bibnamefont {{Lloro}}},
  \bibinfo {author} {\bibfnamefont {N.}~\bibnamefont {{Man}}}, \bibinfo
  {author} {\bibfnamefont {D.}~\bibnamefont {{Mance}}}, \bibinfo {author}
  {\bibfnamefont {J.}~\bibnamefont {{Martino}}}, \bibinfo {author}
  {\bibfnamefont {I.}~\bibnamefont {{Mateos}}}, \bibinfo {author}
  {\bibfnamefont {K.}~\bibnamefont {{McKenzie}}}, \bibinfo {author}
  {\bibfnamefont {S.~T.}\ \bibnamefont {{McWilliams}}}, \bibinfo {author}
  {\bibfnamefont {C.}~\bibnamefont {{Miller}}}, \bibinfo {author}
  {\bibfnamefont {G.}~\bibnamefont {{Mueller}}}, \bibinfo {author}
  {\bibfnamefont {G.}~\bibnamefont {{Nardini}}}, \bibinfo {author}
  {\bibfnamefont {G.}~\bibnamefont {{Nelemans}}}, \bibinfo {author}
  {\bibfnamefont {M.}~\bibnamefont {{Nofrarias}}}, \bibinfo {author}
  {\bibfnamefont {A.}~\bibnamefont {{Petiteau}}}, \bibinfo {author}
  {\bibfnamefont {P.}~\bibnamefont {{Pivato}}}, \bibinfo {author}
  {\bibfnamefont {E.}~\bibnamefont {{Plagnol}}}, \bibinfo {author}
  {\bibfnamefont {E.}~\bibnamefont {{Porter}}}, \bibinfo {author}
  {\bibfnamefont {J.}~\bibnamefont {{Reiche}}}, \bibinfo {author}
  {\bibfnamefont {D.}~\bibnamefont {{Robertson}}}, \bibinfo {author}
  {\bibfnamefont {N.}~\bibnamefont {{Robertson}}}, \bibinfo {author}
  {\bibfnamefont {E.}~\bibnamefont {{Rossi}}}, \bibinfo {author} {\bibfnamefont
  {G.}~\bibnamefont {{Russano}}}, \bibinfo {author} {\bibfnamefont
  {B.}~\bibnamefont {{Schutz}}}, \bibinfo {author} {\bibfnamefont
  {A.}~\bibnamefont {{Sesana}}}, \bibinfo {author} {\bibfnamefont
  {D.}~\bibnamefont {{Shoemaker}}}, \bibinfo {author} {\bibfnamefont
  {J.}~\bibnamefont {{Slutsky}}}, \bibinfo {author} {\bibfnamefont {C.~F.}\
  \bibnamefont {{Sopuerta}}}, \bibinfo {author} {\bibfnamefont
  {T.}~\bibnamefont {{Sumner}}}, \bibinfo {author} {\bibfnamefont
  {N.}~\bibnamefont {{Tamanini}}}, \bibinfo {author} {\bibfnamefont
  {I.}~\bibnamefont {{Thorpe}}}, \bibinfo {author} {\bibfnamefont
  {M.}~\bibnamefont {{Troebs}}}, \bibinfo {author} {\bibfnamefont
  {M.}~\bibnamefont {{Vallisneri}}}, \bibinfo {author} {\bibfnamefont
  {A.}~\bibnamefont {{Vecchio}}}, \bibinfo {author} {\bibfnamefont
  {D.}~\bibnamefont {{Vetrugno}}}, \bibinfo {author} {\bibfnamefont
  {S.}~\bibnamefont {{Vitale}}}, \bibinfo {author} {\bibfnamefont
  {M.}~\bibnamefont {{Volonteri}}}, \bibinfo {author} {\bibfnamefont
  {G.}~\bibnamefont {{Wanner}}}, \bibinfo {author} {\bibfnamefont
  {H.}~\bibnamefont {{Ward}}}, \bibinfo {author} {\bibfnamefont
  {P.}~\bibnamefont {{Wass}}}, \bibinfo {author} {\bibfnamefont
  {W.}~\bibnamefont {{Weber}}}, \bibinfo {author} {\bibfnamefont
  {J.}~\bibnamefont {{Ziemer}}}, \ and\ \bibinfo {author} {\bibfnamefont
  {P.}~\bibnamefont {{Zweifel}}},\ }\href@noop {} {\bibfield  {journal}
  {\bibinfo  {journal} {arXiv e-prints}\ ,\ \bibinfo {eid} {arXiv:1702.00786}}
  (\bibinfo {year} {2017})},\ \Eprint {http://arxiv.org/abs/1702.00786}
  {arXiv:1702.00786 [astro-ph.IM]} \BibitemShut {NoStop}%
\bibitem [{\citenamefont {Baibhav}\ \emph {et~al.}(2019)\citenamefont
  {Baibhav}, \citenamefont {Berti}, \citenamefont {Gerosa}, \citenamefont
  {Mapelli}, \citenamefont {Giacobbo}, \citenamefont {Bouffanais},\ and\
  \citenamefont {Di~Carlo}}]{Baibhav:2019gxm}%
  \BibitemOpen
  \bibfield  {author} {\bibinfo {author} {\bibfnamefont {V.}~\bibnamefont
  {Baibhav}}, \bibinfo {author} {\bibfnamefont {E.}~\bibnamefont {Berti}},
  \bibinfo {author} {\bibfnamefont {D.}~\bibnamefont {Gerosa}}, \bibinfo
  {author} {\bibfnamefont {M.}~\bibnamefont {Mapelli}}, \bibinfo {author}
  {\bibfnamefont {N.}~\bibnamefont {Giacobbo}}, \bibinfo {author}
  {\bibfnamefont {Y.}~\bibnamefont {Bouffanais}}, \ and\ \bibinfo {author}
  {\bibfnamefont {U.~N.}\ \bibnamefont {Di~Carlo}},\ }\href {\doibase
  10.1103/PhysRevD.100.064060} {\bibfield  {journal} {\bibinfo  {journal}
  {Phys. Rev.}\ }\textbf {\bibinfo {volume} {D100}},\ \bibinfo {pages} {064060}
  (\bibinfo {year} {2019})},\ \Eprint {http://arxiv.org/abs/1906.04197}
  {arXiv:1906.04197 [gr-qc]} \BibitemShut {NoStop}%
\bibitem [{\citenamefont {Maggiore}\ \emph {et~al.}(2020)\citenamefont
  {Maggiore} \emph {et~al.}}]{Maggiore:2019uih}%
  \BibitemOpen
  \bibfield  {author} {\bibinfo {author} {\bibfnamefont {M.}~\bibnamefont
  {Maggiore}} \emph {et~al.},\ }\href {\doibase 10.1088/1475-7516/2020/03/050}
  {\bibfield  {journal} {\bibinfo  {journal} {JCAP}\ }\textbf {\bibinfo
  {volume} {03}},\ \bibinfo {pages} {050} (\bibinfo {year} {2020})},\ \Eprint
  {http://arxiv.org/abs/1912.02622} {arXiv:1912.02622 [astro-ph.CO]}
  \BibitemShut {NoStop}%
\bibitem [{\citenamefont {Kalogera}\ \emph {et~al.}(2021)\citenamefont
  {Kalogera} \emph {et~al.}}]{Kalogera:2021bya}%
  \BibitemOpen
  \bibfield  {author} {\bibinfo {author} {\bibfnamefont {V.}~\bibnamefont
  {Kalogera}} \emph {et~al.},\ }\href@noop {} {\  (\bibinfo {year} {2021})},\
  \Eprint {http://arxiv.org/abs/2111.06990} {arXiv:2111.06990 [gr-qc]}
  \BibitemShut {NoStop}%
\bibitem [{\citenamefont {De~Luca}\ \emph
  {et~al.}(2020{\natexlab{b}})\citenamefont {De~Luca}, \citenamefont
  {Franciolini}, \citenamefont {Pani},\ and\ \citenamefont
  {Riotto}}]{DeLuca:2020bjf}%
  \BibitemOpen
  \bibfield  {author} {\bibinfo {author} {\bibfnamefont {V.}~\bibnamefont
  {De~Luca}}, \bibinfo {author} {\bibfnamefont {G.}~\bibnamefont
  {Franciolini}}, \bibinfo {author} {\bibfnamefont {P.}~\bibnamefont {Pani}}, \
  and\ \bibinfo {author} {\bibfnamefont {A.}~\bibnamefont {Riotto}},\ }\href
  {\doibase 10.1088/1475-7516/2020/04/052} {\bibfield  {journal} {\bibinfo
  {journal} {JCAP}\ }\textbf {\bibinfo {volume} {04}},\ \bibinfo {pages} {052}
  (\bibinfo {year} {2020}{\natexlab{b}})},\ \Eprint
  {http://arxiv.org/abs/2003.02778} {arXiv:2003.02778 [astro-ph.CO]}
  \BibitemShut {NoStop}%
\bibitem [{\citenamefont {Ricotti}\ \emph {et~al.}(2008)\citenamefont
  {Ricotti}, \citenamefont {Ostriker},\ and\ \citenamefont
  {Mack}}]{Ricotti:2007au}%
  \BibitemOpen
  \bibfield  {author} {\bibinfo {author} {\bibfnamefont {M.}~\bibnamefont
  {Ricotti}}, \bibinfo {author} {\bibfnamefont {J.~P.}\ \bibnamefont
  {Ostriker}}, \ and\ \bibinfo {author} {\bibfnamefont {K.~J.}\ \bibnamefont
  {Mack}},\ }\href {\doibase 10.1086/587831} {\bibfield  {journal} {\bibinfo
  {journal} {Astrophys. J.}\ }\textbf {\bibinfo {volume} {680}},\ \bibinfo
  {pages} {829} (\bibinfo {year} {2008})},\ \Eprint
  {http://arxiv.org/abs/0709.0524} {arXiv:0709.0524 [astro-ph]} \BibitemShut
  {NoStop}%
\bibitem [{\citenamefont {Ali-Ha\"\i{}moud}(2018)}]{Ali-Haimoud:2018dau}%
  \BibitemOpen
  \bibfield  {author} {\bibinfo {author} {\bibfnamefont {Y.}~\bibnamefont
  {Ali-Ha\"\i{}moud}},\ }\href {\doibase 10.1103/PhysRevLett.121.081304}
  {\bibfield  {journal} {\bibinfo  {journal} {Phys. Rev. Lett.}\ }\textbf
  {\bibinfo {volume} {121}},\ \bibinfo {pages} {081304} (\bibinfo {year}
  {2018})},\ \Eprint {http://arxiv.org/abs/1805.05912} {arXiv:1805.05912
  [astro-ph.CO]} \BibitemShut {NoStop}%
\bibitem [{\citenamefont {Desjacques}\ and\ \citenamefont
  {Riotto}(2018)}]{Desjacques:2018wuu}%
  \BibitemOpen
  \bibfield  {author} {\bibinfo {author} {\bibfnamefont {V.}~\bibnamefont
  {Desjacques}}\ and\ \bibinfo {author} {\bibfnamefont {A.}~\bibnamefont
  {Riotto}},\ }\href {\doibase 10.1103/PhysRevD.98.123533} {\bibfield
  {journal} {\bibinfo  {journal} {Phys. Rev. D}\ }\textbf {\bibinfo {volume}
  {98}},\ \bibinfo {pages} {123533} (\bibinfo {year} {2018})},\ \Eprint
  {http://arxiv.org/abs/1806.10414} {arXiv:1806.10414 [astro-ph.CO]}
  \BibitemShut {NoStop}%
\bibitem [{\citenamefont {Ballesteros}\ \emph {et~al.}(2018)\citenamefont
  {Ballesteros}, \citenamefont {Serpico},\ and\ \citenamefont
  {Taoso}}]{Ballesteros:2018swv}%
  \BibitemOpen
  \bibfield  {author} {\bibinfo {author} {\bibfnamefont {G.}~\bibnamefont
  {Ballesteros}}, \bibinfo {author} {\bibfnamefont {P.~D.}\ \bibnamefont
  {Serpico}}, \ and\ \bibinfo {author} {\bibfnamefont {M.}~\bibnamefont
  {Taoso}},\ }\href {\doibase 10.1088/1475-7516/2018/10/043} {\bibfield
  {journal} {\bibinfo  {journal} {JCAP}\ }\textbf {\bibinfo {volume} {10}},\
  \bibinfo {pages} {043} (\bibinfo {year} {2018})},\ \Eprint
  {http://arxiv.org/abs/1807.02084} {arXiv:1807.02084 [astro-ph.CO]}
  \BibitemShut {NoStop}%
\bibitem [{\citenamefont {Moradinezhad~Dizgah}\ \emph
  {et~al.}(2019)\citenamefont {Moradinezhad~Dizgah}, \citenamefont
  {Franciolini},\ and\ \citenamefont {Riotto}}]{MoradinezhadDizgah:2019wjf}%
  \BibitemOpen
  \bibfield  {author} {\bibinfo {author} {\bibfnamefont {A.}~\bibnamefont
  {Moradinezhad~Dizgah}}, \bibinfo {author} {\bibfnamefont {G.}~\bibnamefont
  {Franciolini}}, \ and\ \bibinfo {author} {\bibfnamefont {A.}~\bibnamefont
  {Riotto}},\ }\href {\doibase 10.1088/1475-7516/2019/11/001} {\bibfield
  {journal} {\bibinfo  {journal} {JCAP}\ }\textbf {\bibinfo {volume} {11}},\
  \bibinfo {pages} {001} (\bibinfo {year} {2019})},\ \Eprint
  {http://arxiv.org/abs/1906.08978} {arXiv:1906.08978 [astro-ph.CO]}
  \BibitemShut {NoStop}%
\bibitem [{\citenamefont {De~Luca}\ \emph
  {et~al.}(2020{\natexlab{c}})\citenamefont {De~Luca}, \citenamefont
  {Franciolini},\ and\ \citenamefont {Riotto}}]{DeLuca:2020ioi}%
  \BibitemOpen
  \bibfield  {author} {\bibinfo {author} {\bibfnamefont {V.}~\bibnamefont
  {De~Luca}}, \bibinfo {author} {\bibfnamefont {G.}~\bibnamefont
  {Franciolini}}, \ and\ \bibinfo {author} {\bibfnamefont {A.}~\bibnamefont
  {Riotto}},\ }\href {\doibase 10.1016/j.physletb.2020.135550} {\bibfield
  {journal} {\bibinfo  {journal} {Phys. Lett.}\ }\textbf {\bibinfo {volume}
  {B807}},\ \bibinfo {pages} {135550} (\bibinfo {year} {2020}{\natexlab{c}})},\
  \Eprint {http://arxiv.org/abs/2001.04371} {arXiv:2001.04371 [astro-ph.CO]}
  \BibitemShut {NoStop}%
\bibitem [{\citenamefont {Nakamura}\ \emph {et~al.}(1997)\citenamefont
  {Nakamura}, \citenamefont {Sasaki}, \citenamefont {Tanaka},\ and\
  \citenamefont {Thorne}}]{Nakamura:1997sm}%
  \BibitemOpen
  \bibfield  {author} {\bibinfo {author} {\bibfnamefont {T.}~\bibnamefont
  {Nakamura}}, \bibinfo {author} {\bibfnamefont {M.}~\bibnamefont {Sasaki}},
  \bibinfo {author} {\bibfnamefont {T.}~\bibnamefont {Tanaka}}, \ and\ \bibinfo
  {author} {\bibfnamefont {K.~S.}\ \bibnamefont {Thorne}},\ }\href {\doibase
  10.1086/310886} {\bibfield  {journal} {\bibinfo  {journal} {Astrophys. J.
  Lett.}\ }\textbf {\bibinfo {volume} {487}},\ \bibinfo {pages} {L139}
  (\bibinfo {year} {1997})},\ \Eprint {http://arxiv.org/abs/astro-ph/9708060}
  {arXiv:astro-ph/9708060} \BibitemShut {NoStop}%
\bibitem [{\citenamefont {Ioka}\ \emph {et~al.}(1998)\citenamefont {Ioka},
  \citenamefont {Chiba}, \citenamefont {Tanaka},\ and\ \citenamefont
  {Nakamura}}]{Ioka:1998nz}%
  \BibitemOpen
  \bibfield  {author} {\bibinfo {author} {\bibfnamefont {K.}~\bibnamefont
  {Ioka}}, \bibinfo {author} {\bibfnamefont {T.}~\bibnamefont {Chiba}},
  \bibinfo {author} {\bibfnamefont {T.}~\bibnamefont {Tanaka}}, \ and\ \bibinfo
  {author} {\bibfnamefont {T.}~\bibnamefont {Nakamura}},\ }\href {\doibase
  10.1103/PhysRevD.58.063003} {\bibfield  {journal} {\bibinfo  {journal} {Phys.
  Rev. D}\ }\textbf {\bibinfo {volume} {58}},\ \bibinfo {pages} {063003}
  (\bibinfo {year} {1998})},\ \Eprint {http://arxiv.org/abs/astro-ph/9807018}
  {arXiv:astro-ph/9807018} \BibitemShut {NoStop}%
\bibitem [{\citenamefont {Cholis}\ \emph {et~al.}(2016)\citenamefont {Cholis},
  \citenamefont {Kovetz}, \citenamefont {Ali-Ha\"\i{}moud}, \citenamefont
  {Bird}, \citenamefont {Kamionkowski}, \citenamefont {Mu\~noz},\ and\
  \citenamefont {Raccanelli}}]{Cholis:2016kqi}%
  \BibitemOpen
  \bibfield  {author} {\bibinfo {author} {\bibfnamefont {I.}~\bibnamefont
  {Cholis}}, \bibinfo {author} {\bibfnamefont {E.~D.}\ \bibnamefont {Kovetz}},
  \bibinfo {author} {\bibfnamefont {Y.}~\bibnamefont {Ali-Ha\"\i{}moud}},
  \bibinfo {author} {\bibfnamefont {S.}~\bibnamefont {Bird}}, \bibinfo {author}
  {\bibfnamefont {M.}~\bibnamefont {Kamionkowski}}, \bibinfo {author}
  {\bibfnamefont {J.~B.}\ \bibnamefont {Mu\~noz}}, \ and\ \bibinfo {author}
  {\bibfnamefont {A.}~\bibnamefont {Raccanelli}},\ }\href {\doibase
  10.1103/PhysRevD.94.084013} {\bibfield  {journal} {\bibinfo  {journal} {Phys.
  Rev. D}\ }\textbf {\bibinfo {volume} {94}},\ \bibinfo {pages} {084013}
  (\bibinfo {year} {2016})},\ \Eprint {http://arxiv.org/abs/1606.07437}
  {arXiv:1606.07437 [astro-ph.HE]} \BibitemShut {NoStop}%
\bibitem [{\citenamefont {Raidal}\ \emph {et~al.}(2017)\citenamefont {Raidal},
  \citenamefont {Vaskonen},\ and\ \citenamefont {Veerm\"ae}}]{Raidal:2017mfl}%
  \BibitemOpen
  \bibfield  {author} {\bibinfo {author} {\bibfnamefont {M.}~\bibnamefont
  {Raidal}}, \bibinfo {author} {\bibfnamefont {V.}~\bibnamefont {Vaskonen}}, \
  and\ \bibinfo {author} {\bibfnamefont {H.}~\bibnamefont {Veerm\"ae}},\ }\href
  {\doibase 10.1088/1475-7516/2017/09/037} {\bibfield  {journal} {\bibinfo
  {journal} {JCAP}\ }\textbf {\bibinfo {volume} {09}},\ \bibinfo {pages} {037}
  (\bibinfo {year} {2017})},\ \Eprint {http://arxiv.org/abs/1707.01480}
  {arXiv:1707.01480 [astro-ph.CO]} \BibitemShut {NoStop}%
\bibitem [{\citenamefont {De~Luca}\ \emph
  {et~al.}(2020{\natexlab{d}})\citenamefont {De~Luca}, \citenamefont
  {Desjacques}, \citenamefont {Franciolini},\ and\ \citenamefont
  {Riotto}}]{DeLuca:2020jug}%
  \BibitemOpen
  \bibfield  {author} {\bibinfo {author} {\bibfnamefont {V.}~\bibnamefont
  {De~Luca}}, \bibinfo {author} {\bibfnamefont {V.}~\bibnamefont {Desjacques}},
  \bibinfo {author} {\bibfnamefont {G.}~\bibnamefont {Franciolini}}, \ and\
  \bibinfo {author} {\bibfnamefont {A.}~\bibnamefont {Riotto}},\ }\href
  {\doibase 10.1088/1475-7516/2020/11/028} {\bibfield  {journal} {\bibinfo
  {journal} {JCAP}\ }\textbf {\bibinfo {volume} {11}},\ \bibinfo {pages} {028}
  (\bibinfo {year} {2020}{\natexlab{d}})},\ \Eprint
  {http://arxiv.org/abs/2009.04731} {arXiv:2009.04731 [astro-ph.CO]}
  \BibitemShut {NoStop}%
\bibitem [{\citenamefont {Schneider}\ \emph {et~al.}(2000)\citenamefont
  {Schneider}, \citenamefont {Ferrara}, \citenamefont {Ciardi}, \citenamefont
  {Ferrari},\ and\ \citenamefont {Matarrese}}]{Schneider:1999us}%
  \BibitemOpen
  \bibfield  {author} {\bibinfo {author} {\bibfnamefont {R.}~\bibnamefont
  {Schneider}}, \bibinfo {author} {\bibfnamefont {A.}~\bibnamefont {Ferrara}},
  \bibinfo {author} {\bibfnamefont {B.}~\bibnamefont {Ciardi}}, \bibinfo
  {author} {\bibfnamefont {V.}~\bibnamefont {Ferrari}}, \ and\ \bibinfo
  {author} {\bibfnamefont {S.}~\bibnamefont {Matarrese}},\ }\href {\doibase
  10.1046/j.1365-8711.2000.03596.x} {\bibfield  {journal} {\bibinfo  {journal}
  {Mon. Not. Roy. Astron. Soc.}\ }\textbf {\bibinfo {volume} {317}},\ \bibinfo
  {pages} {385} (\bibinfo {year} {2000})},\ \Eprint
  {http://arxiv.org/abs/astro-ph/9909419} {arXiv:astro-ph/9909419} \BibitemShut
  {NoStop}%
\bibitem [{\citenamefont {Schneider}\ \emph {et~al.}(2002)\citenamefont
  {Schneider}, \citenamefont {Ferrara}, \citenamefont {Natarajan},\ and\
  \citenamefont {Omukai}}]{Schneider:2001bu}%
  \BibitemOpen
  \bibfield  {author} {\bibinfo {author} {\bibfnamefont {R.}~\bibnamefont
  {Schneider}}, \bibinfo {author} {\bibfnamefont {A.}~\bibnamefont {Ferrara}},
  \bibinfo {author} {\bibfnamefont {P.}~\bibnamefont {Natarajan}}, \ and\
  \bibinfo {author} {\bibfnamefont {K.}~\bibnamefont {Omukai}},\ }\href
  {\doibase 10.1086/339917} {\bibfield  {journal} {\bibinfo  {journal}
  {Astrophys. J.}\ }\textbf {\bibinfo {volume} {571}},\ \bibinfo {pages} {30}
  (\bibinfo {year} {2002})},\ \Eprint {http://arxiv.org/abs/astro-ph/0111341}
  {arXiv:astro-ph/0111341} \BibitemShut {NoStop}%
\bibitem [{\citenamefont {Schneider}\ \emph {et~al.}(2003)\citenamefont
  {Schneider}, \citenamefont {Ferrara}, \citenamefont {Salvaterra},
  \citenamefont {Omukai},\ and\ \citenamefont {Bromm}}]{Schneider:2003em}%
  \BibitemOpen
  \bibfield  {author} {\bibinfo {author} {\bibfnamefont {R.}~\bibnamefont
  {Schneider}}, \bibinfo {author} {\bibfnamefont {A.}~\bibnamefont {Ferrara}},
  \bibinfo {author} {\bibfnamefont {R.}~\bibnamefont {Salvaterra}}, \bibinfo
  {author} {\bibfnamefont {K.}~\bibnamefont {Omukai}}, \ and\ \bibinfo {author}
  {\bibfnamefont {V.}~\bibnamefont {Bromm}},\ }\href {\doibase
  10.1038/nature01579} {\bibfield  {journal} {\bibinfo  {journal} {Nature}\
  }\textbf {\bibinfo {volume} {422}},\ \bibinfo {pages} {869} (\bibinfo {year}
  {2003})},\ \Eprint {http://arxiv.org/abs/astro-ph/0304254}
  {arXiv:astro-ph/0304254} \BibitemShut {NoStop}%
\bibitem [{\citenamefont {Bromm}(2006)}]{Bromm:2005ep}%
  \BibitemOpen
  \bibfield  {author} {\bibinfo {author} {\bibfnamefont {V.}~\bibnamefont
  {Bromm}},\ }\href {\doibase 10.1086/500799} {\bibfield  {journal} {\bibinfo
  {journal} {Astrophys. J.}\ }\textbf {\bibinfo {volume} {642}},\ \bibinfo
  {pages} {382} (\bibinfo {year} {2006})},\ \Eprint
  {http://arxiv.org/abs/astro-ph/0509303} {arXiv:astro-ph/0509303} \BibitemShut
  {NoStop}%
\bibitem [{\citenamefont {de~Souza}\ \emph {et~al.}(2011)\citenamefont
  {de~Souza}, \citenamefont {Yoshida},\ and\ \citenamefont
  {Ioka}}]{deSouza:2011ea}%
  \BibitemOpen
  \bibfield  {author} {\bibinfo {author} {\bibfnamefont {R.~S.}\ \bibnamefont
  {de~Souza}}, \bibinfo {author} {\bibfnamefont {N.}~\bibnamefont {Yoshida}}, \
  and\ \bibinfo {author} {\bibfnamefont {K.}~\bibnamefont {Ioka}},\ }\href
  {\doibase 10.1051/0004-6361/201117242} {\bibfield  {journal} {\bibinfo
  {journal} {Astron. Astrophys.}\ }\textbf {\bibinfo {volume} {533}},\ \bibinfo
  {pages} {A32} (\bibinfo {year} {2011})},\ \Eprint
  {http://arxiv.org/abs/1105.2395} {arXiv:1105.2395 [astro-ph.CO]} \BibitemShut
  {NoStop}%
\bibitem [{\citenamefont {Koushiappas}\ and\ \citenamefont
  {Loeb}(2017)}]{Koushiappas:2017kqm}%
  \BibitemOpen
  \bibfield  {author} {\bibinfo {author} {\bibfnamefont {S.~M.}\ \bibnamefont
  {Koushiappas}}\ and\ \bibinfo {author} {\bibfnamefont {A.}~\bibnamefont
  {Loeb}},\ }\href {\doibase 10.1103/PhysRevLett.119.221104} {\bibfield
  {journal} {\bibinfo  {journal} {Phys. Rev. Lett.}\ }\textbf {\bibinfo
  {volume} {119}},\ \bibinfo {pages} {221104} (\bibinfo {year} {2017})},\
  \Eprint {http://arxiv.org/abs/1708.07380} {arXiv:1708.07380 [astro-ph.CO]}
  \BibitemShut {NoStop}%
\bibitem [{\citenamefont {Mocz}\ \emph {et~al.}(2020)\citenamefont {Mocz} \emph
  {et~al.}}]{Mocz:2019uyd}%
  \BibitemOpen
  \bibfield  {author} {\bibinfo {author} {\bibfnamefont {P.}~\bibnamefont
  {Mocz}} \emph {et~al.},\ }\href {\doibase 10.1093/mnras/staa738} {\bibfield
  {journal} {\bibinfo  {journal} {Mon. Not. Roy. Astron. Soc.}\ }\textbf
  {\bibinfo {volume} {494}},\ \bibinfo {pages} {2027} (\bibinfo {year}
  {2020})},\ \Eprint {http://arxiv.org/abs/1911.05746} {arXiv:1911.05746
  [astro-ph.CO]} \BibitemShut {NoStop}%
\bibitem [{\citenamefont {Liu}\ and\ \citenamefont
  {Bromm}(2020{\natexlab{a}})}]{Liu:2020ufc}%
  \BibitemOpen
  \bibfield  {author} {\bibinfo {author} {\bibfnamefont {B.}~\bibnamefont
  {Liu}}\ and\ \bibinfo {author} {\bibfnamefont {V.}~\bibnamefont {Bromm}},\
  }\href {\doibase 10.1093/mnras/staa1362} {\bibfield  {journal} {\bibinfo
  {journal} {Mon. Not. Roy. Astron. Soc.}\ }\textbf {\bibinfo {volume} {495}},\
  \bibinfo {pages} {2475} (\bibinfo {year} {2020}{\natexlab{a}})},\ \Eprint
  {http://arxiv.org/abs/2003.00065} {arXiv:2003.00065 [astro-ph.CO]}
  \BibitemShut {NoStop}%
\bibitem [{\citenamefont {Trenti}\ and\ \citenamefont
  {Stiavelli}(2009)}]{Trenti:2009cj}%
  \BibitemOpen
  \bibfield  {author} {\bibinfo {author} {\bibfnamefont {M.}~\bibnamefont
  {Trenti}}\ and\ \bibinfo {author} {\bibfnamefont {M.}~\bibnamefont
  {Stiavelli}},\ }\href {\doibase 10.1088/0004-637X/694/2/879} {\bibfield
  {journal} {\bibinfo  {journal} {Astrophys. J.}\ }\textbf {\bibinfo {volume}
  {694}},\ \bibinfo {pages} {879} (\bibinfo {year} {2009})},\ \Eprint
  {http://arxiv.org/abs/0901.0711} {arXiv:0901.0711 [astro-ph.CO]} \BibitemShut
  {NoStop}%
\bibitem [{\citenamefont {Tornatore}\ \emph {et~al.}(2007)\citenamefont
  {Tornatore}, \citenamefont {Borgani}, \citenamefont {Dolag},\ and\
  \citenamefont {Matteucci}}]{Tornatore:2007ds}%
  \BibitemOpen
  \bibfield  {author} {\bibinfo {author} {\bibfnamefont {L.}~\bibnamefont
  {Tornatore}}, \bibinfo {author} {\bibfnamefont {S.}~\bibnamefont {Borgani}},
  \bibinfo {author} {\bibfnamefont {K.}~\bibnamefont {Dolag}}, \ and\ \bibinfo
  {author} {\bibfnamefont {F.}~\bibnamefont {Matteucci}},\ }\href {\doibase
  10.1111/j.1365-2966.2007.12070.x} {\bibfield  {journal} {\bibinfo  {journal}
  {Mon. Not. Roy. Astron. Soc.}\ }\textbf {\bibinfo {volume} {382}},\ \bibinfo
  {pages} {1050} (\bibinfo {year} {2007})},\ \Eprint
  {http://arxiv.org/abs/0705.1921} {arXiv:0705.1921 [astro-ph]} \BibitemShut
  {NoStop}%
\bibitem [{\citenamefont {Kinugawa}\ \emph {et~al.}(2014)\citenamefont
  {Kinugawa}, \citenamefont {Inayoshi}, \citenamefont {Hotokezaka},
  \citenamefont {Nakauchi},\ and\ \citenamefont {Nakamura}}]{Kinugawa:2014zha}%
  \BibitemOpen
  \bibfield  {author} {\bibinfo {author} {\bibfnamefont {T.}~\bibnamefont
  {Kinugawa}}, \bibinfo {author} {\bibfnamefont {K.}~\bibnamefont {Inayoshi}},
  \bibinfo {author} {\bibfnamefont {K.}~\bibnamefont {Hotokezaka}}, \bibinfo
  {author} {\bibfnamefont {D.}~\bibnamefont {Nakauchi}}, \ and\ \bibinfo
  {author} {\bibfnamefont {T.}~\bibnamefont {Nakamura}},\ }\href {\doibase
  10.1093/mnras/stu1022} {\bibfield  {journal} {\bibinfo  {journal} {Mon. Not.
  Roy. Astron. Soc.}\ }\textbf {\bibinfo {volume} {442}},\ \bibinfo {pages}
  {2963} (\bibinfo {year} {2014})},\ \Eprint {http://arxiv.org/abs/1402.6672}
  {arXiv:1402.6672 [astro-ph.HE]} \BibitemShut {NoStop}%
\bibitem [{\citenamefont {Kinugawa}\ \emph {et~al.}(2016)\citenamefont
  {Kinugawa}, \citenamefont {Miyamoto}, \citenamefont {Kanda},\ and\
  \citenamefont {Nakamura}}]{Kinugawa:2015nla}%
  \BibitemOpen
  \bibfield  {author} {\bibinfo {author} {\bibfnamefont {T.}~\bibnamefont
  {Kinugawa}}, \bibinfo {author} {\bibfnamefont {A.}~\bibnamefont {Miyamoto}},
  \bibinfo {author} {\bibfnamefont {N.}~\bibnamefont {Kanda}}, \ and\ \bibinfo
  {author} {\bibfnamefont {T.}~\bibnamefont {Nakamura}},\ }\href {\doibase
  10.1093/mnras/stv2624} {\bibfield  {journal} {\bibinfo  {journal} {Mon. Not.
  Roy. Astron. Soc.}\ }\textbf {\bibinfo {volume} {456}},\ \bibinfo {pages}
  {1093} (\bibinfo {year} {2016})},\ \Eprint {http://arxiv.org/abs/1505.06962}
  {arXiv:1505.06962 [astro-ph.SR]} \BibitemShut {NoStop}%
\bibitem [{\citenamefont {Hartwig}\ \emph {et~al.}(2016)\citenamefont
  {Hartwig}, \citenamefont {Volonteri}, \citenamefont {Bromm}, \citenamefont
  {Klessen}, \citenamefont {Barausse}, \citenamefont {Magg},\ and\
  \citenamefont {Stacy}}]{Hartwig:2016nde}%
  \BibitemOpen
  \bibfield  {author} {\bibinfo {author} {\bibfnamefont {T.}~\bibnamefont
  {Hartwig}}, \bibinfo {author} {\bibfnamefont {M.}~\bibnamefont {Volonteri}},
  \bibinfo {author} {\bibfnamefont {V.}~\bibnamefont {Bromm}}, \bibinfo
  {author} {\bibfnamefont {R.~S.}\ \bibnamefont {Klessen}}, \bibinfo {author}
  {\bibfnamefont {E.}~\bibnamefont {Barausse}}, \bibinfo {author}
  {\bibfnamefont {M.}~\bibnamefont {Magg}}, \ and\ \bibinfo {author}
  {\bibfnamefont {A.}~\bibnamefont {Stacy}},\ }\href {\doibase
  10.1093/mnrasl/slw074} {\bibfield  {journal} {\bibinfo  {journal} {Mon. Not.
  Roy. Astron. Soc.}\ }\textbf {\bibinfo {volume} {460}},\ \bibinfo {pages}
  {L74} (\bibinfo {year} {2016})},\ \Eprint {http://arxiv.org/abs/1603.05655}
  {arXiv:1603.05655 [astro-ph.GA]} \BibitemShut {NoStop}%
\bibitem [{\citenamefont {Belczynski}\ \emph {et~al.}(2017)\citenamefont
  {Belczynski}, \citenamefont {Ryu}, \citenamefont {Perna}, \citenamefont
  {Berti}, \citenamefont {Tanaka},\ and\ \citenamefont
  {Bulik}}]{Belczynski:2016ieo}%
  \BibitemOpen
  \bibfield  {author} {\bibinfo {author} {\bibfnamefont {K.}~\bibnamefont
  {Belczynski}}, \bibinfo {author} {\bibfnamefont {T.}~\bibnamefont {Ryu}},
  \bibinfo {author} {\bibfnamefont {R.}~\bibnamefont {Perna}}, \bibinfo
  {author} {\bibfnamefont {E.}~\bibnamefont {Berti}}, \bibinfo {author}
  {\bibfnamefont {T.~L.}\ \bibnamefont {Tanaka}}, \ and\ \bibinfo {author}
  {\bibfnamefont {T.}~\bibnamefont {Bulik}},\ }\href {\doibase
  10.1093/mnras/stx1759} {\bibfield  {journal} {\bibinfo  {journal} {Mon. Not.
  Roy. Astron. Soc.}\ }\textbf {\bibinfo {volume} {471}},\ \bibinfo {pages}
  {4702} (\bibinfo {year} {2017})},\ \Eprint {http://arxiv.org/abs/1612.01524}
  {arXiv:1612.01524 [astro-ph.HE]} \BibitemShut {NoStop}%
\bibitem [{\citenamefont {Inayoshi}\ \emph {et~al.}(2017)\citenamefont
  {Inayoshi}, \citenamefont {Hirai}, \citenamefont {Kinugawa},\ and\
  \citenamefont {Hotokezaka}}]{Inayoshi:2017mrs}%
  \BibitemOpen
  \bibfield  {author} {\bibinfo {author} {\bibfnamefont {K.}~\bibnamefont
  {Inayoshi}}, \bibinfo {author} {\bibfnamefont {R.}~\bibnamefont {Hirai}},
  \bibinfo {author} {\bibfnamefont {T.}~\bibnamefont {Kinugawa}}, \ and\
  \bibinfo {author} {\bibfnamefont {K.}~\bibnamefont {Hotokezaka}},\ }\href
  {\doibase 10.1093/mnras/stx757} {\bibfield  {journal} {\bibinfo  {journal}
  {Mon. Not. Roy. Astron. Soc.}\ }\textbf {\bibinfo {volume} {468}},\ \bibinfo
  {pages} {5020} (\bibinfo {year} {2017})},\ \Eprint
  {http://arxiv.org/abs/1701.04823} {arXiv:1701.04823 [astro-ph.HE]}
  \BibitemShut {NoStop}%
\bibitem [{\citenamefont {Liu}\ and\ \citenamefont
  {Bromm}(2020{\natexlab{b}})}]{Liu:2020lmi}%
  \BibitemOpen
  \bibfield  {author} {\bibinfo {author} {\bibfnamefont {B.}~\bibnamefont
  {Liu}}\ and\ \bibinfo {author} {\bibfnamefont {V.}~\bibnamefont {Bromm}},\
  }\href {\doibase 10.3847/2041-8213/abc552} {\bibfield  {journal} {\bibinfo
  {journal} {Astrophys. J. Lett.}\ }\textbf {\bibinfo {volume} {903}},\
  \bibinfo {pages} {L40} (\bibinfo {year} {2020}{\natexlab{b}})},\ \Eprint
  {http://arxiv.org/abs/2009.11447} {arXiv:2009.11447 [astro-ph.GA]}
  \BibitemShut {NoStop}%
\bibitem [{\citenamefont {Kinugawa}\ \emph {et~al.}(2020)\citenamefont
  {Kinugawa}, \citenamefont {Nakamura},\ and\ \citenamefont
  {Nakano}}]{Kinugawa:2020ego}%
  \BibitemOpen
  \bibfield  {author} {\bibinfo {author} {\bibfnamefont {T.}~\bibnamefont
  {Kinugawa}}, \bibinfo {author} {\bibfnamefont {T.}~\bibnamefont {Nakamura}},
  \ and\ \bibinfo {author} {\bibfnamefont {H.}~\bibnamefont {Nakano}},\ }\href
  {\doibase 10.1093/mnras/staa2511} {\bibfield  {journal} {\bibinfo  {journal}
  {Mon. Not. Roy. Astron. Soc.}\ }\textbf {\bibinfo {volume} {498}},\ \bibinfo
  {pages} {3946} (\bibinfo {year} {2020})},\ \Eprint
  {http://arxiv.org/abs/2005.09795} {arXiv:2005.09795 [astro-ph.HE]}
  \BibitemShut {NoStop}%
\bibitem [{\citenamefont {Tanikawa}\ \emph {et~al.}(2021)\citenamefont
  {Tanikawa}, \citenamefont {Susa}, \citenamefont {Yoshida}, \citenamefont
  {Trani},\ and\ \citenamefont {Kinugawa}}]{Tanikawa:2020cca}%
  \BibitemOpen
  \bibfield  {author} {\bibinfo {author} {\bibfnamefont {A.}~\bibnamefont
  {Tanikawa}}, \bibinfo {author} {\bibfnamefont {H.}~\bibnamefont {Susa}},
  \bibinfo {author} {\bibfnamefont {T.}~\bibnamefont {Yoshida}}, \bibinfo
  {author} {\bibfnamefont {A.~A.}\ \bibnamefont {Trani}}, \ and\ \bibinfo
  {author} {\bibfnamefont {T.}~\bibnamefont {Kinugawa}},\ }\href {\doibase
  10.3847/1538-4357/abe40d} {\bibfield  {journal} {\bibinfo  {journal}
  {Astrophys. J.}\ }\textbf {\bibinfo {volume} {910}},\ \bibinfo {pages} {30}
  (\bibinfo {year} {2021})},\ \Eprint {http://arxiv.org/abs/2008.01890}
  {arXiv:2008.01890 [astro-ph.HE]} \BibitemShut {NoStop}%
\bibitem [{\citenamefont {Singh}\ \emph {et~al.}(2021)\citenamefont {Singh},
  \citenamefont {Bulik}, \citenamefont {Belczynski},\ and\ \citenamefont
  {Askar}}]{Singh:2021zah}%
  \BibitemOpen
  \bibfield  {author} {\bibinfo {author} {\bibfnamefont {N.}~\bibnamefont
  {Singh}}, \bibinfo {author} {\bibfnamefont {T.}~\bibnamefont {Bulik}},
  \bibinfo {author} {\bibfnamefont {K.}~\bibnamefont {Belczynski}}, \ and\
  \bibinfo {author} {\bibfnamefont {A.}~\bibnamefont {Askar}},\ }\href@noop {}
  {\  (\bibinfo {year} {2021})},\ \Eprint {http://arxiv.org/abs/2112.04058}
  {arXiv:2112.04058 [astro-ph.HE]} \BibitemShut {NoStop}%
\bibitem [{\citenamefont {Ng}\ \emph {et~al.}(2021)\citenamefont {Ng},
  \citenamefont {Chen}, \citenamefont {Goncharov}, \citenamefont {Dupletsa},
  \citenamefont {Borhanian}, \citenamefont {Branchesi}, \citenamefont {Harms},
  \citenamefont {Maggiore}, \citenamefont {Sathyaprakash},\ and\ \citenamefont
  {Vitale}}]{Ng:2021sqn}%
  \BibitemOpen
  \bibfield  {author} {\bibinfo {author} {\bibfnamefont {K.~K.~Y.}\
  \bibnamefont {Ng}}, \bibinfo {author} {\bibfnamefont {S.}~\bibnamefont
  {Chen}}, \bibinfo {author} {\bibfnamefont {B.}~\bibnamefont {Goncharov}},
  \bibinfo {author} {\bibfnamefont {U.}~\bibnamefont {Dupletsa}}, \bibinfo
  {author} {\bibfnamefont {S.}~\bibnamefont {Borhanian}}, \bibinfo {author}
  {\bibfnamefont {M.}~\bibnamefont {Branchesi}}, \bibinfo {author}
  {\bibfnamefont {J.}~\bibnamefont {Harms}}, \bibinfo {author} {\bibfnamefont
  {M.}~\bibnamefont {Maggiore}}, \bibinfo {author} {\bibfnamefont {B.~S.}\
  \bibnamefont {Sathyaprakash}}, \ and\ \bibinfo {author} {\bibfnamefont
  {S.}~\bibnamefont {Vitale}},\ }\href@noop {} {\  (\bibinfo {year} {2021})},\
  \Eprint {http://arxiv.org/abs/2108.07276} {arXiv:2108.07276 [astro-ph.CO]}
  \BibitemShut {NoStop}%
\bibitem [{\citenamefont {Shandera}\ \emph {et~al.}(2018)\citenamefont
  {Shandera}, \citenamefont {Jeong},\ and\ \citenamefont
  {Gebhardt}}]{Shandera:2018xkn}%
  \BibitemOpen
  \bibfield  {author} {\bibinfo {author} {\bibfnamefont {S.}~\bibnamefont
  {Shandera}}, \bibinfo {author} {\bibfnamefont {D.}~\bibnamefont {Jeong}}, \
  and\ \bibinfo {author} {\bibfnamefont {H.~S.~G.}\ \bibnamefont {Gebhardt}},\
  }\href {\doibase 10.1103/PhysRevLett.120.241102} {\bibfield  {journal}
  {\bibinfo  {journal} {Phys. Rev. Lett.}\ }\textbf {\bibinfo {volume} {120}},\
  \bibinfo {pages} {241102} (\bibinfo {year} {2018})},\ \Eprint
  {http://arxiv.org/abs/1802.08206} {arXiv:1802.08206 [astro-ph.CO]}
  \BibitemShut {NoStop}%
\bibitem [{\citenamefont {Cardoso}\ and\ \citenamefont
  {Pani}(2019)}]{Cardoso:2019rvt}%
  \BibitemOpen
  \bibfield  {author} {\bibinfo {author} {\bibfnamefont {V.}~\bibnamefont
  {Cardoso}}\ and\ \bibinfo {author} {\bibfnamefont {P.}~\bibnamefont {Pani}},\
  }\href {\doibase 10.1007/s41114-019-0020-4} {\bibfield  {journal} {\bibinfo
  {journal} {Living Rev. Rel.}\ }\textbf {\bibinfo {volume} {22}},\ \bibinfo
  {pages} {4} (\bibinfo {year} {2019})},\ \Eprint
  {http://arxiv.org/abs/1904.05363} {arXiv:1904.05363 [gr-qc]} \BibitemShut
  {NoStop}%
\bibitem [{\citenamefont {Guo}\ \emph {et~al.}(2019)\citenamefont {Guo},
  \citenamefont {Sinha},\ and\ \citenamefont {Sun}}]{Guo:2019sns}%
  \BibitemOpen
  \bibfield  {author} {\bibinfo {author} {\bibfnamefont {H.-K.}\ \bibnamefont
  {Guo}}, \bibinfo {author} {\bibfnamefont {K.}~\bibnamefont {Sinha}}, \ and\
  \bibinfo {author} {\bibfnamefont {C.}~\bibnamefont {Sun}},\ }\href {\doibase
  10.1088/1475-7516/2019/09/032} {\bibfield  {journal} {\bibinfo  {journal}
  {JCAP}\ }\textbf {\bibinfo {volume} {09}},\ \bibinfo {pages} {032} (\bibinfo
  {year} {2019})},\ \Eprint {http://arxiv.org/abs/1904.07871} {arXiv:1904.07871
  [hep-ph]} \BibitemShut {NoStop}%
\bibitem [{\citenamefont {Barsanti}\ \emph {et~al.}(2021)\citenamefont
  {Barsanti}, \citenamefont {De~Luca}, \citenamefont {Maselli},\ and\
  \citenamefont {Pani}}]{Barsanti:2021ydd}%
  \BibitemOpen
  \bibfield  {author} {\bibinfo {author} {\bibfnamefont {S.}~\bibnamefont
  {Barsanti}}, \bibinfo {author} {\bibfnamefont {V.}~\bibnamefont {De~Luca}},
  \bibinfo {author} {\bibfnamefont {A.}~\bibnamefont {Maselli}}, \ and\
  \bibinfo {author} {\bibfnamefont {P.}~\bibnamefont {Pani}},\ }\href@noop {}
  {\  (\bibinfo {year} {2021})},\ \Eprint {http://arxiv.org/abs/2109.02170}
  {arXiv:2109.02170 [gr-qc]} \BibitemShut {NoStop}%
\bibitem [{\citenamefont {Bramante}\ \emph {et~al.}(2018)\citenamefont
  {Bramante}, \citenamefont {Linden},\ and\ \citenamefont
  {Tsai}}]{Bramante:2017ulk}%
  \BibitemOpen
  \bibfield  {author} {\bibinfo {author} {\bibfnamefont {J.}~\bibnamefont
  {Bramante}}, \bibinfo {author} {\bibfnamefont {T.}~\bibnamefont {Linden}}, \
  and\ \bibinfo {author} {\bibfnamefont {Y.-D.}\ \bibnamefont {Tsai}},\ }\href
  {\doibase 10.1103/PhysRevD.97.055016} {\bibfield  {journal} {\bibinfo
  {journal} {Phys. Rev. D}\ }\textbf {\bibinfo {volume} {97}},\ \bibinfo
  {pages} {055016} (\bibinfo {year} {2018})},\ \Eprint
  {http://arxiv.org/abs/1706.00001} {arXiv:1706.00001 [hep-ph]} \BibitemShut
  {NoStop}%
\bibitem [{\citenamefont {Takhistov}\ \emph {et~al.}(2021)\citenamefont
  {Takhistov}, \citenamefont {Fuller},\ and\ \citenamefont
  {Kusenko}}]{Takhistov:2020vxs}%
  \BibitemOpen
  \bibfield  {author} {\bibinfo {author} {\bibfnamefont {V.}~\bibnamefont
  {Takhistov}}, \bibinfo {author} {\bibfnamefont {G.~M.}\ \bibnamefont
  {Fuller}}, \ and\ \bibinfo {author} {\bibfnamefont {A.}~\bibnamefont
  {Kusenko}},\ }\href {\doibase 10.1103/PhysRevLett.126.071101} {\bibfield
  {journal} {\bibinfo  {journal} {Phys. Rev. Lett.}\ }\textbf {\bibinfo
  {volume} {126}},\ \bibinfo {pages} {071101} (\bibinfo {year} {2021})},\
  \Eprint {http://arxiv.org/abs/2008.12780} {arXiv:2008.12780 [astro-ph.HE]}
  \BibitemShut {NoStop}%
\bibitem [{\citenamefont {Dasgupta}\ \emph {et~al.}(2021)\citenamefont
  {Dasgupta}, \citenamefont {Laha},\ and\ \citenamefont
  {Ray}}]{Dasgupta:2020mqg}%
  \BibitemOpen
  \bibfield  {author} {\bibinfo {author} {\bibfnamefont {B.}~\bibnamefont
  {Dasgupta}}, \bibinfo {author} {\bibfnamefont {R.}~\bibnamefont {Laha}}, \
  and\ \bibinfo {author} {\bibfnamefont {A.}~\bibnamefont {Ray}},\ }\href
  {\doibase 10.1103/PhysRevLett.126.141105} {\bibfield  {journal} {\bibinfo
  {journal} {Phys. Rev. Lett.}\ }\textbf {\bibinfo {volume} {126}},\ \bibinfo
  {pages} {141105} (\bibinfo {year} {2021})},\ \Eprint
  {http://arxiv.org/abs/2009.01825} {arXiv:2009.01825 [astro-ph.HE]}
  \BibitemShut {NoStop}%
\bibitem [{\citenamefont {Giffin}\ \emph {et~al.}(2021)\citenamefont {Giffin},
  \citenamefont {Lloyd}, \citenamefont {McDermott},\ and\ \citenamefont
  {Profumo}}]{Giffin:2021kgb}%
  \BibitemOpen
  \bibfield  {author} {\bibinfo {author} {\bibfnamefont {P.}~\bibnamefont
  {Giffin}}, \bibinfo {author} {\bibfnamefont {J.}~\bibnamefont {Lloyd}},
  \bibinfo {author} {\bibfnamefont {S.~D.}\ \bibnamefont {McDermott}}, \ and\
  \bibinfo {author} {\bibfnamefont {S.}~\bibnamefont {Profumo}},\ }\href@noop
  {} {\  (\bibinfo {year} {2021})},\ \Eprint {http://arxiv.org/abs/2105.06504}
  {arXiv:2105.06504 [hep-ph]} \BibitemShut {NoStop}%
\bibitem [{\citenamefont {Fasano}\ \emph {et~al.}(2020)\citenamefont {Fasano},
  \citenamefont {Wong}, \citenamefont {Maselli}, \citenamefont {Berti},
  \citenamefont {Ferrari},\ and\ \citenamefont
  {Sathyaprakash}}]{Fasano:2020eum}%
  \BibitemOpen
  \bibfield  {author} {\bibinfo {author} {\bibfnamefont {M.}~\bibnamefont
  {Fasano}}, \bibinfo {author} {\bibfnamefont {K.~W.~K.}\ \bibnamefont {Wong}},
  \bibinfo {author} {\bibfnamefont {A.}~\bibnamefont {Maselli}}, \bibinfo
  {author} {\bibfnamefont {E.}~\bibnamefont {Berti}}, \bibinfo {author}
  {\bibfnamefont {V.}~\bibnamefont {Ferrari}}, \ and\ \bibinfo {author}
  {\bibfnamefont {B.~S.}\ \bibnamefont {Sathyaprakash}},\ }\href {\doibase
  10.1103/PhysRevD.102.023025} {\bibfield  {journal} {\bibinfo  {journal}
  {Phys. Rev. D}\ }\textbf {\bibinfo {volume} {102}},\ \bibinfo {pages}
  {023025} (\bibinfo {year} {2020})},\ \Eprint
  {http://arxiv.org/abs/2005.01726} {arXiv:2005.01726 [astro-ph.HE]}
  \BibitemShut {NoStop}%
\bibitem [{\citenamefont {Hofmann}\ \emph {et~al.}(2016)\citenamefont
  {Hofmann}, \citenamefont {Barausse},\ and\ \citenamefont
  {Rezzolla}}]{Hofmann:2016yih}%
  \BibitemOpen
  \bibfield  {author} {\bibinfo {author} {\bibfnamefont {F.}~\bibnamefont
  {Hofmann}}, \bibinfo {author} {\bibfnamefont {E.}~\bibnamefont {Barausse}}, \
  and\ \bibinfo {author} {\bibfnamefont {L.}~\bibnamefont {Rezzolla}},\ }\href
  {\doibase 10.3847/2041-8205/825/2/L19} {\bibfield  {journal} {\bibinfo
  {journal} {Astrophys. J. Lett.}\ }\textbf {\bibinfo {volume} {825}},\
  \bibinfo {pages} {L19} (\bibinfo {year} {2016})},\ \Eprint
  {http://arxiv.org/abs/1605.01938} {arXiv:1605.01938 [gr-qc]} \BibitemShut
  {NoStop}%
\bibitem [{\citenamefont {Bardeen}\ \emph {et~al.}(1986)\citenamefont
  {Bardeen}, \citenamefont {Bond}, \citenamefont {Kaiser},\ and\ \citenamefont
  {Szalay}}]{bbks}%
  \BibitemOpen
  \bibfield  {author} {\bibinfo {author} {\bibfnamefont {J.~M.}\ \bibnamefont
  {Bardeen}}, \bibinfo {author} {\bibfnamefont {J.}~\bibnamefont {Bond}},
  \bibinfo {author} {\bibfnamefont {N.}~\bibnamefont {Kaiser}}, \ and\ \bibinfo
  {author} {\bibfnamefont {A.}~\bibnamefont {Szalay}},\ }\href {\doibase
  10.1086/164143} {\bibfield  {journal} {\bibinfo  {journal} {Astrophys. J.}\
  }\textbf {\bibinfo {volume} {304}},\ \bibinfo {pages} {15} (\bibinfo {year}
  {1986})}\BibitemShut {NoStop}%
\bibitem [{\citenamefont {De~Luca}\ \emph {et~al.}(2019)\citenamefont
  {De~Luca}, \citenamefont {Desjacques}, \citenamefont {Franciolini},
  \citenamefont {Malhotra},\ and\ \citenamefont {Riotto}}]{DeLuca:2019buf}%
  \BibitemOpen
  \bibfield  {author} {\bibinfo {author} {\bibfnamefont {V.}~\bibnamefont
  {De~Luca}}, \bibinfo {author} {\bibfnamefont {V.}~\bibnamefont {Desjacques}},
  \bibinfo {author} {\bibfnamefont {G.}~\bibnamefont {Franciolini}}, \bibinfo
  {author} {\bibfnamefont {A.}~\bibnamefont {Malhotra}}, \ and\ \bibinfo
  {author} {\bibfnamefont {A.}~\bibnamefont {Riotto}},\ }\href {\doibase
  10.1088/1475-7516/2019/05/018} {\bibfield  {journal} {\bibinfo  {journal}
  {JCAP}\ }\textbf {\bibinfo {volume} {05}},\ \bibinfo {pages} {018} (\bibinfo
  {year} {2019})},\ \Eprint {http://arxiv.org/abs/1903.01179} {arXiv:1903.01179
  [astro-ph.CO]} \BibitemShut {NoStop}%
\bibitem [{\citenamefont {Mirbabayi}\ \emph {et~al.}(2020)\citenamefont
  {Mirbabayi}, \citenamefont {Gruzinov},\ and\ \citenamefont
  {Noreña}}]{Mirbabayi:2019uph}%
  \BibitemOpen
  \bibfield  {author} {\bibinfo {author} {\bibfnamefont {M.}~\bibnamefont
  {Mirbabayi}}, \bibinfo {author} {\bibfnamefont {A.}~\bibnamefont {Gruzinov}},
  \ and\ \bibinfo {author} {\bibfnamefont {J.}~\bibnamefont {Noreña}},\ }\href
  {\doibase 10.1088/1475-7516/2020/03/017} {\bibfield  {journal} {\bibinfo
  {journal} {JCAP}\ }\textbf {\bibinfo {volume} {2003}},\ \bibinfo {pages}
  {017} (\bibinfo {year} {2020})},\ \Eprint {http://arxiv.org/abs/1901.05963}
  {arXiv:1901.05963 [astro-ph.CO]} \BibitemShut {NoStop}%
\bibitem [{\citenamefont {Ricotti}(2007)}]{Ricotti:2007jk}%
  \BibitemOpen
  \bibfield  {author} {\bibinfo {author} {\bibfnamefont {M.}~\bibnamefont
  {Ricotti}},\ }\href {\doibase 10.1086/516562} {\bibfield  {journal} {\bibinfo
   {journal} {Astrophys. J.}\ }\textbf {\bibinfo {volume} {662}},\ \bibinfo
  {pages} {53} (\bibinfo {year} {2007})},\ \Eprint
  {http://arxiv.org/abs/0706.0864} {arXiv:0706.0864 [astro-ph]} \BibitemShut
  {NoStop}%
\bibitem [{\citenamefont {Hasinger}(2020)}]{Hasinger:2020ptw}%
  \BibitemOpen
  \bibfield  {author} {\bibinfo {author} {\bibfnamefont {G.}~\bibnamefont
  {Hasinger}},\ }\href@noop {} {\  (\bibinfo {year} {2020})},\ \Eprint
  {http://arxiv.org/abs/2003.05150} {arXiv:2003.05150 [astro-ph.CO]}
  \BibitemShut {NoStop}%
\bibitem [{\citenamefont {Oh}\ and\ \citenamefont {Haiman}(2003)}]{Oh:2003pm}%
  \BibitemOpen
  \bibfield  {author} {\bibinfo {author} {\bibfnamefont {S.~P.}\ \bibnamefont
  {Oh}}\ and\ \bibinfo {author} {\bibfnamefont {Z.}~\bibnamefont {Haiman}},\
  }\href {\doibase 10.1046/j.1365-2966.2003.07103.x} {\bibfield  {journal}
  {\bibinfo  {journal} {Mon. Not. Roy. Astron. Soc.}\ }\textbf {\bibinfo
  {volume} {346}},\ \bibinfo {pages} {456} (\bibinfo {year} {2003})},\ \Eprint
  {http://arxiv.org/abs/astro-ph/0307135} {arXiv:astro-ph/0307135 [astro-ph]}
  \BibitemShut {NoStop}%
\bibitem [{\citenamefont {De~Luca}\ \emph
  {et~al.}(2020{\natexlab{e}})\citenamefont {De~Luca}, \citenamefont
  {Franciolini}, \citenamefont {Pani},\ and\ \citenamefont
  {Riotto}}]{DeLuca:2020fpg}%
  \BibitemOpen
  \bibfield  {author} {\bibinfo {author} {\bibfnamefont {V.}~\bibnamefont
  {De~Luca}}, \bibinfo {author} {\bibfnamefont {G.}~\bibnamefont
  {Franciolini}}, \bibinfo {author} {\bibfnamefont {P.}~\bibnamefont {Pani}}, \
  and\ \bibinfo {author} {\bibfnamefont {A.}~\bibnamefont {Riotto}},\ }\href
  {\doibase 10.1103/PhysRevD.102.043505} {\bibfield  {journal} {\bibinfo
  {journal} {Phys. Rev. D}\ }\textbf {\bibinfo {volume} {102}},\ \bibinfo
  {pages} {043505} (\bibinfo {year} {2020}{\natexlab{e}})},\ \Eprint
  {http://arxiv.org/abs/2003.12589} {arXiv:2003.12589 [astro-ph.CO]}
  \BibitemShut {NoStop}%
\bibitem [{\citenamefont {Harada}\ \emph {et~al.}(2017)\citenamefont {Harada},
  \citenamefont {Yoo}, \citenamefont {Kohri},\ and\ \citenamefont
  {Nakao}}]{Harada:2017fjm}%
  \BibitemOpen
  \bibfield  {author} {\bibinfo {author} {\bibfnamefont {T.}~\bibnamefont
  {Harada}}, \bibinfo {author} {\bibfnamefont {C.-M.}\ \bibnamefont {Yoo}},
  \bibinfo {author} {\bibfnamefont {K.}~\bibnamefont {Kohri}}, \ and\ \bibinfo
  {author} {\bibfnamefont {K.-I.}\ \bibnamefont {Nakao}},\ }\href {\doibase
  10.1103/PhysRevD.96.083517} {\bibfield  {journal} {\bibinfo  {journal} {Phys.
  Rev. D}\ }\textbf {\bibinfo {volume} {96}},\ \bibinfo {pages} {083517}
  (\bibinfo {year} {2017})},\ \bibinfo {note} {[Erratum: Phys.Rev.D 99, 069904
  (2019)]},\ \Eprint {http://arxiv.org/abs/1707.03595} {arXiv:1707.03595
  [gr-qc]} \BibitemShut {NoStop}%
\bibitem [{\citenamefont {Flores}\ and\ \citenamefont
  {Kusenko}(2021)}]{Flores:2021tmc}%
  \BibitemOpen
  \bibfield  {author} {\bibinfo {author} {\bibfnamefont {M.~M.}\ \bibnamefont
  {Flores}}\ and\ \bibinfo {author} {\bibfnamefont {A.}~\bibnamefont
  {Kusenko}},\ }\href {\doibase 10.1103/PhysRevD.104.063008} {\bibfield
  {journal} {\bibinfo  {journal} {Phys. Rev. D}\ }\textbf {\bibinfo {volume}
  {104}},\ \bibinfo {pages} {063008} (\bibinfo {year} {2021})},\ \Eprint
  {http://arxiv.org/abs/2106.03237} {arXiv:2106.03237 [astro-ph.CO]}
  \BibitemShut {NoStop}%
\bibitem [{\citenamefont {Dvali}\ \emph {et~al.}(2021)\citenamefont {Dvali},
  \citenamefont {K\"uhnel},\ and\ \citenamefont {Zantedeschi}}]{Dvali:2021byy}%
  \BibitemOpen
  \bibfield  {author} {\bibinfo {author} {\bibfnamefont {G.}~\bibnamefont
  {Dvali}}, \bibinfo {author} {\bibfnamefont {F.}~\bibnamefont {K\"uhnel}}, \
  and\ \bibinfo {author} {\bibfnamefont {M.}~\bibnamefont {Zantedeschi}},\
  }\href@noop {} {\  (\bibinfo {year} {2021})},\ \Eprint
  {http://arxiv.org/abs/2108.09471} {arXiv:2108.09471 [hep-ph]} \BibitemShut
  {NoStop}%
\bibitem [{\citenamefont {Eroshenko}(2021)}]{Eroshenko:2021sez}%
  \BibitemOpen
  \bibfield  {author} {\bibinfo {author} {\bibfnamefont {Y.~N.}\ \bibnamefont
  {Eroshenko}},\ }\href@noop {} {\  (\bibinfo {year} {2021})},\ \Eprint
  {http://arxiv.org/abs/2111.03403} {arXiv:2111.03403 [astro-ph.CO]}
  \BibitemShut {NoStop}%
\bibitem [{\citenamefont {De~Luca}\ \emph
  {et~al.}(2021{\natexlab{d}})\citenamefont {De~Luca}, \citenamefont
  {Franciolini}, \citenamefont {Kehagias}, \citenamefont {Pani},\ and\
  \citenamefont {Riotto}}]{DeLuca:2021pls}%
  \BibitemOpen
  \bibfield  {author} {\bibinfo {author} {\bibfnamefont {V.}~\bibnamefont
  {De~Luca}}, \bibinfo {author} {\bibfnamefont {G.}~\bibnamefont
  {Franciolini}}, \bibinfo {author} {\bibfnamefont {A.}~\bibnamefont
  {Kehagias}}, \bibinfo {author} {\bibfnamefont {P.}~\bibnamefont {Pani}}, \
  and\ \bibinfo {author} {\bibfnamefont {A.}~\bibnamefont {Riotto}},\
  }\href@noop {} {\  (\bibinfo {year} {2021}{\natexlab{d}})},\ \Eprint
  {http://arxiv.org/abs/2112.02534} {arXiv:2112.02534 [astro-ph.CO]}
  \BibitemShut {NoStop}%
\bibitem [{\citenamefont {Wang}\ and\ \citenamefont
  {Nitz}(2021)}]{Wang:2021qsu}%
  \BibitemOpen
  \bibfield  {author} {\bibinfo {author} {\bibfnamefont {Y.-F.}\ \bibnamefont
  {Wang}}\ and\ \bibinfo {author} {\bibfnamefont {A.~H.}\ \bibnamefont
  {Nitz}},\ }\href {\doibase 10.3847/1538-4357/abe939} {\bibfield  {journal}
  {\bibinfo  {journal} {Astrophys. J.}\ }\textbf {\bibinfo {volume} {912}},\
  \bibinfo {pages} {53} (\bibinfo {year} {2021})},\ \Eprint
  {http://arxiv.org/abs/2101.12269} {arXiv:2101.12269 [astro-ph.HE]}
  \BibitemShut {NoStop}%
\bibitem [{\citenamefont {Kavanagh}\ \emph {et~al.}(2018)\citenamefont
  {Kavanagh}, \citenamefont {Gaggero},\ and\ \citenamefont
  {Bertone}}]{Kavanagh:2018ggo}%
  \BibitemOpen
  \bibfield  {author} {\bibinfo {author} {\bibfnamefont {B.~J.}\ \bibnamefont
  {Kavanagh}}, \bibinfo {author} {\bibfnamefont {D.}~\bibnamefont {Gaggero}}, \
  and\ \bibinfo {author} {\bibfnamefont {G.}~\bibnamefont {Bertone}},\ }\href
  {\doibase 10.1103/PhysRevD.98.023536} {\bibfield  {journal} {\bibinfo
  {journal} {Phys. Rev. D}\ }\textbf {\bibinfo {volume} {98}},\ \bibinfo
  {pages} {023536} (\bibinfo {year} {2018})},\ \Eprint
  {http://arxiv.org/abs/1805.09034} {arXiv:1805.09034 [astro-ph.CO]}
  \BibitemShut {NoStop}%
\bibitem [{\citenamefont {Liu}\ \emph {et~al.}(2019{\natexlab{b}})\citenamefont
  {Liu}, \citenamefont {Guo},\ and\ \citenamefont {Cai}}]{Liu:2018ess}%
  \BibitemOpen
  \bibfield  {author} {\bibinfo {author} {\bibfnamefont {L.}~\bibnamefont
  {Liu}}, \bibinfo {author} {\bibfnamefont {Z.-K.}\ \bibnamefont {Guo}}, \ and\
  \bibinfo {author} {\bibfnamefont {R.-G.}\ \bibnamefont {Cai}},\ }\href
  {\doibase 10.1103/PhysRevD.99.063523} {\bibfield  {journal} {\bibinfo
  {journal} {Phys. Rev. D}\ }\textbf {\bibinfo {volume} {99}},\ \bibinfo
  {pages} {063523} (\bibinfo {year} {2019}{\natexlab{b}})},\ \Eprint
  {http://arxiv.org/abs/1812.05376} {arXiv:1812.05376 [astro-ph.CO]}
  \BibitemShut {NoStop}%
\bibitem [{\citenamefont {Peters}\ and\ \citenamefont
  {Mathews}(1963{\natexlab{a}})}]{Peters:1963ux}%
  \BibitemOpen
  \bibfield  {author} {\bibinfo {author} {\bibfnamefont {P.~C.}\ \bibnamefont
  {Peters}}\ and\ \bibinfo {author} {\bibfnamefont {J.}~\bibnamefont
  {Mathews}},\ }\href {\doibase 10.1103/PhysRev.131.435} {\bibfield  {journal}
  {\bibinfo  {journal} {Phys. Rev.}\ }\textbf {\bibinfo {volume} {131}},\
  \bibinfo {pages} {435} (\bibinfo {year} {1963}{\natexlab{a}})}\BibitemShut
  {NoStop}%
\bibitem [{\citenamefont {Peters}(1964)}]{Peters:1964zz}%
  \BibitemOpen
  \bibfield  {author} {\bibinfo {author} {\bibfnamefont {P.~C.}\ \bibnamefont
  {Peters}},\ }\href {\doibase 10.1103/PhysRev.136.B1224} {\bibfield  {journal}
  {\bibinfo  {journal} {Phys. Rev.}\ }\textbf {\bibinfo {volume} {136}},\
  \bibinfo {pages} {B1224} (\bibinfo {year} {1964})}\BibitemShut {NoStop}%
\bibitem [{\citenamefont {Mandel}(2021)}]{Mandel:2021fra}%
  \BibitemOpen
  \bibfield  {author} {\bibinfo {author} {\bibfnamefont {I.}~\bibnamefont
  {Mandel}},\ }\href@noop {} {\  (\bibinfo {year} {2021})},\ \Eprint
  {http://arxiv.org/abs/2110.09254} {arXiv:2110.09254 [astro-ph.HE]}
  \BibitemShut {NoStop}%
\bibitem [{\citenamefont {Maggiore}(2007)}]{Maggiore:2007ulw}%
  \BibitemOpen
  \bibfield  {author} {\bibinfo {author} {\bibfnamefont {M.}~\bibnamefont
  {Maggiore}},\ }\href@noop {} {\emph {\bibinfo {title} {{Gravitational Waves.
  Vol. 1: Theory and Experiments}}}},\ Oxford Master Series in Physics\
  (\bibinfo  {publisher} {Oxford University Press},\ \bibinfo {year}
  {2007})\BibitemShut {NoStop}%
\bibitem [{\citenamefont {Pannarale}\ \emph {et~al.}(2015)\citenamefont
  {Pannarale}, \citenamefont {Berti}, \citenamefont {Kyutoku}, \citenamefont
  {Lackey},\ and\ \citenamefont {Shibata}}]{Pannarale:2015jia}%
  \BibitemOpen
  \bibfield  {author} {\bibinfo {author} {\bibfnamefont {F.}~\bibnamefont
  {Pannarale}}, \bibinfo {author} {\bibfnamefont {E.}~\bibnamefont {Berti}},
  \bibinfo {author} {\bibfnamefont {K.}~\bibnamefont {Kyutoku}}, \bibinfo
  {author} {\bibfnamefont {B.~D.}\ \bibnamefont {Lackey}}, \ and\ \bibinfo
  {author} {\bibfnamefont {M.}~\bibnamefont {Shibata}},\ }\href {\doibase
  10.1103/PhysRevD.92.081504} {\bibfield  {journal} {\bibinfo  {journal} {Phys.
  Rev. D}\ }\textbf {\bibinfo {volume} {92}},\ \bibinfo {pages} {081504}
  (\bibinfo {year} {2015})},\ \Eprint {http://arxiv.org/abs/1509.06209}
  {arXiv:1509.06209 [gr-qc]} \BibitemShut {NoStop}%
\bibitem [{\citenamefont {{Kowalska}}\ \emph {et~al.}(2011)\citenamefont
  {{Kowalska}}, \citenamefont {{Bulik}}, \citenamefont {{Belczynski}},
  \citenamefont {{Dominik}},\ and\ \citenamefont
  {{Gondek-Rosinska}}}]{2011A&A...527A..70K}%
  \BibitemOpen
  \bibfield  {author} {\bibinfo {author} {\bibfnamefont {I.}~\bibnamefont
  {{Kowalska}}}, \bibinfo {author} {\bibfnamefont {T.}~\bibnamefont {{Bulik}}},
  \bibinfo {author} {\bibfnamefont {K.}~\bibnamefont {{Belczynski}}}, \bibinfo
  {author} {\bibfnamefont {M.}~\bibnamefont {{Dominik}}}, \ and\ \bibinfo
  {author} {\bibfnamefont {D.}~\bibnamefont {{Gondek-Rosinska}}},\ }\href
  {\doibase 10.1051/0004-6361/201015777} {\bibfield  {journal} {\bibinfo
  {journal} {\aap}\ }\textbf {\bibinfo {volume} {527}},\ \bibinfo {eid} {A70}
  (\bibinfo {year} {2011})},\ \Eprint {http://arxiv.org/abs/1010.0511}
  {arXiv:1010.0511 [astro-ph.CO]} \BibitemShut {NoStop}%
\bibitem [{\citenamefont {Kyutoku}\ and\ \citenamefont
  {Seto}(2016)}]{Kyutoku:2016ppx}%
  \BibitemOpen
  \bibfield  {author} {\bibinfo {author} {\bibfnamefont {K.}~\bibnamefont
  {Kyutoku}}\ and\ \bibinfo {author} {\bibfnamefont {N.}~\bibnamefont {Seto}},\
  }\href {\doibase 10.1093/mnras/stw1767} {\bibfield  {journal} {\bibinfo
  {journal} {Mon. Not. Roy. Astron. Soc.}\ }\textbf {\bibinfo {volume} {462}},\
  \bibinfo {pages} {2177} (\bibinfo {year} {2016})},\ \Eprint
  {http://arxiv.org/abs/1606.02298} {arXiv:1606.02298 [astro-ph.HE]}
  \BibitemShut {NoStop}%
\bibitem [{\citenamefont {Samsing}\ \emph {et~al.}(2014)\citenamefont
  {Samsing}, \citenamefont {MacLeod},\ and\ \citenamefont
  {Ramirez-Ruiz}}]{Samsing:2013kua}%
  \BibitemOpen
  \bibfield  {author} {\bibinfo {author} {\bibfnamefont {J.}~\bibnamefont
  {Samsing}}, \bibinfo {author} {\bibfnamefont {M.}~\bibnamefont {MacLeod}}, \
  and\ \bibinfo {author} {\bibfnamefont {E.}~\bibnamefont {Ramirez-Ruiz}},\
  }\href {\doibase 10.1088/0004-637X/784/1/71} {\bibfield  {journal} {\bibinfo
  {journal} {Astrophys. J.}\ }\textbf {\bibinfo {volume} {784}},\ \bibinfo
  {pages} {71} (\bibinfo {year} {2014})},\ \Eprint
  {http://arxiv.org/abs/1308.2964} {arXiv:1308.2964 [astro-ph.HE]} \BibitemShut
  {NoStop}%
\bibitem [{\citenamefont {Nishizawa}\ \emph {et~al.}(2016)\citenamefont
  {Nishizawa}, \citenamefont {Berti}, \citenamefont {Klein},\ and\
  \citenamefont {Sesana}}]{Nishizawa:2016jji}%
  \BibitemOpen
  \bibfield  {author} {\bibinfo {author} {\bibfnamefont {A.}~\bibnamefont
  {Nishizawa}}, \bibinfo {author} {\bibfnamefont {E.}~\bibnamefont {Berti}},
  \bibinfo {author} {\bibfnamefont {A.}~\bibnamefont {Klein}}, \ and\ \bibinfo
  {author} {\bibfnamefont {A.}~\bibnamefont {Sesana}},\ }\href {\doibase
  10.1103/PhysRevD.94.064020} {\bibfield  {journal} {\bibinfo  {journal} {Phys.
  Rev. D}\ }\textbf {\bibinfo {volume} {94}},\ \bibinfo {pages} {064020}
  (\bibinfo {year} {2016})},\ \Eprint {http://arxiv.org/abs/1605.01341}
  {arXiv:1605.01341 [gr-qc]} \BibitemShut {NoStop}%
\bibitem [{\citenamefont {Breivik}\ \emph {et~al.}(2016)\citenamefont
  {Breivik}, \citenamefont {Rodriguez}, \citenamefont {Larson}, \citenamefont
  {Kalogera},\ and\ \citenamefont {Rasio}}]{Breivik:2016ddj}%
  \BibitemOpen
  \bibfield  {author} {\bibinfo {author} {\bibfnamefont {K.}~\bibnamefont
  {Breivik}}, \bibinfo {author} {\bibfnamefont {C.~L.}\ \bibnamefont
  {Rodriguez}}, \bibinfo {author} {\bibfnamefont {S.~L.}\ \bibnamefont
  {Larson}}, \bibinfo {author} {\bibfnamefont {V.}~\bibnamefont {Kalogera}}, \
  and\ \bibinfo {author} {\bibfnamefont {F.~A.}\ \bibnamefont {Rasio}},\ }\href
  {\doibase 10.3847/2041-8205/830/1/L18} {\bibfield  {journal} {\bibinfo
  {journal} {Astrophys. J. Lett.}\ }\textbf {\bibinfo {volume} {830}},\
  \bibinfo {pages} {L18} (\bibinfo {year} {2016})},\ \Eprint
  {http://arxiv.org/abs/1606.09558} {arXiv:1606.09558 [astro-ph.GA]}
  \BibitemShut {NoStop}%
\bibitem [{\citenamefont {Nishizawa}\ \emph {et~al.}(2017)\citenamefont
  {Nishizawa}, \citenamefont {Sesana}, \citenamefont {Berti},\ and\
  \citenamefont {Klein}}]{Nishizawa:2016eza}%
  \BibitemOpen
  \bibfield  {author} {\bibinfo {author} {\bibfnamefont {A.}~\bibnamefont
  {Nishizawa}}, \bibinfo {author} {\bibfnamefont {A.}~\bibnamefont {Sesana}},
  \bibinfo {author} {\bibfnamefont {E.}~\bibnamefont {Berti}}, \ and\ \bibinfo
  {author} {\bibfnamefont {A.}~\bibnamefont {Klein}},\ }\href {\doibase
  10.1093/mnras/stw2993} {\bibfield  {journal} {\bibinfo  {journal} {Mon. Not.
  Roy. Astron. Soc.}\ }\textbf {\bibinfo {volume} {465}},\ \bibinfo {pages}
  {4375} (\bibinfo {year} {2017})},\ \Eprint {http://arxiv.org/abs/1606.09295}
  {arXiv:1606.09295 [astro-ph.HE]} \BibitemShut {NoStop}%
\bibitem [{\citenamefont {Zevin}\ \emph {et~al.}(2021)\citenamefont {Zevin},
  \citenamefont {Romero-Shaw}, \citenamefont {Kremer}, \citenamefont {Thrane},\
  and\ \citenamefont {Lasky}}]{Zevin:2021rtf}%
  \BibitemOpen
  \bibfield  {author} {\bibinfo {author} {\bibfnamefont {M.}~\bibnamefont
  {Zevin}}, \bibinfo {author} {\bibfnamefont {I.~M.}\ \bibnamefont
  {Romero-Shaw}}, \bibinfo {author} {\bibfnamefont {K.}~\bibnamefont {Kremer}},
  \bibinfo {author} {\bibfnamefont {E.}~\bibnamefont {Thrane}}, \ and\ \bibinfo
  {author} {\bibfnamefont {P.~D.}\ \bibnamefont {Lasky}},\ }\href@noop {} {\
  (\bibinfo {year} {2021})},\ \Eprint {http://arxiv.org/abs/2106.09042}
  {arXiv:2106.09042 [astro-ph.HE]} \BibitemShut {NoStop}%
\bibitem [{\citenamefont {Jedamzik}(2020)}]{Jedamzik:2020ypm}%
  \BibitemOpen
  \bibfield  {author} {\bibinfo {author} {\bibfnamefont {K.}~\bibnamefont
  {Jedamzik}},\ }\href {\doibase 10.1088/1475-7516/2020/09/022} {\bibfield
  {journal} {\bibinfo  {journal} {JCAP}\ }\textbf {\bibinfo {volume} {09}},\
  \bibinfo {pages} {022} (\bibinfo {year} {2020})},\ \Eprint
  {http://arxiv.org/abs/2006.11172} {arXiv:2006.11172 [astro-ph.CO]}
  \BibitemShut {NoStop}%
\bibitem [{\citenamefont {Trashorras}\ \emph {et~al.}(2021)\citenamefont
  {Trashorras}, \citenamefont {Garc\'\i{}a-Bellido},\ and\ \citenamefont
  {Nesseris}}]{Trashorras:2020mwn}%
  \BibitemOpen
  \bibfield  {author} {\bibinfo {author} {\bibfnamefont {M.}~\bibnamefont
  {Trashorras}}, \bibinfo {author} {\bibfnamefont {J.}~\bibnamefont
  {Garc\'\i{}a-Bellido}}, \ and\ \bibinfo {author} {\bibfnamefont
  {S.}~\bibnamefont {Nesseris}},\ }\href {\doibase 10.3390/universe7010018}
  {\bibfield  {journal} {\bibinfo  {journal} {Universe}\ }\textbf {\bibinfo
  {volume} {7}},\ \bibinfo {pages} {18} (\bibinfo {year} {2021})},\ \Eprint
  {http://arxiv.org/abs/2006.15018} {arXiv:2006.15018 [astro-ph.CO]}
  \BibitemShut {NoStop}%
\bibitem [{\citenamefont {Kilic}\ \emph {et~al.}(2007)\citenamefont {Kilic},
  \citenamefont {Allende~Prieto}, \citenamefont {Brown},\ and\ \citenamefont
  {Koester}}]{Kilic:2006as}%
  \BibitemOpen
  \bibfield  {author} {\bibinfo {author} {\bibfnamefont {M.}~\bibnamefont
  {Kilic}}, \bibinfo {author} {\bibfnamefont {C.}~\bibnamefont
  {Allende~Prieto}}, \bibinfo {author} {\bibfnamefont {W.~R.}\ \bibnamefont
  {Brown}}, \ and\ \bibinfo {author} {\bibfnamefont {D.}~\bibnamefont
  {Koester}},\ }\href {\doibase 10.1086/514327} {\bibfield  {journal} {\bibinfo
   {journal} {Astrophys. J.}\ }\textbf {\bibinfo {volume} {660}},\ \bibinfo
  {pages} {1451} (\bibinfo {year} {2007})},\ \Eprint
  {http://arxiv.org/abs/astro-ph/0611498} {arXiv:astro-ph/0611498} \BibitemShut
  {NoStop}%
\bibitem [{\citenamefont {Strobel}\ \emph {et~al.}(1999)\citenamefont
  {Strobel}, \citenamefont {Schaab},\ and\ \citenamefont
  {Weigel}}]{Strobel:1999vn}%
  \BibitemOpen
  \bibfield  {author} {\bibinfo {author} {\bibfnamefont {K.}~\bibnamefont
  {Strobel}}, \bibinfo {author} {\bibfnamefont {C.}~\bibnamefont {Schaab}}, \
  and\ \bibinfo {author} {\bibfnamefont {M.~K.}\ \bibnamefont {Weigel}},\
  }\href@noop {} {\bibfield  {journal} {\bibinfo  {journal} {Astron.
  Astrophys.}\ }\textbf {\bibinfo {volume} {350}},\ \bibinfo {pages} {497}
  (\bibinfo {year} {1999})},\ \Eprint {http://arxiv.org/abs/astro-ph/9908132}
  {arXiv:astro-ph/9908132} \BibitemShut {NoStop}%
\bibitem [{\citenamefont {Lattimer}(2012)}]{Lattimer:2012nd}%
  \BibitemOpen
  \bibfield  {author} {\bibinfo {author} {\bibfnamefont {J.~M.}\ \bibnamefont
  {Lattimer}},\ }\href {\doibase 10.1146/annurev-nucl-102711-095018} {\bibfield
   {journal} {\bibinfo  {journal} {Ann. Rev. Nucl. Part. Sci.}\ }\textbf
  {\bibinfo {volume} {62}},\ \bibinfo {pages} {485} (\bibinfo {year} {2012})},\
  \Eprint {http://arxiv.org/abs/1305.3510} {arXiv:1305.3510 [nucl-th]}
  \BibitemShut {NoStop}%
\bibitem [{\citenamefont {Silva}\ \emph {et~al.}(2016)\citenamefont {Silva},
  \citenamefont {Sotani},\ and\ \citenamefont {Berti}}]{Silva:2016myw}%
  \BibitemOpen
  \bibfield  {author} {\bibinfo {author} {\bibfnamefont {H.~O.}\ \bibnamefont
  {Silva}}, \bibinfo {author} {\bibfnamefont {H.}~\bibnamefont {Sotani}}, \
  and\ \bibinfo {author} {\bibfnamefont {E.}~\bibnamefont {Berti}},\ }\href
  {\doibase 10.1093/mnras/stw969} {\bibfield  {journal} {\bibinfo  {journal}
  {Mon. Not. Roy. Astron. Soc.}\ }\textbf {\bibinfo {volume} {459}},\ \bibinfo
  {pages} {4378} (\bibinfo {year} {2016})},\ \Eprint
  {http://arxiv.org/abs/1601.03407} {arXiv:1601.03407 [astro-ph.HE]}
  \BibitemShut {NoStop}%
\bibitem [{\citenamefont {Suwa}\ \emph {et~al.}(2018)\citenamefont {Suwa},
  \citenamefont {Yoshida}, \citenamefont {Shibata}, \citenamefont {Umeda},\
  and\ \citenamefont {Takahashi}}]{Suwa:2018uni}%
  \BibitemOpen
  \bibfield  {author} {\bibinfo {author} {\bibfnamefont {Y.}~\bibnamefont
  {Suwa}}, \bibinfo {author} {\bibfnamefont {T.}~\bibnamefont {Yoshida}},
  \bibinfo {author} {\bibfnamefont {M.}~\bibnamefont {Shibata}}, \bibinfo
  {author} {\bibfnamefont {H.}~\bibnamefont {Umeda}}, \ and\ \bibinfo {author}
  {\bibfnamefont {K.}~\bibnamefont {Takahashi}},\ }\href {\doibase
  10.1093/mnras/sty2460} {\bibfield  {journal} {\bibinfo  {journal} {Mon. Not.
  Roy. Astron. Soc.}\ }\textbf {\bibinfo {volume} {481}},\ \bibinfo {pages}
  {3305} (\bibinfo {year} {2018})},\ \Eprint {http://arxiv.org/abs/1808.02328}
  {arXiv:1808.02328 [astro-ph.HE]} \BibitemShut {NoStop}%
\bibitem [{\citenamefont {Liebling}\ and\ \citenamefont
  {Palenzuela}(2012)}]{Liebling:2012fv}%
  \BibitemOpen
  \bibfield  {author} {\bibinfo {author} {\bibfnamefont {S.~L.}\ \bibnamefont
  {Liebling}}\ and\ \bibinfo {author} {\bibfnamefont {C.}~\bibnamefont
  {Palenzuela}},\ }\href {\doibase 10.12942/lrr-2012-6} {\bibfield  {journal}
  {\bibinfo  {journal} {Living Rev. Rel.}\ }\textbf {\bibinfo {volume} {15}},\
  \bibinfo {pages} {6} (\bibinfo {year} {2012})},\ \Eprint
  {http://arxiv.org/abs/1202.5809} {arXiv:1202.5809 [gr-qc]} \BibitemShut
  {NoStop}%
\bibitem [{\citenamefont {Cardoso}\ \emph {et~al.}(2017)\citenamefont
  {Cardoso}, \citenamefont {Franzin}, \citenamefont {Maselli}, \citenamefont
  {Pani},\ and\ \citenamefont {Raposo}}]{Cardoso:2017cfl}%
  \BibitemOpen
  \bibfield  {author} {\bibinfo {author} {\bibfnamefont {V.}~\bibnamefont
  {Cardoso}}, \bibinfo {author} {\bibfnamefont {E.}~\bibnamefont {Franzin}},
  \bibinfo {author} {\bibfnamefont {A.}~\bibnamefont {Maselli}}, \bibinfo
  {author} {\bibfnamefont {P.}~\bibnamefont {Pani}}, \ and\ \bibinfo {author}
  {\bibfnamefont {G.}~\bibnamefont {Raposo}},\ }\href {\doibase
  10.1103/PhysRevD.95.084014} {\bibfield  {journal} {\bibinfo  {journal} {Phys.
  Rev. D}\ }\textbf {\bibinfo {volume} {95}},\ \bibinfo {pages} {084014}
  (\bibinfo {year} {2017})},\ \bibinfo {note} {[Addendum: Phys.Rev.D 95, 089901
  (2017)]},\ \Eprint {http://arxiv.org/abs/1701.01116} {arXiv:1701.01116
  [gr-qc]} \BibitemShut {NoStop}%
\bibitem [{\citenamefont {Ruffini}\ and\ \citenamefont
  {Bonazzola}(1969)}]{Ruffini:1969qy}%
  \BibitemOpen
  \bibfield  {author} {\bibinfo {author} {\bibfnamefont {R.}~\bibnamefont
  {Ruffini}}\ and\ \bibinfo {author} {\bibfnamefont {S.}~\bibnamefont
  {Bonazzola}},\ }\href {\doibase 10.1103/PhysRev.187.1767} {\bibfield
  {journal} {\bibinfo  {journal} {Phys. Rev.}\ }\textbf {\bibinfo {volume}
  {187}},\ \bibinfo {pages} {1767} (\bibinfo {year} {1969})}\BibitemShut
  {NoStop}%
\bibitem [{\citenamefont {Colpi}\ \emph {et~al.}(1986)\citenamefont {Colpi},
  \citenamefont {Shapiro},\ and\ \citenamefont {Wasserman}}]{Colpi:1986ye}%
  \BibitemOpen
  \bibfield  {author} {\bibinfo {author} {\bibfnamefont {M.}~\bibnamefont
  {Colpi}}, \bibinfo {author} {\bibfnamefont {S.~L.}\ \bibnamefont {Shapiro}},
  \ and\ \bibinfo {author} {\bibfnamefont {I.}~\bibnamefont {Wasserman}},\
  }\href {\doibase 10.1103/PhysRevLett.57.2485} {\bibfield  {journal} {\bibinfo
   {journal} {Phys. Rev. Lett.}\ }\textbf {\bibinfo {volume} {57}},\ \bibinfo
  {pages} {2485} (\bibinfo {year} {1986})}\BibitemShut {NoStop}%
\bibitem [{\citenamefont {Binnington}\ and\ \citenamefont
  {Poisson}(2009)}]{Binnington:2009bb}%
  \BibitemOpen
  \bibfield  {author} {\bibinfo {author} {\bibfnamefont {T.}~\bibnamefont
  {Binnington}}\ and\ \bibinfo {author} {\bibfnamefont {E.}~\bibnamefont
  {Poisson}},\ }\href {\doibase 10.1103/PhysRevD.80.084018} {\bibfield
  {journal} {\bibinfo  {journal} {Phys. Rev. D}\ }\textbf {\bibinfo {volume}
  {80}},\ \bibinfo {pages} {084018} (\bibinfo {year} {2009})},\ \Eprint
  {http://arxiv.org/abs/0906.1366} {arXiv:0906.1366 [gr-qc]} \BibitemShut
  {NoStop}%
\bibitem [{\citenamefont {Damour}\ and\ \citenamefont
  {Nagar}(2009)}]{Damour:2009vw}%
  \BibitemOpen
  \bibfield  {author} {\bibinfo {author} {\bibfnamefont {T.}~\bibnamefont
  {Damour}}\ and\ \bibinfo {author} {\bibfnamefont {A.}~\bibnamefont {Nagar}},\
  }\href {\doibase 10.1103/PhysRevD.80.084035} {\bibfield  {journal} {\bibinfo
  {journal} {Phys. Rev. D}\ }\textbf {\bibinfo {volume} {80}},\ \bibinfo
  {pages} {084035} (\bibinfo {year} {2009})},\ \Eprint
  {http://arxiv.org/abs/0906.0096} {arXiv:0906.0096 [gr-qc]} \BibitemShut
  {NoStop}%
\bibitem [{\citenamefont {Damour}\ and\ \citenamefont
  {Lecian}(2009)}]{Damour:2009va}%
  \BibitemOpen
  \bibfield  {author} {\bibinfo {author} {\bibfnamefont {T.}~\bibnamefont
  {Damour}}\ and\ \bibinfo {author} {\bibfnamefont {O.~M.}\ \bibnamefont
  {Lecian}},\ }\href {\doibase 10.1103/PhysRevD.80.044017} {\bibfield
  {journal} {\bibinfo  {journal} {Phys. Rev. D}\ }\textbf {\bibinfo {volume}
  {80}},\ \bibinfo {pages} {044017} (\bibinfo {year} {2009})},\ \Eprint
  {http://arxiv.org/abs/0906.3003} {arXiv:0906.3003 [gr-qc]} \BibitemShut
  {NoStop}%
\bibitem [{\citenamefont {Pani}\ \emph
  {et~al.}(2015{\natexlab{a}})\citenamefont {Pani}, \citenamefont {Gualtieri},
  \citenamefont {Maselli},\ and\ \citenamefont {Ferrari}}]{Pani:2015hfa}%
  \BibitemOpen
  \bibfield  {author} {\bibinfo {author} {\bibfnamefont {P.}~\bibnamefont
  {Pani}}, \bibinfo {author} {\bibfnamefont {L.}~\bibnamefont {Gualtieri}},
  \bibinfo {author} {\bibfnamefont {A.}~\bibnamefont {Maselli}}, \ and\
  \bibinfo {author} {\bibfnamefont {V.}~\bibnamefont {Ferrari}},\ }\href
  {\doibase 10.1103/PhysRevD.92.024010} {\bibfield  {journal} {\bibinfo
  {journal} {Phys. Rev. D}\ }\textbf {\bibinfo {volume} {92}},\ \bibinfo
  {pages} {024010} (\bibinfo {year} {2015}{\natexlab{a}})},\ \Eprint
  {http://arxiv.org/abs/1503.07365} {arXiv:1503.07365 [gr-qc]} \BibitemShut
  {NoStop}%
\bibitem [{\citenamefont {Pani}\ \emph
  {et~al.}(2015{\natexlab{b}})\citenamefont {Pani}, \citenamefont {Gualtieri},\
  and\ \citenamefont {Ferrari}}]{Pani:2015nua}%
  \BibitemOpen
  \bibfield  {author} {\bibinfo {author} {\bibfnamefont {P.}~\bibnamefont
  {Pani}}, \bibinfo {author} {\bibfnamefont {L.}~\bibnamefont {Gualtieri}}, \
  and\ \bibinfo {author} {\bibfnamefont {V.}~\bibnamefont {Ferrari}},\ }\href
  {\doibase 10.1103/PhysRevD.92.124003} {\bibfield  {journal} {\bibinfo
  {journal} {Phys. Rev. D}\ }\textbf {\bibinfo {volume} {92}},\ \bibinfo
  {pages} {124003} (\bibinfo {year} {2015}{\natexlab{b}})},\ \Eprint
  {http://arxiv.org/abs/1509.02171} {arXiv:1509.02171 [gr-qc]} \BibitemShut
  {NoStop}%
\bibitem [{\citenamefont {G\"urlebeck}(2015)}]{Gurlebeck:2015xpa}%
  \BibitemOpen
  \bibfield  {author} {\bibinfo {author} {\bibfnamefont {N.}~\bibnamefont
  {G\"urlebeck}},\ }\href {\doibase 10.1103/PhysRevLett.114.151102} {\bibfield
  {journal} {\bibinfo  {journal} {Phys. Rev. Lett.}\ }\textbf {\bibinfo
  {volume} {114}},\ \bibinfo {pages} {151102} (\bibinfo {year} {2015})},\
  \Eprint {http://arxiv.org/abs/1503.03240} {arXiv:1503.03240 [gr-qc]}
  \BibitemShut {NoStop}%
\bibitem [{\citenamefont {Porto}(2016)}]{Porto:2016zng}%
  \BibitemOpen
  \bibfield  {author} {\bibinfo {author} {\bibfnamefont {R.~A.}\ \bibnamefont
  {Porto}},\ }\href {\doibase 10.1002/prop.201600064} {\bibfield  {journal}
  {\bibinfo  {journal} {Fortsch. Phys.}\ }\textbf {\bibinfo {volume} {64}},\
  \bibinfo {pages} {723} (\bibinfo {year} {2016})},\ \Eprint
  {http://arxiv.org/abs/1606.08895} {arXiv:1606.08895 [gr-qc]} \BibitemShut
  {NoStop}%
\bibitem [{\citenamefont {Le~Tiec}\ and\ \citenamefont
  {Casals}(2021)}]{LeTiec:2020spy}%
  \BibitemOpen
  \bibfield  {author} {\bibinfo {author} {\bibfnamefont {A.}~\bibnamefont
  {Le~Tiec}}\ and\ \bibinfo {author} {\bibfnamefont {M.}~\bibnamefont
  {Casals}},\ }\href {\doibase 10.1103/PhysRevLett.126.131102} {\bibfield
  {journal} {\bibinfo  {journal} {Phys. Rev. Lett.}\ }\textbf {\bibinfo
  {volume} {126}},\ \bibinfo {pages} {131102} (\bibinfo {year} {2021})},\
  \Eprint {http://arxiv.org/abs/2007.00214} {arXiv:2007.00214 [gr-qc]}
  \BibitemShut {NoStop}%
\bibitem [{\citenamefont {Chia}(2020)}]{Chia:2020yla}%
  \BibitemOpen
  \bibfield  {author} {\bibinfo {author} {\bibfnamefont {H.~S.}\ \bibnamefont
  {Chia}},\ }\href@noop {} {\  (\bibinfo {year} {2020})},\ \Eprint
  {http://arxiv.org/abs/2010.07300} {arXiv:2010.07300 [gr-qc]} \BibitemShut
  {NoStop}%
\bibitem [{\citenamefont {Le~Tiec}\ \emph {et~al.}(2021)\citenamefont
  {Le~Tiec}, \citenamefont {Casals},\ and\ \citenamefont
  {Franzin}}]{LeTiec:2020bos}%
  \BibitemOpen
  \bibfield  {author} {\bibinfo {author} {\bibfnamefont {A.}~\bibnamefont
  {Le~Tiec}}, \bibinfo {author} {\bibfnamefont {M.}~\bibnamefont {Casals}}, \
  and\ \bibinfo {author} {\bibfnamefont {E.}~\bibnamefont {Franzin}},\ }\href
  {\doibase 10.1103/PhysRevD.103.084021} {\bibfield  {journal} {\bibinfo
  {journal} {Phys. Rev. D}\ }\textbf {\bibinfo {volume} {103}},\ \bibinfo
  {pages} {084021} (\bibinfo {year} {2021})},\ \Eprint
  {http://arxiv.org/abs/2010.15795} {arXiv:2010.15795 [gr-qc]} \BibitemShut
  {NoStop}%
\bibitem [{\citenamefont {Hui}\ \emph {et~al.}(2021)\citenamefont {Hui},
  \citenamefont {Joyce}, \citenamefont {Penco}, \citenamefont {Santoni},\ and\
  \citenamefont {Solomon}}]{Hui:2020xxx}%
  \BibitemOpen
  \bibfield  {author} {\bibinfo {author} {\bibfnamefont {L.}~\bibnamefont
  {Hui}}, \bibinfo {author} {\bibfnamefont {A.}~\bibnamefont {Joyce}}, \bibinfo
  {author} {\bibfnamefont {R.}~\bibnamefont {Penco}}, \bibinfo {author}
  {\bibfnamefont {L.}~\bibnamefont {Santoni}}, \ and\ \bibinfo {author}
  {\bibfnamefont {A.~R.}\ \bibnamefont {Solomon}},\ }\href {\doibase
  10.1088/1475-7516/2021/04/052} {\bibfield  {journal} {\bibinfo  {journal}
  {JCAP}\ }\textbf {\bibinfo {volume} {04}},\ \bibinfo {pages} {052} (\bibinfo
  {year} {2021})},\ \Eprint {http://arxiv.org/abs/2010.00593} {arXiv:2010.00593
  [hep-th]} \BibitemShut {NoStop}%
\bibitem [{\citenamefont {Charalambous}\ \emph
  {et~al.}(2021{\natexlab{a}})\citenamefont {Charalambous}, \citenamefont
  {Dubovsky},\ and\ \citenamefont {Ivanov}}]{Charalambous:2021kcz}%
  \BibitemOpen
  \bibfield  {author} {\bibinfo {author} {\bibfnamefont {P.}~\bibnamefont
  {Charalambous}}, \bibinfo {author} {\bibfnamefont {S.}~\bibnamefont
  {Dubovsky}}, \ and\ \bibinfo {author} {\bibfnamefont {M.~M.}\ \bibnamefont
  {Ivanov}},\ }\href {\doibase 10.1103/PhysRevLett.127.101101} {\bibfield
  {journal} {\bibinfo  {journal} {Phys. Rev. Lett.}\ }\textbf {\bibinfo
  {volume} {127}},\ \bibinfo {pages} {101101} (\bibinfo {year}
  {2021}{\natexlab{a}})},\ \Eprint {http://arxiv.org/abs/2103.01234}
  {arXiv:2103.01234 [hep-th]} \BibitemShut {NoStop}%
\bibitem [{\citenamefont {Charalambous}\ \emph
  {et~al.}(2021{\natexlab{b}})\citenamefont {Charalambous}, \citenamefont
  {Dubovsky},\ and\ \citenamefont {Ivanov}}]{Charalambous:2021mea}%
  \BibitemOpen
  \bibfield  {author} {\bibinfo {author} {\bibfnamefont {P.}~\bibnamefont
  {Charalambous}}, \bibinfo {author} {\bibfnamefont {S.}~\bibnamefont
  {Dubovsky}}, \ and\ \bibinfo {author} {\bibfnamefont {M.~M.}\ \bibnamefont
  {Ivanov}},\ }\href {\doibase 10.1007/JHEP05(2021)038} {\bibfield  {journal}
  {\bibinfo  {journal} {JHEP}\ }\textbf {\bibinfo {volume} {05}},\ \bibinfo
  {pages} {038} (\bibinfo {year} {2021}{\natexlab{b}})},\ \Eprint
  {http://arxiv.org/abs/2102.08917} {arXiv:2102.08917 [hep-th]} \BibitemShut
  {NoStop}%
\bibitem [{\citenamefont {Pere\~niguez}\ and\ \citenamefont
  {Cardoso}(2021)}]{Pereniguez:2021xcj}%
  \BibitemOpen
  \bibfield  {author} {\bibinfo {author} {\bibfnamefont {D.}~\bibnamefont
  {Pere\~niguez}}\ and\ \bibinfo {author} {\bibfnamefont {V.}~\bibnamefont
  {Cardoso}},\ }\href@noop {} {\  (\bibinfo {year} {2021})},\ \Eprint
  {http://arxiv.org/abs/2112.08400} {arXiv:2112.08400 [gr-qc]} \BibitemShut
  {NoStop}%
\bibitem [{\citenamefont {Flanagan}\ and\ \citenamefont
  {Hinderer}(2008)}]{Flanagan:2007ix}%
  \BibitemOpen
  \bibfield  {author} {\bibinfo {author} {\bibfnamefont {E.~E.}\ \bibnamefont
  {Flanagan}}\ and\ \bibinfo {author} {\bibfnamefont {T.}~\bibnamefont
  {Hinderer}},\ }\href {\doibase 10.1103/PhysRevD.77.021502} {\bibfield
  {journal} {\bibinfo  {journal} {Phys. Rev. D}\ }\textbf {\bibinfo {volume}
  {77}},\ \bibinfo {pages} {021502} (\bibinfo {year} {2008})},\ \Eprint
  {http://arxiv.org/abs/0709.1915} {arXiv:0709.1915 [astro-ph]} \BibitemShut
  {NoStop}%
\bibitem [{\citenamefont {Vines}\ \emph {et~al.}(2011)\citenamefont {Vines},
  \citenamefont {Flanagan},\ and\ \citenamefont {Hinderer}}]{Vines:2011ud}%
  \BibitemOpen
  \bibfield  {author} {\bibinfo {author} {\bibfnamefont {J.}~\bibnamefont
  {Vines}}, \bibinfo {author} {\bibfnamefont {E.~E.}\ \bibnamefont {Flanagan}},
  \ and\ \bibinfo {author} {\bibfnamefont {T.}~\bibnamefont {Hinderer}},\
  }\href {\doibase 10.1103/PhysRevD.83.084051} {\bibfield  {journal} {\bibinfo
  {journal} {Phys. Rev. D}\ }\textbf {\bibinfo {volume} {83}},\ \bibinfo
  {pages} {084051} (\bibinfo {year} {2011})},\ \Eprint
  {http://arxiv.org/abs/1101.1673} {arXiv:1101.1673 [gr-qc]} \BibitemShut
  {NoStop}%
\bibitem [{\citenamefont {Poisson}\ and\ \citenamefont
  {Will}(1953)}]{PoissonWill}%
  \BibitemOpen
  \bibfield  {author} {\bibinfo {author} {\bibfnamefont {E.}~\bibnamefont
  {Poisson}}\ and\ \bibinfo {author} {\bibfnamefont {C.}~\bibnamefont {Will}},\
  }\href@noop {} {\emph {\bibinfo {title} {{Gravity: Newtonian, Post-Newtonian,
  Relativistic}}}}\ (\bibinfo  {publisher} {Cambridge University Press},\
  \bibinfo {address} {Cambridge, UK},\ \bibinfo {year} {1953})\BibitemShut
  {NoStop}%
\bibitem [{\citenamefont {Cardoso}\ and\ \citenamefont
  {Duque}(2020)}]{Cardoso:2019upw}%
  \BibitemOpen
  \bibfield  {author} {\bibinfo {author} {\bibfnamefont {V.}~\bibnamefont
  {Cardoso}}\ and\ \bibinfo {author} {\bibfnamefont {F.}~\bibnamefont
  {Duque}},\ }\href {\doibase 10.1103/PhysRevD.101.064028} {\bibfield
  {journal} {\bibinfo  {journal} {Phys. Rev. D}\ }\textbf {\bibinfo {volume}
  {101}},\ \bibinfo {pages} {064028} (\bibinfo {year} {2020})},\ \Eprint
  {http://arxiv.org/abs/1912.07616} {arXiv:1912.07616 [gr-qc]} \BibitemShut
  {NoStop}%
\bibitem [{\citenamefont {De~Luca}\ and\ \citenamefont
  {Pani}(2021)}]{DeLuca:2021ite}%
  \BibitemOpen
  \bibfield  {author} {\bibinfo {author} {\bibfnamefont {V.}~\bibnamefont
  {De~Luca}}\ and\ \bibinfo {author} {\bibfnamefont {P.}~\bibnamefont {Pani}},\
  }\href {\doibase 10.1088/1475-7516/2021/08/032} {\bibfield  {journal}
  {\bibinfo  {journal} {JCAP}\ }\textbf {\bibinfo {volume} {08}},\ \bibinfo
  {pages} {032} (\bibinfo {year} {2021})},\ \Eprint
  {http://arxiv.org/abs/2106.14428} {arXiv:2106.14428 [gr-qc]} \BibitemShut
  {NoStop}%
\bibitem [{\citenamefont {Alvi}(2001)}]{Alvi:2001mx}%
  \BibitemOpen
  \bibfield  {author} {\bibinfo {author} {\bibfnamefont {K.}~\bibnamefont
  {Alvi}},\ }\href {\doibase 10.1103/PhysRevD.64.104020} {\bibfield  {journal}
  {\bibinfo  {journal} {Phys. Rev. D}\ }\textbf {\bibinfo {volume} {64}},\
  \bibinfo {pages} {104020} (\bibinfo {year} {2001})},\ \Eprint
  {http://arxiv.org/abs/gr-qc/0107080} {arXiv:gr-qc/0107080} \BibitemShut
  {NoStop}%
\bibitem [{\citenamefont {Maselli}\ \emph {et~al.}(2018)\citenamefont
  {Maselli}, \citenamefont {Pani}, \citenamefont {Cardoso}, \citenamefont
  {Abdelsalhin}, \citenamefont {Gualtieri},\ and\ \citenamefont
  {Ferrari}}]{Maselli:2017cmm}%
  \BibitemOpen
  \bibfield  {author} {\bibinfo {author} {\bibfnamefont {A.}~\bibnamefont
  {Maselli}}, \bibinfo {author} {\bibfnamefont {P.}~\bibnamefont {Pani}},
  \bibinfo {author} {\bibfnamefont {V.}~\bibnamefont {Cardoso}}, \bibinfo
  {author} {\bibfnamefont {T.}~\bibnamefont {Abdelsalhin}}, \bibinfo {author}
  {\bibfnamefont {L.}~\bibnamefont {Gualtieri}}, \ and\ \bibinfo {author}
  {\bibfnamefont {V.}~\bibnamefont {Ferrari}},\ }\href {\doibase
  10.1103/PhysRevLett.120.081101} {\bibfield  {journal} {\bibinfo  {journal}
  {Phys. Rev. Lett.}\ }\textbf {\bibinfo {volume} {120}},\ \bibinfo {pages}
  {081101} (\bibinfo {year} {2018})},\ \Eprint
  {http://arxiv.org/abs/1703.10612} {arXiv:1703.10612 [gr-qc]} \BibitemShut
  {NoStop}%
\bibitem [{\citenamefont {Vallisneri}(2008)}]{Vallisneri:2007ev}%
  \BibitemOpen
  \bibfield  {author} {\bibinfo {author} {\bibfnamefont {M.}~\bibnamefont
  {Vallisneri}},\ }\href {\doibase 10.1103/PhysRevD.77.042001} {\bibfield
  {journal} {\bibinfo  {journal} {Phys. Rev. D}\ }\textbf {\bibinfo {volume}
  {77}},\ \bibinfo {pages} {042001} (\bibinfo {year} {2008})},\ \Eprint
  {http://arxiv.org/abs/gr-qc/0703086} {arXiv:gr-qc/0703086} \BibitemShut
  {NoStop}%
\bibitem [{\citenamefont {Robson}\ \emph {et~al.}(2019)\citenamefont {Robson},
  \citenamefont {Cornish},\ and\ \citenamefont {Liu}}]{Robson:2018ifk}%
  \BibitemOpen
  \bibfield  {author} {\bibinfo {author} {\bibfnamefont {T.}~\bibnamefont
  {Robson}}, \bibinfo {author} {\bibfnamefont {N.~J.}\ \bibnamefont {Cornish}},
  \ and\ \bibinfo {author} {\bibfnamefont {C.}~\bibnamefont {Liu}},\ }\href
  {\doibase 10.1088/1361-6382/ab1101} {\bibfield  {journal} {\bibinfo
  {journal} {Class. Quant. Grav.}\ }\textbf {\bibinfo {volume} {36}},\ \bibinfo
  {pages} {105011} (\bibinfo {year} {2019})},\ \Eprint
  {http://arxiv.org/abs/1803.01944} {arXiv:1803.01944 [astro-ph.HE]}
  \BibitemShut {NoStop}%
\bibitem [{\citenamefont {Kaiser}\ and\ \citenamefont
  {McWilliams}(2021)}]{Kaiser:2020tlg}%
  \BibitemOpen
  \bibfield  {author} {\bibinfo {author} {\bibfnamefont {A.~R.}\ \bibnamefont
  {Kaiser}}\ and\ \bibinfo {author} {\bibfnamefont {S.~T.}\ \bibnamefont
  {McWilliams}},\ }\href {\doibase 10.1088/1361-6382/abd4f6} {\bibfield
  {journal} {\bibinfo  {journal} {Class. Quant. Grav.}\ }\textbf {\bibinfo
  {volume} {38}},\ \bibinfo {pages} {055009} (\bibinfo {year} {2021})},\
  \Eprint {http://arxiv.org/abs/2010.02135} {arXiv:2010.02135 [gr-qc]}
  \BibitemShut {NoStop}%
\bibitem [{\citenamefont {Abbott}\ \emph {et~al.}(2020)\citenamefont {Abbott}
  \emph {et~al.}}]{LIGOScientific:2019eut}%
  \BibitemOpen
  \bibfield  {author} {\bibinfo {author} {\bibfnamefont {B.~P.}\ \bibnamefont
  {Abbott}} \emph {et~al.} (\bibinfo {collaboration} {LIGO Scientific,
  Virgo}),\ }\href {\doibase 10.1088/1361-6382/ab5f7c} {\bibfield  {journal}
  {\bibinfo  {journal} {Class. Quant. Grav.}\ }\textbf {\bibinfo {volume}
  {37}},\ \bibinfo {pages} {045006} (\bibinfo {year} {2020})},\ \Eprint
  {http://arxiv.org/abs/1908.01012} {arXiv:1908.01012 [gr-qc]} \BibitemShut
  {NoStop}%
\bibitem [{\citenamefont {Pacilio}\ \emph {et~al.}(2021)\citenamefont
  {Pacilio}, \citenamefont {Maselli}, \citenamefont {Fasano},\ and\
  \citenamefont {Pani}}]{Pacilio:2021jmq}%
  \BibitemOpen
  \bibfield  {author} {\bibinfo {author} {\bibfnamefont {C.}~\bibnamefont
  {Pacilio}}, \bibinfo {author} {\bibfnamefont {A.}~\bibnamefont {Maselli}},
  \bibinfo {author} {\bibfnamefont {M.}~\bibnamefont {Fasano}}, \ and\ \bibinfo
  {author} {\bibfnamefont {P.}~\bibnamefont {Pani}},\ }\href@noop {} {\
  (\bibinfo {year} {2021})},\ \Eprint {http://arxiv.org/abs/2104.10035}
  {arXiv:2104.10035 [gr-qc]} \BibitemShut {NoStop}%
\bibitem [{\citenamefont {Favata}(2014)}]{Favata:2013rwa}%
  \BibitemOpen
  \bibfield  {author} {\bibinfo {author} {\bibfnamefont {M.}~\bibnamefont
  {Favata}},\ }\href {\doibase 10.1103/PhysRevLett.112.101101} {\bibfield
  {journal} {\bibinfo  {journal} {Phys. Rev. Lett.}\ }\textbf {\bibinfo
  {volume} {112}},\ \bibinfo {pages} {101101} (\bibinfo {year} {2014})},\
  \Eprint {http://arxiv.org/abs/1310.8288} {arXiv:1310.8288 [gr-qc]}
  \BibitemShut {NoStop}%
\bibitem [{\citenamefont {Abbott}\ \emph
  {et~al.}(2017{\natexlab{a}})\citenamefont {Abbott} \emph
  {et~al.}}]{LIGOScientific:2017vwq}%
  \BibitemOpen
  \bibfield  {author} {\bibinfo {author} {\bibfnamefont {B.~P.}\ \bibnamefont
  {Abbott}} \emph {et~al.} (\bibinfo {collaboration} {LIGO Scientific,
  Virgo}),\ }\href {\doibase 10.1103/PhysRevLett.119.161101} {\bibfield
  {journal} {\bibinfo  {journal} {Phys. Rev. Lett.}\ }\textbf {\bibinfo
  {volume} {119}},\ \bibinfo {pages} {161101} (\bibinfo {year}
  {2017}{\natexlab{a}})},\ \Eprint {http://arxiv.org/abs/1710.05832}
  {arXiv:1710.05832 [gr-qc]} \BibitemShut {NoStop}%
\bibitem [{\citenamefont {Favata}\ \emph {et~al.}(2021)\citenamefont {Favata},
  \citenamefont {Kim}, \citenamefont {Arun}, \citenamefont {Kim},\ and\
  \citenamefont {Lee}}]{Favata:2021vhw}%
  \BibitemOpen
  \bibfield  {author} {\bibinfo {author} {\bibfnamefont {M.}~\bibnamefont
  {Favata}}, \bibinfo {author} {\bibfnamefont {C.}~\bibnamefont {Kim}},
  \bibinfo {author} {\bibfnamefont {K.~G.}\ \bibnamefont {Arun}}, \bibinfo
  {author} {\bibfnamefont {J.}~\bibnamefont {Kim}}, \ and\ \bibinfo {author}
  {\bibfnamefont {H.~W.}\ \bibnamefont {Lee}},\ }\href@noop {} {\  (\bibinfo
  {year} {2021})},\ \Eprint {http://arxiv.org/abs/2108.05861} {arXiv:2108.05861
  [gr-qc]} \BibitemShut {NoStop}%
\bibitem [{\citenamefont {Lower}\ \emph {et~al.}(2018)\citenamefont {Lower},
  \citenamefont {Thrane}, \citenamefont {Lasky},\ and\ \citenamefont
  {Smith}}]{Lower:2018seu}%
  \BibitemOpen
  \bibfield  {author} {\bibinfo {author} {\bibfnamefont {M.~E.}\ \bibnamefont
  {Lower}}, \bibinfo {author} {\bibfnamefont {E.}~\bibnamefont {Thrane}},
  \bibinfo {author} {\bibfnamefont {P.~D.}\ \bibnamefont {Lasky}}, \ and\
  \bibinfo {author} {\bibfnamefont {R.}~\bibnamefont {Smith}},\ }\href
  {\doibase 10.1103/PhysRevD.98.083028} {\bibfield  {journal} {\bibinfo
  {journal} {Phys. Rev.}\ }\textbf {\bibinfo {volume} {D98}},\ \bibinfo {pages}
  {083028} (\bibinfo {year} {2018})},\ \Eprint
  {http://arxiv.org/abs/1806.05350} {arXiv:1806.05350 [astro-ph.HE]}
  \BibitemShut {NoStop}%
\bibitem [{\citenamefont {Romero-Shaw}\ \emph {et~al.}(2021)\citenamefont
  {Romero-Shaw}, \citenamefont {Lasky},\ and\ \citenamefont
  {Thrane}}]{Romero-Shaw:2021ual}%
  \BibitemOpen
  \bibfield  {author} {\bibinfo {author} {\bibfnamefont {I.~M.}\ \bibnamefont
  {Romero-Shaw}}, \bibinfo {author} {\bibfnamefont {P.~D.}\ \bibnamefont
  {Lasky}}, \ and\ \bibinfo {author} {\bibfnamefont {E.}~\bibnamefont
  {Thrane}},\ }\href {\doibase 10.3847/2041-8213/ac3138} {\bibfield  {journal}
  {\bibinfo  {journal} {Astrophys. J. Lett.}\ }\textbf {\bibinfo {volume}
  {921}},\ \bibinfo {pages} {L31} (\bibinfo {year} {2021})},\ \Eprint
  {http://arxiv.org/abs/2108.01284} {arXiv:2108.01284 [astro-ph.HE]}
  \BibitemShut {NoStop}%
\bibitem [{\citenamefont {Romero-Shaw}\ \emph {et~al.}(2019)\citenamefont
  {Romero-Shaw}, \citenamefont {Lasky},\ and\ \citenamefont
  {Thrane}}]{10.1093/mnras/stz2996}%
  \BibitemOpen
  \bibfield  {author} {\bibinfo {author} {\bibfnamefont {I.~M.}\ \bibnamefont
  {Romero-Shaw}}, \bibinfo {author} {\bibfnamefont {P.~D.}\ \bibnamefont
  {Lasky}}, \ and\ \bibinfo {author} {\bibfnamefont {E.}~\bibnamefont
  {Thrane}},\ }\href {\doibase 10.1093/mnras/stz2996} {\bibfield  {journal}
  {\bibinfo  {journal} {Monthly Notices of the Royal Astronomical Society}\
  }\textbf {\bibinfo {volume} {490}},\ \bibinfo {pages} {5210} (\bibinfo {year}
  {2019})},\ \Eprint
  {http://arxiv.org/abs/https://academic.oup.com/mnras/article-pdf/490/4/5210/30725872/stz2996.pdf}
  {https://academic.oup.com/mnras/article-pdf/490/4/5210/30725872/stz2996.pdf}
  \BibitemShut {NoStop}%
\bibitem [{\citenamefont {O'Shea}\ and\ \citenamefont
  {Kumar}(2021)}]{OShea:2021ugg}%
  \BibitemOpen
  \bibfield  {author} {\bibinfo {author} {\bibfnamefont {E.}~\bibnamefont
  {O'Shea}}\ and\ \bibinfo {author} {\bibfnamefont {P.}~\bibnamefont {Kumar}},\
  }\href@noop {} {\  (\bibinfo {year} {2021})},\ \Eprint
  {http://arxiv.org/abs/2107.07981} {arXiv:2107.07981 [astro-ph.HE]}
  \BibitemShut {NoStop}%
\bibitem [{\citenamefont {Romero-Shaw}\ \emph {et~al.}(2020)\citenamefont
  {Romero-Shaw}, \citenamefont {Lasky}, \citenamefont {Thrane},\ and\
  \citenamefont {Bustillo}}]{Romero-Shaw:2020thy}%
  \BibitemOpen
  \bibfield  {author} {\bibinfo {author} {\bibfnamefont {I.~M.}\ \bibnamefont
  {Romero-Shaw}}, \bibinfo {author} {\bibfnamefont {P.~D.}\ \bibnamefont
  {Lasky}}, \bibinfo {author} {\bibfnamefont {E.}~\bibnamefont {Thrane}}, \
  and\ \bibinfo {author} {\bibfnamefont {J.~C.}\ \bibnamefont {Bustillo}},\
  }\href {\doibase 10.3847/2041-8213/abbe26} {\bibfield  {journal} {\bibinfo
  {journal} {Astrophys. J. Lett.}\ }\textbf {\bibinfo {volume} {903}},\
  \bibinfo {pages} {L5} (\bibinfo {year} {2020})},\ \Eprint
  {http://arxiv.org/abs/2009.04771} {arXiv:2009.04771 [astro-ph.HE]}
  \BibitemShut {NoStop}%
\bibitem [{\citenamefont {Gayathri}\ \emph {et~al.}(2020)\citenamefont
  {Gayathri}, \citenamefont {Healy}, \citenamefont {Lange}, \citenamefont
  {O'Brien}, \citenamefont {Szczepanczyk}, \citenamefont {Bartos},
  \citenamefont {Campanelli}, \citenamefont {Klimenko}, \citenamefont
  {Lousto},\ and\ \citenamefont {O'Shaughnessy}}]{Gayathri:2020coq}%
  \BibitemOpen
  \bibfield  {author} {\bibinfo {author} {\bibfnamefont {V.}~\bibnamefont
  {Gayathri}}, \bibinfo {author} {\bibfnamefont {J.}~\bibnamefont {Healy}},
  \bibinfo {author} {\bibfnamefont {J.}~\bibnamefont {Lange}}, \bibinfo
  {author} {\bibfnamefont {B.}~\bibnamefont {O'Brien}}, \bibinfo {author}
  {\bibfnamefont {M.}~\bibnamefont {Szczepanczyk}}, \bibinfo {author}
  {\bibfnamefont {I.}~\bibnamefont {Bartos}}, \bibinfo {author} {\bibfnamefont
  {M.}~\bibnamefont {Campanelli}}, \bibinfo {author} {\bibfnamefont
  {S.}~\bibnamefont {Klimenko}}, \bibinfo {author} {\bibfnamefont
  {C.}~\bibnamefont {Lousto}}, \ and\ \bibinfo {author} {\bibfnamefont
  {R.}~\bibnamefont {O'Shaughnessy}},\ }\href@noop {} {\  (\bibinfo {year}
  {2020})},\ \Eprint {http://arxiv.org/abs/2009.05461} {arXiv:2009.05461
  [astro-ph.HE]} \BibitemShut {NoStop}%
\bibitem [{\citenamefont {Gamba}\ \emph {et~al.}(2021)\citenamefont {Gamba},
  \citenamefont {Breschi}, \citenamefont {Carullo}, \citenamefont {Rettegno},
  \citenamefont {Albanesi}, \citenamefont {Bernuzzi},\ and\ \citenamefont
  {Nagar}}]{Gamba:2021gap}%
  \BibitemOpen
  \bibfield  {author} {\bibinfo {author} {\bibfnamefont {R.}~\bibnamefont
  {Gamba}}, \bibinfo {author} {\bibfnamefont {M.}~\bibnamefont {Breschi}},
  \bibinfo {author} {\bibfnamefont {G.}~\bibnamefont {Carullo}}, \bibinfo
  {author} {\bibfnamefont {P.}~\bibnamefont {Rettegno}}, \bibinfo {author}
  {\bibfnamefont {S.}~\bibnamefont {Albanesi}}, \bibinfo {author}
  {\bibfnamefont {S.}~\bibnamefont {Bernuzzi}}, \ and\ \bibinfo {author}
  {\bibfnamefont {A.}~\bibnamefont {Nagar}},\ }\href@noop {} {\  (\bibinfo
  {year} {2021})},\ \Eprint {http://arxiv.org/abs/2106.05575} {arXiv:2106.05575
  [gr-qc]} \BibitemShut {NoStop}%
\bibitem [{\citenamefont {Hoy}\ \emph {et~al.}(2021)\citenamefont {Hoy},
  \citenamefont {Mills},\ and\ \citenamefont {Fairhurst}}]{Hoy:2021dqg}%
  \BibitemOpen
  \bibfield  {author} {\bibinfo {author} {\bibfnamefont {C.}~\bibnamefont
  {Hoy}}, \bibinfo {author} {\bibfnamefont {C.}~\bibnamefont {Mills}}, \ and\
  \bibinfo {author} {\bibfnamefont {S.}~\bibnamefont {Fairhurst}},\ }\href@noop
  {} {\  (\bibinfo {year} {2021})},\ \Eprint {http://arxiv.org/abs/2111.10455}
  {arXiv:2111.10455 [gr-qc]} \BibitemShut {NoStop}%
\bibitem [{\citenamefont {Abbott}\ \emph
  {et~al.}(2017{\natexlab{b}})\citenamefont {Abbott} \emph
  {et~al.}}]{LIGOScientific:2017ync}%
  \BibitemOpen
  \bibfield  {author} {\bibinfo {author} {\bibfnamefont {B.~P.}\ \bibnamefont
  {Abbott}} \emph {et~al.} (\bibinfo {collaboration} {LIGO Scientific, Virgo,
  Fermi GBM, INTEGRAL, IceCube, AstroSat Cadmium Zinc Telluride Imager Team,
  IPN, Insight-Hxmt, ANTARES, Swift, AGILE Team, 1M2H Team, Dark Energy Camera
  GW-EM, DES, DLT40, GRAWITA, Fermi-LAT, ATCA, ASKAP, Las Cumbres Observatory
  Group, OzGrav, DWF (Deeper Wider Faster Program), AST3, CAASTRO, VINROUGE,
  MASTER, J-GEM, GROWTH, JAGWAR, CaltechNRAO, TTU-NRAO, NuSTAR, Pan-STARRS,
  MAXI Team, TZAC Consortium, KU, Nordic Optical Telescope, ePESSTO, GROND,
  Texas Tech University, SALT Group, TOROS, BOOTES, MWA, CALET, IKI-GW
  Follow-up, H.E.S.S., LOFAR, LWA, HAWC, Pierre Auger, ALMA, Euro VLBI Team, Pi
  of Sky, Chandra Team at McGill University, DFN, ATLAS Telescopes, High Time
  Resolution Universe Survey, RIMAS, RATIR, SKA South Africa/MeerKAT}),\ }\href
  {\doibase 10.3847/2041-8213/aa91c9} {\bibfield  {journal} {\bibinfo
  {journal} {Astrophys. J. Lett.}\ }\textbf {\bibinfo {volume} {848}},\
  \bibinfo {pages} {L12} (\bibinfo {year} {2017}{\natexlab{b}})},\ \Eprint
  {http://arxiv.org/abs/1710.05833} {arXiv:1710.05833 [astro-ph.HE]}
  \BibitemShut {NoStop}%
\bibitem [{\citenamefont {Abbott}\ \emph
  {et~al.}(2019{\natexlab{b}})\citenamefont {Abbott} \emph
  {et~al.}}]{LIGOScientific:2019kan}%
  \BibitemOpen
  \bibfield  {author} {\bibinfo {author} {\bibfnamefont {B.~P.}\ \bibnamefont
  {Abbott}} \emph {et~al.} (\bibinfo {collaboration} {LIGO Scientific,
  Virgo}),\ }\href {\doibase 10.1103/PhysRevLett.123.161102} {\bibfield
  {journal} {\bibinfo  {journal} {Phys. Rev. Lett.}\ }\textbf {\bibinfo
  {volume} {123}},\ \bibinfo {pages} {161102} (\bibinfo {year}
  {2019}{\natexlab{b}})},\ \Eprint {http://arxiv.org/abs/1904.08976}
  {arXiv:1904.08976 [astro-ph.CO]} \BibitemShut {NoStop}%
\bibitem [{\citenamefont {Nitz}\ and\ \citenamefont
  {Wang}(2021{\natexlab{a}})}]{Nitz:2021mzz}%
  \BibitemOpen
  \bibfield  {author} {\bibinfo {author} {\bibfnamefont {A.~H.}\ \bibnamefont
  {Nitz}}\ and\ \bibinfo {author} {\bibfnamefont {Y.-F.}\ \bibnamefont
  {Wang}},\ }\href {\doibase 10.3847/1538-4357/ac01d9} {\  (\bibinfo {year}
  {2021}{\natexlab{a}}),\ 10.3847/1538-4357/ac01d9},\ \Eprint
  {http://arxiv.org/abs/2102.00868} {arXiv:2102.00868 [astro-ph.HE]}
  \BibitemShut {NoStop}%
\bibitem [{\citenamefont {Nitz}\ and\ \citenamefont
  {Wang}(2021{\natexlab{b}})}]{Nitz:2021vqh}%
  \BibitemOpen
  \bibfield  {author} {\bibinfo {author} {\bibfnamefont {A.~H.}\ \bibnamefont
  {Nitz}}\ and\ \bibinfo {author} {\bibfnamefont {Y.-F.}\ \bibnamefont
  {Wang}},\ }\href {\doibase 10.1103/PhysRevLett.127.151101} {\bibfield
  {journal} {\bibinfo  {journal} {Phys. Rev. Lett.}\ }\textbf {\bibinfo
  {volume} {127}},\ \bibinfo {pages} {151101} (\bibinfo {year}
  {2021}{\natexlab{b}})},\ \Eprint {http://arxiv.org/abs/2106.08979}
  {arXiv:2106.08979 [astro-ph.HE]} \BibitemShut {NoStop}%
\bibitem [{\citenamefont {Abbott}\ \emph
  {et~al.}(2021{\natexlab{d}})\citenamefont {Abbott} \emph
  {et~al.}}]{LIGOScientific:2021job}%
  \BibitemOpen
  \bibfield  {author} {\bibinfo {author} {\bibfnamefont {R.}~\bibnamefont
  {Abbott}} \emph {et~al.} (\bibinfo {collaboration} {LIGO Scientific, VIRGO,
  KAGRA}),\ }\href@noop {} {\  (\bibinfo {year} {2021}{\natexlab{d}})},\
  \Eprint {http://arxiv.org/abs/2109.12197} {arXiv:2109.12197 [astro-ph.CO]}
  \BibitemShut {NoStop}%
\bibitem [{\citenamefont {Graham}\ \emph {et~al.}(2020)\citenamefont {Graham}
  \emph {et~al.}}]{Graham:2020gwr}%
  \BibitemOpen
  \bibfield  {author} {\bibinfo {author} {\bibfnamefont {M.~J.}\ \bibnamefont
  {Graham}} \emph {et~al.},\ }\href {\doibase 10.1103/PhysRevLett.124.251102}
  {\bibfield  {journal} {\bibinfo  {journal} {Phys. Rev. Lett.}\ }\textbf
  {\bibinfo {volume} {124}},\ \bibinfo {pages} {251102} (\bibinfo {year}
  {2020})},\ \Eprint {http://arxiv.org/abs/2006.14122} {arXiv:2006.14122
  [astro-ph.HE]} \BibitemShut {NoStop}%
\bibitem [{\citenamefont {Tsai}\ \emph {et~al.}(2020)\citenamefont {Tsai},
  \citenamefont {Palmese}, \citenamefont {Profumo},\ and\ \citenamefont
  {Jeltema}}]{Tsai:2020hpi}%
  \BibitemOpen
  \bibfield  {author} {\bibinfo {author} {\bibfnamefont {Y.-D.}\ \bibnamefont
  {Tsai}}, \bibinfo {author} {\bibfnamefont {A.}~\bibnamefont {Palmese}},
  \bibinfo {author} {\bibfnamefont {S.}~\bibnamefont {Profumo}}, \ and\
  \bibinfo {author} {\bibfnamefont {T.}~\bibnamefont {Jeltema}},\ }\href
  {\doibase 10.1088/1475-7516/2021/10/019} {\  (\bibinfo {year} {2020}),\
  10.1088/1475-7516/2021/10/019},\ \Eprint {http://arxiv.org/abs/2007.03686}
  {arXiv:2007.03686 [astro-ph.HE]} \BibitemShut {NoStop}%
\bibitem [{\citenamefont {Sasaki}\ \emph {et~al.}(2021)\citenamefont {Sasaki},
  \citenamefont {Takhistov}, \citenamefont {Vardanyan},\ and\ \citenamefont
  {Zhang}}]{Sasaki:2021iuc}%
  \BibitemOpen
  \bibfield  {author} {\bibinfo {author} {\bibfnamefont {M.}~\bibnamefont
  {Sasaki}}, \bibinfo {author} {\bibfnamefont {V.}~\bibnamefont {Takhistov}},
  \bibinfo {author} {\bibfnamefont {V.}~\bibnamefont {Vardanyan}}, \ and\
  \bibinfo {author} {\bibfnamefont {Y.-l.}\ \bibnamefont {Zhang}},\ }\href@noop
  {} {\  (\bibinfo {year} {2021})},\ \Eprint {http://arxiv.org/abs/2110.09509}
  {arXiv:2110.09509 [astro-ph.CO]} \BibitemShut {NoStop}%
\bibitem [{\citenamefont {Kritos}\ and\ \citenamefont
  {Cholis}(2021)}]{Kritos:2021yty}%
  \BibitemOpen
  \bibfield  {author} {\bibinfo {author} {\bibfnamefont {K.}~\bibnamefont
  {Kritos}}\ and\ \bibinfo {author} {\bibfnamefont {I.}~\bibnamefont
  {Cholis}},\ }\href {\doibase 10.1103/PhysRevD.104.043004} {\bibfield
  {journal} {\bibinfo  {journal} {Phys. Rev. D}\ }\textbf {\bibinfo {volume}
  {104}},\ \bibinfo {pages} {043004} (\bibinfo {year} {2021})},\ \Eprint
  {http://arxiv.org/abs/2104.02073} {arXiv:2104.02073 [astro-ph.GA]}
  \BibitemShut {NoStop}%
\bibitem [{\citenamefont {Pankow}\ \emph {et~al.}(2017)\citenamefont {Pankow},
  \citenamefont {Sampson}, \citenamefont {Perri}, \citenamefont {Chase},
  \citenamefont {Coughlin}, \citenamefont {Zevin},\ and\ \citenamefont
  {Kalogera}}]{Pankow:2016udj}%
  \BibitemOpen
  \bibfield  {author} {\bibinfo {author} {\bibfnamefont {C.}~\bibnamefont
  {Pankow}}, \bibinfo {author} {\bibfnamefont {L.}~\bibnamefont {Sampson}},
  \bibinfo {author} {\bibfnamefont {L.}~\bibnamefont {Perri}}, \bibinfo
  {author} {\bibfnamefont {E.}~\bibnamefont {Chase}}, \bibinfo {author}
  {\bibfnamefont {S.}~\bibnamefont {Coughlin}}, \bibinfo {author}
  {\bibfnamefont {M.}~\bibnamefont {Zevin}}, \ and\ \bibinfo {author}
  {\bibfnamefont {V.}~\bibnamefont {Kalogera}},\ }\href {\doibase
  10.3847/1538-4357/834/2/154} {\bibfield  {journal} {\bibinfo  {journal}
  {Astrophys. J.}\ }\textbf {\bibinfo {volume} {834}},\ \bibinfo {pages} {154}
  (\bibinfo {year} {2017})},\ \Eprint {http://arxiv.org/abs/1610.05633}
  {arXiv:1610.05633 [astro-ph.HE]} \BibitemShut {NoStop}%
\bibitem [{\citenamefont {Vitale}\ \emph {et~al.}(2017)\citenamefont {Vitale},
  \citenamefont {Gerosa}, \citenamefont {Haster}, \citenamefont
  {Chatziioannou},\ and\ \citenamefont {Zimmerman}}]{Vitale:2017cfs}%
  \BibitemOpen
  \bibfield  {author} {\bibinfo {author} {\bibfnamefont {S.}~\bibnamefont
  {Vitale}}, \bibinfo {author} {\bibfnamefont {D.}~\bibnamefont {Gerosa}},
  \bibinfo {author} {\bibfnamefont {C.-J.}\ \bibnamefont {Haster}}, \bibinfo
  {author} {\bibfnamefont {K.}~\bibnamefont {Chatziioannou}}, \ and\ \bibinfo
  {author} {\bibfnamefont {A.}~\bibnamefont {Zimmerman}},\ }\href {\doibase
  10.1103/PhysRevLett.119.251103} {\bibfield  {journal} {\bibinfo  {journal}
  {Phys. Rev. Lett.}\ }\textbf {\bibinfo {volume} {119}},\ \bibinfo {pages}
  {251103} (\bibinfo {year} {2017})},\ \Eprint
  {http://arxiv.org/abs/1707.04637} {arXiv:1707.04637 [gr-qc]} \BibitemShut
  {NoStop}%
\bibitem [{\citenamefont {Mandel}\ and\ \citenamefont
  {Fragos}(2020)}]{Mandel:2020lhv}%
  \BibitemOpen
  \bibfield  {author} {\bibinfo {author} {\bibfnamefont {I.}~\bibnamefont
  {Mandel}}\ and\ \bibinfo {author} {\bibfnamefont {T.}~\bibnamefont
  {Fragos}},\ }\href {\doibase 10.3847/2041-8213/ab8e41} {\bibfield  {journal}
  {\bibinfo  {journal} {Astrophys. J. Lett.}\ }\textbf {\bibinfo {volume}
  {895}},\ \bibinfo {pages} {L28} (\bibinfo {year} {2020})},\ \Eprint
  {http://arxiv.org/abs/2004.09288} {arXiv:2004.09288 [astro-ph.HE]}
  \BibitemShut {NoStop}%
\bibitem [{\citenamefont {Gerosa}\ \emph {et~al.}(2020)\citenamefont {Gerosa},
  \citenamefont {Vitale},\ and\ \citenamefont {Berti}}]{Gerosa:2020bjb}%
  \BibitemOpen
  \bibfield  {author} {\bibinfo {author} {\bibfnamefont {D.}~\bibnamefont
  {Gerosa}}, \bibinfo {author} {\bibfnamefont {S.}~\bibnamefont {Vitale}}, \
  and\ \bibinfo {author} {\bibfnamefont {E.}~\bibnamefont {Berti}},\ }\href
  {\doibase 10.1103/PhysRevLett.125.101103} {\bibfield  {journal} {\bibinfo
  {journal} {Phys. Rev. Lett.}\ }\textbf {\bibinfo {volume} {125}},\ \bibinfo
  {pages} {101103} (\bibinfo {year} {2020})},\ \Eprint
  {http://arxiv.org/abs/2005.04243} {arXiv:2005.04243 [astro-ph.HE]}
  \BibitemShut {NoStop}%
\bibitem [{\citenamefont {Zevin}\ \emph {et~al.}(2020)\citenamefont {Zevin},
  \citenamefont {Berry}, \citenamefont {Coughlin}, \citenamefont
  {Chatziioannou},\ and\ \citenamefont {Vitale}}]{Zevin:2020gxf}%
  \BibitemOpen
  \bibfield  {author} {\bibinfo {author} {\bibfnamefont {M.}~\bibnamefont
  {Zevin}}, \bibinfo {author} {\bibfnamefont {C.~P.~L.}\ \bibnamefont {Berry}},
  \bibinfo {author} {\bibfnamefont {S.}~\bibnamefont {Coughlin}}, \bibinfo
  {author} {\bibfnamefont {K.}~\bibnamefont {Chatziioannou}}, \ and\ \bibinfo
  {author} {\bibfnamefont {S.}~\bibnamefont {Vitale}},\ }\href {\doibase
  10.3847/2041-8213/aba8ef} {\bibfield  {journal} {\bibinfo  {journal}
  {Astrophys. J. Lett.}\ }\textbf {\bibinfo {volume} {899}},\ \bibinfo {pages}
  {L17} (\bibinfo {year} {2020})},\ \Eprint {http://arxiv.org/abs/2006.11293}
  {arXiv:2006.11293 [astro-ph.HE]} \BibitemShut {NoStop}%
\bibitem [{\citenamefont {Callister}\ \emph {et~al.}(2021)\citenamefont
  {Callister}, \citenamefont {Haster}, \citenamefont {Ng}, \citenamefont
  {Vitale},\ and\ \citenamefont {Farr}}]{Callister:2021fpo}%
  \BibitemOpen
  \bibfield  {author} {\bibinfo {author} {\bibfnamefont {T.~A.}\ \bibnamefont
  {Callister}}, \bibinfo {author} {\bibfnamefont {C.-J.}\ \bibnamefont
  {Haster}}, \bibinfo {author} {\bibfnamefont {K.~K.~Y.}\ \bibnamefont {Ng}},
  \bibinfo {author} {\bibfnamefont {S.}~\bibnamefont {Vitale}}, \ and\ \bibinfo
  {author} {\bibfnamefont {W.~M.}\ \bibnamefont {Farr}},\ }\href {\doibase
  10.3847/2041-8213/ac2ccc} {\bibfield  {journal} {\bibinfo  {journal}
  {Astrophys. J. Lett.}\ }\textbf {\bibinfo {volume} {922}},\ \bibinfo {pages}
  {L5} (\bibinfo {year} {2021})},\ \Eprint {http://arxiv.org/abs/2106.00521}
  {arXiv:2106.00521 [astro-ph.HE]} \BibitemShut {NoStop}%
\bibitem [{\citenamefont {Franciolini}\ and\ \citenamefont
  {Pani}(2022)}]{Franciolini:2022iaa}%
  \BibitemOpen
  \bibfield  {author} {\bibinfo {author} {\bibfnamefont {G.}~\bibnamefont
  {Franciolini}}\ and\ \bibinfo {author} {\bibfnamefont {P.}~\bibnamefont
  {Pani}},\ }\href@noop {} {\  (\bibinfo {year} {2022})},\ \Eprint
  {http://arxiv.org/abs/2201.13098} {arXiv:2201.13098 [astro-ph.HE]}
  \BibitemShut {NoStop}%
\bibitem [{\citenamefont {Vitale}\ and\ \citenamefont
  {Evans}(2017)}]{Vitale:2016icu}%
  \BibitemOpen
  \bibfield  {author} {\bibinfo {author} {\bibfnamefont {S.}~\bibnamefont
  {Vitale}}\ and\ \bibinfo {author} {\bibfnamefont {M.}~\bibnamefont {Evans}},\
  }\href {\doibase 10.1103/PhysRevD.95.064052} {\bibfield  {journal} {\bibinfo
  {journal} {Phys. Rev. D}\ }\textbf {\bibinfo {volume} {95}},\ \bibinfo
  {pages} {064052} (\bibinfo {year} {2017})},\ \Eprint
  {http://arxiv.org/abs/1610.06917} {arXiv:1610.06917 [gr-qc]} \BibitemShut
  {NoStop}%
\bibitem [{\citenamefont {Moore}\ and\ \citenamefont
  {Yunes}(2020)}]{Moore:2019vjj}%
  \BibitemOpen
  \bibfield  {author} {\bibinfo {author} {\bibfnamefont {B.}~\bibnamefont
  {Moore}}\ and\ \bibinfo {author} {\bibfnamefont {N.}~\bibnamefont {Yunes}},\
  }\href {\doibase 10.1088/1361-6382/ab7963} {\bibfield  {journal} {\bibinfo
  {journal} {Class. Quant. Grav.}\ }\textbf {\bibinfo {volume} {37}},\ \bibinfo
  {pages} {225015} (\bibinfo {year} {2020})},\ \Eprint
  {http://arxiv.org/abs/1910.01680} {arXiv:1910.01680 [gr-qc]} \BibitemShut
  {NoStop}%
\bibitem [{\citenamefont {Sesana}(2016)}]{Sesana:2016ljz}%
  \BibitemOpen
  \bibfield  {author} {\bibinfo {author} {\bibfnamefont {A.}~\bibnamefont
  {Sesana}},\ }\href {\doibase 10.1103/PhysRevLett.116.231102} {\bibfield
  {journal} {\bibinfo  {journal} {Phys. Rev. Lett.}\ }\textbf {\bibinfo
  {volume} {116}},\ \bibinfo {pages} {231102} (\bibinfo {year} {2016})},\
  \Eprint {http://arxiv.org/abs/1602.06951} {arXiv:1602.06951 [gr-qc]}
  \BibitemShut {NoStop}%
\bibitem [{\citenamefont {Wong}\ \emph {et~al.}(2018)\citenamefont {Wong},
  \citenamefont {Kovetz}, \citenamefont {Cutler},\ and\ \citenamefont
  {Berti}}]{Wong:2018uwb}%
  \BibitemOpen
  \bibfield  {author} {\bibinfo {author} {\bibfnamefont {K.~W.~K.}\
  \bibnamefont {Wong}}, \bibinfo {author} {\bibfnamefont {E.~D.}\ \bibnamefont
  {Kovetz}}, \bibinfo {author} {\bibfnamefont {C.}~\bibnamefont {Cutler}}, \
  and\ \bibinfo {author} {\bibfnamefont {E.}~\bibnamefont {Berti}},\ }\href
  {\doibase 10.1103/PhysRevLett.121.251102} {\bibfield  {journal} {\bibinfo
  {journal} {Phys. Rev. Lett.}\ }\textbf {\bibinfo {volume} {121}},\ \bibinfo
  {pages} {251102} (\bibinfo {year} {2018})},\ \Eprint
  {http://arxiv.org/abs/1808.08247} {arXiv:1808.08247 [astro-ph.HE]}
  \BibitemShut {NoStop}%
\bibitem [{\citenamefont {Cutler}\ \emph {et~al.}(2019)\citenamefont {Cutler}
  \emph {et~al.}}]{Cutler:2019krq}%
  \BibitemOpen
  \bibfield  {author} {\bibinfo {author} {\bibfnamefont {C.}~\bibnamefont
  {Cutler}} \emph {et~al.},\ }\href@noop {} {\  (\bibinfo {year} {2019})},\
  \Eprint {http://arxiv.org/abs/1903.04069} {arXiv:1903.04069 [astro-ph.HE]}
  \BibitemShut {NoStop}%
\bibitem [{\citenamefont {Gerosa}\ \emph {et~al.}(2019)\citenamefont {Gerosa},
  \citenamefont {Ma}, \citenamefont {Wong}, \citenamefont {Berti},
  \citenamefont {O'Shaughnessy}, \citenamefont {Chen},\ and\ \citenamefont
  {Belczynski}}]{Gerosa:2019dbe}%
  \BibitemOpen
  \bibfield  {author} {\bibinfo {author} {\bibfnamefont {D.}~\bibnamefont
  {Gerosa}}, \bibinfo {author} {\bibfnamefont {S.}~\bibnamefont {Ma}}, \bibinfo
  {author} {\bibfnamefont {K.~W.~K.}\ \bibnamefont {Wong}}, \bibinfo {author}
  {\bibfnamefont {E.}~\bibnamefont {Berti}}, \bibinfo {author} {\bibfnamefont
  {R.}~\bibnamefont {O'Shaughnessy}}, \bibinfo {author} {\bibfnamefont
  {Y.}~\bibnamefont {Chen}}, \ and\ \bibinfo {author} {\bibfnamefont
  {K.}~\bibnamefont {Belczynski}},\ }\href {\doibase
  10.1103/PhysRevD.99.103004} {\bibfield  {journal} {\bibinfo  {journal} {Phys.
  Rev. D}\ }\textbf {\bibinfo {volume} {99}},\ \bibinfo {pages} {103004}
  (\bibinfo {year} {2019})},\ \Eprint {http://arxiv.org/abs/1902.00021}
  {arXiv:1902.00021 [astro-ph.HE]} \BibitemShut {NoStop}%
\bibitem [{\citenamefont {Moore}\ \emph {et~al.}(2019)\citenamefont {Moore},
  \citenamefont {Gerosa},\ and\ \citenamefont {Klein}}]{Moore:2019pke}%
  \BibitemOpen
  \bibfield  {author} {\bibinfo {author} {\bibfnamefont {C.~J.}\ \bibnamefont
  {Moore}}, \bibinfo {author} {\bibfnamefont {D.}~\bibnamefont {Gerosa}}, \
  and\ \bibinfo {author} {\bibfnamefont {A.}~\bibnamefont {Klein}},\ }\href
  {\doibase 10.1093/mnrasl/slz104} {\bibfield  {journal} {\bibinfo  {journal}
  {Mon. Not. Roy. Astron. Soc.}\ }\textbf {\bibinfo {volume} {488}},\ \bibinfo
  {pages} {L94} (\bibinfo {year} {2019})},\ \Eprint
  {http://arxiv.org/abs/1905.11998} {arXiv:1905.11998 [astro-ph.HE]}
  \BibitemShut {NoStop}%
\bibitem [{\citenamefont {Ewing}\ \emph {et~al.}(2021)\citenamefont {Ewing},
  \citenamefont {Sachdev}, \citenamefont {Borhanian},\ and\ \citenamefont
  {Sathyaprakash}}]{Ewing:2020brd}%
  \BibitemOpen
  \bibfield  {author} {\bibinfo {author} {\bibfnamefont {B.}~\bibnamefont
  {Ewing}}, \bibinfo {author} {\bibfnamefont {S.}~\bibnamefont {Sachdev}},
  \bibinfo {author} {\bibfnamefont {S.}~\bibnamefont {Borhanian}}, \ and\
  \bibinfo {author} {\bibfnamefont {B.~S.}\ \bibnamefont {Sathyaprakash}},\
  }\href {\doibase 10.1103/PhysRevD.103.023025} {\bibfield  {journal} {\bibinfo
   {journal} {Phys. Rev. D}\ }\textbf {\bibinfo {volume} {103}},\ \bibinfo
  {pages} {023025} (\bibinfo {year} {2021})},\ \Eprint
  {http://arxiv.org/abs/2011.03036} {arXiv:2011.03036 [gr-qc]} \BibitemShut
  {NoStop}%
\bibitem [{\citenamefont {Buscicchio}\ \emph {et~al.}(2021)\citenamefont
  {Buscicchio}, \citenamefont {Klein}, \citenamefont {Roebber}, \citenamefont
  {Moore}, \citenamefont {Gerosa}, \citenamefont {Finch},\ and\ \citenamefont
  {Vecchio}}]{Buscicchio:2021dph}%
  \BibitemOpen
  \bibfield  {author} {\bibinfo {author} {\bibfnamefont {R.}~\bibnamefont
  {Buscicchio}}, \bibinfo {author} {\bibfnamefont {A.}~\bibnamefont {Klein}},
  \bibinfo {author} {\bibfnamefont {E.}~\bibnamefont {Roebber}}, \bibinfo
  {author} {\bibfnamefont {C.~J.}\ \bibnamefont {Moore}}, \bibinfo {author}
  {\bibfnamefont {D.}~\bibnamefont {Gerosa}}, \bibinfo {author} {\bibfnamefont
  {E.}~\bibnamefont {Finch}}, \ and\ \bibinfo {author} {\bibfnamefont
  {A.}~\bibnamefont {Vecchio}},\ }\href {\doibase 10.1103/PhysRevD.104.044065}
  {\bibfield  {journal} {\bibinfo  {journal} {Phys. Rev. D}\ }\textbf {\bibinfo
  {volume} {104}},\ \bibinfo {pages} {044065} (\bibinfo {year} {2021})},\
  \Eprint {http://arxiv.org/abs/2106.05259} {arXiv:2106.05259 [astro-ph.HE]}
  \BibitemShut {NoStop}%
\bibitem [{web()}]{webpage}%
  \BibitemOpen
  \href@noop {} {}\bibinfo {note}
  {\noindent\url{https://web.uniroma1.it/gmunu/}}\BibitemShut {NoStop}%
\bibitem [{\citenamefont {Yunes}\ and\ \citenamefont
  {Berti}(2008)}]{Yunes:2008tw}%
  \BibitemOpen
  \bibfield  {author} {\bibinfo {author} {\bibfnamefont {N.}~\bibnamefont
  {Yunes}}\ and\ \bibinfo {author} {\bibfnamefont {E.}~\bibnamefont {Berti}},\
  }\href {\doibase 10.1103/PhysRevD.77.124006} {\bibfield  {journal} {\bibinfo
  {journal} {Phys. Rev. D}\ }\textbf {\bibinfo {volume} {77}},\ \bibinfo
  {pages} {124006} (\bibinfo {year} {2008})},\ \bibinfo {note} {[Erratum:
  Phys.Rev.D 83, 109901 (2011)]},\ \Eprint {http://arxiv.org/abs/0803.1853}
  {arXiv:0803.1853 [gr-qc]} \BibitemShut {NoStop}%
\bibitem [{\citenamefont {Moore}\ \emph {et~al.}(2016)\citenamefont {Moore},
  \citenamefont {Favata}, \citenamefont {Arun},\ and\ \citenamefont
  {Mishra}}]{Moore:2016qxz}%
  \BibitemOpen
  \bibfield  {author} {\bibinfo {author} {\bibfnamefont {B.}~\bibnamefont
  {Moore}}, \bibinfo {author} {\bibfnamefont {M.}~\bibnamefont {Favata}},
  \bibinfo {author} {\bibfnamefont {K.~G.}\ \bibnamefont {Arun}}, \ and\
  \bibinfo {author} {\bibfnamefont {C.~K.}\ \bibnamefont {Mishra}},\ }\href
  {\doibase 10.1103/PhysRevD.93.124061} {\bibfield  {journal} {\bibinfo
  {journal} {Phys. Rev. D}\ }\textbf {\bibinfo {volume} {93}},\ \bibinfo
  {pages} {124061} (\bibinfo {year} {2016})},\ \Eprint
  {http://arxiv.org/abs/1605.00304} {arXiv:1605.00304 [gr-qc]} \BibitemShut
  {NoStop}%
\bibitem [{\citenamefont {Buonanno}\ \emph {et~al.}(2009)\citenamefont
  {Buonanno}, \citenamefont {Iyer}, \citenamefont {Ochsner}, \citenamefont
  {Pan},\ and\ \citenamefont {Sathyaprakash}}]{PhysRevD.80.084043}%
  \BibitemOpen
  \bibfield  {author} {\bibinfo {author} {\bibfnamefont {A.}~\bibnamefont
  {Buonanno}}, \bibinfo {author} {\bibfnamefont {B.~R.}\ \bibnamefont {Iyer}},
  \bibinfo {author} {\bibfnamefont {E.}~\bibnamefont {Ochsner}}, \bibinfo
  {author} {\bibfnamefont {Y.}~\bibnamefont {Pan}}, \ and\ \bibinfo {author}
  {\bibfnamefont {B.~S.}\ \bibnamefont {Sathyaprakash}},\ }\href {\doibase
  10.1103/PhysRevD.80.084043} {\bibfield  {journal} {\bibinfo  {journal} {Phys.
  Rev. D}\ }\textbf {\bibinfo {volume} {80}},\ \bibinfo {pages} {084043}
  (\bibinfo {year} {2009})}\BibitemShut {NoStop}%
\bibitem [{\citenamefont {Kidder}\ \emph {et~al.}(1993)\citenamefont {Kidder},
  \citenamefont {Will},\ and\ \citenamefont {Wiseman}}]{Kidder:1992fr}%
  \BibitemOpen
  \bibfield  {author} {\bibinfo {author} {\bibfnamefont {L.~E.}\ \bibnamefont
  {Kidder}}, \bibinfo {author} {\bibfnamefont {C.~M.}\ \bibnamefont {Will}}, \
  and\ \bibinfo {author} {\bibfnamefont {A.~G.}\ \bibnamefont {Wiseman}},\
  }\href {\doibase 10.1103/PhysRevD.47.R4183} {\bibfield  {journal} {\bibinfo
  {journal} {Phys. Rev. D}\ }\textbf {\bibinfo {volume} {47}},\ \bibinfo
  {pages} {R4183} (\bibinfo {year} {1993})},\ \Eprint
  {http://arxiv.org/abs/gr-qc/9211025} {arXiv:gr-qc/9211025} \BibitemShut
  {NoStop}%
\bibitem [{\citenamefont {Poisson}(1993)}]{PhysRevD.48.1860}%
  \BibitemOpen
  \bibfield  {author} {\bibinfo {author} {\bibfnamefont {E.}~\bibnamefont
  {Poisson}},\ }\href {\doibase 10.1103/PhysRevD.48.1860} {\bibfield  {journal}
  {\bibinfo  {journal} {Phys. Rev. D}\ }\textbf {\bibinfo {volume} {48}},\
  \bibinfo {pages} {1860} (\bibinfo {year} {1993})}\BibitemShut {NoStop}%
\bibitem [{\citenamefont {Kidder}(1995)}]{Kidder:1995zr}%
  \BibitemOpen
  \bibfield  {author} {\bibinfo {author} {\bibfnamefont {L.~E.}\ \bibnamefont
  {Kidder}},\ }\href {\doibase 10.1103/PhysRevD.52.821} {\bibfield  {journal}
  {\bibinfo  {journal} {Phys. Rev. D}\ }\textbf {\bibinfo {volume} {52}},\
  \bibinfo {pages} {821} (\bibinfo {year} {1995})},\ \Eprint
  {http://arxiv.org/abs/gr-qc/9506022} {arXiv:gr-qc/9506022} \BibitemShut
  {NoStop}%
\bibitem [{\citenamefont {Mikoczi}\ \emph {et~al.}(2005)\citenamefont
  {Mikoczi}, \citenamefont {Vasuth},\ and\ \citenamefont
  {Gergely}}]{Mikoczi:2005dn}%
  \BibitemOpen
  \bibfield  {author} {\bibinfo {author} {\bibfnamefont {B.}~\bibnamefont
  {Mikoczi}}, \bibinfo {author} {\bibfnamefont {M.}~\bibnamefont {Vasuth}}, \
  and\ \bibinfo {author} {\bibfnamefont {L.~A.}\ \bibnamefont {Gergely}},\
  }\href {\doibase 10.1103/PhysRevD.71.124043} {\bibfield  {journal} {\bibinfo
  {journal} {Phys. Rev. D}\ }\textbf {\bibinfo {volume} {71}},\ \bibinfo
  {pages} {124043} (\bibinfo {year} {2005})},\ \Eprint
  {http://arxiv.org/abs/astro-ph/0504538} {arXiv:astro-ph/0504538} \BibitemShut
  {NoStop}%
\bibitem [{\citenamefont {Poisson}(1998)}]{Poisson:1997ha}%
  \BibitemOpen
  \bibfield  {author} {\bibinfo {author} {\bibfnamefont {E.}~\bibnamefont
  {Poisson}},\ }\href {\doibase 10.1103/PhysRevD.57.5287} {\bibfield  {journal}
  {\bibinfo  {journal} {Phys. Rev. D}\ }\textbf {\bibinfo {volume} {57}},\
  \bibinfo {pages} {5287} (\bibinfo {year} {1998})},\ \Eprint
  {http://arxiv.org/abs/gr-qc/9709032} {arXiv:gr-qc/9709032} \BibitemShut
  {NoStop}%
\bibitem [{\citenamefont {Gergely}(2000)}]{Gergely:1999pd}%
  \BibitemOpen
  \bibfield  {author} {\bibinfo {author} {\bibfnamefont {L.~A.}\ \bibnamefont
  {Gergely}},\ }\href {\doibase 10.1103/PhysRevD.61.024035} {\bibfield
  {journal} {\bibinfo  {journal} {Phys. Rev. D}\ }\textbf {\bibinfo {volume}
  {61}},\ \bibinfo {pages} {024035} (\bibinfo {year} {2000})},\ \Eprint
  {http://arxiv.org/abs/gr-qc/9911082} {arXiv:gr-qc/9911082} \BibitemShut
  {NoStop}%
\bibitem [{\citenamefont {Blanchet}\ \emph {et~al.}(2006)\citenamefont
  {Blanchet}, \citenamefont {Buonanno},\ and\ \citenamefont
  {Faye}}]{PhysRevD.74.104034}%
  \BibitemOpen
  \bibfield  {author} {\bibinfo {author} {\bibfnamefont {L.}~\bibnamefont
  {Blanchet}}, \bibinfo {author} {\bibfnamefont {A.}~\bibnamefont {Buonanno}},
  \ and\ \bibinfo {author} {\bibfnamefont {G.}~\bibnamefont {Faye}},\ }\href
  {\doibase 10.1103/PhysRevD.74.104034} {\bibfield  {journal} {\bibinfo
  {journal} {Phys. Rev. D}\ }\textbf {\bibinfo {volume} {74}},\ \bibinfo
  {pages} {104034} (\bibinfo {year} {2006})}\BibitemShut {NoStop}%
\bibitem [{\citenamefont {Mishra}\ \emph {et~al.}(2016)\citenamefont {Mishra},
  \citenamefont {Kela}, \citenamefont {Arun},\ and\ \citenamefont
  {Faye}}]{Mishra:2016whh}%
  \BibitemOpen
  \bibfield  {author} {\bibinfo {author} {\bibfnamefont {C.~K.}\ \bibnamefont
  {Mishra}}, \bibinfo {author} {\bibfnamefont {A.}~\bibnamefont {Kela}},
  \bibinfo {author} {\bibfnamefont {K.~G.}\ \bibnamefont {Arun}}, \ and\
  \bibinfo {author} {\bibfnamefont {G.}~\bibnamefont {Faye}},\ }\href {\doibase
  10.1103/PhysRevD.93.084054} {\bibfield  {journal} {\bibinfo  {journal} {Phys.
  Rev. D}\ }\textbf {\bibinfo {volume} {93}},\ \bibinfo {pages} {084054}
  (\bibinfo {year} {2016})},\ \Eprint {http://arxiv.org/abs/1601.05588}
  {arXiv:1601.05588 [gr-qc]} \BibitemShut {NoStop}%
\bibitem [{\citenamefont {Peters}\ and\ \citenamefont
  {Mathews}(1963{\natexlab{b}})}]{PhysRev.131.435}%
  \BibitemOpen
  \bibfield  {author} {\bibinfo {author} {\bibfnamefont {P.~C.}\ \bibnamefont
  {Peters}}\ and\ \bibinfo {author} {\bibfnamefont {J.}~\bibnamefont
  {Mathews}},\ }\href {\doibase 10.1103/PhysRev.131.435} {\bibfield  {journal}
  {\bibinfo  {journal} {Phys. Rev.}\ }\textbf {\bibinfo {volume} {131}},\
  \bibinfo {pages} {435} (\bibinfo {year} {1963}{\natexlab{b}})}\BibitemShut
  {NoStop}%
\bibitem [{\citenamefont {{Junker}}\ and\ \citenamefont
  {{Schaefer}}(1992)}]{1992MNRAS.254..146J}%
  \BibitemOpen
  \bibfield  {author} {\bibinfo {author} {\bibfnamefont {W.}~\bibnamefont
  {{Junker}}}\ and\ \bibinfo {author} {\bibfnamefont {G.}~\bibnamefont
  {{Schaefer}}},\ }\href {\doibase 10.1093/mnras/254.1.146} {\bibfield
  {journal} {\bibinfo  {journal} {\mnras}\ }\textbf {\bibinfo {volume} {254}},\
  \bibinfo {pages} {146} (\bibinfo {year} {1992})}\BibitemShut {NoStop}%
\bibitem [{\citenamefont {Gopakumar}\ \emph {et~al.}(1997)\citenamefont
  {Gopakumar}, \citenamefont {Iyer},\ and\ \citenamefont
  {Iyer}}]{Gopakumar:1997ng}%
  \BibitemOpen
  \bibfield  {author} {\bibinfo {author} {\bibfnamefont {A.}~\bibnamefont
  {Gopakumar}}, \bibinfo {author} {\bibfnamefont {B.~R.}\ \bibnamefont {Iyer}},
  \ and\ \bibinfo {author} {\bibfnamefont {S.}~\bibnamefont {Iyer}},\ }\href
  {\doibase 10.1103/PhysRevD.57.6562} {\bibfield  {journal} {\bibinfo
  {journal} {Phys. Rev. D}\ }\textbf {\bibinfo {volume} {55}},\ \bibinfo
  {pages} {6030} (\bibinfo {year} {1997})},\ \bibinfo {note} {[Erratum:
  Phys.Rev.D 57, 6562 (1998)]},\ \Eprint {http://arxiv.org/abs/gr-qc/9703075}
  {arXiv:gr-qc/9703075} \BibitemShut {NoStop}%
\bibitem [{\citenamefont {Arun}\ \emph
  {et~al.}(2008{\natexlab{a}})\citenamefont {Arun}, \citenamefont {Blanchet},
  \citenamefont {Iyer},\ and\ \citenamefont {Qusailah}}]{Arun:2007rg}%
  \BibitemOpen
  \bibfield  {author} {\bibinfo {author} {\bibfnamefont {K.~G.}\ \bibnamefont
  {Arun}}, \bibinfo {author} {\bibfnamefont {L.}~\bibnamefont {Blanchet}},
  \bibinfo {author} {\bibfnamefont {B.~R.}\ \bibnamefont {Iyer}}, \ and\
  \bibinfo {author} {\bibfnamefont {M.~S.~S.}\ \bibnamefont {Qusailah}},\
  }\href {\doibase 10.1103/PhysRevD.77.064034} {\bibfield  {journal} {\bibinfo
  {journal} {Phys. Rev. D}\ }\textbf {\bibinfo {volume} {77}},\ \bibinfo
  {pages} {064034} (\bibinfo {year} {2008}{\natexlab{a}})},\ \Eprint
  {http://arxiv.org/abs/0711.0250} {arXiv:0711.0250 [gr-qc]} \BibitemShut
  {NoStop}%
\bibitem [{\citenamefont {Arun}\ \emph
  {et~al.}(2008{\natexlab{b}})\citenamefont {Arun}, \citenamefont {Blanchet},
  \citenamefont {Iyer},\ and\ \citenamefont {Qusailah}}]{Arun:2007sg}%
  \BibitemOpen
  \bibfield  {author} {\bibinfo {author} {\bibfnamefont {K.~G.}\ \bibnamefont
  {Arun}}, \bibinfo {author} {\bibfnamefont {L.}~\bibnamefont {Blanchet}},
  \bibinfo {author} {\bibfnamefont {B.~R.}\ \bibnamefont {Iyer}}, \ and\
  \bibinfo {author} {\bibfnamefont {M.~S.~S.}\ \bibnamefont {Qusailah}},\
  }\href {\doibase 10.1103/PhysRevD.77.064035} {\bibfield  {journal} {\bibinfo
  {journal} {Phys. Rev. D}\ }\textbf {\bibinfo {volume} {77}},\ \bibinfo
  {pages} {064035} (\bibinfo {year} {2008}{\natexlab{b}})},\ \Eprint
  {http://arxiv.org/abs/0711.0302} {arXiv:0711.0302 [gr-qc]} \BibitemShut
  {NoStop}%
\bibitem [{\citenamefont {Arun}\ \emph {et~al.}(2009)\citenamefont {Arun},
  \citenamefont {Blanchet}, \citenamefont {Iyer},\ and\ \citenamefont
  {Sinha}}]{Arun:2009mc}%
  \BibitemOpen
  \bibfield  {author} {\bibinfo {author} {\bibfnamefont {K.~G.}\ \bibnamefont
  {Arun}}, \bibinfo {author} {\bibfnamefont {L.}~\bibnamefont {Blanchet}},
  \bibinfo {author} {\bibfnamefont {B.~R.}\ \bibnamefont {Iyer}}, \ and\
  \bibinfo {author} {\bibfnamefont {S.}~\bibnamefont {Sinha}},\ }\href
  {\doibase 10.1103/PhysRevD.80.124018} {\bibfield  {journal} {\bibinfo
  {journal} {Phys. Rev. D}\ }\textbf {\bibinfo {volume} {80}},\ \bibinfo
  {pages} {124018} (\bibinfo {year} {2009})},\ \Eprint
  {http://arxiv.org/abs/0908.3854} {arXiv:0908.3854 [gr-qc]} \BibitemShut
  {NoStop}%
\bibitem [{\citenamefont {Ade}\ \emph {et~al.}(2016)\citenamefont {Ade} \emph
  {et~al.}}]{Planck:2015fie}%
  \BibitemOpen
  \bibfield  {author} {\bibinfo {author} {\bibfnamefont {P.~A.~R.}\
  \bibnamefont {Ade}} \emph {et~al.} (\bibinfo {collaboration} {Planck}),\
  }\href {\doibase 10.1051/0004-6361/201525830} {\bibfield  {journal} {\bibinfo
   {journal} {Astron. Astrophys.}\ }\textbf {\bibinfo {volume} {594}},\
  \bibinfo {pages} {A13} (\bibinfo {year} {2016})},\ \Eprint
  {http://arxiv.org/abs/1502.01589} {arXiv:1502.01589 [astro-ph.CO]}
  \BibitemShut {NoStop}%
\bibitem [{\citenamefont {Finn}(1992)}]{PhysRevD.46.5236}%
  \BibitemOpen
  \bibfield  {author} {\bibinfo {author} {\bibfnamefont {L.~S.}\ \bibnamefont
  {Finn}},\ }\href {\doibase 10.1103/PhysRevD.46.5236} {\bibfield  {journal}
  {\bibinfo  {journal} {Phys. Rev. D}\ }\textbf {\bibinfo {volume} {46}},\
  \bibinfo {pages} {5236} (\bibinfo {year} {1992})}\BibitemShut {NoStop}%
\bibitem [{\citenamefont {Finn}\ and\ \citenamefont
  {Chernoff}(1993)}]{PhysRevD.47.2198}%
  \BibitemOpen
  \bibfield  {author} {\bibinfo {author} {\bibfnamefont {L.~S.}\ \bibnamefont
  {Finn}}\ and\ \bibinfo {author} {\bibfnamefont {D.~F.}\ \bibnamefont
  {Chernoff}},\ }\href {\doibase 10.1103/PhysRevD.47.2198} {\bibfield
  {journal} {\bibinfo  {journal} {Phys. Rev. D}\ }\textbf {\bibinfo {volume}
  {47}},\ \bibinfo {pages} {2198} (\bibinfo {year} {1993})}\BibitemShut
  {NoStop}%
\bibitem [{\citenamefont {Cutler}\ and\ \citenamefont
  {Flanagan}(1994)}]{Cutler:1994ys}%
  \BibitemOpen
  \bibfield  {author} {\bibinfo {author} {\bibfnamefont {C.}~\bibnamefont
  {Cutler}}\ and\ \bibinfo {author} {\bibfnamefont {E.~E.}\ \bibnamefont
  {Flanagan}},\ }\href {\doibase 10.1103/PhysRevD.49.2658} {\bibfield
  {journal} {\bibinfo  {journal} {Phys. Rev. D}\ }\textbf {\bibinfo {volume}
  {49}},\ \bibinfo {pages} {2658} (\bibinfo {year} {1994})},\ \Eprint
  {http://arxiv.org/abs/gr-qc/9402014} {arXiv:gr-qc/9402014} \BibitemShut
  {NoStop}%
\bibitem [{\citenamefont {Poisson}\ and\ \citenamefont
  {Will}(1995)}]{Poisson:1995ef}%
  \BibitemOpen
  \bibfield  {author} {\bibinfo {author} {\bibfnamefont {E.}~\bibnamefont
  {Poisson}}\ and\ \bibinfo {author} {\bibfnamefont {C.~M.}\ \bibnamefont
  {Will}},\ }\href {\doibase 10.1103/PhysRevD.52.848} {\bibfield  {journal}
  {\bibinfo  {journal} {Phys. Rev. D}\ }\textbf {\bibinfo {volume} {52}},\
  \bibinfo {pages} {848} (\bibinfo {year} {1995})},\ \Eprint
  {http://arxiv.org/abs/gr-qc/9502040} {arXiv:gr-qc/9502040} \BibitemShut
  {NoStop}%
\bibitem [{\citenamefont {Berti}\ \emph {et~al.}(2005)\citenamefont {Berti},
  \citenamefont {Buonanno},\ and\ \citenamefont {Will}}]{Berti:2004bd}%
  \BibitemOpen
  \bibfield  {author} {\bibinfo {author} {\bibfnamefont {E.}~\bibnamefont
  {Berti}}, \bibinfo {author} {\bibfnamefont {A.}~\bibnamefont {Buonanno}}, \
  and\ \bibinfo {author} {\bibfnamefont {C.~M.}\ \bibnamefont {Will}},\ }\href
  {\doibase 10.1103/PhysRevD.71.084025} {\bibfield  {journal} {\bibinfo
  {journal} {Phys. Rev. D}\ }\textbf {\bibinfo {volume} {71}},\ \bibinfo
  {pages} {084025} (\bibinfo {year} {2005})},\ \Eprint
  {http://arxiv.org/abs/gr-qc/0411129} {arXiv:gr-qc/0411129} \BibitemShut
  {NoStop}%
\bibitem [{\citenamefont {Ajith}\ and\ \citenamefont
  {Bose}(2009)}]{Ajith:2009fz}%
  \BibitemOpen
  \bibfield  {author} {\bibinfo {author} {\bibfnamefont {P.}~\bibnamefont
  {Ajith}}\ and\ \bibinfo {author} {\bibfnamefont {S.}~\bibnamefont {Bose}},\
  }\href {\doibase 10.1103/PhysRevD.79.084032} {\bibfield  {journal} {\bibinfo
  {journal} {Phys. Rev. D}\ }\textbf {\bibinfo {volume} {79}},\ \bibinfo
  {pages} {084032} (\bibinfo {year} {2009})},\ \Eprint
  {http://arxiv.org/abs/0901.4936} {arXiv:0901.4936 [gr-qc]} \BibitemShut
  {NoStop}%
\bibitem [{\citenamefont {Rodriguez}\ \emph {et~al.}(2013)\citenamefont
  {Rodriguez}, \citenamefont {Farr}, \citenamefont {Farr},\ and\ \citenamefont
  {Mandel}}]{Rodriguez:2013mla}%
  \BibitemOpen
  \bibfield  {author} {\bibinfo {author} {\bibfnamefont {C.~L.}\ \bibnamefont
  {Rodriguez}}, \bibinfo {author} {\bibfnamefont {B.}~\bibnamefont {Farr}},
  \bibinfo {author} {\bibfnamefont {W.~M.}\ \bibnamefont {Farr}}, \ and\
  \bibinfo {author} {\bibfnamefont {I.}~\bibnamefont {Mandel}},\ }\href
  {\doibase 10.1103/PhysRevD.88.084013} {\bibfield  {journal} {\bibinfo
  {journal} {Phys. Rev. D}\ }\textbf {\bibinfo {volume} {88}},\ \bibinfo
  {pages} {084013} (\bibinfo {year} {2013})},\ \Eprint
  {http://arxiv.org/abs/1308.1397} {arXiv:1308.1397 [astro-ph.IM]} \BibitemShut
  {NoStop}%
\bibitem [{\citenamefont {and}(2010)}]{Harry_2010}%
  \BibitemOpen
  \bibfield  {author} {\bibinfo {author} {\bibfnamefont {G.~M.~H.}\
  \bibnamefont {and}},\ }\href {\doibase 10.1088/0264-9381/27/8/084006}
  {\bibfield  {journal} {\bibinfo  {journal} {Classical and Quantum Gravity}\
  }\textbf {\bibinfo {volume} {27}},\ \bibinfo {pages} {084006} (\bibinfo
  {year} {2010})}\BibitemShut {NoStop}%
\bibitem [{\citenamefont {Ajith}(2011)}]{Ajith:2011ec}%
  \BibitemOpen
  \bibfield  {author} {\bibinfo {author} {\bibfnamefont {P.}~\bibnamefont
  {Ajith}},\ }\href {\doibase 10.1103/PhysRevD.84.084037} {\bibfield  {journal}
  {\bibinfo  {journal} {Phys. Rev. D}\ }\textbf {\bibinfo {volume} {84}},\
  \bibinfo {pages} {084037} (\bibinfo {year} {2011})},\ \Eprint
  {http://arxiv.org/abs/1107.1267} {arXiv:1107.1267 [gr-qc]} \BibitemShut
  {NoStop}%
\bibitem [{\citenamefont {Maselli}\ \emph {et~al.}(2021)\citenamefont
  {Maselli}, \citenamefont {Franchini}, \citenamefont {Gualtieri},
  \citenamefont {Sotiriou}, \citenamefont {Barsanti},\ and\ \citenamefont
  {Pani}}]{Maselli:2021men}%
  \BibitemOpen
  \bibfield  {author} {\bibinfo {author} {\bibfnamefont {A.}~\bibnamefont
  {Maselli}}, \bibinfo {author} {\bibfnamefont {N.}~\bibnamefont {Franchini}},
  \bibinfo {author} {\bibfnamefont {L.}~\bibnamefont {Gualtieri}}, \bibinfo
  {author} {\bibfnamefont {T.~P.}\ \bibnamefont {Sotiriou}}, \bibinfo {author}
  {\bibfnamefont {S.}~\bibnamefont {Barsanti}}, \ and\ \bibinfo {author}
  {\bibfnamefont {P.}~\bibnamefont {Pani}},\ }\href@noop {} {\  (\bibinfo
  {year} {2021})},\ \Eprint {http://arxiv.org/abs/2106.11325} {arXiv:2106.11325
  [gr-qc]} \BibitemShut {NoStop}%
\bibitem [{\citenamefont {{Bardeen}}\ \emph {et~al.}(1972)\citenamefont
  {{Bardeen}}, \citenamefont {{Press}},\ and\ \citenamefont
  {{Teukolsky}}}]{1972ApJ...178..347B}%
  \BibitemOpen
  \bibfield  {author} {\bibinfo {author} {\bibfnamefont {J.~M.}\ \bibnamefont
  {{Bardeen}}}, \bibinfo {author} {\bibfnamefont {W.~H.}\ \bibnamefont
  {{Press}}}, \ and\ \bibinfo {author} {\bibfnamefont {S.~A.}\ \bibnamefont
  {{Teukolsky}}},\ }\href {\doibase 10.1086/151796} {\bibfield  {journal}
  {\bibinfo  {journal} {\apj}\ }\textbf {\bibinfo {volume} {178}},\ \bibinfo
  {pages} {347} (\bibinfo {year} {1972})}\BibitemShut {NoStop}%
\bibitem [{\citenamefont {Husa}\ \emph {et~al.}(2016)\citenamefont {Husa},
  \citenamefont {Khan}, \citenamefont {Hannam}, \citenamefont {P\"urrer},
  \citenamefont {Ohme}, \citenamefont {Jim\'enez~Forteza},\ and\ \citenamefont
  {Boh\'e}}]{Husa:2015iqa}%
  \BibitemOpen
  \bibfield  {author} {\bibinfo {author} {\bibfnamefont {S.}~\bibnamefont
  {Husa}}, \bibinfo {author} {\bibfnamefont {S.}~\bibnamefont {Khan}}, \bibinfo
  {author} {\bibfnamefont {M.}~\bibnamefont {Hannam}}, \bibinfo {author}
  {\bibfnamefont {M.}~\bibnamefont {P\"urrer}}, \bibinfo {author}
  {\bibfnamefont {F.}~\bibnamefont {Ohme}}, \bibinfo {author} {\bibfnamefont
  {X.}~\bibnamefont {Jim\'enez~Forteza}}, \ and\ \bibinfo {author}
  {\bibfnamefont {A.}~\bibnamefont {Boh\'e}},\ }\href {\doibase
  10.1103/PhysRevD.93.044006} {\bibfield  {journal} {\bibinfo  {journal} {Phys.
  Rev. D}\ }\textbf {\bibinfo {volume} {93}},\ \bibinfo {pages} {044006}
  (\bibinfo {year} {2016})},\ \Eprint {http://arxiv.org/abs/1508.07250}
  {arXiv:1508.07250 [gr-qc]} \BibitemShut {NoStop}%
\end{thebibliography}%

\end{document}